\documentclass[smallcondensed]{svjour3}

\usepackage{graphicx}
\usepackage{epstopdf}
\usepackage{longtable}
\usepackage{url}
\usepackage{lscape}
\usepackage{natbib}
\usepackage{color}
\usepackage{deluxetable}
\usepackage{amsmath}
\usepackage{hyperref}
\usepackage{amssymb}

\usepackage{times}
\hypersetup{
    unicode=true,          % non-Latin characters in AcrobatÕs bookmarks
   hyperindex=true,
   colorlinks=true,       % false: boxed links; true: colored links
   linkcolor=red,          % color of internal links
   citecolor=blue,        % color of links to bibliography
   filecolor=magenta,      % color of file links
    urlcolor=cyan           % color of external links
}
\usepackage[all]{hypcap}
\defcitealias{Binggeli:1988}{BST88}
\defcitealias{Felten:1977}{Felten\ 1977}
\defcitealias{sandage:1979}{STY79}
\defcitealias{Efstathiou:1988}{EEP88}
\defcitealias{Marshall:1983}{M83}
\defcitealias{Heyl:1997}{H97}
\defcitealias{Willmer:1997}{W97}
\defcitealias{Takeuchi:2000}{TYI00}
\defcitealias{Choloniewski:1987}{C87}
\defcitealias{Caditz:1993}{CP93}
\defcitealias{Ilbert:2004}{I04}

\begin{document}

%TITLE INPUT HERE
\title{Shedding Light on the Galaxy Luminosity Function}

\author{Russell Johnston}
\institute{University of the Western Cape, Modderdam Road, Bellville 7535, Cape Town, South Africa \\ 
South African Astronomical Observatory (SAAO), Observatory Road, Cape Town 7925, South Africa \\
 University of Glasgow, Kelvin Building, University Avenue, Glasgow, Scotland, UK G12 8QQ  \\
\email{rwi.johnston@gmail.com}}
\date{Accepted 26 August 2011}

\maketitle

%ABSTRACT INPUT HERE
\begin{abstract}From as early as the 1930s, astronomers have tried to quantify the statistical nature of the evolution and large-scale structure of galaxies by studying their luminosity distribution as a function of redshift - known as the galaxy luminosity function (LF).  Accurately  constructing  the  LF remains a popular and yet  tricky pursuit in modern observational cosmology where  the presence of observational selection effects due to e.g. detection thresholds in apparent magnitude, colour, surface brightness or some combination thereof can render any given galaxy survey {\it incomplete} and thus introduce bias into the LF.

Over the last 70 years there have been numerous sophisticated statistical approaches devised to tackle these issues; all have advantages  -- but not one is perfect.  This review  takes a broad historical look at the key statistical tools that have been developed over this period, discussing their relative merits and highlighting  any significant  extensions and modifications.  In addition, the  more generalised methods that have emerged within the last few years are examined.  These methods  propose a more  rigorous statistical framework within which to determine the LF  compared to  some of  the more traditional methods. I also look at how photometric redshift estimations are being incorporated into the LF methodology as well as considering the construction of bivariate LFs. Finally, I review  the ongoing development of completeness estimators  which  test some of the fundamental assumptions going into LF estimators and can be powerful   probes of any residual systematic effects inherent   magnitude-redshift data.

\end{abstract}
%ALL SECTIONS INPUT HERE
%%
\section{Introduction}\label{sec:intro}
Understanding the origins and growth of structure that form the galaxies we observe today  is one of the many driving forces behind current cosmological research. The luminosity function (LF), denoted by $\Phi(L)$ (in units of ergs s$^{-1}$ Mpc$^{-3}$), provides one of the most fundamental tools to probe the distribution of galaxies over cosmological time.  It describes the relative number of galaxies of different luminosities by counting them  in a representative volume of the Universe which then measures the comoving number density of galaxies per unit luminosity, $L$,  such that
\begin{equation}
dN=\Phi(L)dLdV.
\end{equation}
where $dN$ is the observed number of galaxies within a  luminosity range [$L, L+dL$].  When working in  luminosities it is common practice  to apply $\log$ intervals of $L$.  The quantity $\Phi(L)$ can be normalised such that
\begin{equation}
\int\limits_{0}^\infty  {\Phi (L)dL = \rho },
\end{equation}
where $\rho$ is the number of objects per unit volume $V$, and thus $\Phi(L)dL$ gives the number density of objects within a given luminosity range. In general, the density function can be defined by $\rho(\mathbf{x})$, where $\mathbf{x}$ represents the 3D spatial cartesian co-ordinates such that  the total number $N$ of objects per unit volume (Mpc$^{-3}$) is
\begin{equation}
N = \int_V {\rho (\mathbf{x})}\,d\mathbf{x}
\end{equation}
However, it is common place to compute $\rho$ from the measured redshifts $z$ and angular co-ordinates.  Thus, for a given sample within a respective minimum and maximum redshift range $z_{\min}$ and $z_{\max}$ and solid angle $\Omega$ at a distance $r$ it is possible to compute,

\begin{equation}
N = \int\limits_{z_{\min } }^{z_{\max } } {\rho (z)} \frac{{dV'}}
{{dz}}dz,
\end{equation}
where $\rho(z)$ is now the density as a function of redshift and  $dV'\equiv\Omega r^2 dr$ is a solid angle-integrated differential volume element \citep[see e.g.][]{Choloniewski:1987}.

The LF  provides us with a robust handle to compare the difference between different sets of galaxies i.e. at different redshifts, galaxy types, environment etc...  It allows us  to assess the statistical nature of galaxy formation and evolution and indeed, it seems that as soon as a new survey is carried out one of  the first actions  is to compute the  LF,  see e.g. \cite{Blanton:2001AJ....121.2358B,Fried:2001AA...367..788F,Norberg:2002b,Im:2002ApJ...571..136I,Blanton:2003ApJ...592..819B,Liske:2003,Wolf:2003AA...401...73W,Croom:2004MNRAS.349.1397C,Richards:2006,Ilbert:2006AA...453..809I,Faber:2007,Bouwens:2008,Siana:2008,Crawford:2009,Zucca:2009,Haberzettl:2009,Montero:2009MNRAS.399.1106M,Croom:2009MNRAS.399.1755C,Willott:2010AJ....139..906W,Rodighiero:2010AA...515A...8R,Eales:2010arXiv1005.2189E,Tilvi:2010ApJ...721.1853T,Hill:2010MNRAS.404.1215H} to highlight  just a small fraction of studies over the last ten years. This is perhaps indicative of  the continuing popularity and relevance of this area of research.

The study of  galaxy LFs is now a vast subject area  spanning a   broad range of wavelengths and probing  back to the earliest galaxies.  Therefore,  it should be noted that this review is led with  a  strong emphasis on areas of work that have driven the development of the  statistical methodology of LF estimation.  Consequently,  there will   be areas of research that have not been cited, and so apologies are given in advance.  Instead, the most relevant extensions and variations of the traditional approaches are examined in detail  as well as considering the most  recent  generalised  statistical advances for LF estimation.  Furthermore, I explore the branch of astro-statistics concerning completeness estimators  that, at their core, represent a test of the validity of the assumption of  separability between the LF, $\phi(M)$ and the density function $\rho(z)$, that is inherent in the LF methodology. I also discuss how such estimators have been used to constrain evolutionary models.  

The format of this article will be as follows. The remainder of this section discusses some of the parameterisations of the LF that have led to the maximum likelihood estimators.  There is then a brief introduction to  the traditional  non-parametric methods before moving to  \S~\ref{sec:motivation} where the historical  driving forces  within astronomy and cosmology are considered both from  an observational and theoretical/numerical point of view.    In  \S~\ref{sec:assumptions} some of the critical underlying assumptions inherent to most LF estimators are discussed  before beginning the  review  in \S~\ref{sec:MLE} with the maximum likelihood parametric approach. The focus is then shifted  toward the myriad of   non-parametric methods in \S~\ref{sec:nonparam} before looking at how both the MLE and non-parametric methods have been used to construct bivariate luminosity functions (BLF) in \S~\ref{sec:bivariate},  as well as incorporating  photometric redshift estimation into current LF methodology in \S~\ref{sec:photoz}. In \S~\ref{sec:comp_review} the results of  papers which have compared the traditional estimators are reviewed.  I  then  discuss in detail some of the emerging methods that have been developed in recent years in \S~\ref{sec:emerging}.   \S~\ref{sec:indep} explores the tests of independence and completeness estimators that probe the separability assumption which underpins most  LF estimators.  This then leads to a discussion in \S~\ref{sec:applications} on some  astrophysical applications for which accurate LF estimation has been crucially important. \S~\ref{sec:conclusions} closes the review with final a summary and  discussion.

\subsection{Parameterising the luminosity function}
The processes by which one can estimate the LF vary greatly.  A popular  parametric method developed \citep*{sandage:1979}  is  based on the  Maximum Likelihood Estimator (MLE) where a parametric form  of the LF is assumed.  The most common of these models is that of  the Press-Schechter function named after  William Press and Paul Schechter \citep{Press:1974,Schechter:1976} who originally derived it in the form of  the  mass  function during their studies of structure formation and evolution. It is typically written in the form given by,
\begin{equation}\label{Eq:LF_schechter}
\Phi (L)dL = \phi^*\left( {\frac{L}{{L_* }}} \right)^{\alpha } \exp \left( {\frac{{ - L}}{{L_* }}} \right)\frac{{dL}}
{{L_* }},
\end{equation}
where, $\phi^*$ is a normalisation factor defining the overall density of galaxies, usually quoted in units  of  $h^3$Mpc$^{-3}$, and $L_*$ is the characteristic luminosity. The quantity  $\alpha$ defines the faint-end slope of the LF and  is typically negative, implying large numbers of galaxies with faint luminosities.  LF studies of the local Universe ($z\lesssim0.2$) have estimated close to $\alpha\sim-1.0$ which could imply that  there may be an infinite number of faint galaxies. However, integrating Equation~\ref{Eq:LF_schechter} over luminosity provides us with  the total luminosity density, $j_L$ (in solar luminosity units, $h$ L$_\odot$ Mpc$^{-3}$) and ensures the total luminosity density remains finite,
\begin{equation}	
j_L  = \int_{L_{\lim}}^\infty  {L\,\Phi (L)\;dL}  = \phi _* L_* \Gamma (2 + \alpha,L/L_*),
\end{equation}
where $\Gamma(a,b)$ is the incomplete gamma function defined as
\begin{equation}
\Gamma (a,b) = \int\limits_b^\infty  {t^{(a - 1)} e^{ - b} dt}.
\end{equation}
See e.g. \cite{Norberg:2002b} and \cite{Blanton:2001AJ....121.2358B} for estimates of $j_L$ from the respective Two Degree Field Galaxy Redshift Survey (2dFGRS)  and the Sloan Digital Sky Survey (SDSS) redshift surveys respectively.  A recent paper by \cite{Hill:2010MNRAS.404.1215H} combines the survey data from the Millennium Galaxy Catalogue \citep[MGC][]{Liske:2003,Driver:2005},  SDSS \citep[][]{York:2000,Adelman-McCarthy:2007} and the UKIRT Infrared Deep Sky Survey Large Area  Survey \citep[UKIDSS LAS][]{Lawrence:2007,Warren:2007MNRAS.375..213W} to determine LFs and luminosity densities over a broad range of wavelengths (UV to NIR) and thus probe the cosmic spectral energy distribution at $z<0.1$. 

By assuming the  mass-to-light ratio relation, it is also possible to use $j_L$  to compute the total contribution from stars and galaxies to the mean mass density of the Universe $\bar\rho$ \citep[e.g.][]{Loveday:1992ApJ...390..338L}.

\vspace{5mm}
\noindent
At this point it should be noted that it is common practice to convert Equation~\ref{Eq:LF_schechter} from  absolute luminosities $L$ to absolute magnitudes $M$ via the simple relation
\begin{equation}
\frac{L}{L_*}=10^{-0.4(M-M_*)}.
\end{equation}
This leads to the equivalent  expression of the Schechter LF,
\begin{align}\label{Eq:schM}
\Phi (M) &= 0.4\ln (10)\phi ^* \frac{{\left( {10^{0.4(M^*  - M)} } \right)^{^{(\alpha  + 1)} } }}{{e^{10^{0.4(M^*  - M)} } }}.
\end{align}
In a similar way for $\Phi(L)$ the LF in terms of magnitudes $M$ is usually plotted in log space i.e $\log[\Phi(M)]$ vs. $M$. For a galaxy with an  observed apparent magnitude $m$, it is then straightforward to determine its absolute magnitude $M$ by
\begin{equation}
M=m-5\log_{10}(d_L)-25 - A_g(l,b) - K(z) - E(z),
\end{equation}
provided that the corrections for galactic extinction, $A_g(l,b)$,   $K$-correction (see \S~\ref{sec:kcorr}), $K(z)$,  and evolution (see \S~\ref{sec:evolution}), $E(z)$, are well understood. The quantity $d_L$ is the luminosity distance, which, by invoking the standard $\Lambda$CDM cosmology, is defined as,
\begin{equation}\label{Eq:lumdist}
d_{L}=(1+z)\left(\frac{c}{H_{0}}\right)\int^{z}_0
\frac{dz}{\sqrt{(1+z)^{3}\Omega_{m}+\Omega_{\Lambda}}},
\end{equation}
where $\Omega_{m}$ and $\Omega_{\Lambda}$ represent  the present-day dimensionless matter density and cosmological energy density constant respectively. $z$ is the redshift  of the object, $c$ is the speed of light and $H_0$ is the Hubble constant.

Although the Schechter form of the LF  has been very successful as a generic fit to a wide variety of survey data, it has also been shown that   galaxy surveys sampled in the  infra-red have yielded LFs that do not seem to fit the standard Press-Schechter formalism.  For example, in \cite{Lawrence:1986} the following power law analytical form was fitted to data obtained from the Infrared Astronomical Satellite (IRAS),
\begin{equation}
\phi (L) = \frac{{d\Phi }}
{{dL}} = \phi_* L^{1 - \beta_1 } \left( {1 + \frac{L}
{{L_* \beta_2 }}} \right)^{ - \beta_2 } ,
\end{equation}
where $\beta_1$ and $\beta_2$ define the slopes of the two power laws.  In \cite{Saunders:1990}  a log-Gaussian form  was adopted for a survey also using the IRAS given by
\begin{equation}\label{equ:LFSaunders}
\phi (L) = \frac{{d\Phi }}
{{dL}} = \phi_* \left( {\frac{L}
{{L_* }}} \right)^{1 - \gamma } \exp \left[ { - \frac{1}
{{2\sigma ^2 }}\log_{10} ^2 \left( {1 + \frac{L}
{{L_* }}} \right)} \right],
\end{equation}
where $\gamma$ is analogous to $\alpha$ in the Schechter formalism.  \cite{Sanders:2003AJ....126.1607S} instead fitted a broken power-law of the form
\begin{equation}
\phi(L)\propto L^\alpha,
\end{equation}
to  IRAS 60 $\mu$m bright galaxy sample.  This form of the LF has more recently been adopted by \cite{Magnelli:2009AA...496...57M, Magnelli:2011AA...528A..35M} when probing the infrared LF out to $z=2.3$ using $Spitzer$ observations.

 A study of Early-Type galaxies within the Sloan Digital Sky Survey (SDSS) by \cite{Bernardi:2003b} found that a Gaussian function of the following form best described the data:
\begin{equation}
\phi(M)=\frac{\phi_*}{\sqrt{2\pi\sigma^2_M}}\exp\left( -\frac{[M_i-M_*]^2}{2\sigma^2_M }\right).
\end{equation}
For the study of quasi-stellar objects (QSO) \cite{Boyle:1988ASPC....2....1B,Boyle:1988MNRAS.235..935B} demonstrated that  a two-power law parameterisation is a reasonable fit to the data,
\begin{equation}\label{equ_LFQSO}
\Phi (M) = \frac{{\Phi ^* }}{{10^{0.4[M - M^*](\alpha  + 1)}  + 10^{0.4[M - M^*](\beta  + 1)} }},
\end{equation}
with a break at $M^*$ and a bright-end slope  $\alpha$ steeper than  the faint-end slope  $\beta$.
\subsection{Toward more robust estimates of  the luminosity function}
Complementary to the parametric approach are the  non-parametric methods which do not  require any underlying assumption of the parametric form of the LF. The traditional approaches for this class of estimator are the classical number count test \citep[e.g][]{Hubble:1936,Christensen:1975}, the \cite{Schmidt:1968} $1/V_{\max}$ estimator, the $\phi/\Phi$ method \citep[e.g.][]{Turner:1979}, and the \cite{lynden:1971} $C^-$ method.
There is also the non-parametric counterpart of the MLE developed by \citet*{Efstathiou:1988} and often referred to as the Step-wise Maximum Likelihood method (SWML).    A summary of all the traditional methods and their extensions is show in Table~1 at the end of \S~\ref{sec:photoz}.

More recently, this area of work has seen a renaissance where  more generalised techniques have emerged in an attempt  to provide a more rigorous statistical footing.  One such approach by \cite{Schafer:2007} developed a semi-parametric approach that differs crucially from the above methods by not explicitly assuming separability between the luminosity function, $\phi(M)$, and  the density function, $\rho(z)$ (see \S~\ref{sec:sep} for further discussion of separability).  Alternatively, work by  \cite{Andreon:2006MNRAS.369..969A}  and \cite{Kelly:2008}   developed approaches rooted within a Bayesian framework to estimate the LF.  For  the bivariate LF (BLF),  \cite{Takeuchi2010MNRAS.406.1830T} applies the {\it copula} \footnote{The copula  is a function used to  {\it join} multivariate distribution functions to their one-dimensional marginal distribution function and is particularly useful for variables with co-dependence.} to construct the far-ultraviolet (FUV) - far-infrared (FIR) BLF.  A non-parametric method by \cite{Borgne:2009AA...504..727L} has been developed and applied to  multi-wavelength high redshift IR and sub-mm data.  The method differs from other non-parametric estimators by applying an inversion technique to galaxy counts that does not require the use of redshift information. With no assumption on the parametric form of the LF, the method uses an assumed set of  spectral energy distribution (SED) templates to explore the range of possible LFs  for the observed number counts.

All these diverse approaches  perhaps underline the difficulty in its nature where intrinsic bias can often lead to conflicting results.   As a result  it is quite often the case that several methods are applied to a given sample to  help achieve consistency and quantify any bias found.   For example,   \cite{Ilbert:2005} developed their  ``Algorithm for Luminosity Function" (ALF) tool that implements  the   $1/V_{\max}$, $C^+$ (a modified version of Lynden-Bell's $C^-$ method),  SWML and the STY estimators applied to the VIMOS-VLT DEEP survey data \citep[see also][]{Zucca:2006AA...455..879Z,Ilbert:2006AA...453..809I}.  More recently, these estimators were applied together to the zCOSMOS survey \citep[][]{Zucca:2009}.

There have been several reviews of the LF  methodology as it has developed over the years. The earliest of these was by \cite{Felten:1977} who  performed nine determinations of the LF  using variations of the {\it classical} method.  \citetalias{Binggeli:1988} gave a comprehensive review of all non-parametric and parametric methods that had been developed up to 1988.  However,  key papers by  \cite{Heyl:1997},  \cite{Willmer:1997}, \citet*{Takeuchi:2000}  and \cite{Ilbert:2004} have made direct comparisons of the  traditional methods with the use of  real and simulated survey  data.  For the first time a rigorous  exploration into  their relative merits  was made by applying the major LF estimators to  different scenarios  from e.g. homogeneous samples to ones  with strong clustering properties, density evolution, varying LFs, observational bias from $K$-corrections etc. (see \S~\ref{sec:comp_review}).  For a nice introduction to LF estimators with numerous examples, the reader is also directed to Chapters 6 and  7 of \cite{wall:2003}.

%section 1
\section{Motivation}\label{sec:motivation}
We do not yet have a  complete theory that fully describes how galaxies form and evolve into the structures we observe today, and any model of formation and evolution must match the observed data. Since the fledgling redshifts surveys during the  1970s, the mapping of the sky  has turned into a thriving industry where the largest survey to date, the Sloan Digital Sky Survey, has imaged in excess of  270 million galaxies with approximately a  million of these having  spectroscopic redshifts.   As as a result, luminosity function studies have been a vital tool with which to analyse  these surveys, since they allow us to probe the evolutionary processes of extragalactic sources as a function of redshift and thus place powerful constraints on current galaxy formation and evolutionary models.   

At around the same time  redshift surveys came to fruition, the theoretical framework was being laid for how we currently model galaxy formation and evolution.  Moreover, as computational technology  came of age, the field of numerical cosmology saw rapid  development providing state-of-the-art N-body simulations based  on our current knowledge of these physical processes and  observations. 
\subsection{Constructing a model  of galaxy evolution}
The current standard  Concordance model of cosmology represents the most concise model to date,  combining  astronomical observations with  theoretical predictions to explain the origins, evolution, structure and dynamics  of the Universe.  The origin of the Concordance model is rooted in the Copernican principle -  a fundamental assumption proposed by Nicolaus Copernicus in the 16th century that states we do not occupy  a privileged position in the Universe.  This concept was generalised into what is termed, the Cosmological Principle, in which we assume that the Universe is both homogeneous and isotropic.  In 1922  Alexander Friedman provided a formal description in terms of a metric that was  later independently improved upon by Howard Robertson, Arthur Walker and Georges Lema\^{i}tre into what is now referred to as the Friedmann-Lema\^{i}tre-Robertson-Walker metric, or more simply - the FLRW metric. 

In its current, simplest form, the standard model is known as $\Lambda$ Cold Dark Matter ($\Lambda$CDM).  Whilst dark matter particles  have not yet been directly detected, there are a number of independent observations derived from high velocity dispersions of galaxies observed  in clusters \citep{Zwicky:1933AcHPh...6..110Z} and   of  flat galaxy rotation curves \citep{Rubin:1980ApJ...238..471R,2001ARA&A..39..137S}, which support its inferred existence.  The  $\Lambda$ term refers to the non-zero cosmological constant in   general relativity theory which implies the Universe is currently undergoing a period of cosmic acceleration \citep[see e.g.][]{Peebles:1988ApJ...325L..17P,Steinhardt:1999PhRvD..59l3504S}. The recent observational techniques derived from distance measurements of  Type Ia supernovae by \cite{Riess:1998} and \cite{Perlmutter:1999} have provided evidence of this acceleration;  later studies from baryonic acoustic oscillations \citep{Eisenstein:2005ApJ...633..560E}, the integrated Sachs-Wolf (ISW)  effect  \citep{Giannantonio:2008PhRvD..77l3520G} and weak lensing \citep{Schrabback:2010AA...516A..63S} have helped strengthen support of the $\Lambda$CDM model.  

By the 1960s a theory of galaxy formation was proposed by  \citet{Eggen:1962}  known as  {\it monolithic collapse} (or the `top-down' scenario) in which galaxies originate from large regions of primordial baryonic gas. This baryonic mass then collapses  to form  stars within the central  region  allowing the most massive galaxies to form first.  Seminal work by James Peebles \citep[see e.g.][]{Peebles:1970AJ.....75...13P,Gunn:1972ApJ...176....1G,Peebles:1973PASJ...25..291P,Peebles:1974ApJ...189L..51P,Peebles:1980} established the theoretical underpinnings for structure formation  known as the gravitational instability paradigm.   In essence, this theory states that small scale density fluctuations in the early Universe grew by gravitational instability into the structures we see today. This led to the establishment of the   {\it Hierarchical clustering} model (or `bottom-up' scenario) in which larger objects evolve from mergers with smaller objects \citep[see e.g.][]{Searle:1978ApJ...225..357S}.  The bottom-up  scenario  is now the  more favoured of the two models and is fundamental to  galaxy formation models, ultimately leading to the theoretical and numerical  modelling  applied in current N-body simulations. 

The work by \cite{Press:1974} demonstrated how  the structure of halos in the early Universe could form  through gravitational condensation in a density field using a   Gaussian random field of gas-like particles. Thus, the  first analytical treatment  of the mass function was derived.    \cite{White:1978MNRAS.183..341W} developed these ideas  formulating  a more sophisticated  model of galaxy formation in which  dark matter halos formed hierarchically and consequently baryons cooled and condensed at
their centres to form galaxies.  Thus, small proto-galaxies form early in the history of the Universe and  through merging processes build up into larger galaxies we see today.

The production  of early  N-body codes of  the 1960s and 1970s  \citep[e.g.][]{Aarseth:1963MNRAS.126..223A,Gingold:1977MNRAS.181..375G,Lucy:1977AJ.....82.1013L}   began to see rapid development during the 1980s.   \cite{Aarseth:1979ApJ...228..664A} was one of the first to develop simulations of galaxy clustering. However, as  the Cold Dark Matter model \citep{Blumenthal:1984Natur.311..517B} garnered momentum,  \cite{Davis:1985ApJ...292..371D} provided the first simulations within the hierarchical  clustering CDM framework. Over the next two decades hydrodynamics were integrated into simulations that would include  processes such as star formation, feedback and gas cooling \citep[see also e.g.][highlighting just a few]{Fall:1980MNRAS.193..189F,White:1987Natur.330..451W,Schaeffer:1988AA...203..273S,White:1991ApJ...379...52W,Katz:1991ApJ...377..365K, Navarro:1991ApJ...380..320N,Cen:1992ApJ...399L.113C,Katz:1993ApJ...412..455K,Cole:1994MNRAS.271..781C,Lacey:1994MNRAS.271..676L,Navarro:1994MNRAS.267L...1N,Springel:2001NewA....6...79S,Springel:2005MNRAS.364.1105S,Kere:2005MNRAS.363....2K,Bower:2006MNRAS.370..645B,DeLucia:2007MNRAS.375....2D,Bertone:2007MNRAS.379.1143B,Weinberg:2008ApJ...678....6W,Dav:2008MNRAS.391..110D,2009MNRAS.395..160K}. 

However, despite the tremendous achievements made in the development of cosmological simulations, there  remains  a number of discrepancies between astronomical observations  and what has been simulated within  a  $\Lambda$CDM framework. This has led some to question  the validity of the current concordance model.  Such discrepancies include the lack of direct detection of dark matter; the missing satellite problem,  where observations of the local Universe have found orders of magnitude less dwarf galaxies than predicted by simulations \citep[e.g.][]{1999ApJ...522...82K,1999ApJ...524L..19M,2004MNRAS.353..639W,2010AdAst2010E...8K}; the core-cusp problem, in which  observations indicate a constant dark matter density in the inner parts of galaxies  \citep{2010AdAst2010E...5D}, whilst simulations prefer a power-law cusp as originally developed by \cite{Navarro:2004MNRAS.349.1039N}; and finally, the angular momentum problem, in which  $\Lambda$CDM simulations  have had difficulties reproducing disk-dominated and bulgeless galaxies. 

Whilst the existence of dark matter remains an illusive quantity, possible solutions to addressing the remaining issues may lie with not only in  improving  simulations with higher resolutions and more rigorous treatment of physical mechanism within galaxies (e.g. supernovae feedback), but also obtaining higher resolution observations from  next generation telescopes \citep[see e.g.][for recent developments]{Gnedin:2002MNRAS.333..299G,Governato:2004ApJ...607..688G,Robertson:2004ApJ...606...32R, Simon:2007ApJ...670..313S,2007MNRAS.374.1479G,Simon:2007ApJ...670..313S,2009Natur.461...66M}.  Of course there exists the possibility that maybe as these improvements are made  we may  require an entirely different model beyond  $\Lambda$CDM.
\subsection{The rise and rise of redshift surveys}

\subsubsection{Measuring redshifts}
The difference between redshifts obtained spectroscopically or photometrically essentially reduces to the precision of the methods.  Spectroscopy  is  the most precise and, consequently, the most popular way to  measure redshifts. The technique  requires  identification of  spectral lines (typically emission lines) and record their wavelength $\lambda$.   The  relative shift of a these lines compared to their known position measured in the laboratory ($\lambda_o$)  allows us to measure the redshift.

The acquisition of redshift data in this way  can be described as a two-stage process. One firstly images galaxies in a region of space using low resolution photometry and then targets these galaxies to perform high resolution spectroscopy requiring long integration times to achieve sufficiently high signal-to-noise.  Redshift surveys such as the  Two Degree Field Galaxy Redshift Survey (2dFGRS) and the Sloan Digital Sky Survey (SDSS) used multi-fibre spectrographs which could, respectively, record up to 400 and 600 redshifts simultaneously. 

Alternatively, the use of photometry, offers a much quicker way to estimate redshifts to a much greater depth and has thus  gained in popularity in recent times.  However, the precision to which they are currently measured  remains poor, with a typical uncertainty in the range $0.05 \lesssim \delta z \lesssim 0.1$. Moreover,  effects from e.g. absorption due to galactic extinction and the Lyman-alpha forest  can contribute to systematics \citep[see e.g.][]{Massarotti:2001AA...380..425M}.  The techniques of photometric estimation of redshifts can be traced back to \cite{Baum:1962} who  first developed a method using multi-band photometry in 9 filters for elliptical galaxies.  Further work by  \citet{Loh:1986a}  and \cite{Connolly:1995} saw this area develop into two respective  techniques: the {\it template fitting methods} and {\it empirical fitting methods}. Template fitting requires a library of theoretical or empirical SEDs to be generated coupled with spectroscopic redshifts (for calibration purposes) which are then fitted to the observed colours of the galaxies, where redshift is a fitted parameter  \citep[see e.g.][]{Bolzonella:2000}. Empirical fitting  methods, whilst similar, instead use empirical relations with a  training set of  galaxies with spectroscopically obtained redshifts  \citep[see also e.g.][]{Brunner:1997,Wang:1998}. In the case of \cite{Connolly:1995}, for example, they used linear regression where the redshift was assumed to be a linear or quadratic function of the magnitudes. 

Both  methods have been improved upon by the incorporation of, for example, Bayesian inference  \citep{Kodama:1999,Benitez:2000,Stabenau:2008,Wolf:2009}, nearest neighbour weighting schemes  \citep[e.g.][]{Ball:2008,Lima:2008}, boosted decision trees \citep{Gerdes:2010ApJ...715..823G}, and  artificial neural networks \citep{Firth:2003MNRAS.339.1195F,Vanzella:2004AA...423..761V,Collister:2007MNRAS.375...68C,Oyaizu:2008ApJ...674..768O,Yeche:2010AA...523A..14Y}. Alternatively, others have attempted more generalised approaches that incorporate aspects of both the traditional methods \citep[e.g.][]{Budavri:2009}.

Improvements on photometry have also assisted the case for photo-z. For example, the COMBO-17 photometric redshift survey produced multi-colour data in a total of 17 optical filters - five broad-band filters ({\it UBVRI}) and 12 medium-band covering a wavelength range of 400 to 930 nm \citep[see e.g.][]{Wolf:2004AA...421..913W}. Having this many filters allowed significant increases in resolution improving  galaxy redshift accuracy to $\delta z \sim 0.03$ \citep{Wolf:2003AA...401...73W,Bonfield:2010MNRAS.405..987B}.

\begin{figure}\label{fig:cfa}
    \begin{center}
      \includegraphics[width=1.0\textwidth]{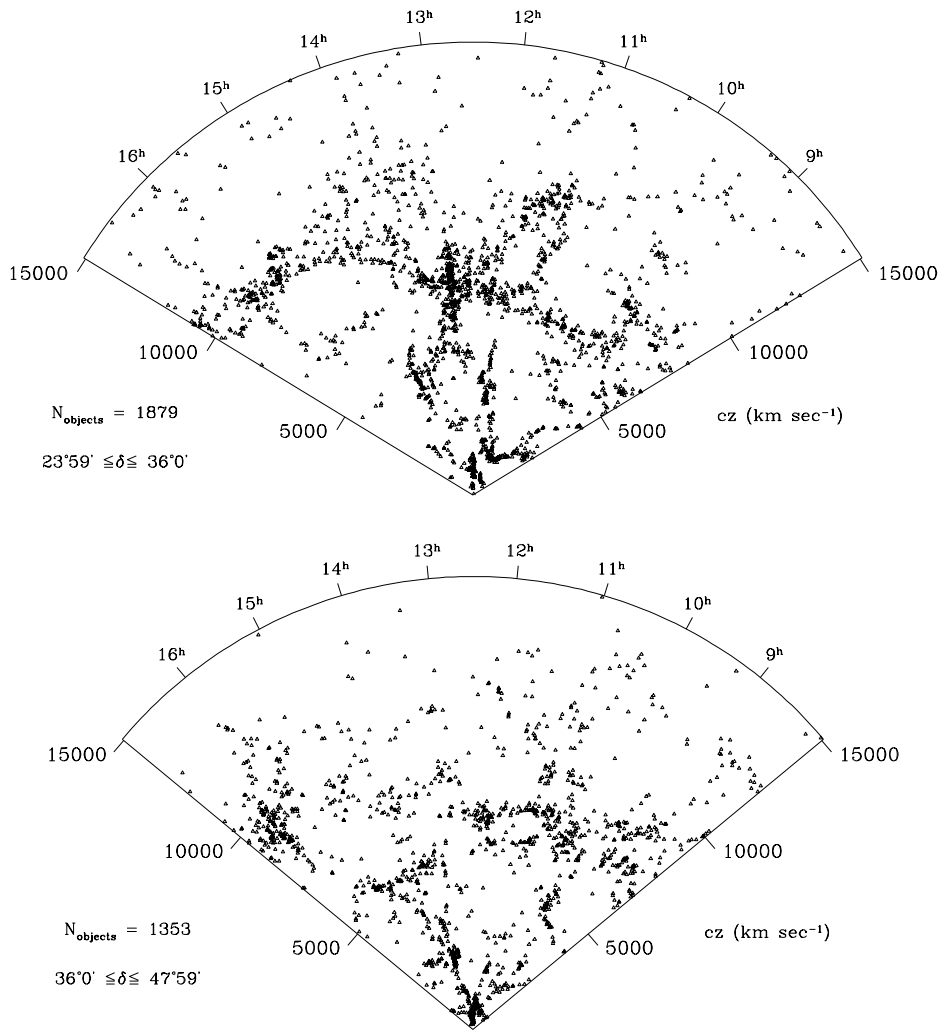}\\
      \caption[Redshift cones of the CFA survey]{\small Extract strips from the CfA redshift surveys. The strip on the sky was 6 degrees wide and 130 degrees long with our origin begin at the apex of the wedge.  Image courtesy of the Smithsonian Astrophysical Observatory Emilio \cite{Falco:1999}.}        
      \label{fig:cfa}
    \end{center}
  \end{figure}
\subsubsection{Redshift surveys}
Galaxy redshift surveys have played, and  will continue to play a vital role in our understanding of the formation, evolution and distribution of galaxies in the Universe.   Prior to the 1970's,  models of the structure of the Universe were based on the observed distribution of galaxies projected onto the plane of the sky.   Whilst early pioneers had already identified the clustering nature of galaxies from 2-D samples \citep[e.g.][]{Hubble:1936,Charlier:1922}, it would take the move to three-dimensional data-sets before the wider astronomy community would accept these claims.  As a consequence, this  required the measurement of redshifts on a much grander scale.  To make a large enough survey where redshifts of thousands of galaxies could be measured would require a lot of dedicated telescope time and funding. Nevertheless, it was in 1977 that these investments were made and dedicated redshift surveys began.  

The first major breakthroughs in mapping large-scale structure began with the CfA survey which ran from 1977 to 1982 \citep{Huchra:1983} and measured spectroscopic redshifts for a total of 2401 galaxies out to a limiting apparent magnitude of $m_{\lim}\le14.5$~mag.  This survey represented the first large area maps of large-scale structure in the nearby universe and confirmed the 3-D clustering properties of galaxies already proposed a little over 50 years previously.  CfA2 \citep{Geller:1989} continued the survey measuring a total of 18,000 redshifts out to 15,000 kms$^{-1}$ and $m_{\lim}\le15.5$~mag (see Figure~\ref{fig:cfa}).   With survey data provided by the Southern Sky Redshift Survey (SSRS/SSRS2) \cite{Costa:1998AJ....116....1D,daCosta:1994ApJ...424L...1D}, CfA was  extended to include the southern hemisphere.  Despite this tremendous achievement, spectrographic technological constraints allowed only one galaxy at a time to be observed, making the whole process extremely time consuming.   However,  the technology developments during the 1980s provided the  first multi-object fibre spectrographs that allowed between 20 and 200 galaxies to be observed simultaneously during one exposure.  
 \begin{figure}\label{fig:PSCZ}
    \begin{center}
        \includegraphics[width=1.\textwidth]{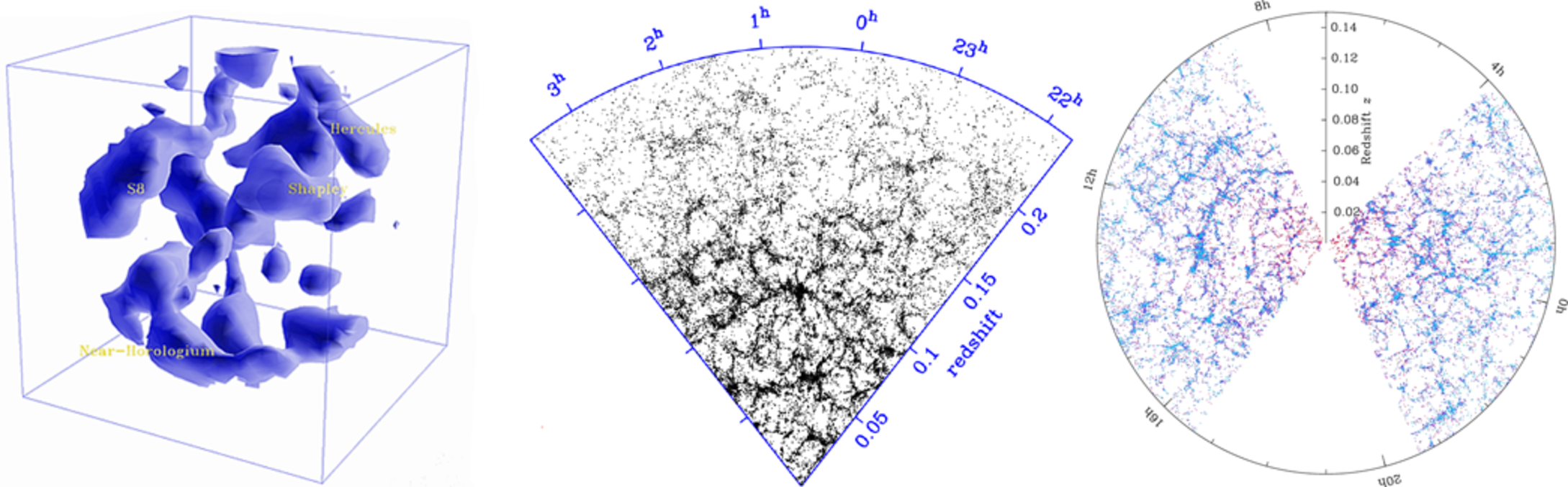}        
      \caption{\small Left -- 3D representation of the of the PSCz survey (Image courtesy of Dr. Luis Teodoro). Middle -- The 2dF galaxy redshift survey (2dFGRS)  final data release
	showing one of the two wedges (Image courtesy of \href{http://msowww.anu.edu.au/2dFGRS/}{http://msowww.anu.edu.au/2dFGRS/}). Right - The Sloan Digital Sky Survey (SDSS) (Image courtesy of \href{http://www.sdss.org}{http://www.sdss.org}).}
      \label{fig:PSCZ}
    \end{center}
  \end{figure}

A new era of space-based telescopes began with the Infrared  Astronomical  Satellite (IRAS), launched 1983.  The  IRAS Point Source Catalogue (PSCz) redshift survey  ran from 1992 to 1995 and  mapping 15,411 galaxies over 84\% of the sky out to 0.6 Jy in the far-IR (60 $\mu$m)  \citep{Saunders:2000} marking the first all-sky survey (see Figure~\ref{fig:PSCZ} left).  The Hubble Space Telescope (HST) was launched in 1990 and the Hubble Deep Field-North (HDF-N) survey in 1995 \citep{Williams:1996} was the next landmark,  allowing unprecedented detail of faint galaxy populations to  a magnitude of $m_V=30$~mag out to high redshift.  In 1998 the follow-up survey, HDF-South, sampled a random field in the southern hemisphere sky with equal success  \citep[see e.g.][]{Fernandez-Soto:1999}. The more recent Hubble Ultra Deep Field  \citep[HUDF][]{Beckwith:2006} ran from 2003 to 2004 and has allowed evolutionary LF constraints on the very faintest galaxies to redshifts reaching the end of the epoch of re-ionisation, $z_{photo}\sim6$ \citep[e.g.][]{Ferguson:2000ARAA..38..667F,Kneib:2004ApJ...607..697K,Bouwens:2009ApJ...690.1764B,Zheng:2009ApJ...697.1907Z,Finkelstein:2010ApJ...719.1250F,McLure:2010MNRAS.403..960M,Oesch:2010ApJ...725L.150O}.

Throughout  the 1990s there were a number of surveys that paved the way to constrain the local LF out to $z\sim0.2$ such as the  Stromlo-APM Redshift Survey (S-APM) \citep{Loveday:1992ApJ...390..338L}, Las Campanas (LCRS)  \citep{Lin:1996ApJ...464...60L} and the ESO Slice Project (ESP) \citep{Zucca:1997AA...326..477Z}. However,  the next milestone from ground-based telescopes was in the form of the Two Degree Field Galaxy Redshift Survey (2dFGRS)  \citep{Colless:1998}.  This survey ran from 1998 to 2003 and used the multi-fibre spectrograph on the Anglo-Australian Telescope (AAT), which could measure up to  400 galaxy redshifts simultaneously.   The photometry was taken from the Automatic Plate Measuring (APM) scans of the UK Schmidt Telescope (UKST) plates with measured magnitudes out to $m_{b_J}^{\lim}=19.45$~mag.  The 2dFGRS team recovered a total of 245,591 redshifts, 220,000 of which were galaxies  (see Figure~\ref{fig:PSCZ} middle).   With this increase in instrumentation precision and the vast number of objects catalogued, the scientific goals became equally ambitious.  Some of 2dFGRS goals included measuring the power spectrum of the galaxy distribution on scales up to few hundred M$\rm{pc^{-1}}$, determining the galaxy LF, clustering amplitude and the mean star-formation rate out to a redshift $z\sim0.5$.

At around the same time as the 2dFGRS another team was carrying  out a survey called the Two Micron All Sky Survey (2MASS) \citep{Skrutskie:2006AJ....131.1163S}. This saw the return of a near-infrared full sky survey and was the first all-sky photometric survey of galaxies brighter than $m_K=13.5$~mag cataloguing approximately 100,000 galaxies.

However, it is the Sloan Digital Sky Survey  (SDSS) that takes the prize as the most ambitious  survey to date by mapping a quarter of the entire sky  to a median spectroscopic redshift $z_{\rm spec} \sim 0.2$ using multi-band photometry with unprecedented accuracy \citep[e.g.][]{Abazajian:2003AJ.126.2081A,Adelman-McCarthy:2006,Abazajian:2009ApJS..182..543A}.  Using the dedicated, 2.5-meter telescope on Apache Point, New Mexico, USA and multi-fibre spectrographs, SDSS has imaged over 200 million galaxies and obtained just over 1 million spectroscopic redshifts. The combination of 2dFGRS and SDSS has provided the most accurate maps of the nearby Universe placing strong  constraints on the LF out to $z\sim0.2$ \citep{Norberg:2002b,Blanton:2003ApJ...592..819B,Montero:2009MNRAS.399.1106M}.

Exploring galaxy evolution out to $z\sim1.0$ and beyond has been and continues to be  explored with surveys such as the Canada-France Redshift Survey (CFRS) \citep{Lilly:1995ApJ...455..108L}, Autofib I \& II \citep{Ellis:1996MNRAS.280..235E,Heyl:1997}, the Canadian Network for Observational Cosmology survey (CNOC1 \& 2) \citep{Lin:1997ApJ...475..494L,Lin:1999}, COMBO-17 \citep{Wolf:2003AA...401...73W},VIMOS-VLT Deep Survey  (VVDS) \citep[e.g.][]{Ilbert:2005}, the Deep Extragalactic Evolutionary  Probe 2 (DEEP2)  \citep[e.g.][]{Willmer:2006ApJ...647..853W} and the more recent zCOSMOS \citep{Lilly:2007,Zucca:2009}.  All  have been instrumental in constraining  the statistical nature of the evolutionary  processes of early-type to  late-type galaxies at intermediate redshifts through the  study of  LF as a function of color \citep[e.g.][]{Bell:2004ApJ...608..752B}.

There appears to be no sign of a slowing down of  redshift surveys. In fact, future projects such as the Dark Energy Survey and those involving the  Large Synoptic Survey Telescope (LSST), the Square Kilometre Array  demonstrator telescopes, ASKAP  and MeerKAT,  and the James Web Space Telescope (JWST) will provide the next generation of surveys with data-sets  orders of magnitude larger, with greater wavelength coverage, and probing both fainter and more distant galaxies.  
%end of file			%section 2
\section{K-corrections, Evolution and Completeness}\label{sec:assumptions}
This section begins with a brief overview of  the $K$-correction and the methods adopted to constrain  source evolution for a given population of galaxies.  The final part discusses the two important fundamental assumptions common to {\it all} the traditional LF estimators, namely, completeness of observed data and separability between the luminosity and density functions. %
\subsection{The $K$-correction}\label{sec:kcorr}
The modelling of  galaxy evolution and $K$-correction is a vital part of most galaxy survey analysis, which, if not accounted for properly, can adversely affect accurate determination of the LF. 
The use of $K$-correction can be traced backed to early 20th century pioneering observers such as \cite{Hubble:1936b} and \cite{Humason:1956}.  The  observed wavelength from a galaxy is different  from the one that was emitted due to cosmological redshift, $z$.  The $K$-correction allows us to transform from the observed wavelength, $\lambda_o$ when measured through a particular filter (or bandpass) at $z$, into the emitted wavelength, $\lambda_e$ in the rest frame at $z=0$. Following the derivation from  \cite{Hogg:2002} \citep[see also e.g.][]{Oke:1968ApJ...154...21O}   we consider an object with an apparent magnitude $m_R$ that has been observed in the $R$ bandpass. However, it is desirable to obtain the absolute magnitude $M_Q$ in the rest-frame bandpass $Q$.  Therefore, we firstly consider the relation between the corresponding emitted frequency, $\nu_e$ and the observed frequency, $\nu_o$, given by
\begin{equation}
\nu_e=(1+z)\nu_o,
\end{equation}
where $z$ is redshift at which the source object was observed.  Hogg {\it et al.} then go onto  show that the corresponding $K$-correction can be determined by
\begin{equation}
K =  - 2.5\log _{10} \left[ {\left[ {1 + z} \right]\frac{{\int {\frac{{d\nu _o }}{{\nu _o }}f_\nu  (\nu _o )S_R(\nu _o )} \int {\frac{{d\nu _e }}{{\nu _e }}g_\nu ^Q (\nu _e )S_Q(\nu _e )} }}{{\int {\frac{{d\nu _o }}{{\nu _o }}g_\nu ^R (\nu _o )S_R(\nu _o )\int {\frac{{d\nu _e }}{{\nu _e }}f_\nu  \left( {\frac{{\nu _e }}{{1 + z}}} \right)S_Q(\nu _e )} } }}} \right],
\end{equation}
where $S$ is often  referred to as the transmission (or sensitivity)  function of the detector for which at each frequency $\nu$ is the mean contribution of a photon  of frequency $\nu$ to the output signal from the detector in the respective bandpass. The quantity  $g$ is the spectral density of flux  for a zero-magnitude or {\it standard} source.
Thus determining the $K$-corrected absolute magnitude in the rest frame of the object is simply given by
\begin{equation}
M_Q=m_R-5\log(d_L)-K,
\end{equation}
where  $d_L$ is the cosmology dependent luminosity distance already defined in Equation~\ref{Eq:lumdist}.
\subsection{Evolution}\label{sec:evolution}
From a purely practical point of view, a  frustrating aspect of studying galaxy evolution resides in our inability to trace the individual evolutionary processes of a galaxy through cosmological time. Since we do not have a concise  and complete theory of galaxy formation and evolution, we can instead examine the   statistical  properties of populations of galaxies at different epochs  \citep[see e.g.][for detailed studies]{1993MNRAS.263..123R, Marzke:1994AJ....108..437M,Lilly:1995ApJ...455..108L,Willmer:2006ApJ...647..853W,Heyl:1997,Faber:2007}.  In this section two different approaches regularly applied to constrain evolution are outlined.  

 \begin{figure*}
     \begin{center}
      \includegraphics[width=1.0\textwidth]{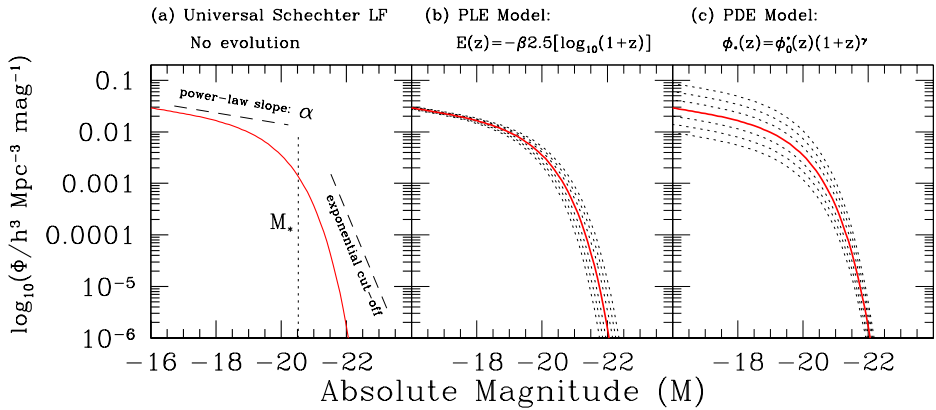}
           \caption{\small The characteristic  shape of the Schechter luminosity function.  The LF is typically presented in terms of absolute magnitudes $M$ with the number density on the y-axis as the  $\log_{10}(\Phi)$.  For this example the LF has been modelled using constrained LF parameter values based on  the 2dFGRS results. In  panel (a) this implies: $\alpha=-1.21$, $M_*=-19.61$ and $\phi_*=1.61\times10^{-2}h$ Mpc$^{-3}$. The steepness of the faint-end slope is determined by the $\alpha$ parameter  and the characteristic magnitude, $M_*$, indicates the `knee' of the LF. Panel (b) shows the effect of introducing a pure luminosity evolution (PLE) model. In terms of magnitudes this model is shown as $E(z)$ and is dependent on the evolutionary parameter $\beta$. The red line shows the LF from panel (a) and the dotted lines  (from left to right on the plot) represents the same LF for a range of $\beta$ = 1.5, 1.0, 0.5,  -0.5, -1.0  and -1.5, at a redshift of $z=0.2$.  In panel (c) a pure density evolution (PDE) model is alternatively introduced in to the LF.  As with panel (b), the dotted lines (from top to bottom on the plot) represent LFs for the range of $\gamma$ = 6.0, 4.0, 2.0 and -2.0, -4.0 and -6.0 at the same redshift. }        
      \label{fig:LF1}
    \end{center}
  \end{figure*}
  \vspace{2mm}
\noindent

The first method adopts  a strict parametric form  based on some physical assumptions regarding our current understanding of  luminosity evolution and/or number (mergers) evolution \citep[see e.g.][]{Wall:1980MNRAS.193..683W,Phillipps:1995MNRAS.274..832P,Heyl:1997,Driver:2005,Wall:2008MNRAS.383..435W,Prescott:2009MNRAS.397...90P,Croom:2009MNRAS.399.1755C,Rodighiero:2010AA...515A...8R}.  In the Pure Luminosity Evolution (PLE) scenario it is  assumed that massive galaxies were assembled and formed most of their stars at high redshift and have evolved without merging i.e. not convolved number evolution. The correction applied to account for this form of evolution is redshift and galaxy-type dependent and typically takes the form
\begin{equation}\label{Eq:PLE_L}
L_* (z) = L_*(0)(1 + z)^\beta  
\end{equation}
where $\beta$ is the evolution parameter and although it is galaxy-type dependent, a global correction is often adopted.  Applying this correction to magnitudes, the evolution correction $E(z)$ is the written as
\begin{align}
M_* &=M_*(0) - E(z)  \\
E(z) &= -\beta\ 2.5\log_{10}(1+z) \label{equ:ev_m}
\end{align}
For the study of quasi-stellar objects (QSOs) in the 2df QSO redshift survey, \cite{Croom:2004MNRAS.349.1397C}  instead, characterised luminosity evolution as a second-order polynomial luminosity given by
\begin{align}\label{equ:croom}
L_*(z) &=L_*(0)10^{k_1z+k_2z^2} \\
M_*(z) &=M_*(0) - 2.5(k_1z+k_2z^2)
\end{align}
where the model requires a two parameter fit given by $k_1$ and $k_2$.

Pure Density (Number) Evolution (PDE) assumes that galaxies were more numerous in the past but have since merged. The treatment of this type of evolution can be modelled in a similar way as PLE,
\begin{equation}\label{Eq:PDE}
\phi_* (z) = \phi_0^* (z)(1 + z)^\gamma,
\end{equation}
where the parameter, $\gamma$, is the number evolution parameter.   By assuming  either of the above models (or  a combination of both) one can apply this technique to observables and simultaneously constrain both the LF and $E(z)$ parameters via a  maximum likelihood technique.  Whilst this has proved to be a popular approach  it should be noted that the wrong evolutionary model may be adopted in the analysis yielding questionable results.

The second alternative approach is termed as   a {\it free-form} technique that, as the name suggests,  requires no  strict parametric assumption for the evolutionary model.  An early method by  \citet{Robertson:1978MNRAS.182..617R,Robertson:1980MNRAS.190..143R} developed an iterative procedure for exploring evolution of the radio luminosity function (RLF) whereby   $E(z)$ was sampled over multiple $\log(z)$ slices until convergence.  Whilst this approach allowed a free-form analysis  in redshift, assumptions regarding luminosity evolution were still required and there was no built-in measure of  the range of evolution allowed by the data  \citep[see e.g.][for applications and extensions]{Zawislak:1986MNRAS.222..487Z,Zawislak:1987ApSS.139..305Z,Zawislak:1990ApSS.172...89Z}.  \cite{Peacock:1981MNRAS.196..611P} and \cite{Peacock:1985MNRAS.217..601P} developed  a fully free-form methodology that instead assumes a  series expansion for the evolutionary model and employs a $\chi^2$ minimisation technique to constrain its  coefficients.   A more recent development in this area by \cite{Dye:2010MNRAS.tmp..610D} extends the ideas of Peacock and Gull by   introducing a reconstruction technique that discretises the  functions of redshift and luminosity and  employs an adaptive gridding  in ($L-z$) space  to constrain evolution. 

Recent observational techniques \cite[e.g.][]{Blanton:2003ApJ...594..186B,Bell:2004ApJ...608..752B} have identified a bimodal distribution in color across a broad range of redshifts allowing clear segregation of early- (red) and late- (blue star forming) type galaxies.  This has allowed deeper exploration into the evolutionary properties of these populations beyond the simple e.g  PLE approach (see \S~\ref{sec:appintz} for further discussion).

Figure~\ref{fig:LF1} shows a simple example of a typical Schechter LF modelled on results from the 2dFGRS \cite{Norberg:2002b} paper.  In panel (a) $\Phi(M)$ is plotted in log space.  The input parameters are, $\alpha=-1.21$, $M_*=-19.61$ and $\phi_*=1.61\times10^{-2}h$ Mpc$^{-3}$. The panel indicates the various components that make up the shape of the LF i.e. the power-law slope at the faint end the steepness of which is governed by $\alpha$;  $M_*$, which characterises the turnover from the bright-end to the faint end; and the exponential cut-off which constrains the bright end of the LF. Panel (b) then shows how pure luminosity evolution (PLE)  affects the shape of the LF by incorporating the PLE model from Equation~\ref{equ:ev_m} into the Schechter function for a range of $\beta$ given by [1.5, 1.0, 0.5,  -0.5, -1.0 , -1.5] at a fixed redshift of $z=0.2$.  These are indicated, respectively, from left to right on the plot by the dotted lines. The red solid line shows the Universal LF from panel (a).  As one would expect, the LF is affected only in its position in magnitude and is independent of any shift in number density.   Panel (c) now shows the case where only pure density evolution (PDE) is present by adopting Equation~\ref{Eq:PDE}  for a range of $\gamma$ given by [6.0, 4.0, 2.0 and -2.0, -4.0, -6.0] as, respectively, indicated by the dotted lines from top to bottom in the plot.  It is clear that the shape of the LF is now dominated by shifts in the $\phi$ direction.

\subsection{Completeness,  separability and the selection function }\label{sec:completeness}
In general, when compiling a magnitude-redshift catalogue, we would like to be able to quantify in some way, how close we are to having a representative sample of the underlying distribution of galaxies. However, there are a number of constraints preventing us from observing all objects in the sky. This is termed  a measure of {\it completeness} for a given survey.  The ability to  interpret and measure it accurately is not  trivial.  There are many diverse  contributing  sources  of {\it in}-completeness that have to be corrected for and understood to construct  accurately  the LF.    Nevertheless, an assumption of LF estimation is  one of  {\it completeness} of   a volume- or flux-limited survey.    For clarity, I now discuss two general interpretations of  completeness  that commonly  appear in the observational cosmology literature:

\subsubsection{Redshift completeness} 
This can be described by the process of  combining photometric and spectroscopic galaxy survey catalogues.  In the most general sense, an observer will measure apparent magnitudes, $m$, of galaxies in a portion of the sky out to a faint limiting apparent magnitude, $m^{\rm f}_{\lim}$ imposed by the physical limitations of the telescope.  CCD instrumentation can be  affected by  pixel saturation due to very bright 
objects. This imposes a bright limit to the survey, $m^{\rm b}_{\lim}$. At this stage the catalogue consists of measured magnitudes and sky positions only, without their 3D spatial redshift distribution. Therefore, each galaxy is then targeted to obtain a measure of its redshift.  The most accurate approach is by multi-fibre spectroscopy, where optical fibres are positioned on a plate which has holes drilled at the positions of the sources measured from the photometry. However, a drawback of measuring redshifts in this way arises from spatial limitations. For example, in a region that has a high density of objects, a lot of galaxies may be missed since there is a physical spatial limit on how close the fibres can be placed i.e. {\it  fibre collisions}. Therefore, redshift completeness can be presented as the percentage of successfully measured redshifts over a list of targeted galaxies  within a survey.  Whilst this is in itself an important contributing factor to the overall completeness picture,  we can describe the overarching completeness in terms of {\it magnitude completeness}.

\subsubsection{Magnitude completeness} 
At this stage we can now ask the question, {\it how do we know our percentage of successfully matched targets is complete up to (or within)  the apparent magnitude limit(s),} $m_{\lim}^{\rm f}$  (and $m_{\lim}^{\rm b}$)? Some of the contributing factors that affect this specific  type of completeness can be summarised as:  galaxies that are missed because they are located close to bright stars or lie close to the edge or defected part of the CCD image; the wrong $K$-corrections applied; the wrong evolutionary model adopted;  galaxies with the same surface brightness  that may or may not be detected depending on their shape and overall extent i.e.  a compact object is more likely to have enough pixels above the detection limit than a very diffuse galaxy of the same brightness; and adverse effects from a varying magnitude limit over a photographic plate or CCD image.

Cosmological  surface brightness dimming of objects due to the expansion of the universe has been examined by e.g.  \cite{Lanzetta:2002ApJ...570..492L}.  By studying  the distribution of  star-formation rates as a function of redshift in the Hubble Deep Field  they demonstrated that such an effect would bias derived quantities such as the luminosity density.

An effect  explored by \cite{Ilbert:2004} that can potentially bias the shape of the global LF  arises from  a wide range of $K$-corrections being applied across different  galaxy types.  Such an effect  results in a varying galaxy-type dependent absolute magnitude limit where certain galaxy populations will not be detectable out to the full extent of the magnitude limit of the survey.  This form of incompleteness is particularly crucial toward  the faint end of the LF and  discussed in greater detail in \S~\ref{sec:comp_review}.

A more fundamental form of  incompleteness arises from the observational limitations of a telescope and is often referred to as {\it Malmquist Bias} \citep[see e.g.][]{Hendry:1990A&A...237..275H}.  As one images out to higher redshifts only intrinsically bright sources will be observed.  Since bright objects at large distances are rare, one observes a decrease in the  number density of imaged objects as a function of  redshift.     A way to quantify this effect is to compute  the {\it selection function}, the probability that a galaxy at a redshift $z$ will be included in  a given magnitude- (or flux) limited survey.  Thus, in the simplest scenario, the selection function $S(z)$ may be expressed as

\begin{equation}
S(z) = \frac{{\int_{ - \infty }^{M_{\lim } (z)} {\Phi (M)\;dM} }}{{\int_{ - \infty }^\infty  {\Phi (M)\;dM} }}
\end{equation}
where $M_{\lim}$ is the absolute magnitude limit of the survey.  In practical terms this observational effect implies we are sampling less of the underlying distribution of galaxies at increasing redshifts. To correct for this effect it is common to apply a  weighting scheme by incorporating the inverse of the selection function in the LF estimation such that
\begin{equation}
w\propto\frac{1}{S(z)}.
\end{equation}
More generally, both magnitude and redshift completeness definitions can now be grouped in terms of, the probability that a galaxy of apparent magnitude, $m$, is observable. 
\subsubsection{Separability}\label{sec:sep}
Weighting galaxies via the selection function introduces a very large assumption that is  central to {\it all} the traditional methods for constructing the LF - {\it separability} between the probability densities $\phi(M)$ and $\rho(z)$.  This directly implies that the absolute magnitudes $M$ (or luminosities $L$) are statistically independent from  their spatial distribution and thus the LF has a Universal form.  As such, the joint probability density of $M$ and $z$, $P(M,z)$,  can be expressed in a separable form as the product of the two univariate distributions such that
\begin{equation}\label{dP_1}
P(M,z)\propto\phi(M)\rho(z)
\end{equation}
The assumption of separability between the probability densities, can thus be thought of as an extension to the idea of magnitude completeness.  In this scenario it is assumed that the absolute magnitudes, $M$,  have also been corrected, where deemed necessary, for any luminosity and/or density evolution (Equations~\ref{Eq:PLE_L} and  \ref{Eq:PDE}, respectively), galactic extinction and $K$-correction.   In terms of the LF, these corrections can also be thought of as additional contributing factors to completeness.  Therefore,  only when such corrections have been accurately made can the separability assumption be valid.
%

			%section 3
 \section{The Parametric Maximum Likelihood Estimator}\label{sec:MLE}
This review begins with the maximum likelihood estimator (MLE).  As a statistical tool, the MLE is by no means a recent development.  Its origins can be traced as far back as e.g. \cite{Bernoulli:1769,Bernoulli:1778}\nocite{Kendall:1961} through to  R. A. Fisher  who provided a more formal derivation of the MLE in his seminal papers \cite{Fisher:1912, Fisher:1922}.  For a comprehensive historical review see e.g.  \cite{Aldrich:1997} and \cite{Stigler2008arXiv0804.2996S}.  In terms of its application within the context of observational cosmology  it was \citet*{sandage:1979}, hereafter \citetalias{sandage:1979}, who were the first to see it as powerful approach to  galaxy  LF estimation. This is a parametric technique which  therefore assumes an analytical form for the LF and thus eliminates the need for  the binning of data (as usually required by most non-parametric methods).    The equally popular non-parametric counterpart of the MLE, the step-wise maximum likelihood (SWML), is discussed in \S~\ref{sec:swml}.

The derivation of the MLE begins by considering  $x$, a continuous random variable, that is described by a probability distribution function (PDF) given by
\begin{equation}
f(x;\theta),
\end{equation}
where $\theta$ represent the parameter we wish to estimate. As we shall see, in practice $\theta$  represents more than one parameter of the LF.  If $x$  represents our observed data then the likelihood function, $\mathcal{L}$, can be written as

\begin{equation}
f(x_1, x_2, ..., x_N|\theta_1, \theta_2, ..., \theta_k)=\mathcal{L}= \prod\limits_{i = 1}^{N} {f(x_i;  \theta_1, \theta_2, ..., \theta_k)} 
\end{equation}
where  $x_i$ are $N$ independent observations.  It is often the case that the likelihood function is expressed in terms of the logarithmic likelihood such that
\begin{equation}
\ln\{\mathcal{L}\}=\sum\limits_{i = 1}^N {\ln f(x_i ;\theta _1 ,} \theta _2 ,...,\theta _k )
\end{equation}
Constraining the $\theta_1, \theta_2, ..., \theta_k$ parameters is, in principle, a straightforward matter of maximising the likelihood function $\mathcal{L}(\theta)$ or $\ln\{\mathcal{L}(\theta)\}$ such that
\begin{equation}
\frac{{\partial (\mathcal{L} \ \mbox{or} \ \ln\{\mathcal{L}\})}}
{{\partial \theta _j }} = 0,\quad j = 1,2,...,k
\end{equation}
\\
\noindent In the context of  estimating the parameters of the LF we consider a galaxy at redshift $z$ for which we can define the cumulative luminosity function (CLF) and thus determine the probability that the galaxy will have an absolute magnitude brighter than $M$ as
\begin{equation}
p(M|z) = \frac{{\int\limits_{ - \infty }^M {\phi (M')\rho(z)f(M')dM'} }}{{\int\limits_{ - \infty }^\infty  {\phi (M')\rho(z)f(M')dM'} }},
\end{equation}
where $\rho(z)$ is the density function for the redshift distribution, $f(M')$ is the completeness function which for a 100$\%$ complete survey would be

\begin{equation}
f(M')=
\begin{cases}
1, &M^{\rm bright}_{\rm lim}\le M' \le M^{\rm faint}_{\rm lim} \\ \\
0, &\mbox{otherwise}.
\end{cases}
\end{equation}
It follows that the probability density for detected galaxies is given by the partial derivative of $P(M,z)$ with respect to $M$,
\begin{equation}
p(M_i ,z_i ) = \frac{{\partial p(M,z)}}{{\partial M}}=\frac{{\phi (M_i )}}{{\int\limits_{M_{{\rm{faint}}(z_i )} }^{M_{{\rm{bright}}(z_i )} } {\phi (M')dM'} }}
\end{equation}

Note that the density functions have cancelled thus rendering the technique insensitive to density inhomogeneities. Finally, the likelihood is maximised to give
\begin{equation}\label{equ:like}
\mathcal{L} = \prod\limits_{i = 1}^{N } {p(M_i ,z_i ).} 
\end{equation}
Most commonly a Schechter function is assumed where the parameters that we wish to estimate are $\alpha$ and  $M_*$  as defined in Equation~\ref{Eq:schM} on page~\pageref{Eq:schM}.  
\subsection{Normalisation, goodness of fit \& error estimates} \label{sec:mle_norm}
Although the MLE method has become more popular  than other traditional non-parametric methods there are aspects not to be overlooked. This approach does not determine the normalisation parameter  $\phi^*$ of the LF and consequently  has to be estimated  by independent means.  The approach originally described by \citet{Davis:1982}  and later adopted by e.g. \cite{Loveday:1992ApJ...390..338L, Lin:1996ApJ...464...60L,Willmer:1997,Springel:1998,Blanton:2003AJ.125.2276B,Montero:2009MNRAS.399.1106M} incorporates   a minimum variance density estimator to determine the mean density of objects.  The method can be summarised as follows. The normalisation can be cast in terms of
\begin{equation}
\bar n = \frac{{\sum\nolimits_{j = 1}^{N_{\rm gal} } {w(z_j )} }}{{\int {dVS(z)w(z)} }},
\end{equation}
where $N_{gal}$ is the number of galaxies in the sample,  $w(z)$ is a weighting function for each galaxy  defined by the inverse of the second moment of the two-point correlation function given by
\begin{equation}
w(z)=\frac{1}{1+\bar n J_3 S(z)},
\end{equation}
and $S(z)$ is the selection function for the survey defined within a maximum and minimum redshift range,
\begin{equation}
S(z) = \frac{\int_{L_{\min}(z) }^{L_{\max}(z) } {dL\,\Phi (L,z)}}{\int_{L_{\min} }^{L_{\max}} {dL\,\Phi (L,z)}}, 
\end{equation}
where the quantity $J_3$ is the integral of the correlation function given by
\begin{equation}
J_3  = \int_0^\infty  {dr\;r^2 \xi (r)}.
\end{equation}
The normalisation is then calculated iteratively and the error can be computed by
\begin{equation}
\left\langle {\delta \bar n^2 } \right\rangle ^{1/2}  = \left[ {\frac{{\bar n}}{{\int {dV\phi (z)w(z)} }}} \right]^{1/2} .
\end{equation}
\citet*{Davis:1982}   and \citet{Willmer:1997} point out that whilst this method is robust it can return a biased estimate if the survey sample has significant inhomogeneities.  In a more recent paper by \cite{Hill:2010MNRAS.404.1215H} it was commented that further bias may be introduced due to incompleteness at higher redshifts  resulting in over weighting of  $\phi^*$. For more exploration into this and other normalisation methods, see \citet{Willmer:1997}.

In terms of the goodness-of-fit of the adopted  parametric form of the LF, this too, as highlighted by \cite{Springel:1998}, is not built into the MLE and, therefore, has to be assessed independently. For survey samples that may not so obviously  be described by a Schechter function, caution should be taken as this implies that nearly any functional form could be made to fit a given data-set.  Furthermore, the nature of the method effectively determines the slope of the LF at any point.  One can, however,  apply a simple $\chi^2$ minimisation test  to probe the goodness-of-fit.  Aside from this, if the survey sample is not complete near the apparent magnitude limit sources close to the limit will be underestimated thus making the slope of the LF underestimated \citep{Saunders:1990}.  

A standard approach for estimating the  relative error on the LF parameters $M_*$ and $\alpha$ was adopted by  \cite{Efstathiou:1988}. This involves jointly varying these parameters around the maximum likelihood value to find where the likelihood increases by the $\beta$-point of the $\chi^2$ distribution; we have
\begin{equation}
\ln\mathcal{L} = \ln\mathcal{L}_{\max} - \frac{1}{2}\chi^2_\beta(N)
\end{equation}
where $N$ is the number of  degrees of freedom corresponding to $\Delta\chi^2=2.30$ for $1\sigma$ limit ($68.3\%$ confidence interval) and $\Delta\chi^2$=6.17 for $2\sigma$ limit ($95\%$ confidence interval).

\subsection{Further extensions}\label{sec:mle_ext}
The \citetalias{sandage:1979} method remains one of the most widely applied LF  estimators to date and as a result has been modified over the years.
For example, \citet{Marshall:1983} (hereafter, \citetalias{Marshall:1983}) extended its use for quasars by simultaneously fitting evolution parameters  with the luminosity function parameters.  For this  they test both pure density and pure luminosity models.  In their analysis the probability distribution in the likelihood for the observables is described instead by Poisson probabilities.  The luminosity and redshift space ($L-z$) distribution is gridded such that the likelihood is  defined as the product of the probabilities of observing  either 1 or 0 quasars in each cell such that
\begin{equation}\label{equ:marshal}
\mathcal{L}=\prod\limits_i^N {\lambda (z_i ,L_i )dzdL\,e^{ - \lambda (z_i ,L_i )dzdL} } \prod\limits_j^N {e^{ - \lambda (z_j ,L_j )dzdL} },
\end{equation}
where the quantity $\lambda (z_i ,L_i )dzdL$ represents the expected number of objects in each cell in the $L-z$ plane.  The $j$ index takes into account cells where no objects were observed. This form of the likelihood has  proved popular and has been widely applied. \cite{Choloniewski:1986}  adopted the method and applied it to the CfA survey data. More recent examples by  \cite{Boyle2000MNRAS.317.1014B}  studied the quasar LF in the 2df-QSO survey and \cite{Wall:2008MNRAS.383..435W} for exploring a sub-millimetre galaxy sample from the GOODS-N survey. As I will discuss in more detail in \S~\ref{sec:photoz}, \cite{Christlein2009MNRAS.400..429C} also drew on this approach when incorporating photometric redshift estimates into the MLE.

\cite{Saunders:1990} used the  \citetalias{sandage:1979} approach not only to  constrain the LF but instead integrate over the comoving volume to determine the  radial  density field. In this way no knowledge of the LF is required.   They demonstrated that by parameterising the radial density function $\rho(\bf|\underset{\raise0.3em\hbox{$\smash{\scriptscriptstyle-}$}}{r} |$) they can fit it as a step function and obtain the variation on the MLE as
\begin{equation}
\mathcal{L} = \prod\limits_{i = 1}^{N} {\frac{{\rho(z_i )}}{\int{\rho(z_i )(dV/dz)dz}}.} 
\end{equation}

\cite{Heyl:1997} generalised  \citetalias{sandage:1979} by constructing  a statistical framework to  explore how the LF evolves with redshift.  This generalisation also allowed for the  combination of multiple  samples  (see also \cite{avni:1980} extension of $V/V_{\max}$ in  \S~\ref{sec:vmax}).

For the study of the LF as applied to galaxy clusters \citet*{Andreon:2005MNRAS.360..727A} extend \citetalias{sandage:1979} by developing a likelihood approach that retains the normalisation and accounts for Poisson fluctuations and is cast in the context of two density probability functions: one describing the {\it signal} (the cluster LF), the other accounting for {\it background} galaxy counts from the observations of many individual events (the galaxies luminosities), without knowledge of which event is the signal and which is background. Given $j$ data-sets of e.g. cluster 1, cluster 2... etc.. each comprising of $N_j$ galaxies, the likelihood is maximised such that
\begin{equation}
\ln \mathcal{L} = \sum\limits_{clusters,\,j} {\left( {\sum\limits_{galaxies,\,i} {\ln \;p_i  - s_j } } \right)},
\end{equation}
where $p_i=p(m_i)$ is the probability of the $i$th galaxy of the $j$th cluster to have an apparent magnitude $m_i$. The quantity $s$ is the expected number of galaxies given the model, evaluated by
\begin{equation}
s = \int_{m^{\rm b}_{{\lim},j} }^{m^{\rm f}_{{\lim } ,j}} {\phi (m)dm},
\end{equation}
where $m^{\rm b}_{\lim}$ and $m^{\rm f}_{\lim}$ are the respective bright and faint limiting magnitudes of the $j$th cluster data in this example. In their analysis the LF $\phi$ is modelled as a convolved power-law and Schechter function. The goodness-of-fit was determined adopting the $\chi^2$ test.  In \cite{Andreon:2006MNRAS.369..969A} and \cite{Andreon:2006MNRAS.372...60A} they extend this approach within a bayesian framework and apply a  Markov Chain Monte Carlo algorithm to constrain the LF parameters \citep[MCMC][]{Metropolis:1953JChPh..21.1087M}.  See also e.g. \cite{Andreon:2008MNRAS.385..979A,Andreon:2010MNRAS.407..263A} for further applications of the method.

In  \S~\ref{sec:bivariate}   further applications of the MLE to bivariate distributions are discussed in greater detail.			%section 4
\section{The Traditional Non-Parametric Approaches}\label{sec:nonparam}
One of the main difficulties in constructing an accurate LF from flux-limited survey samples is the issue of completeness. I have discussed some of the major problems with galaxy detection.   However, in general terms we are hindered observationally by the notorious {\it Malmquist bias} effect. This effect means we are  biased to observe intrinsically brighter objects at higher redshifts and observe only the fainter objects over smaller volumes nearby.   In this section I will discuss all the various non-parametric  weighting schemes that have been  devised  over the years to correct for this  bias.
\subsection{The $classical$ approach}\label{sec:classical}
The $classical$ method, as coined by \cite{Felten:1977}, represents the first rudimentary binned number count approach to determining the LF and was initially developed and applied  by e.g. \cite{Hubble:1936}, \cite{VandenBergh:1961}, \cite{Kiang:1961}, and \cite{Shapiro:1971}.  However, as pointed out in \citetalias{Binggeli:1988} the method was not formally introduced until the publications of \cite{Christensen:1975}, \cite{Schechter:1976} and \cite{Felten:1977}.

The underlying assumption of the method is that the distribution of sources within the data-set in question is spatially homogeneous i.e. with no strong large-scale clustering. From this starting point we count the number of galaxies $N$ within a volume $V$ such that
\begin{equation}\label{Eq:classical}
\Phi \equiv \frac{N}{V}
\end{equation}
The volume, $V(M)$,  is  calculated for the maximum distance that each galaxy with an absolute magnitude, $M_i$, could have and still remain in the sample.  As an example, \cite{Felten:1977} applies the following expression  (neglecting $K$-corrections)  to calculate the volume,
\begin{align}
V(M) = \frac{4}{3}\pi \exp &[0.6(m_{\lim }  - M_i  - 25)] \\ \nonumber
&\times \left[ {E_2 (0.6\alpha \ln 10) - \frac{{E_2 (0.6\alpha \ln 10\csc b_{\min } )}}{{\csc b_{\min } }}} \right]
\end{align}
where $m_{\rm lim}$ is the apparent magnitude limit of the survey, $\alpha$ and $b_{\min}$ are related to the directional-dependent galactic absorption calculation, and $E_2(x)$ is a second exponential integral  \citep*[][Chap 5]{Abramowitz:1964}.

The number of galaxies, $N$,  within the absolute magnitude limits of the survey, $M$, is binned into a histogram \citepalias[see][]{Felten:1977,Binggeli:1988} with each bin divided by $V(M)$ to convert the histogram to units of $\rm {mag^{-1}}$$\rm {Mpc^{-3}}$ and return a  differential estimate of the LF, $\phi (M)$.

Whilst this method is relatively straightforward to apply, its basic assumption of homogeneity is well understood to be a handicap.  At the time when galaxy surveys were shallow  it was common practice to exclude clustered regions such as the Virgo cluster and members of the Local Group to try and avoid biasing in the shape and thus the parameters of the LF \citep{Felten:1977}.  Also, \cite{Felten:1976}  showed mathematically that the {\it classical} test gives a biased estimate of the LF and provides expressions for the fractional bias and the variance of $\Phi$.
\subsection{The $1/V_{\max}$ and $V/V_{\rm max}$ estimators}\label{sec:vmax}
\begin{figure} \label{fig:vmax}   
   \begin{center}
      \includegraphics[width=0.5\textwidth]{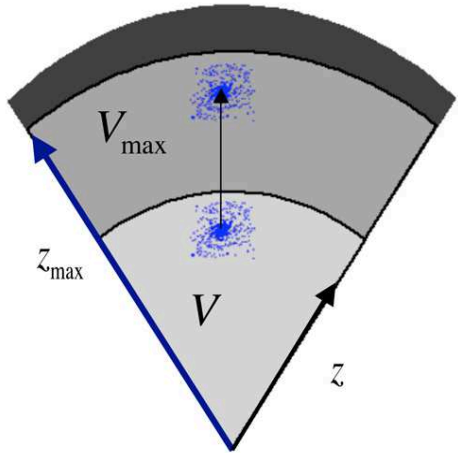}
              \caption{\small The construction of the traditional \cite{Schmidt:1968} $V/{V}_{\rm max}$ test.  The basic construction of the method considers the ratio between the volume $V$, in which a galaxy is observed to the volume $V_{\max}$, the maximum volume the said galaxy could occupy and still be observed.}
       \label{fig:vmax}  
    \end{center}
  \end{figure}
A natural development from the $classical$ approach is now the famous $V/{V}_{\rm max}$ test first described by \cite{Kafka:1967,Rowan-Robinson1968MNRAS.138..445R} but more formally detailed and applied in the much celebrated paper by \cite{Schmidt:1968} to assess the uniformity and the cosmological evolution of quasars at high redshift  \citep[see also][]{Schmidt:1972,Schmidt:1976}.   As with the {\it classical} method, $V/{V}_{\rm max}$ assumes spatial homogeneity.  $V/{V}_{\rm max}$ is essentially a completeness estimator and  Fig.~\ref{fig:vmax} illustrates its construction.  The basic principle of the test is simple and is defined by considering two volumes:
\begin{itemize}
\item $V$, the volume of the sphere of radius $R$, where $R$ is the distance at which  a galaxy was actually detected, compared to 
\item $V_{\rm max}$, the maximum volume within which a galaxy could have been detected and still remain in the catalogue in question.   Thus, $V_{\max}$ is the volume enclosed at maximum redshift, $z_{\max}$ at which the galaxy in question  could still have been observed. 
\end{itemize}
Assuming that the distribution of objects within the survey sample is homogeneous, then it follows that the value ${V}/{V}_{\rm max}$  is expected to be uniformly distributed in the interval [0,1]. Thus, for a complete sample with no evolution ${V}/{V}_{\rm max}$ has expectation value
\begin{equation}
\small
\left< \frac{V}{V_{\rm max}} \right> = \frac{1}{2},
\end{equation}
with an often quoted statistical uncertainty of $1/(12N)^{1/2}$, where $N$ is the total number of objects in the sample \citep*[see e.g.][] {Hudson:1991}.  In reality, the value calculated from  $V/V_{\max}$ for a survey will deviate from $1/2$.  By how much the value deviates from $1/2$ is usually considered to be either a signature of incompleteness (e.g. Malmquist bias) and/or an indication of evolution: a value that is greater than $1/2$ would imply a density evolution where galaxies were more numerous in the past, where as a value less than $1/2$ would imply that galaxies were less numerous in the past. 

In the same paper, Schmidt also outlined a variation of this statistic that could be used to estimate the  LF under the condition where the maximum distance $r_{\max}$ an object could have and still be included in the sample was independent of its direction,
\begin{equation}\label{Eq:schmidt}
\Phi = \sum\limits_{i = 1}^N {\frac{1}{{V_{\max,i}. }}} 
\end{equation}
Once again it was \cite{Felten:1976} who would dub this estimator as the `Schmidt's estimator'. The comoving volume can be determined by evaluating,
\begin{equation}
V_{\max} = \frac{c}
{{H_0 }}\int_\Omega  {\int_{z_{\min } }^{z_{\max} } {\frac{{d_L^2 }}
{{(1 + z)^2 A(z)}}dz\;d\Omega } },
\end{equation}
where $\Omega$ is the solid angle of the survey, $d_L$ is the luminosity distance as given by Equation~\ref{Eq:lumdist},  and $z_{\min}$ and $z_{\max}$ is redshift range of the sample.  The quantity $A(z)$ is related to the  transverse comoving distance and defined as
\begin{equation}
A(z)\equiv\sqrt{\Omega_m(1+z)^3+\Omega_k(1+z)^2+\Omega_\Lambda},
\end{equation}
where $\Omega_m$, $\Omega_k$ and $\Omega_\Lambda$ are, respectively, the matter density, curvature and cosmological energy density  constants.

If one assumes Poisson fluctuations and homogeneity within each bin then a standard approach \citep[see e.g.][]{Condon:1989ApJ...338...13C} to error estimation is simply computing the rms on each bin,
\begin{equation}
\sigma_i=\left[\sum\limits_{i=1}^N{\frac{1}{V_{\max,i}}}^2 \right]^{1/2}.
\end{equation}
As we shall see in the following section, if these assumptions breakdown \cite{Eales:1993}  provided an more rigorous modified approach to error estimation. 
\subsubsection{Development and variations of the  $V_{\rm max}$ estimator}\label{sec:vmax_ext}

Although strictly speaking $V/V_{\max}$ is a completeness estimator I have included its development in this section as it is so intimately linked with the  $1/{V}_{\rm max}$ estimator for constructing the LF.  

Since its inception,  $1/{V}_{\rm max}$ has remained a popular estimator for  determining luminosity functions and as a probe of evolution, most likely due to its simplicity and ease of implementation.  However, as we will see in \S~\ref{sec:comp_review} a survey sample with strong clustering properties will bias the slope of the recovered LF. Nevertheless, the  $V_{\max}$ estimators have evolved,  been improved and refined over the years to accommodate the many different  types of survey that have steadily grown in size and complexity.  
Selected below is a summary of some of the most significant developments of the $V_{\max}$ method.
\\
\\ \noindent
\cite{Huchra:1973} were the first to extend its use to galaxies from the Markarian lists I to IV \citep*[see][]{Markarian:1967,Markarian:1969a, Markarian:1969b,Markarian:1971} and perform $V/{V}_{\rm max}$ as a completeness test whilst including the Virgo Cluster and the Local Group.  They showed that the effects of including such clusters had a minimal impact on their results.   Furthermore, they calculated the space density $\Phi(M)$ via Schmidt's $1/V_{\max}$ estimator, where they summed  over all galaxies within absolute magnitude intervals.
\\
\\ \noindent
\cite{Felten:1976} made an extensive comparison of  $1/V_{\max}$ with the {\it classical} test.  This paper derives a generalised version of  $1/V_{\max}$  between absolute magnitude ranges $M_1<M<M_2$ to give
\begin{equation}
\int\limits_{M_1 }^{M_2 } {\phi(M)\  dM = \sum\limits_{i = 1}^N {\frac{1}{{V_i(M_i) }}} } 
\end{equation}
and shows that it is superior to that of the $classical$ estimator by being an unbiased estimator. 
\begin{figure*}\label{fig:avni}   
   \begin{center}
      \includegraphics[width=0.6\textwidth]{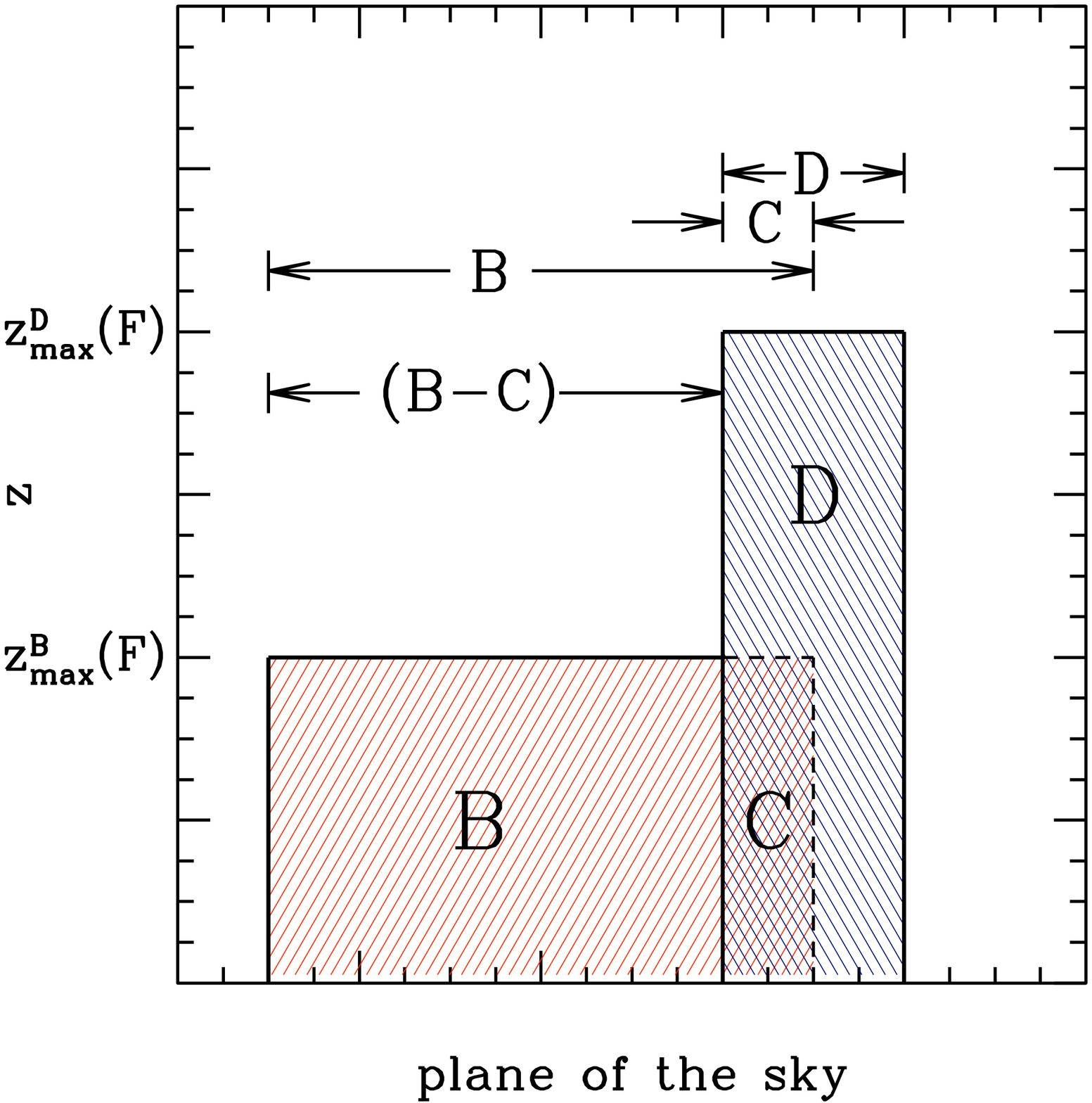} \\ 
      \vspace{5mm}
      \includegraphics[width=0.6\textwidth]{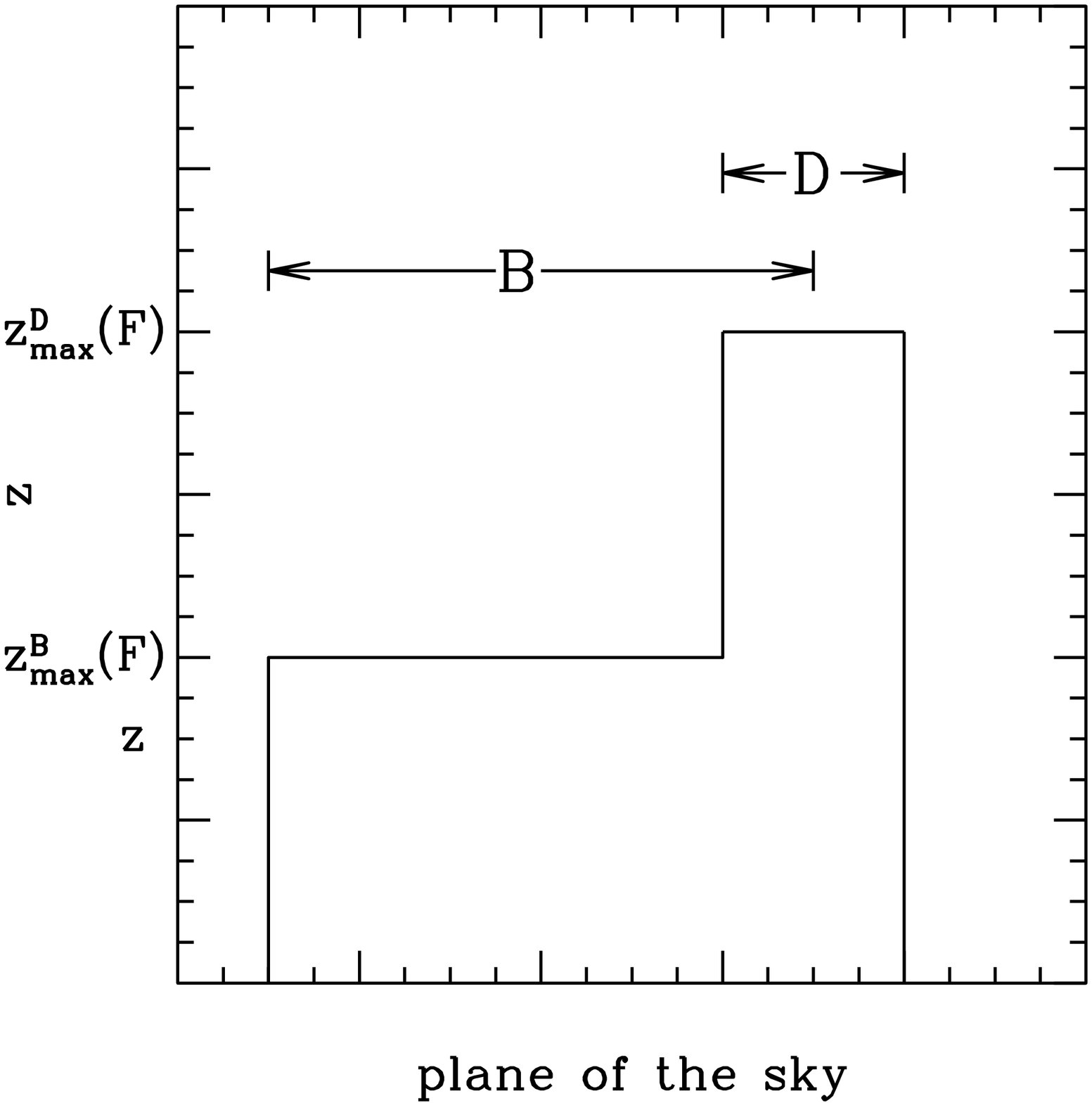} \\  
      \vspace{5mm}
      \includegraphics[width=0.6\textwidth]{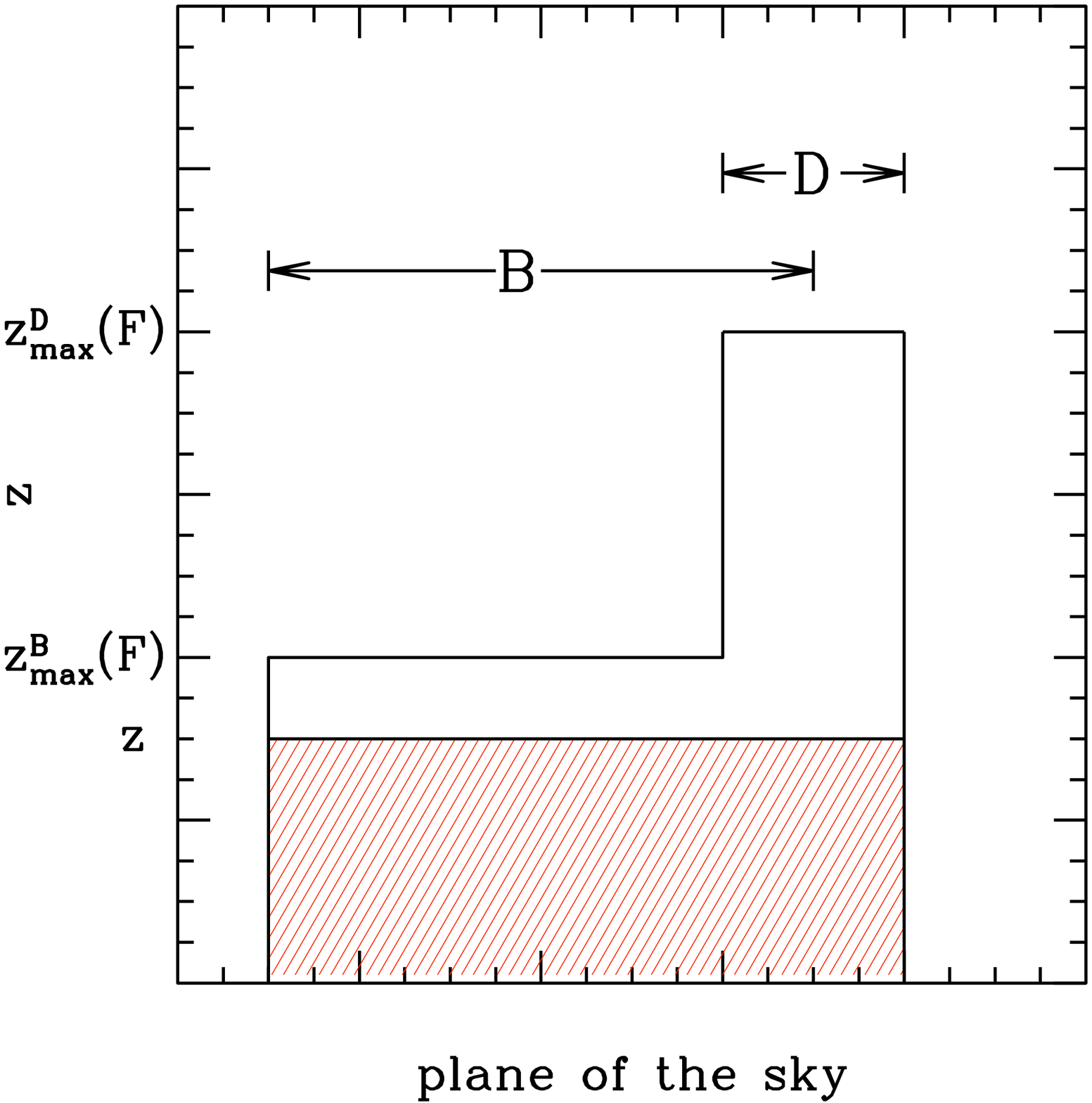}
              \caption {\small \cite{avni:1980}:  The top panel shows the construction of the generalised $V/{V}_{\rm max}$  for two incoherent  overlapping samples (B and D) into two region independent samples, (B-C) and D (see Equation~\ref{Eq:avni_vmax}). The middle panel shows the construction of $V_e/V_a$ for a coherent sample constructed from two overlapping samples (B and D).   The shaded region in the bottom panel represents $V_e$ for the case $z \le z^{\rm B}_{\rm max}(F)$ in Equation~\ref{Eq:Ve}. }
       \label{fig:avni}   
    \end{center}
  \end{figure*}
\\
\\ \noindent
\cite{avni:1980}  generalised  $V/{V}_{\rm max}$ for multiple samples for two distinct cases:
\begin{enumerate}
\item{Firstly, for combining independent multiple samples  that are still physically separated.}
\item{Secondly, for combining independent samples in which the individual samples are overlapping.}
\end{enumerate}
In the first scenario they consider complete `incoherent'  samples which do not overlap on the plane of the sky that could either be initially non-overlapping, or could be constructed from overlapping samples as illustrated on the  top panel in Fig.~\ref{fig:avni}.  The term `incoherent' refers to combining samples in which one remembers for each object its parent sample.

In this particular case the $V/{V}_{\rm max}$ statistic can be constructed from overlapping samples dividing the space into two distinct volumes  (B-C) and D.  For this method they show that a combined sample average of $V'/{V'}_{\rm max}$ is given by
\begin{equation}\label{Eq:avni_vmax}
\left\langle {\frac{{V'}}{{V'_{\max } }}} \right\rangle _{\rm B - C,D}  = \frac{{N_{\rm B - C} }}{{N_{\rm B - C}  + N_{\rm D} }}\left\langle {\frac{{V'}}{{V'_{\max } }}} \right\rangle _{\rm B - C}  + \frac{{N_{\rm D} }}{{N_{\rm B - C}  + N_{\rm D} }}\left\langle {\frac{{V'}}{{V'_{\max } }}} \right\rangle _{\rm D} 
\end{equation}
where $V'$ represents the density-weighted volume, $N_{\rm B-C}$ is the number of objects in sample (B-C) and $N_{\rm D}$ is the number of objects in sample D.

The second scenario considers the simultaneous analysis of independent complete  `coherent'  samples in which the individual samples $are$ physically joined and a new statistic, $V_e/{V}_{a}$, is constructed (see illustration in the middle panel of Fig.~\ref{fig:avni}).  By this description, `coherent'  refers to the method of combining independent samples.   Here, a new variable $V'_a$ is defined as the density-weighted volume $available$ to an object for being included in the coherent sample.  This new volume is defined as
\begin{equation}
V'_a [F_i ] = \frac{{\Omega _{\rm B - C} }}{{4\pi }}V'[z_{\max }^{\rm B} (F_i )] + \frac{{\Omega _{\rm D} }}{{4\pi }}V'[z_{\max }^{\rm D} (F_i )]
\end{equation}
where $\Omega _{\rm B - C}$ and $\Omega _{\rm D}$ are the solid angles subtended on the sky and $F_i$ is flux of the object. 
The second  new variable $V'_e$ is defined as the density-weighted volume enclosed by an object in the coherent sample and is given by
\begin{equation}\label{Eq:Ve}
V'_e[z_i,F_i]=
\begin{cases}
\frac{{\Omega _{\rm B - C} }}{{4\pi }}V'(z_i ) + \frac{{\Omega _{\rm D} }}{{4\pi }}V'(z_i ), & z_i  \le z_{\max }^{\rm B} (F_i) \\ \\
\frac{{\Omega _{\rm B - C} }}{{4\pi }}V'[z_{\max }^{\rm B} (F_i )] + \frac{{\Omega _{\rm D} }}{{4\pi }}V'(z_i ), & z_i  > z_{\max }^{\rm B} (F_i ) 
\end{cases}
\end{equation}
This first case in Equation~\ref{Eq:Ve} is illustrated in the bottom panel of Fig.~\ref{fig:avni}.  This leads to the sample average of  $V'_e/{V'}_{a}$ being defined as
\begin{equation}
\left\langle {\frac{{V'_e }}{{V'_a }}} \right\rangle  = \frac{1}{{N_T }}\sum\limits_i{\left\{ {\frac{{V'_e [z_i ,F_i ]}}{{V'_a [F_i]}}} \right\}} 
\end{equation}
where $N_T$ is the total number of objects in the two combined samples.
\\
\\ \noindent
\cite{Hudson:1991} recast  $V/{V}_{\rm max}$ for analysis of the diameter function of galaxies.  Therefore, for diameter-limited catalogues which have both a maximum and minimum diameter cut-off they show that the completeness test can be written as
\begin{equation}
\frac{V}{{V_{\max } }} = \frac{{\theta ^{ - 3}  - \theta _{\max }^{ - 3} }}{{\theta _{\lim }^{ - 3}  - \theta _{\max }^{ - 3} }},
\end{equation}
where $\theta$ is the major diameter of a given galaxy, $\theta_{\rm lim}$ is the lower diameter limit of the survey and  $\theta_{\rm max}$ is the maximum diameter cut-off of the survey.
\\
\\ \noindent
\cite{Eales:1993} provided a more rigorous and generalised  treatment for estimating errors for surveys sampling smaller volumes than the quasar samples examined in Schmidt's original work. Consequently, the error estimation laid out by Eales takes into account effects from strong clustering of galaxies and shot noise. The key to this approach draws from  \cite{Peebles:1980} 
by incorporating the variance in the number of galaxies $N$ within successive $i$ slices of redshift and absolute magnitude of a parent sample
\begin{equation}
{\rm Var}(N_i)= \left\langle {\left( {N_i  - \left\langle {N_i } \right\rangle } \right)^2 } \right\rangle  = \int {\varphi (r)dV} +
\int\int{\varphi ({\rm \bf r_1} )\varphi ({\rm \bf r_2} )\xi \left( {\left| {\rm \bf r_1  - r_2 } \right|} \right)dV_1 dV_2 },
\end{equation}
where ${\rm \bf r}$ is a position vector and the integrals are over the comoving volume subtended by the solid angle of the sample bounded by the redshift limits. The 2-point correlation function is given by $\xi(r)$, which accounts for the error due to clustering and the selection function, and $\varphi(r)$, taking into account the contribution of shot noise. Eales goes on to show that the total error in the LF is then estimated by
\begin{equation}
\sigma=\phi\left(\frac{{\rm var}(N)^{1/2}}{N}\right)
\end{equation}
In the same paper  the generalised $1/V_{\max}$ estimator introduced by \cite{Felten:1976} is extended to examine the evolution of the LF as a function of redshift.  Similarly, {\cite{Waerbeke:1996}}  looked specifically at the effects of pure luminosity evolutionary models  on QSOs via the $V_{\max}$ estimator to constrain cosmological parameters.
\\
\\ \noindent
\cite{Qin:1997} generalised the now familiar {\it Schmidt}  notation in terms of a new statistic called $n/n_{\rm max}$ that is applicable to {\it any} kind of distribution of objects.  This, therefore, would be an improved measure of the traditional  $V/{V}_{\rm max}$ test where the estimator is weighted differently and the distribution in question is assumed to homogeneous.  This fitting technique demonstrated that if the adopted LF is correct then the distribution of $n/n_{\rm max}$ is uniform on the interval [0,1] ,
\begin{equation}
\frac{{n(M,z)}}{{n_{\max } [M,z_{\max } (M)]}} = \frac{{\int\limits_0^z {\Phi (M,z)dV(z)\quad } }}{{\int\limits_0^{z_{\max } (M)} {\Phi (M,z)dV(z)} }},
\end{equation}
and the authors showed that its expectation value $\langle n/n_{\rm max}\rangle$ is $1/2$. 

Following from this, another statistic, $o/o_{\max}$, based on the cumulative LF and independent from  $n/n_{\rm max}$ was introduced by \cite{Qin:1999}:
\begin{equation}
\frac{{o(M,z)}}{{o_{\max } (z)}} = \frac{{\int\limits_{M_{\min } }^M {\Phi (M,z)dM\quad } }}{{\int\limits_{M_{\min } }^{M_{\max } (z)} {\Phi (M,z)dM} }}.
\end{equation}
This statistic combined with that of \cite{Qin:1997} are designed to provide a sufficient test for any adopted LF form. In the latter paper they apply both estimators to AAT sample data from the UVX survey \citep{Boyle:1990}.
\\
\\ \noindent
\cite{Page:2000} improved the method to take into account  systematic errors introduced for objects close to the flux limit of a survey.  As they point out, for evolutionary studies of galaxies the traditional approach, as extended by \cite{avni:1980} and  \cite{Eales:1993}, is very common \citep[see e.g.][]{Maccacaro:1991,Ellis:1996MNRAS.280..235E} but can distort the apparent evolution of  extragalactic populations.  Through the use of Monte Carlo simulations, with a sample of 10,000 objects and simulating an unevolving two-power law model X-ray LF,   they compare the $1/V_{\max}$ estimation of the differential LF given by
\begin{equation} \label{page2000} 
\phi _{1/V_{\max } } (L,z) = \frac{1}{{\Delta L}}\sum\limits_{i = 1}^N {\frac{1}{{V_{\max ,i}}}},
\end{equation}
to their improved binned approximation of the $\phi_{\rm est}$, which assumes that $\phi$ does not change significantly over the luminosity and redshift intervals $\Delta L$ and $\Delta z$, respectively, and is defined as
\begin{equation}
\phi_{\rm est}  = \frac{N}{{\int\limits_{L_{\min } }^{L_{\max } } \int\limits_{z_{\min } }^{z_{\max } (L)} {(dV)/(dz)dzdL} }},
\end{equation}
where $N$ is the number of objects within some volume-luminosity region.
\\
\\ \noindent
\citet*{Miyaji:2001AA...369...49M} extend the \cite{Page:2000} method for the study of active galactic nuclei (AGN).  To help reduce biases from binning effects they employ  a parametric model to correct for the variation of the LF within each bin. With this weighting scheme,  the binned LF is calculated by
\begin{equation}
\Phi(L_i,z_i)=\Phi(L_i,z_i)^{\rm model}\frac{N_i^{\rm obs}}{N_i^{\rm model}},
\end{equation}
where $L_i$ and $z_i$ represent the luminosity and redshift at the centre of the $i$th bin. $\Phi(L_i,z_i)^{\rm m}$ is the best-fit model evaluated at $L_i$ and $z_i$. $N_i^{\rm obs}$ is the number of observed objects in each bin and $N_i^{\rm m}$ is the number of objects estimated from the best-fit model.  The method still assumes homogeneity within each bin and obviously requires that the model accurately describes the data.  However, it is remarked that Poisson statistics can be used to compute the errors. This approach has been more recently applied by e.g.  \cite{Croom:2009MNRAS.399.1755C} for their studies of quasi-stellar objects (QSOs).
\begin{figure*}\label{fig:tojeiro}   
   \begin{center}
      \includegraphics[width=1.\textwidth]{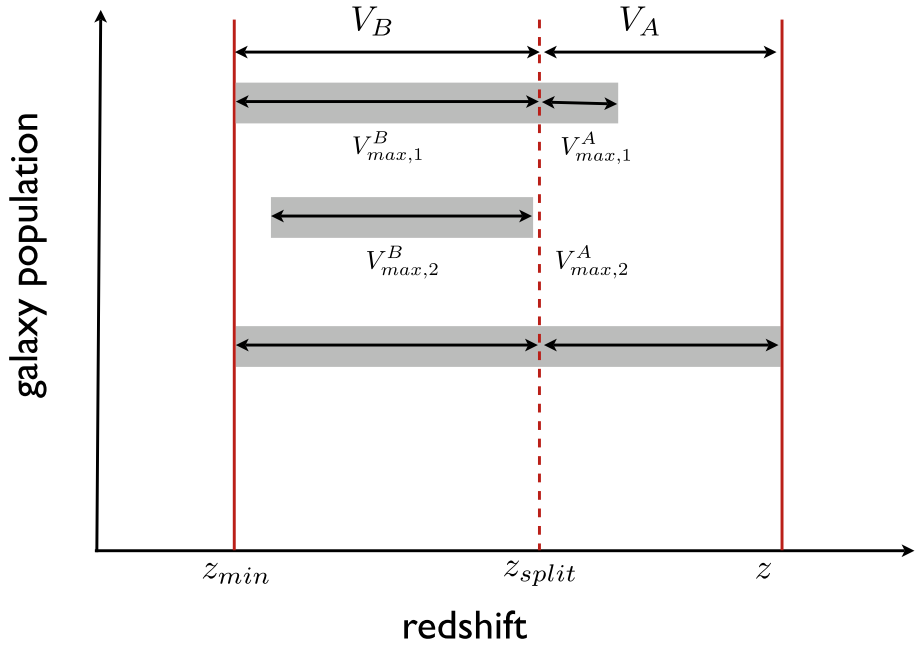} \\ 
              \caption {\small Image courtesy of  \cite{Tojeiro2010MNRAS.405.2534T}.  In the above schematic three populations (i.e galaxies with similar colours and luminosities) are considered.  For $V_A$ in  the  population 1 scenario galaxies are given a weight equal to unity.  However, a galaxy in $V_B$ in this population is down-weighted.   If a galaxy is in the population 2 sample and therefore confined to one slice only it would be given a weight of zero.  The population 3 galaxies are those that can been see throughout the entire survey sample.  This represents a volume-limited sample where the corresponding weight is always unity.}
       \label{fig:tojeiro}   
    \end{center}
  \end{figure*}
\\
\\ \noindent
\cite{Chapman2003ApJ...588..186C} uses $1/V_{\max}$ in order to construct the bivariate luminosity function (BLF) $\Phi(L,C)$, in luminosity $L$ and  colour $C$. This method is discussed in greater detail in \S~\ref{sec:bivariate} which reviews various BLF techniques.
\\
\\ \noindent
To account for low precision photometric redshift estimation  \cite{Sheth:2007} incorporated the $dN/dz_{photo}$ distribution into $V_{\max}$ by adopting a deconvolution technique.  This method will  be detailed in \S~\ref{sec:photoz}. 
\\
\\ \noindent
\cite{Tojeiro2010MNRAS.405.2534T} provide a variation of the traditional $V/V_{\max}$ estimator to probe  evolution of the Sloan Digital Sky Survey (SDSS) DR7 Luminous Red Galaxies (LRGs) \citep[see][for main sample selection]{Eisenstein:2001AJ.122.2267E}.  As pointed out by the authors, recent studies of evolution in LRG samples employs a procedure that constructs pairs of samples  - one being at high redshift and the other at low redshift.  However, matching the individual galaxy properties between the two samples traditionally requires the  removal of  galaxies that could not have been observed due to a varying selection function \citep[see e.g.][]{Wake:2006MNRAS.372..537W}. To overcome this, a weighting scheme is constructed to the two redshift slices: $V_A$ (for high redshift) and $V_B$ (for low redshift).  The weighting scheme down-weights galaxies to keep the total weight of each galaxy population the same in the different redshift slices.  Therefore, each galaxy in $V_A$ or $V_B$ is, respectively, weighted  by
\begin{equation}
w_i  = \frac{{V_A }}{{V_{\max ,i}^A }}\min \left\{ {\frac{{V_{\max ,i}^A }}{{V_A }},\frac{{V_{\max ,i}^B }}{{V_B }}} \right\} \ \  \mbox{or,} \ \ \frac{{V_B }}{{V_{\max ,i}^B }}\min \left\{ {\frac{{V_{\max ,i}^A }}{{V_A }},\frac{{V_{\max ,i}^B }}{{V_B }}} \right\}.
\end{equation}
This is illustrated in Fig.~\ref{fig:tojeiro}.  However, as the authors carefully note, this approach provides a weighting scheme {\it only} and is not a completeness estimator unlike the traditional Schmidt test.  As such, incompleteness may still be inherent in the parent sample that is under test.  Moreover, for a survey constructed by a magnitude-limited sample, the Schmidt estimator would be applied instead.
\\
\\ \noindent
Lastly, there has been a more recent paper by \cite{Cole:2011arXiv1104.0009C} in which he generalises $1/V_{\max}$ with a density corrected  $V_{\rm dc,max}$ estimator that takes into account effects from density fluctuations within a given volume.  The development of such a method was used to provide an algorithm which generates  magnitude-limited random (unclustered) galaxy catalogues, which take into account both the correlation function and galaxy evolution.
 \subsection{The $\rm C^-$ method}\label{sec:cmethod}
 
 As already discussed, the drawback in the use of the $1/V_{\max}$ is the assumption that the distribution of objects is spatially uniform.  The  increase in the number and variety of redshift surveys over the years has confirmed that galaxies have strong clustering properties.  Naturally, this can introduce a bias in constructing the differential LF.  However,  it was not  long before alternative approaches were developed that could circumvent this problem.

 The $C^-$ method was introduced by \cite{lynden:1971}, where he applied it to the quasar data of \cite{Schmidt:1968}. It is a maximum likelihood procedure adapted from the survival function and does not require any binning of data. Instead, the $C^-$ method estimates  the cumulative luminosity function (CLF).   It has the advantage over the $classical$ and $1/V_{\max}$ methods as it does not require any assumptions about the distribution of objects within the data-set.  Furthermore, as remarked by \cite{Petrosian:1992}, all non-parametric methods are essentially variations on the $C^-$ method in the limiting case of having one galaxy per bin.  Its mathematical properties have also been examined rigorously in the statistical literature \citep[see][]{woodroofe:1985,chen:1995}.
 
 In practice the method  recovers the CLF  without normalisation with the use of a weighted sum of Dirac $\delta$-functions (thus assuming no errors in magnitude).  As we shall see, to account for errors in magnitude one can add a smoothing kernel into the procedure.
\begin{figure} \label{fig:cminus}   
   \begin{center}
      \includegraphics[width= 0.9\textwidth]{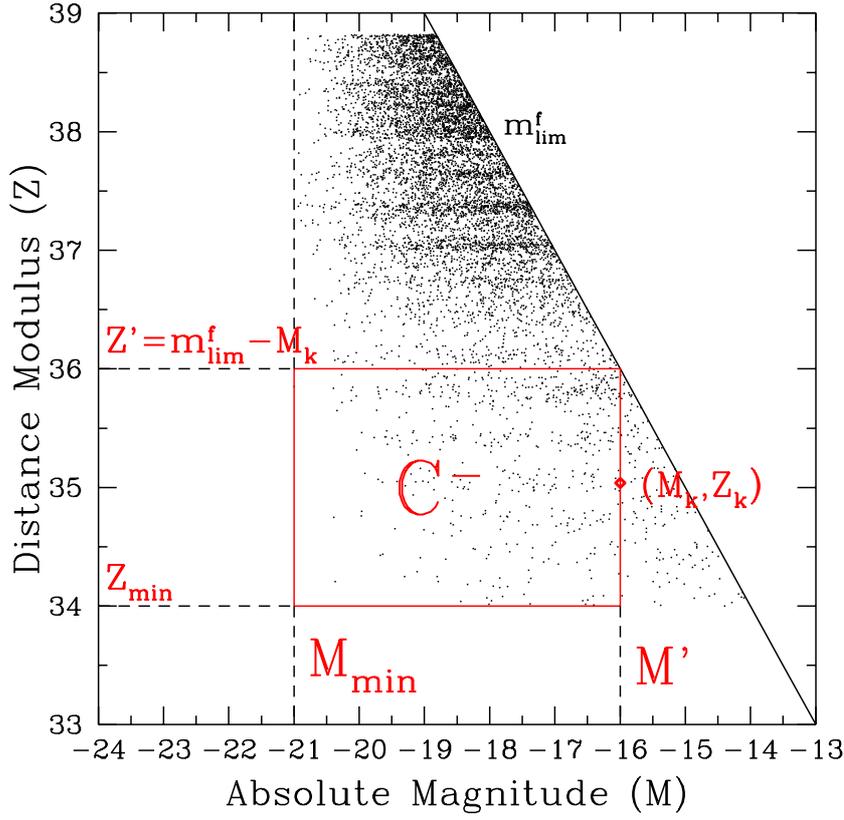}
              \caption {\small The construction of the $C^-$ method introduced by \cite{lynden:1971}. $C_k$ defines a region  $C^-$ for each galaxy  ($M_k,Z_k$) in the sample.  All the galaxies in this region are counted excepted for  ($M_k,Z_k$). $M_{\min}$ is imposed by the brightest galaxy in the survey data.  Similarly, $Z_{\min}$ is the minimum distance modulus defined by  the nearest galaxy at redshift $z_{\min}$.  }
       \label{fig:cminus}  
    \end{center}
  \end{figure}
To understand how the method works let us firstly consider the following.   In an ideal scenario where one is  not restricted by observational constraints such as faint and bright apparent magnitude limits,  constructing the cumulative luminosity function (CLF), $\Phi(M)$ would be  a relatively trivial task.  However, these limitations, in reality, lead us instead to observe a sub-population of galaxies sampling a  CLF which we can refer to as $X(M)$ \citep{Subbarao1996AJ....112..929S,Willmer:1997}.  Thus we find that
\begin{equation}
\frac{d\Phi}{\Phi}>\frac{dX}{X}
\end{equation}
Let us now consider this  observed distribution of galaxies in terms of  the absolute magnitude, $M$, and distance modulus, $Z$, plane $M-Z$, and assume (as with all the methods described so  far) separability between $M$ and $Z$ (see Fig.~\ref{fig:cminus} ignoring the red coloured markings for the moment).  From this figure we can see the sharp faint apparent magnitude  limit $m_{\rm lim}^{\rm f}$ of the survey indicated by the  diagonal line.  We can now write the probability density function of $M$ and $Z$ for the observable population as
\begin{equation}
dP=\rho(Z)dZ\  \phi(M)dM \ \Theta(m_{\rm lim}^{\rm f} - m)
\end{equation}
where,  $\Theta(m_{\rm lim}^{\rm f} - m)$ is a Heaviside function describing the magnitude cut,  $m_{\rm lim}^{\rm f}$ defined as
\begin{equation}
\Theta(x) =    \left \{\begin{array}{ll}
1 &\quad {\rm if} \, \, x \geq 0,\\
0&\quad {\rm if} \, \, x < 0,
\end{array}
\right .	
\end{equation}

However, to determine $\Phi(M)$ given the observed $X(M)$ we can construct a subset region of $X$ we  call $C^-(M)$ where we are  now able to obtain
\begin{equation}
\frac{d\Phi}{\Phi}=\frac{dX}{C^-}
\end{equation}
In this scenario, the integrated CLF can be written in form,
\begin{align}\label{equ:will97}
\Phi (M) = A\exp \left\{ {\int\limits_{ - \infty }^M {\frac{{dX(M)}}{C^-(M)}}} \right\},
\end{align}
% %
where, $A$ is a normalisation factor and $X(M)$ represents the parent set of points within which one constructs the $C^-(M)$ subset. However, we require the differential luminosity and density distribution functions which can be represented by a series of Dirac $\delta$ functions, respectively, given as
\begin{equation}\label{Eq:phiDF}
\phi (M) = \sum\limits_{i = 1}^N {\psi _i \delta (M - M_i ),} 
\end{equation}
\begin{equation} \label{Eq:DDF}
\rho(Z) = \sum\limits_{i = 1}^N {\rho_i \delta (Z - Z_i )},
\end{equation}
where $\psi_i$ and $\rho_i$ are the respective step sizes.  The distance modulus, $Z$, is calculated by
\begin{equation}
Z=m-M=5\log_{10}(d_L) +25 
\end{equation}
where $d_L$ is the luminosity distance to the object and $m$ is the apparent magnitude.  
To then construct the CLF  the $(M_k,Z_k)$ data are firstly  sorted from the brightest to the faintest galaxy  such that $M_{k+1} \ge M_k$ for $k=1,N$, and a region on the plane for each galaxy located at $M_k$ defines the  $C^-(M_k)$ function such that
\begin{equation}
C_k  \equiv C^ -  (M_k), \ \ \ k=1,..., N
\begin{cases}
& M_{\min }  \le M < M', \\ \\
& Z_{\min }  \le Z \le Z'
\end{cases}
\end{equation}
as illustrated in Fig.~\ref{fig:cminus}.  According to  \cite{Jackson:1974}, the superscript `minus' is to emphasise that  the point at ($M_k,Z_k$) is not included when evaluating $C^-(M_k)$.  The coefficients of the  LF are determined from the recursion relation,
\begin{equation}\label{equ:LFcoeff}
\psi _{k + 1}  = \psi _k \frac{{C_k^-(M)  + 1}}{{C_{k+1}^-(M) }},
\end{equation}
Therefore, the  CLF is given by,
\begin{equation}\label{equ:Cm}
\Phi(M_k)=\int\limits_{M_{\min } }^M {\phi (M)}dM= \psi_1 \prod\limits_{i}^{M_k  < M} {\frac{{C_k^-(M)  + 1}}{{C_k ^-(M)}}},
\end{equation}
In Equation~\ref{equ:Cm}  $(C_1+1)/C_1$ is set to unity so  that the product begins with $k=2$  \citep{Choloniewski:1987,Takeuchi:2000}.
\subsubsection{Extensions to the method}\label{sec:cminus_ext}
Although \cite{Jackson:1974} extended the original method to account for the combining of multiple data-sets and deriving suitable error estimates,  the method remained limited to deriving only the shape of the probability density function.  However,  \cite{Choloniewski:1987} revisited and improved the method by not only simplifying it, but by properly normalising the LF and estimating the density evolution of galaxies simultaneously.  However, Choloniewski's improved version has only been applied sporadically over the years \citep[see e.g.][]{Warren:1988ASPC....2...96W,Takeuchi:2000,Cuesta:2003AA...405..917C,Takeuchi:2006AA...448..525T}. A  paper by \cite{Schmitt:1990} extended the method for samples with multi-flux limits.

\cite{Caditz:1993} introduced a smoothing non-parametric method based on a Gaussian kernel, which replaces the $\delta$-function in Equation~\ref{Eq:phiDF} and \ref{Eq:DDF} with
\begin{equation}\label{equ:CP93}
f_{obs} (x) = \frac{1}{{h^d }}\sum\limits_{i = 1}^N {K\left[ {\frac{1}{h}(x - x_i )} \right]} 
\end{equation}
where  $x$ represents the observables, $K(x)$ is the kernel function, $d$ is the number of measured parameters for each object.  The parameter $h$ is a free parameter which defines the magnitude of the smoothing scale. The kernel therefore replaces the $\delta$-function distributions given in Equations~\ref{Eq:phiDF} $\&$ \ref{Eq:DDF} which can otherwise limit the use of $C^-$ towards the faint magnitude limit of a survey.

\cite{Subbarao1996AJ....112..929S} extended the method for photometric redshifts by considering, for each galaxy, the probability distribution in absolute magnitude $M$ resultant from the photometric redshift error.  By adopting a Gaussian error distribution for the function $z^*(m_i,M)$, the redshift for the $i$th galaxy with an apparent magnitude $m_i$, they showed that for a complete magnitude-limited sample the defined region $C^-(M)$ is now given as \\
\begin{equation}\label{equ:subberr}
C^-(M) = 0.5\sum\limits_i {\left[ {{\rm{erfc}}\left( {\frac{{z^* (m_i ,M) - z_i }}{{\sigma _i }}} \right) - {\rm{erfc}}\left( {\frac{{z^* (m_{{\rm{lim}}} ,M) - z_i }}{{\sigma _i }}} \right)} \right],} 
\end{equation}
where $\rm{erfc}$$(x)$ is the complementary error function.

\cite{Willmer:1997} included Lynden-Bell's approach applied  to simulated data and the CfA~-~1 survey \citep{Huchra:1983}  when  comparing various  LF estimators, and is  discussed in more detail in \S~\ref{sec:comp_review}.

\cite{Zucca:1997AA...326..477Z} provided a variant on the original Lynden-Bell version termed the $C^+$ estimator, which they then incorporated into their  ``Algorithm for Luminosity Function" (ALF)  tool  \citep[see e.g][]{Ilbert:2004,Ilbert:2005,Zucca:2009}.  Essentially Equation~\ref{equ:LFcoeff} is redefined such that for each $k^{\rm th}$ galaxy, the contribution to the LF is given by
\begin{equation}\label{equ:zucca}
\psi(M_k) = \frac{{1 - \sum\nolimits_{j = 1}^{k -1} {\psi (M_j )} }}{{C^ +  (M_k )}}
\end{equation}
where,  galaxies for the $\psi(M_j)$ summation are  sorted  from the faintest  to the brightest and $C^+(M_k)$ is the number of galaxies in the region,
\begin{equation}
C^ +  (M_k), \ \ \ k=1,..., N
\begin{cases}
& M < M_k, \\ \\
& Z_{\min }  \le Z \le Z_{\max ,k}(M_k)
\end{cases}
\end{equation}
Thus to obtain a binned version of the LF, the  quantity $\psi(M_k)$ is estimated for each galaxy and the returned values binned in absolute magnitude.
\subsection{The $\phi/\Phi$ method}\label{sec:phioverphi}
Originally introduced by \cite{Turner:1979} and \cite{Kirshner:1979} the $\phi/\Phi$ method (as coined by \citetalias{Binggeli:1988}) is a natural progression from the $classical$ method (\S~\ref{sec:classical}) that  returns a binned estimate of the LF.  For a magnitude-limited sample we calculate the ratio of the number of galaxies in the interval $dM$, $N(dM)$ to the total number of galaxies brighter than $M$, $N(\le M)$ within the maximum volume assuming  a complete sample:
\\
\\
\begin{align}
\frac{{N(dM)}}{{N( \le M)}} = \frac{{dN( \le M)}}{{N( \le M)}} &= \frac{{\phi (M)\rho(z)dM{\kern 1pt} dV}}{{\int\limits_{ - \infty }^M {\phi (M')\rho(z)dM'{\kern 1pt} dV} }} = \frac{{\phi (M)dM}}{{\Phi (M)}} \approx d\ln \Phi (M),\label{Eq:phi_PHI}
\end{align}
 where $\rho(z)$ is the density function and $\Phi(M)$ is the integrated LF.  It is clear from this equation that the density functions cancel, thus rendering the estimator independent of the distribution of galaxies.   This estimator has been further developed slightly  - \cite{Davis:1980}, \cite{Davis:1982} to bin the data in equal distance intervals. However, as shown in  \cite{Lapparent:1989} the approximation in Equation~\ref{Eq:phi_PHI} introduces a bias in the determination of the LF for large $dM$.  To avoid this bias it has been common place to instead assume an analytical form for the LF as in \cite{Turner:1979}, \cite{Kirshner:1979}, \cite{Davis:1982} and   \cite{Lapparent:1989}. So although by its  original construction this estimator is non-parametric, subsequent applications have found its place to be more useful as a parametric one.  %
 \subsection{The step-wise maximum likelihood method}\label{sec:swml}
One of the first embodiments  of the step-wise MLE approach, was presented by \cite{Nicoll:1983} and can be thought of as a binned version of the Lynden-Bell's $C^-$~method.  This was applied in their analysis of chronometrical cosmology and also considers variations of progressive truncation in apparent magnitude as well as multivariate complete samples.

A more advanced version of Nicoll and Segal's  approach by \cite{Choloniewski:1986} applied the same Poisson probability distribution in the MLE of \cite{Marshall:1983} (see section~\ref{sec:MLE}).  In this paper, the data are projected on the absolute magnitude $M$ and distance modulus $Z$ plane and divided into equal sized cells such that
\begin{align}
M&\in[M_{i-1},M_i], \quad i=1,..., A, \\ \nonumber
Z&\in[Z_{j-1},Z_i], \quad j=1,..., B, \\ \nonumber
\end{align}
The likelihood function can be represented as
\begin{equation}
\mathcal{L} = \prod\limits_{i = 1}^A {\prod\limits_{j = 1}^B} {\frac{{e^{ - \lambda _{ij} } \lambda _{ij}^{N_{ij} } }}{{N_{ij} !}}} ,
\end{equation}
where
\begin{equation}
\lambda _{ij}  = \frac{1}
{{\bar n}}\Phi_i(M)\rho_j(Z)dM\ dZ,
\end{equation}
where $N_{ij}$ is the number of galaxies in the $(i,j)$ cell, $\bar n$ is the mean density of the sample, $\Phi(M)$ is  LF, and $\rho(Z)$ is the  density function where the separability between $\Phi(M)$ and $\rho(Z)$ is assumed.
\\
\\
The method which is would become known as the  step-wise maximum likelihood method (SWML) was introduced by  \citet*{Efstathiou:1988} (hereafter \citetalias{Efstathiou:1988}) and represents the non-parametric version of the \citetalias{sandage:1979}  method.  As its name implies the  technique does not depend on an analytical form for $\phi(M)$.  Instead the LF is in effect parameterised as a series of $N_p$ step functions allowing us to define the following initial setup:
\begin{align}
\Phi(M)&=\phi_k, \ \ \ \ \ M_k-\frac{\Delta M}{2} < M < M_k +\frac{\Delta M}{2}, \\  
\mbox{where,} \ k&=1,...,N_p \nonumber
\end{align}
The differential LF can then be expressed as
\begin{equation}\label{Eq:SWMLtophats}
\phi (M) = \sum\limits_{i = 1}^N {\phi _i W(M_i  - M),} 
\end{equation}
 where $W(x)$ represents two window functions,  
 \begin{equation}\label{Eq:SWMLweight}
 W(x)\equiv
 \begin{cases}
 1, &- \frac{{\Delta M}}{2} \le x \le  \frac{{\Delta M}}{2}, \\
 0, & \mbox{otherwise}.
 \end{cases}
 \end{equation}
 Therefore, it can be shown that the expression for the step-wise likelihood is given by
\begin{equation}
\mathcal{L}(\left\{ {\phi _i } \right\}_{i = 1,...,I} |\left\{ {M_k } \right\}_{l = 1,...,K} ) = \prod\limits_{k = 1}^{N_{{\rm{obs}}} } {\frac{{\sum\limits_{l = 1}^K {W(M_l  - M_k )\phi _l } }}{{\sum\limits_{l = 1}^K {\phi _l H(M_{{\rm{lim}}} (z_k ) - M_l )\Delta M} }}}  
  \end{equation}
 where
 \begin{equation}
 H(x)=
 \begin{cases}
 0, &x\le -\Delta M/2, \\
 (x/\Delta M+1/2), &-\Delta M/2\le x\le \Delta M/2, \\
 1, &x \ge \Delta M/2.
 \end{cases}
 \end{equation}
 In \citetalias{Efstathiou:1988} the errors on the LF parameters are assumed to be asymptotically normally distributed giving a covariance matrix,
 \begin{equation}
 {\rm cov}(\phi_k)=[{\rm\bf I}(\phi_k)]^{-1},
  \end{equation}
 where ${\rm\bf I}(\phi_k)$ is the information matrix \citep[see][]{Eadie:1971smep.book.....E}.  \cite{Strauss:1995PhR...261..271S,1997ApJ...477...36K} noted two drawbacks of the method. The first concerns discretisation of the LF using step functions.  A bias is introduced into the selection function due to having discontinuous first derivatives. The authors instead suggest  interpolating through the steps and then calculating the selection function to reduce this bias. The second drawback is the sensitivity of the LF to the choice of bin size. In \cite{1997ApJ...477...36K} they show by example,  that if  the total number of bins is too small, this can dramatically underestimate the faint-end slope of the LF.

\cite{Heyl:1997} extended the use of the SWML method by generalising it in a similar way as \cite{avni:1980} did for $V/V_{\max}$ by combining various surveys with different magnitude limits, coherently.  Moreover, this extension also provided an absolute normalisation and was used to probe density  evolution in the LF by spectral type.  In the same paper, they compare this method against  the $1/V_{\max}$ estimator (see \S~\ref{sec:comp_review}).  

\cite{Springel:1998} also explored evolution and provided a variation of the method that instead models the selection function as a series of linked piecewise power laws as opposed to step functions as in Equation~\ref{Eq:SWMLtophats}.   			%section 5
\section{Bivariate Luminosity Functions}\label{sec:bivariate}

The move beyond the standard LF analysis has seen constructing bivariate LFs (BLF) which have  proven useful to further constrain galaxy formation and evolution theory.   Much of the work described in this section  represent natural progressions of the \citetalias{sandage:1979} and the \citetalias{Efstathiou:1988} MLE estimators that have combined various observables such as  galaxy radius, colour, galaxy diameters and surface brightness with space density. A more general approach recently developed by  \cite{Takeuchi2010MNRAS.406.1830T} shall be discussed in greater detail in \S~\ref{sec:emerging}.
\subsection{Galaxy radius}\label{sec:blf_radius}
 \cite{Choloniewski1985MNRAS.214..197C} constructed a BLF with galaxy radius, $\Phi(M,\log r)$. In this scenario the number of galaxies in the interval $dM d(\log r)dm$ can be written as
\begin{equation}
n(M,\log r,m) = \phi (M,\log r)f(m)\rho (Z)10^{0.6(Z)},
\end{equation}
where $Z=m-M $ is the distance modulus, $f(m)$ is a completeness function which describes selection effects and  $\rho(Z)$ is the density function.  The likelihood function is then defined as
\begin{equation}
\mathcal{L} = \sum\limits_i {\ln \frac{{\phi (M_i )f(m_i )\rho (m_i  - M_i )10^{0.6(m_i  - M_i )} }}{{\int\limits_{ - \infty }^{ + \infty } {dM\int\limits_{ - \infty }^{ + \infty } {dm\;\phi (M)f(m)\rho (m - M)10^{0.6(m - M)} } } }}}  + \sum\limits_i {\ln S(M_i ,\log (r_i ))},
\end{equation}
The maximisation of the  first component of the likelihood gives the shape of the LF $\phi(M)$. However, the second component $S(M,\log r)$ requires performing the MLE to the distribution of galaxies in the $M-\log r$ plane which has the assumed form, 
\begin{equation}
S(M,\log r) = \frac{1}{{\sqrt {2\pi \sigma } }}\exp \left[ { - \frac{{(\log r + aM + b)^2 }}{{2\sigma ^2 }}} \right].
\end{equation}

\subsection{Colour}\label{sec:blf_colour}
\cite{Chapman2003ApJ...588..186C} described a  parametric approach to fit a  bivariate distribution,  $\Phi(L,C)$, in luminosity $L$ and  colour $C$,  that utilises   the $1/V_{\rm max}$ estimator.  This paper is concerned specifically with the local infrared-luminous galaxies from the {\it IRAS} 1.2 Jy sample selected at 60 $\mu$m \citep[see][for full survey description]{Fisher:1995.APJS.100.69}. They use the IR  colour distribution defined as the $R(60,100) \equiv S_{60\mu m}/S_{100\mu m}$ flux ratio (where $S$ refers to the wave band) to explore colour-added evolutionary models.

Recalling that the $V_{\rm max}$ estimator can be expressed as
\begin{equation}
\Phi (L)\Delta L = \sum\limits_i {\frac{1}{{V_{\max, i}}}},
\end{equation}
where $\Delta L$ is the luminosity range within which we have a number density of galaxies, $\Phi(L)$.   To reiterate, $V_{{\rm max}, i}$ represents the maximum volume that the $i$th galaxy can be located in and still be detected. 

They find that the population of sources in the {\it IRAS} sample is represented best by a lognormal distribution  such that
\begin{equation}
G(C) = \exp \left[ { - \frac{1}{2} \times \left( {\frac{{C - C_0 }}{{\sigma _C }}} \right)^2 } \right].
\end{equation}
The colour distribution  is modelled analytically as 
\begin{equation}
R\left( {\frac{{60}}{{100}}} \right)  = C_*  \times \left( {1 + \frac{{L_* }}{{L_{\rm TIR} }}} \right)^{ - \delta } \left( {1 + \frac{{L_{\rm TIR} }}{{L_* }}} \right)^\gamma ,
\end{equation}
where $L_{\rm TIR}$ is defined as the  total infrared luminosity. $\delta$ and $\gamma$ represent the slopes of the faint-end and bright-end of the LF, respectively.

The final BLF convolves the LF, $\Phi_1(L)$, modelled as a two-power law with the colour function, $\Phi_2(C)$, as defined above to give
\begin{align}\nonumber
 \Phi (L,C)dLdC &= \Phi _1 (L) \times \Phi _2 (C)dLdC \\  \nonumber
  &= \rho _*  \times \left( {\frac{L}{{L_* }}} \right)^{(1 - \alpha )}  \times \left( {1 + \frac{L}{{L_* }}} \right)^{ - \beta }  \\  
  &\times \exp \left[ { - \frac{1}{2}\left( {\frac{{C - C_0 }}{{\sigma _C }}} \right)^2 } \right]dLdC,\quad (1 - \alpha ), 
\end{align}
where the $C$ is calculated using the observed {\it IRAS} broad-band fluxes. See also, \cite{Chapin2009MNRAS.393..653C}, \cite{Marsden2010arXiv1010.1176M} for further extensions to this method.

\subsection{Galaxy diameters}\label{sec:swml_biv}
The SWML has now become a very popular method for determining the LF and \citet*{Sodre:1993} extended it  (and the \citetalias{sandage:1979} estimator)  to a bivariate distribution of magnitudes and galaxy diameters \citep[see also][]{Santiago:1996}.  In this variation the conditional probability of finding a galaxy  with a diameter $D$ and absolute magnitude $M$ can be written as
\begin{equation}
P(D,M) = P(M)P(D|M) = P(D)P(M|D).
\end{equation}
The bivariate distribution of diameters and magnitudes is then expressed as
\begin{equation}\label{equ:sodre_dist}
\Psi (D,M)\,{\rm{d}}D{\rm{d}}M = \phi (D)\,\varphi (M|D){\rm{d}}D{\rm{d}}M,
\end{equation}
where $\phi (D)$  is the diameter distribution function  and $\varphi (M|D)$ gives the luminosity distribution for galaxies with diameter $D$.  In terms of the step-wise likelihood, the distribution function of Equation~\ref{equ:sodre_dist} is now parameterised as $N_D$ bins in diameter and $N_M$ bins in absolute magnitude such that,
\begin{equation}
\Psi(D,M)=\Psi_{jk}, \ \ \ \ j=1,...,N_D, \ \ \ \ k=1,...,N_M.
\end{equation}
It can be shown that the log likelihood is given by
\begin{align}
 \ln (\mathcal{L}) &= \sum\limits_{i = 1}^N {\sum\limits_{j = 1}^{N_D } {\sum\limits_{k = 1}^{N_M } {W_{ijk} \ln [\Psi _{jk} C_V (D_i /v_i^* ,M_i  + Z_i )]} } }  \\ 
  &- \sum\limits_{i = 1}^N {\ln \left( {\sum\limits_{i = 1}^{N_D } {\sum\limits_{m = 1}^{N_M } {H_{ilm} \Psi _{lm} \Delta D\Delta M} } } \right) + {\rm{constant}}},
 \end{align}
where
 \begin{equation}
 W(x,y)=
 \begin{cases}
 1, \ \ \mbox{if} &-(\Delta D)/2\le x\le (\Delta D)/2\\
 &\mbox{and}\ -(\Delta M)/2\le y\le (\Delta M)/2 \\
 0, &\mbox{otherwise}.
 \end{cases}
 \end{equation}
In a general way the quantity $C_V$ defines the velocity completeness function $C_V(\theta^m,m)$ as the fraction of galaxies with measured velocities $v$  and with apparent diameters  between $\theta$ and $\theta + {\rm d}\theta$, and apparent magnitudes between $m$ and  $m+{\rm d}m$.   $Z_i$ is  the distance modulus $v_i^*\equiv D_i/\theta^m$.  \cite{Ball2006MNRAS.373..845B} adopted this approach to construct the BLF with surface brightness.

\subsection{Surface brightness}\label{sec:biv_surfdens}
\cite{Cross2002MNRAS.329..579C} \citep[see also][]{Driver:2005}   extended the \citet*{Sodre:1993} approach and that of \citetalias{Efstathiou:1988} to determine the joint luminosity-surface brightness LF and thus recover the bivariate brightness distribution (BBD) using the Millennium Galaxy Catalogue (MGC) survey data.  In this scenario they consider magnitude and surface brightness limited data where  the effective absolute surface brightness is defined as
\begin{equation}
\mu^e=m+2.5\log[2\pi(r^0_e)^2]-10\log(1+z)-k(z)-e(z),
\end{equation}
where $r^0_e$ is the {\it true} observed half-light radius and $k(z)$ and $e(z)$ are the respective $K-$ and evolution corrections. To construct the bivariate SWML for this case they firstly define an observable window function given by
\begin{align}\label{Eq:Driver_O}
 O_{i}(M,\mu^e)=
 \begin{cases}
1, \ \ &\mbox{if} \ \ \ M_{{\rm bright},i}< M< M_{{\rm faint},i}\\ 
 &\mbox{and}\ \ \ \mu^e_{{\rm high},i}<  \mu^e <\mu^e_{{\rm low},i}\\
 0, &\mbox{otherwise}.
 \end{cases}
 \end{align} 
 Taking the \cite{Sodre:1993} approach they define $W_{ijk}$ to weight each galaxy to account for incompleteness such that
 \begin{align}\label{Eq:Driver_W}
 W_{ijk}=
 \begin{cases}
 \frac{N_i}{N_i(Q_z\ge 3)}, \ \ &\mbox{if} \ \ \ M_j-\frac{\Delta M}{2}\le  M_i <M_j+\frac{\Delta M}{2}\\ \\ 
 &\mbox{and}\ \ \ \mu^e_k-\frac{\Delta \mu^e}{2}\le  \mu^e_k <\mu^e_k+\frac{\Delta \mu^e}{2}\\ \\
 0, &\mbox{otherwise},
 \end{cases}
 \end{align}
where $N_i$ is the total number of galaxies lying in the same $m-\mu^e$ bin as galaxy $i$. $Q_z$ defines a redshift quality, where a $Q_z\ge3$ equates to a reliable redshift measurement.  The quantity $N_i(Q_z\ge3)$ is thus the number of galaxies which known redshifts in the same bin.
Finally, to account for the fraction of the $M_j-\mu^e_k$ bin which lies inside the observable window of galaxy $i$ they define the visibility function to be, 
 \begin{equation}\label{Eq:Driver_H}
H_{ijk}  = \frac{1}{{(\Delta M\Delta \mu ^e )}}\int\limits_{M_j  - \Delta M/2}^{M_j  + \Delta M/2} {dM} \int\limits_{\mu _k^e  - \Delta \mu ^e /2}^{\mu _k^e  + \Delta \mu ^e /2} {d\mu ^e O_i (m,\mu ^e ),} 
\end{equation}
They then fit a functional form characterising the joint luminosity-surface brightness (BBD) distribution given by
\begin{align}\label{equ:driverbivlf}
 \Phi (M,\mu ^e ) &= \frac{{0.4\ln 10}}{{\sqrt {2\pi } \sigma _{\mu ^e } }}\phi _* 10^{0.4(M_*  - M)(\alpha  + 1)} e^{ - 10^{0.4(M_*  - M)} }  \\ \nonumber
  &\times \exp \left\{ {\frac{1}{2}\left[ {\frac{{\mu ^e  - \mu ^{e*}  - \beta (M - M_* )}}{{\sigma _{\mu ^e } }}} \right]} \right\},
 \end{align}
where the upper part of Equation~\ref{equ:driverbivlf} has the usual Schechter function parameters and the lower part of the equation has additional parameters, $\mu ^{e*}$, $\sigma _{\mu ^e }$ and $\beta$, to characterise the surface brightness data.

 Section~\ref{sec:emerging}  examines in more detail a new technique to the BLF by \cite{Takeuchi2010MNRAS.406.1830T} which offers a more generalised approach by using the copula to connect two distribution functions and the Pearson correlation coefficient to explore the correlation between the two. 
			%section 6
 \section{Methods to Account for Large Uncertainties in Redshift Estimation}\label{sec:photoz}
The use of  photometric redshifts (photo-z) is playing a  more central role in  probing the cosmological model \citep[see e.g.][]{Blake:2005MNRAS.363.1329B,Banerji:2008MNRAS.386.1219B}. Measurements can be performed  quickly and to high redshift and are therefore specifically suited to large  galaxy surveys.  However, as previously discussed, the resulting precision is substantially worse than measuring them spectroscopically. Consequently, the ability to accurately constrain the LF  is  severely impeded.  In what follows, I discuss  some of the modifications made to the standard LF  estimators  to incorporate photo-z measurements that are not precisely known.

\subsection{Smoothing kernel}
 Already discussed in \S~\ref{sec:cmethod} was the  \cite{Subbarao1996AJ....112..929S}  Gaussian error distribution which was incorporated into the $C^-$ method given by
\begin{equation}
X(M) = 0.5\sum\limits_i {\rm erfc}\left( \frac{z^* (m_i ,M) - z_i}{\sigma _i } \right),
\end{equation}
which led to estimating the $C^-$ function as shown in Equation~\ref{equ:subberr}.  This essentially transforms the traditional Lynden-Bell `discrete' approach into a more `continuous'  one, where errors in the redshift distribution  coupled with $K$-corrections can now be represented by a smoothing function and integrated over the redshift probability distribution for each galaxy.  To assess the errors and explore biases they use bootstrapping Monte Carlo techniques.
\subsection{Deconvolution - a $V_{\max}$ generalisation}\label{sec:photoz_sheth_vmax}
A paper by \cite{Sheth:2007} revisited the Schmidt $V_{\max}$ estimator  and extended its use for surveys with measured photo-z by casting it as a deconvolution problem.  
To begin with the probability of estimating a redshift $z_e$ given its {\it true} value  $z_t$\footnote{The subscript $t$ refers throughout this section as the  {\it true} measured value, which, in this case, would be derived from a spectroscopic redshift} is given by
\begin{equation}
\frac{{dN_e (z_e )}}{{dz_e }} = \int {dz\frac{{dN(z_t)}}{{dz}}p(z_e |z_t)}.
\end{equation}
By now considering the generalised case in which a catalogue has both a minimum limiting volume, $V_{\min}$ for which an object would be too bright to be included in the catalogue, and the usual maximum volume, $V_{\max}$, the number of galaxies, $N$, with $true$ absolute magnitude, $M_t$, in a magnitude-limited  catalogue is given by
\begin{equation}
N(M_t) = \phi(M_t)[V_{\max } (M_t)-V_{\min } (M_t)].
\end{equation}
This equation is consistent within the usual Schmidt framework, where the LF would now be constructed by a $1/[V_{\max } (M_t)-V_{\min } (M_t)]$ weighting scheme. However, we are now required to consider the number of galaxies $N_e$ with an {\it estimated} absolute magnitude $M_e$ which is shown to be given by
\begin{equation}
N_e (M_e ) = \int {dM_t \phi (M_t )\int\limits_{d_L (M_t^{\min } )}^{d_L (M_t^{\max } )} {dd_L } } \frac{{dV_{\rm com} (d_L )}}{{dd_L }} \times p(M_t  - M_e |d_L ,M_t ),
\end{equation}
where $d_L$ is the luminosity distance and $V_{\rm com}$ is the comoving volume.
In order to simplify the process it is assumed that $p(M_t-M_e|d_L,M_t)$ does not depend on $d_L$ and so the number of estimated absolute magnitudes can be reduced to
\begin{equation}
N_e(M_e) = \int {dM\phi (M_t)\mathcal{V}(V_{\max } ,V_{\min } ,M_t)},
\end{equation}
where,
\begin{equation}
\mathcal{V}(V_{\max } ,V_{\min } ,M_t)=\int\limits_{d_L (M_t^{\min } )}^{d_L (M_t^{\max } )} {dd_L \frac{{dV_{\rm com} (d_L )}}
{{dd_L }}p(M_t - M_e|d_L ,M_t)}.
\end{equation}
An iterative  deconvolution algorithm originally developed by \cite{Lucy:1974AJ.....79..745L} is finally applied to estimate the redshift distribution and thus the luminosity function.  In \cite{Rossi:2008} they apply this method via mock catalogues to probe their effectiveness in areas such as the size-luminosity relation which is often distance-dependent \citep[see also][]{Rossi2010MNRAS.401..666R}.  

\subsection{Modifying the maximum likelihood estimator}\label{sec:photoz_mle}
Using Monte Carlos simulations \cite{Chen:2003ApJ...586..745C} demonstrated that  applying the standard \citetalias{sandage:1979} MLE to photo-z data produces an intrinsic bias into the shape of the LF.  This manifests as a steeper slope at the bright end of the LF caused by intrinsic fainter galaxies being scattered more  into the brighter end than brighter galaxies being scattered into the faint end of the LF.   To attempt to reduce this bias they convolve for each photometric redshift of galaxy $i$, the Gaussian kernel with the redshift likelihood function to give  
\begin{equation}
p_i (z - z_i ;z_i ) = \int\limits_0^\infty  {\frac{1}{{\sigma _z \sqrt {2\pi } }}\mathcal{L}_z^i (z'}  - z_i ;z_i )\exp \left[ {\frac{{ - (z - z')^2 }}{{2\sigma _z^2 }}} \right]dz'.
\end{equation}
Thus, they show that the total likelihood function is then represented by 
\begin{align}
\mathcal{L}&=\prod\limits_i P_i (m_i ,z_i ;M_* )  \nonumber \\
&= \prod\limits_i  \int\limits_0^{z_f } {\psi _i p_i (z_i  - z'} ;z_i )dz'\left[ {\Theta \int\limits_{m_{\min } }^{m_{\max } } {\int\limits_0^{z_f } {\psi _i } } p_i (z_i  - z';z_i )dzdm} \right]^{ - 1}, 
\end{align}
where  $\Theta$ is the fraction of the angular area sensitive enough to detect the galaxy $m_i$ at a redshift $z_i$ and,
\begin{equation}
\psi _i  = 10^{0.4\left[ {M_*  - M_i (m_i ,z')} \right](1 + \alpha )} \exp \left( { - 10^{0.4\left[ {M_*  - M_i (m_i ,z')} \right]} } \right).
\end{equation}
Applying this modification to the same simulated data showed an improvement in the recovered input LF with a reduction in the systematic uncertainties of $\sim0.7$~mag.
\\
\\
\noindent
Following from this, and in the same paper as his deconvolution technique for $V_{\rm max}$,  \cite{Sheth:2007} improves upon the work of Chen {\it et al.} by deriving a new  likelihood. To begin with the general likelihood is expressed  as
\begin{equation}
\mathcal{L}({\bf a}) = \prod\limits_i {\frac{{\phi (L_i |z_i,{\bf a})}}{{S(z_i, {\bf a})}}},
\end{equation}
where $S$ is the selection function, ${\bf a}$ represents the parameters that describe the shape of the LF and $\phi (L_i |z_i,{\bf a})$ is the LF at a redshift $z$.

If we now denote  the estimated photometric redshift as $z_e$ and thus the estimated apparent luminosity as $l_e$.  Also, if  $l_t$ and $z_t$ are the respective true apparent luminosity and redshift then  the number $N$ of estimated redshifts within a flux-limited catalogue is shown to be,
\begin{align}
 N(z_e ,{\bf a}) = \int {dz} \frac{{dV_{\rm com} }}{{dz}}\int\limits_{l^{\min}_t }^{l^{\max}_t } {dl_t4\pi d_L^2 (z_t)}   \times \phi (L_t |{\bf a})p(z_e |z_t,l_t),
 \end{align}
where, $V_{\rm com}$ is the comoving volume. The absolute luminosity,  $L_t=l_t4\pi d_L^2(z)$,where  $d_L(z)$ is the luminosity distance. The joint distribution of $z_e$ and $l_e$ is then given by
\begin{align}
 \l_eN(l_e,z_e ,{\bf a}) = \int {dz} \frac{{dV_{\rm com} }}{{dz}}{4\pi d_L^2 (z_t)}  \times \phi (L_t|{\bf a})p(z_e |z_t,l_t).
 \end{align}
Thus the probability distribution by which the likelihood is to be maximised is such that
\begin{equation}
\mathcal{L({\bf a})} = \prod\limits_{i = 1}^{N} \frac{N(l_{i,e},z_{i,e},{\bf a})}{N(z_{i,e},{\bf a})}, 
\end{equation}
\\
\\
\noindent
The modification by \cite{Christlein2009MNRAS.400..429C}  attempts to define a likelihood function similar in approach to that of  \citetalias{Marshall:1983} (as described  in Equation~\ref{equ:marshal}) which, as suggested by the authors, would replace the photometric redshift of Sheth by a more robust observable. Just as \citetalias{Marshall:1983} gridded the luminosity-redshift information, this method  defines a parameter space that encompasses  absolute magnitude $M_0$, SED and redshift $z_0$ of a galaxy.  This so-called {\it photometric space} is treated as an $n$-dimensional flux space where each dimension corresponds to one filter band in which  a flux, $f$, is measured.  This space is gridded into cells such that each cell either contains 0 or 1 galaxies.  Equation~\ref{equ:marshal} can then be written as
\begin{equation}
\mathcal{L} = \prod\limits_i^N {\lambda (f_i )df\,e^{ - \lambda (f_i )df} } \prod\limits_j^{} {e^{ - \lambda (f_j )df} },
\end{equation}
where $\lambda(f_i)df$ is the expectation of the number of galaxies in the $i^{\rm th}$ cell in photometric space.  To map between the LF parameters and photometric space it is assumed that the photometric data are normally distributed centred on the expectation value for a given absolute magnitude $M_0$, SED and redshift $z_0$. Thus $\lambda(f_i)$ is defined as

\begin{equation}
\lambda (f_i ) = \sum\limits_{\rm SED} {\int {dM_0 \int {dz_0 \left( {\frac{{dV_c }}{{dz}}} \right)} \; \frac{{p_\sigma  (f_i |M_0 ;{\rm SED};z_0 )}}{{\prod\nolimits_n {\Delta _n } }}\;} } \Phi (M_0 ;{\rm SED};z_0 |P),
\end{equation}
where $P$ represents the LF parameters, $dV_c/dz$ is the differential comoving volume and $p_\sigma$ represents the fraction of galaxies  that contribute to the galaxy density at a point $f_i$.  The standard  Schechter LF  of the form in Equation~\ref{Eq:schM} was adopted  to which an evolutionary term was added. To constrain a galaxy to a given $z$, SED and $M$ the above equation is modified as
\begin{align}
 \lambda _i  &= \sum\limits_{\rm SED} {\int {dM_0 \int {dz_0 \left( {\frac{{dV_c }}{{dz}}} \right)} \;} } \delta (z_0  - z_t ) \nonumber \\ 
  &\times \delta ({\rm SED - SED}_t ) \times \delta (M - M_t ) \times (M_0 ;{\rm SED};z_0 |P),
 \end{align}
where  $z_t$, ${\rm SED}_t$ and $M_t$ are the {\it true} values to which a given galaxy is constrained. The standard  Schechter LF  of the form in Equation~\ref{Eq:schM} was adopted  which an evolutionary term added.

\begin{landscape}
{\scriptsize
\begin{center}
\begin{longtable}{lccccc}
\caption{\small Summary of the traditional LF estimators.}\label{t:surveys}
\\ \hline \hline
 \\
{\bf Estimator} & {\bf Type} & {\bf Developed by} & {\bf Section} 	& {\bf Pros} & {\bf Cons} \\
\endfirsthead

\multicolumn{3}{c}%
{{\bfseries \tablename\ \thetable{} -- continued from previous page}} 
\\ \hline \hline
 \\{\bf Estimator} & {\bf Type} & {\bf Developed by} & {\bf Section} 	& {\bf Pros} & {\bf Cons} \\
 \\ \hline \\
\endhead

\\ \hline \\ \\ \multicolumn{3}{l}{{Continued on next page}}
\endfoot

\hline \hline
\endlastfoot
\\
\hline
\\
{\bf \underline{Maximum Likelihood (MLE)}}	& Parametric 			&\citet{sandage:1979} 	&\S~\ref{sec:MLE}, Page~\pageref{sec:MLE}	   &No binning required		&$\begin{cases}\mbox{No built-in goodness of fit}  \\ 
																														 					  \mbox{No built-in normalisation} \end{cases}$ \\ 
{\bf Extensions to MLE:}
\\					
Provide normalisation		       				& 	&\cite{Davis:1982}		  						&\S~\ref{sec:mle_norm}, Page~\pageref{sec:mle_norm}				&&\\ 
${\mathcal{L}}$ modelled by Poisson probabilities    & 	&\citet{Marshall:1983} [\citetalias{Marshall:1983}]		&\S~\ref{sec:mle_ext}, Page~\pageref{sec:mle_ext}					&&\\		
Provides error estimates		      				& 	&\cite{Efstathiou:1988}		  					&\S~\ref{sec:mle_norm}, Page~\pageref{sec:mle_norm}				&&\\ 							
Combining multiple data-sets		       			& 	&\cite{Heyl:1997}		  						&\S~\ref{sec:mle_ext}, Page~\pageref{sec:mle_ext}					&&\\ 
Galaxy clusters							     	 &         &\cite{Andreon:2005MNRAS.360..727A} 			&\S~\ref{sec:mle_ext}, Page~\pageref{sec:mle_ext}					&&\\
\mbox{\quad\quad \bf Bivariate LF:} &&&&& \\
\mbox{\quad\quad  Galaxy radius} 			      	 &          &\cite{Choloniewski1985MNRAS.214..197C} 		& \S~\ref{sec:blf_radius}, Page~\pageref{sec:blf_radius} &&\\
\mbox{\quad\quad  Colour (utilises $1/V_{\max}$)}  			&          &\cite{Chapman2003ApJ...588..186C} 		& \S~\ref{sec:blf_colour}, Page~\pageref{sec:blf_colour} &&\\
\mbox{\quad\quad \bf Photo z:} &&&&& \\
\mbox{\quad\quad  Gaussian kernel}  					&          &\cite{Chen:2003ApJ...586..745C} 		& \S~\ref{sec:photoz_mle}, Page~\pageref{sec:photoz_mle} &&\\ 
\mbox{\quad\quad  Recast as conditional probability}  		&          &\cite{Sheth:2007} & \S~\ref{sec:photoz_mle}, Page~\pageref{sec:photoz_mle} &&\\ 
\mbox{\quad\quad  Adopts \citetalias{Marshall:1983} method}  &          &\cite{Christlein2009MNRAS.400..429C} 	& \S~\ref{sec:photoz_mle}, Page~\pageref{sec:photoz_mle} &&\\ \\ \\

{\bf \underline{Classical Method ($\Phi=N/V$)}}	 &Non-Parametric 		&  $\begin{cases}\mbox{\cite{Christensen:1975}} \\ \mbox{\cite{Schechter:1976}} \\ \mbox{\cite{Felten:1977}}\end{cases}$ 	&\S~\ref{sec:classical}, Page~\pageref{sec:classical}				&No assumption of form of LF	&$\begin{cases}\mbox{Biased by clustering} \\ \mbox{Returns bias LF estimate}\end{cases}$ \\ \\ \\

{\bf \underline{${\bf 1/V_{\max}}$ \& ${\bf V/V_{\max}}$}}	&Non-Parametric			& \cite{Schmidt:1968}	&\S~\ref{sec:vmax}, Page~\pageref{sec:vmax} 			&$\begin{cases}\mbox{No assumption of form of LF} \\ \mbox{Easy to apply} \\ \mbox{Many improvements made}	\\ \mbox{Provides test of evolution} \\ 	\mbox{Provides test of completeness} \end{cases}$		&Biased by clustering \\ 

{\bf Extensions to ${\bf V_{\max}}$:}
\\	
Combining multiple data-sets		& 		&\cite{avni:1980}   		&\S~\ref{sec:vmax_ext} &&\\
Diameter limited surveys				& 		&\cite{Hudson:1991}   	&\S~\ref{sec:vmax_ext} &&\\
Reduce clustering and shot noise bias	& 		&\cite{Eales:1993}   		&\S~\ref{sec:vmax_ext} &&\\
$n/n_{\max}$ \& $o/o_{\max}$ variants	& 		&\cite{Qin:1997,Qin:1999}   &\S~\ref{sec:vmax_ext} &&\\
Reduce bias toward flux limit			& 		&\citet*{Page:2000}   		&\S~\ref{sec:vmax_ext} &&\\
Fit  model to reduce bin bias			& 		&\cite{Miyaji:2001AA...369...49M}   		&\S~\ref{sec:vmax_ext} &&\\
Tracing LRG evolution across z-slices			& 		&\cite{Tojeiro2010MNRAS.405.2534T}   		&\S~\ref{sec:vmax_ext} &&\\

\mbox{\quad\quad \bf Bivariate LF:} &&&&& \\
\mbox{\quad\quad  Colour (utilises MLE)} 				& 		&\cite{Chapman2003ApJ...588..186C}   		&\S~\ref{sec:blf_colour}, Page~\pageref{sec:blf_colour} &&\\
\mbox{\quad\quad \bf Photo z:} &&&&& \\
\mbox{\quad\quad  Deconvolution technique}				& 		&\cite{Sheth:2007}   						&\S~\ref{sec:photoz_sheth_vmax}, Page~\pageref{sec:photoz_sheth_vmax} &&\\ \\ \\

{\bf \underline{The ${\bf \rm C^-}$ method}}	& Non-Parametric			& \cite{lynden:1971}	&\S~\ref{sec:cmethod}, Page~\pageref{sec:cmethod} 			&$\begin{cases}\mbox{No parametrisation of the LF required} \\ \mbox{recovers the CLF} \\ \mbox{Independent of clustering effects} \end{cases}$		&No built-in normalisation\\

{\bf Extensions to the $\rm C^-$ method:}\\
Combining multiple data-sets		& 		&\cite{Jackson:1974}   			&\S~\ref{sec:cminus_ext}, Page~\pageref{sec:cminus_ext}&&\\
\mbox{\quad \& derives error estimates}\\
Simplified \& includes normalisation		& 		&\cite{Choloniewski:1987} [\citetalias{Choloniewski:1987}]  		&\S~\ref{sec:cminus_ext}, Page~\pageref{sec:cminus_ext} &Provides normalisation&\\
Smoothing kernel					& 		&\cite{Caditz:1993} [\citetalias{Caditz:1993}]  					&\S~\ref{sec:cminus_ext}, Page~\pageref{sec:cminus_ext} &&\\
Combines methods of \citetalias{Choloniewski:1987} \& \citetalias{Caditz:1993} & 		&\cite{Takeuchi:2000}   		&\S~\ref{sec:takeu_comp}, Page~\pageref{sec:takeu_comp} &&\\ 
$C^+$ variant						& 		&\cite{Zucca:1997AA...326..477Z}   								&\S~\ref{sec:cminus_ext}, Page~\pageref{sec:cminus_ext} &&\\

\mbox{\quad\quad \bf Photo z:} &&&&& \\
\mbox{\quad\quad  Gaussian error distribution}		& 		&\cite{Subbarao1996AJ....112..929S}   	&\S~\ref{sec:cminus_ext}, Page~\pageref{sec:cminus_ext} &&\\ \\ \\

{\bf \underline{The ${\bf \phi/\Phi}$ method}}	& Non-Parametric			& $\begin{cases}\mbox{\cite{Turner:1979}}, \\ \mbox{\cite{Kirshner:1979}} \end{cases}$ 	&\S~\ref{sec:phioverphi}, Page~\pageref{sec:phioverphi} 			&Independent of clustering effects 		&$\begin{cases}\mbox{Biased estimator} \\ \mbox{Correlated errors}\end{cases}$\\ 
{\bf Extensions to ${\bf \phi/\Phi}$:}\\
Adopting parametric form of the LF					& 		&e.g. \cite{Lapparent:1989}   		&\S~\ref{sec:phioverphi}, Page~\pageref{sec:phioverphi} &&\\ \\ \\

{\bf \underline{The Step-Wise Max Likelihood}}	& Non-Parametric 			&$\begin{cases}\mbox{\cite{Nicoll:1983}}, \\ \mbox{\cite{Choloniewski:1986}} \\ \mbox{\cite{Efstathiou:1988}}\end{cases}$ &\S~\ref{sec:swml}, Page~\pageref{sec:swml} 			&$\begin{cases}\mbox{No assumption of form of LF} \\ \mbox{robust estimator} \\ \mbox{Independent of clustering effects} \end{cases}$		&$\begin{cases}\mbox{No built-in normalisation}\\ 
\mbox{selection f$^{\rm n}$ bias}\end{cases}$ \\
{\bf Extensions to the SWML:}\\
Combining multiple data-sets		& 		&\cite{Heyl:1997}   	&\S~\ref{sec:swml}, Page~\pageref{sec:swml} &&\\
Model S(z) as piecewise power laws & 		&\cite{Springel:1998}   	&\S~\ref{sec:swml}, Page~\pageref{sec:swml} &Reduce selection f$^{\rm n}$ bias&\\
Interpolating through steps & 		&\cite{1997ApJ...477...36K}   	&\S~\ref{sec:swml}, Page~\pageref{sec:swml} &Reduce selection f$^{\rm n}$ bias &\\
\mbox{\quad\quad \bf Bivariate LF:} &&&&& \\
\mbox{\quad\quad  Galaxy diameters} 				& 		&\citet*{Sodre:1993}   		&\S~\ref{sec:swml_biv}, Page~\pageref{sec:swml_biv} &&\\
\mbox{\quad\quad  Surface brightness} 				& 		&\citet{Driver:2005}   		&\S~\ref{sec:biv_surfdens}, Page~\pageref{sec:biv_surfdens} &&\\\\
%\mbox{\quad\quad\quad  Space density} 	\\

\end{longtable} 
 \end{center}

}
\end{landscape}
			%section 7
\section{Method comparison review}\label{sec:comp_review}
Thus far all the traditional LF estimators have been summarised along with their most relevant modifications and extensions.  This aim of this section is to highlight  some of the key results from  papers by \cite{Heyl:1997} \citepalias{Heyl:1997}, \cite{Willmer:1997}   \citepalias{Willmer:1997},  \citet*{Takeuchi:2000} \citepalias{Takeuchi:2000} and \cite{Ilbert:2004} \citepalias{Ilbert:2004}, which have provided detailed  comparisons and, therefore, useful insights of selected LF  estimators.  
Such comparisons are  vital and enable us to gain a practical sense of the limitations and any intrinsic bias of these techniques in a more controlled environment.
\subsection{\cite{Heyl:1997}}
 \begin{figure*}\label{fig:heyl97}
    \begin{center}
      \includegraphics[width=0.49\textwidth]{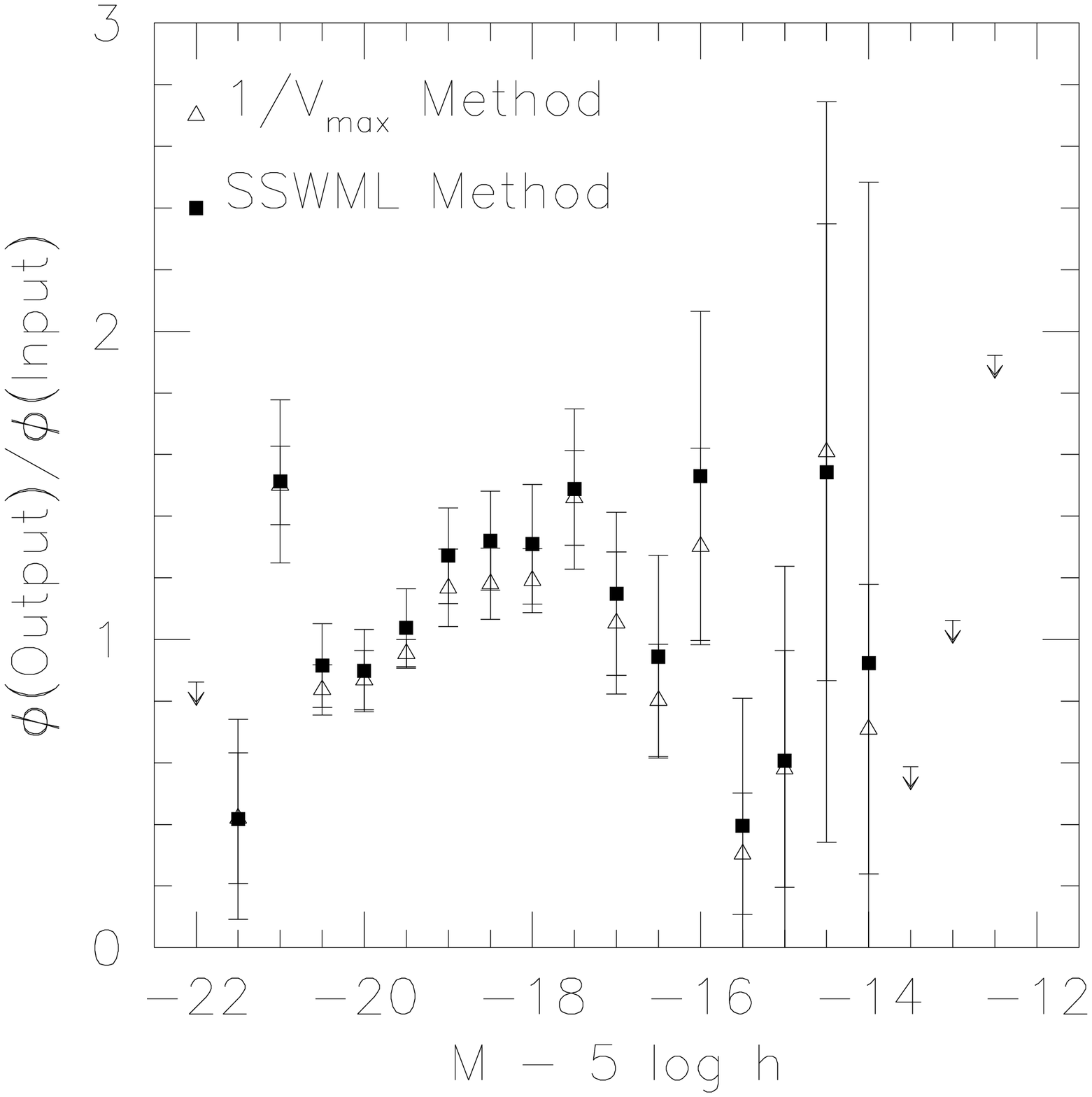} \hfill
       \includegraphics[width=0.49\textwidth]{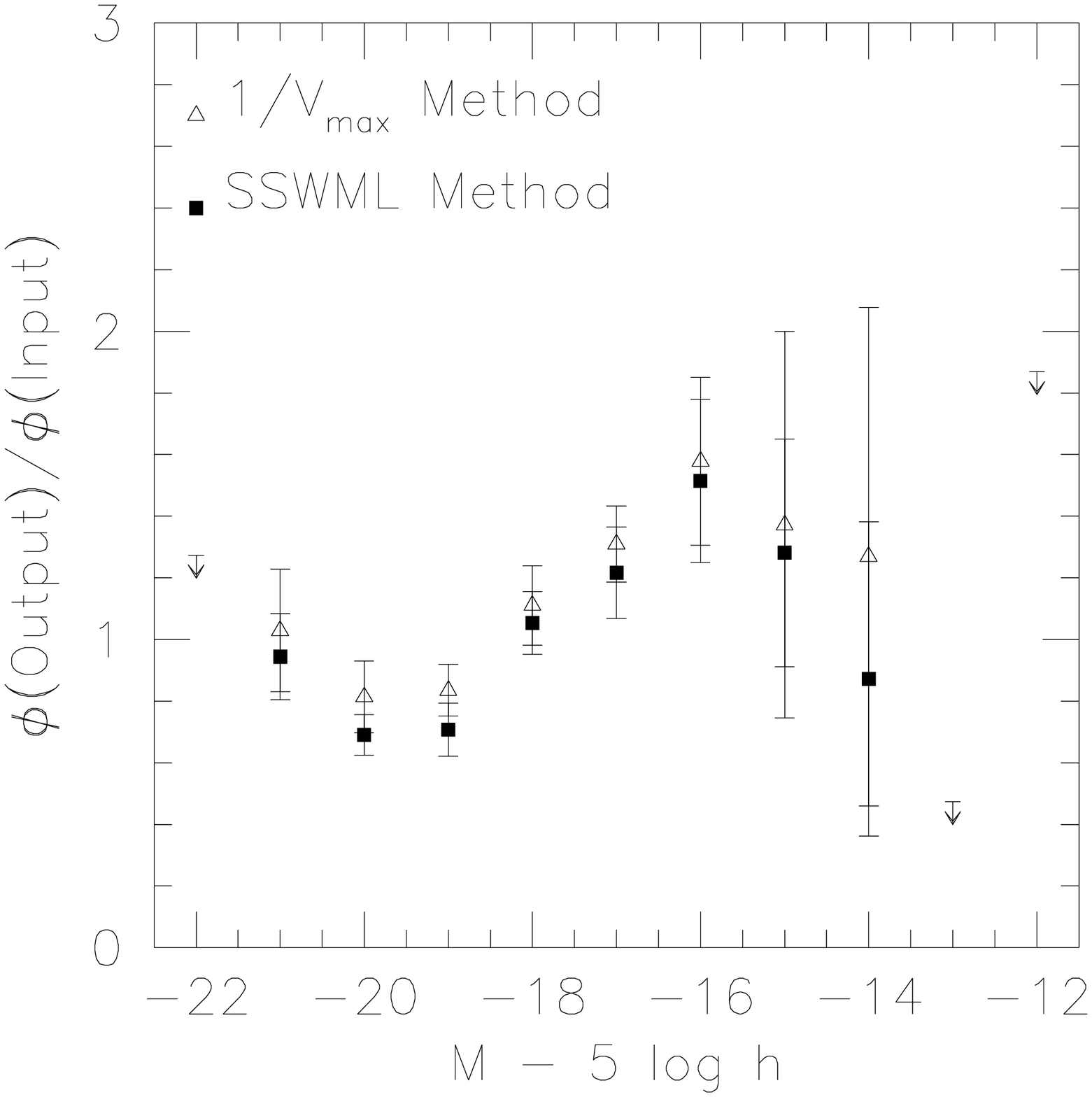} \\
        \includegraphics[width=0.49\textwidth]{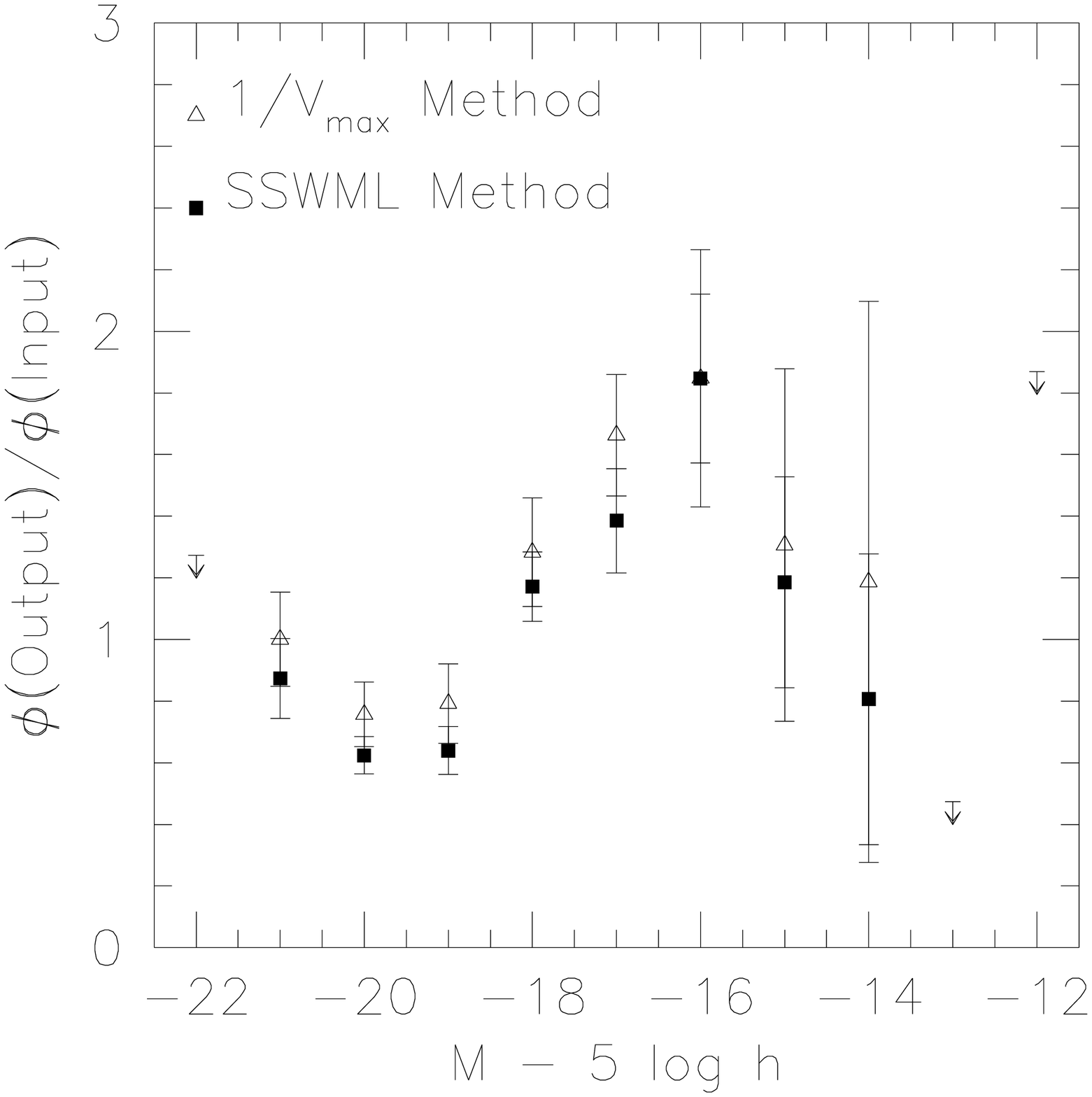} \hfill
         \includegraphics[width=0.49\textwidth]{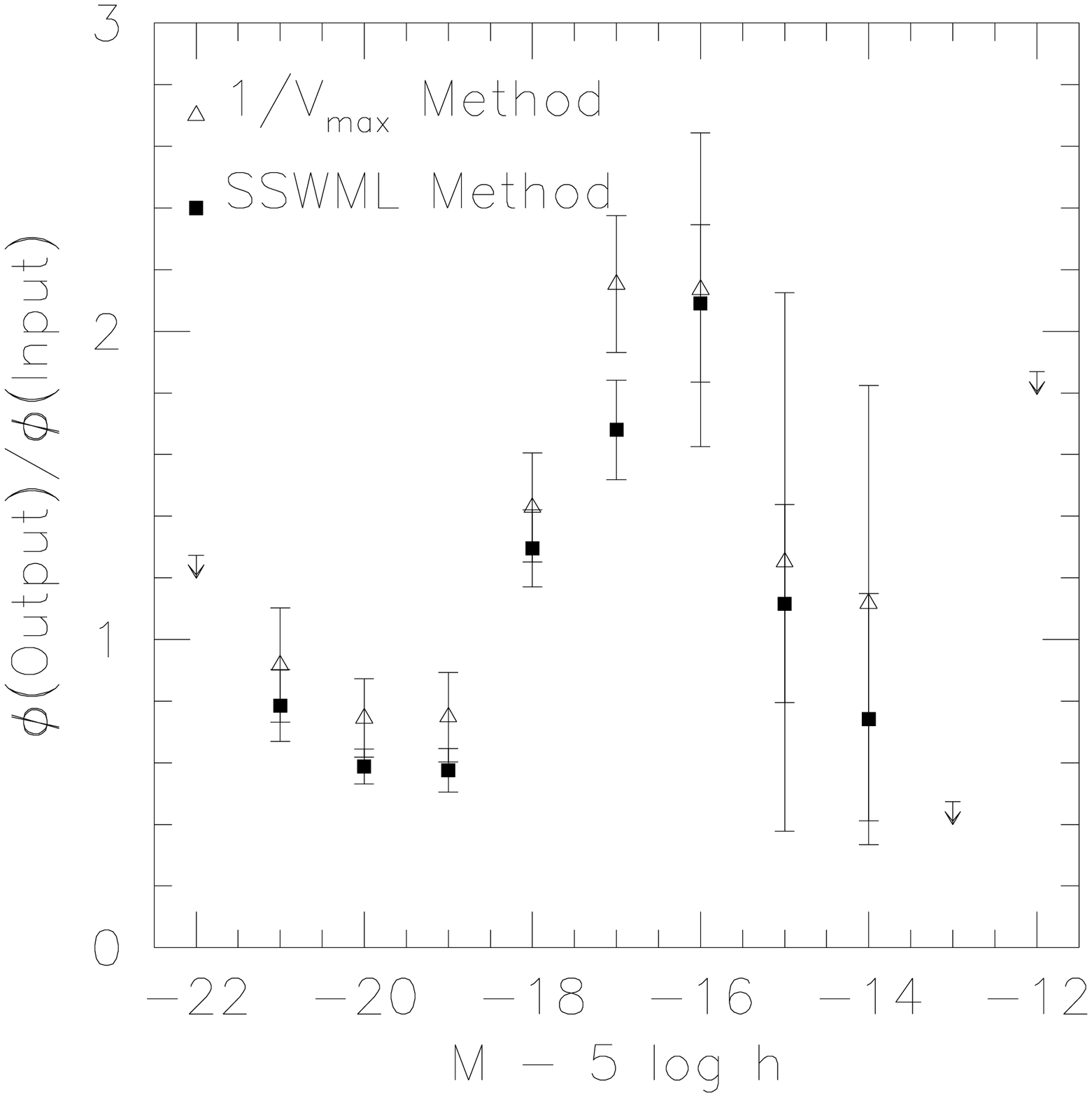}      
      \caption{\small Plots courtesy of \cite{Heyl:1997}.  Selected results from \citetalias{Heyl:1997} (Case II in our text)  where increasing levels of clustering in  the MC samples are shown.  The top-left panel has no clustering, the remaining panels are the result of 45\% (top right), 65\% (bottom left) and 85\% (bottom right) clustering.}        
      \label{fig:heyl97}
    \end{center}
  \end{figure*}
As  already discussed in the previous section, \citetalias{Heyl:1997} extended the \citetalias{Efstathiou:1988} SWML (which was referred to as the SSWML to make the distinction) which could account for multiple samples with varying magnitude limits.   The paper was also concerned with exploring the role of evolution in galaxy  surveys. This discussion is focussed on the first half of  this paper where they compare the SSWML with the $1/V_{\max}$ estimator using simulated data.  The  Monte Carlo (MC) samples drew magnitudes from a Schechter function adopting a  standard cosmology. For simplicity,   $K$-corrections were  not considered.  With this basic set up, the following  scenarios were explored.
\subsubsection{Case I - Density evolution:} In the first scenario they constructed a sample of 1800 galaxies, binned equally over 6 apparent magnitude ranges with an overall magnitude range between $11.0<m<24.0$~mag.  To give a simplistic model of density evolution they doubled the density of galaxies beyond $z=0.2$.  

Their  findings  in this case showed both estimators were very consistent with each other, with the exception of the SSWML returning smoother results.  It was concluded that  neither  indicated any intrinsic  bias with respect to the input LFs.
\subsubsection{Case II - Clustering:}  Potential intrinsic biasing due to clustering (particularly in the $1/V_{\max}$ method) was then explored using three MC samples with over densities built in at $z=0.05$ and with each sample having clustered fractions of 45\%, 65\% and 85\%.   By exploring various bin width sizes they found that for very narrow bin widths both estimators were biased at the faint-end slope of the LF, over predicting the density. However, as  the bin widths increased $1/V_{\max}$ remained unchanged but the SSWML method showed more robustness to the clustering showing a systematic decrease in bias.  For  the three catalogues both estimators did show an increase in steepness of the LF with the SSWML being less affected (see Figure~\ref{fig:heyl97}). It was therefore concluded that the SSWML was a superior estimator to that of $V_{\max}$ for surveys with strong clustering.
\subsection{\cite{Willmer:1997}}
In this paper,  the author compares  LF parameter values derived from the \cite{Choloniewski:1987} version of   Lynden-Bell's $C^-$ method, the MLE of  \citetalias{sandage:1979}, the \cite{Turner:1979}  $\phi/\Phi$ method, the SWML estimators of \cite{Choloniewski:1986}  and \citetalias{Efstathiou:1988}, and finally  Schmidt's $1/V_{\max}$ estimator.  All were applied in two different cases: the first using MC simulations (based on the CfA survey) and the second to the CfA-1 \citep{Huchra:1983} survey data.
\subsubsection{Case 1 - Monte Carlo (MC) simulations}  In this case a  total of  1000 MCs were drawn from a Schechter function \citep{Schechter:1976} using a homogeneous redshift distribution and inputting three different values of $\alpha$ (i.e. $\alpha_{in}=-0.7, -1.1$ and -1.5 ) in order to probe sensitivity of the LF estimators to the inclination of the faint-end slope. Finally, the MCs were designed to have a similar survey geometry as the CfA-1 redshift survey.

It was demonstrated that all  of the estimators under comparison recovered the $\alpha_{in}$ values extremely well with the exception of $1/V_{\max}$ (see Table~2).  Overall the \citetalias{sandage:1979} and $C^-$ methods were shown to be  the most robust providing the best results.  They show that despite having a homogeneous sample, $1/V_{\max}$ indicated bias, giving higher values  for the faint-end slope compared to the other estimators under test. However,  as we shall see in the following section, \citetalias{Takeuchi:2000}  applied the modified  \cite{Eales:1993} $V_{\max}$  variant to a homogeneous sample and demonstrated its consistency with other estimators.  In \cite{Willmer:2006ApJ...647..853W} they apply this variant to the Deep Extragalactic Evolutionary  Probe 2 (DEEP2) survey data to avoid potential bias.

\subsubsection{Case II - The CfA-1 redshift survey} The estimators were then tested against the actual survey data from CfA-1 (see Table~3).  The results indicated that all the estimators gave consistent results with $\alpha=-1.2$ and $M^*=-19.2$~mag with the exception once again of the $V_{\max}$ .  In this case they noted that the density measured was lower than \citetalias{sandage:1979} by a factor of 2.
\begin{deluxetable}{lcccc}
\label{table1}
\small
\tablewidth{0pc}
\tablecaption{Extract from \cite{Willmer:1997} reproduced by permission of the AAS. Median values for recovered parameters $M^*$ and
$\alpha$, derived from 1000 CfA-1 like Monte Carlo simulations.}
\tablehead{
%\colhead{Method}    & 
%\colhead{$\alpha$}  &
\colhead{M$^*$ }    &
\colhead{$\alpha$}  &
\colhead{M$^*$ }    &
\colhead{$\alpha$}  &
\colhead{M$^*$ }    
}
% data for -1.10 -19.10 sim.stat.97.03.30
% data for -1.50 -19.19 sim.stat.97.01.
\startdata 
{\bf Input Values }                             &{\bf -1.50  }                 &  {\bf -19.10}                        &  {\bf-0.70}                  & {\bf-19.10} \nl
SWML                                       & -1.45 $\pm$ 0.14  &  -19.13 $\pm$ 0.11  &   -0.82 $\pm$    0.14  &  -19.19 $\pm$    0.12 \nl
STY                                           & -1.43 $\pm$ 0.06  &  -19.06 $\pm$ 0.07  &   -0.69 $\pm$    0.08  &  -19.10 $\pm$    0.06 \nl
Choloniewski                          & -1.50 $\pm$ 0.08  &  -19.00 $\pm$ 0.11  &   -0.87 $\pm$    0.15  &  -19.08 $\pm$    0.12 \nl
Turner                                       & -1.51 $\pm$ 0.09  &  -19.16 $\pm$ 0.10  &   -0.66 $\pm$    0.11  &  -19.11 $\pm$    0.09 \nl
$1/V_{\max}$$_{\Delta z}$   & -1.67 $\pm$ 0.05  &  -19.14 $\pm$ 0.09  &   -0.94 $\pm$    0.08  &  -19.14 $\pm$    0.07 \nl
$1/V_{\max}$$_{\Delta M}$  & -1.50 $\pm$ 0.06  &  -18.98 $\pm$ 0.07  &   -0.83 $\pm$    0.11  &  -19.04 $\pm$    0.09 \nl
C$^-$                                        & -1.51 $\pm$ 0.07  &  -19.11 $\pm$ 0.08  &   -0.72 $\pm$    0.12  &  -19.10 $\pm$    0.07 \nl
\enddata
\end{deluxetable}

\begin{deluxetable}{llll}
%                   123456789
\small
\tablewidth{0pc}
\tablecaption{Extract from \cite{Willmer:1997} reproduced by permission of the AAS. Schechter function parameters for CfA1 survey, using same limits for all methods}
\tablehead{
\colhead{Method}              & \colhead{$\alpha$}        &
\colhead{M$^*$ }              & \colhead{Notes } \nl
{} & {} & {} & { }
}

\startdata
SWML          & -1.20 $\pm$ 0.03  & -19.30 $\pm$ 0.04 & $\Delta M$ = 0.25\nl
STY           & -1.11 $\pm$ 0.08  & -19.17 $\pm$ 0.08 & { } \nl
Choloniewski  & -1.18 $\pm$ 0.05  & -19.26 $\pm$ 0.07 & $\Delta M$ = 0.25 \nl				       
Turner        & -1.11 $\pm$ 0.06  & -19.32 $\pm$ 0.05 & $\Delta z$ = 500 ${\rm km  \ s^{-1}}$ \nl
1/V$_{\max}$   & -1.59 $\pm$ 0.04  & -19.43 $\pm$ 0.07 & $\Delta M$ = 0.25\nl     			       
1/V$_{\max}$   & -1.70 $\pm$ 0.05  & -19.55 $\pm$ 0.05 & $\Delta z$ = 500 ${\rm km \  s^{-1}}$ \nl				       
C$^-$         & -1.20 $\pm$ 0.01  & -19.21 $\pm$ 0.01 & { } \nl
%\tablecomments{Cols.}
\enddata
\end{deluxetable}

Overall, it was found that he \citetalias{sandage:1979} and the $C^-$ methods gave the best most robust results.  However, they did note that \citetalias{sandage:1979} fit underestimated the faint-end slope whereas the bias observed in $1/V_{\max}$ showed overestimating the faint-end slope.
Therefore, for scenarios where samples have a steeper slope, the $C^-$ method would be a preferred choice of estimator.
\subsection{\citet*{Takeuchi:2000} } \label{sec:takeu_comp}
Undoubtedly this article provides the most rigorous comparisons to date.  Four estimators were put under test.  This first was the \cite{Eales:1993} variation of the $1/V_{\max}$ estimator which traces evolution with redshift. The second method was a variation on  the $C^-$ estimator which they refer to as  the {\it Lynden-Bell--Choloniewski--Catditz--Petrosian (LCCP)} method, which incorporates  the \cite{Choloniewski:1987} extension and the smoothing kernel introduced by \cite{Caditz:1993}  (see Equation~\ref{equ:CP93}).    The third and forth methods were the \cite{Choloniewski:1987} and the \citetalias{Efstathiou:1988} respective variations of the SWML estimators. They consider the following three distinct scenarios.

\subsubsection{Case I - Simulated luminosity functions}The four LF estimators were applied to three different types of  mock catalogue using the following  LFs: 
\begin{itemize}
\item Uniform distribution. 
\item Power-law that increases towards faint magnitudes,
\item Power-law that decreases towards fainter magnitudes,
\item  Gaussian distribution,
\item Schechter function with $\alpha=-1.1$, implying a flat faint-end slope,
\item Schechter function with $\alpha=-1.6$, implying a steep faint-end slope.
\end{itemize}
\noindent
The MC simulations were created with the following spatial density features:
\begin{itemize}
\item A homogeneous distribution.  This was largely to explore any bias that may have been inherent in the $1/V_{\max}$ as originally claimed by \citetalias{Willmer:1997}.  The sample was created neglecting the  $K$-correction,  with a redshift limit of  $z=0.1$ and an apparent magnitude limit $m_{\lim}=13.0$~mag.
\item  Two inhomogeneous samples.  In the first,  half the galaxies in the sample lie within a dense cluster at a distance of 0.8 Mpc with a radius of 0.8 Mpc.  The second mocks an overall under-density with a spherical void  with a radius of 1.6 Mpc at a distance of 0.8 Mpc. Both scenarios have the same overall density as the homogeneous sample.
\end{itemize}
 \begin{figure*}\label{fig:TYI1}
    \begin{center}
      \includegraphics[width=0.49\textwidth]{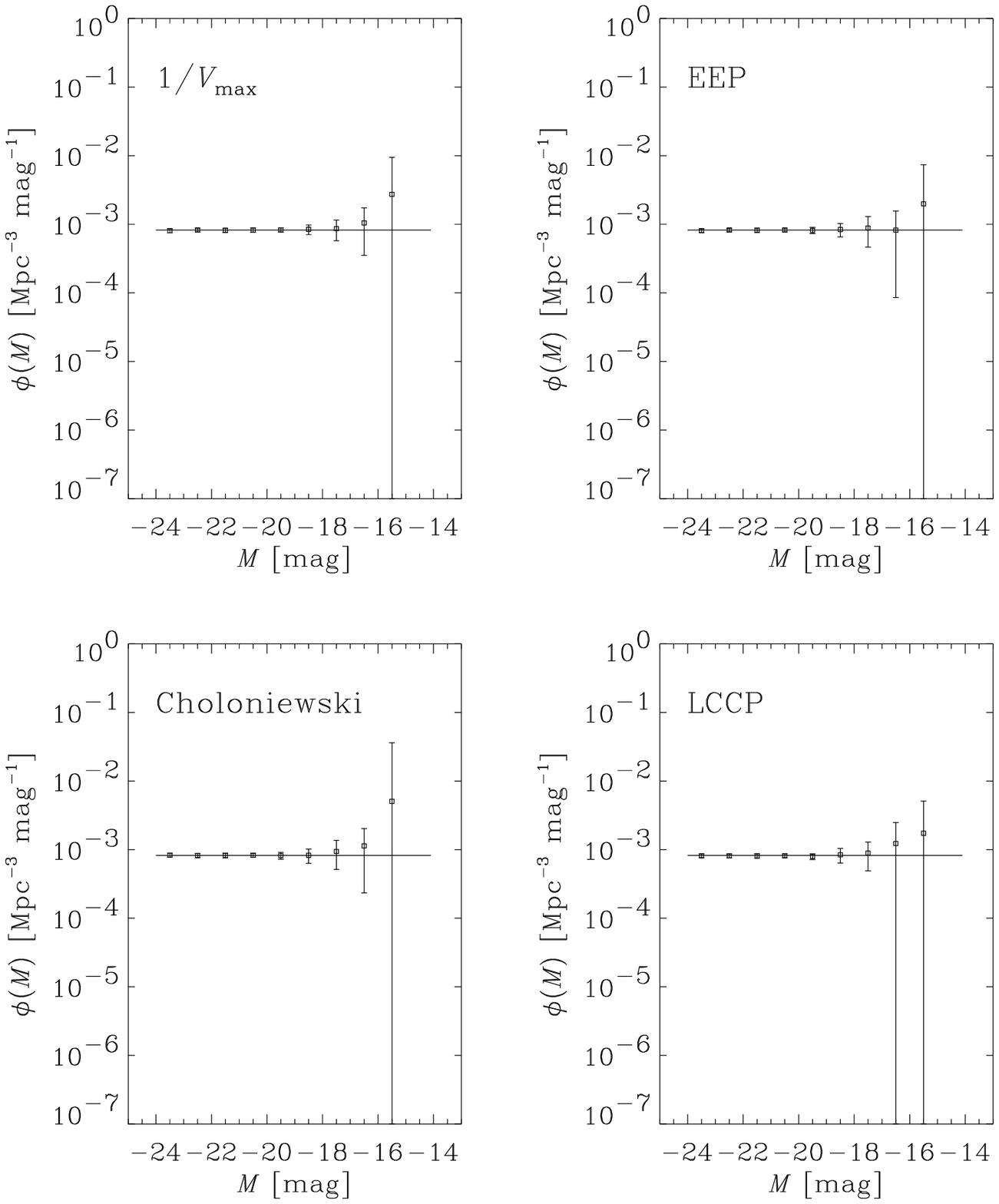} 
       \includegraphics[width=0.49\textwidth]{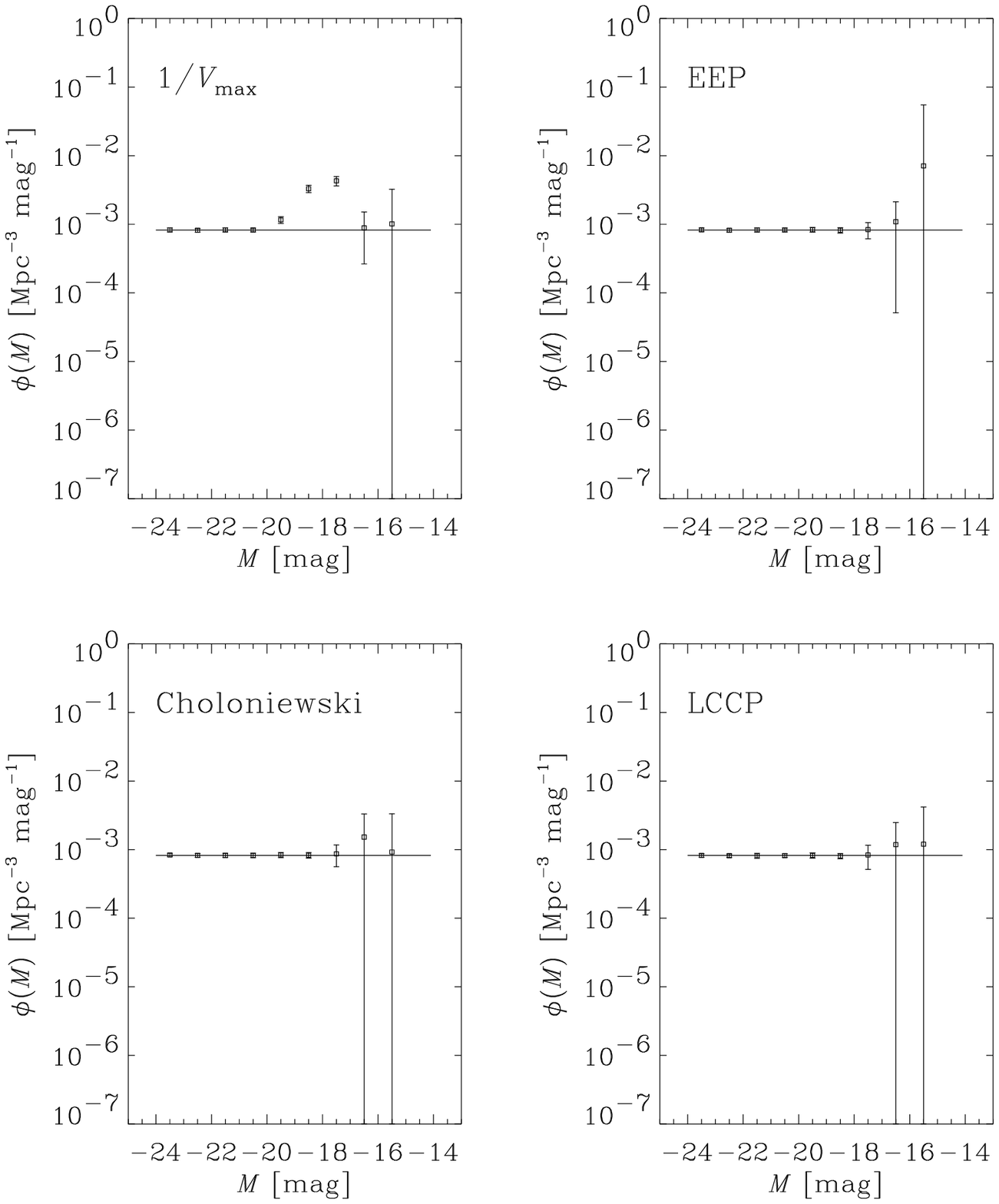}      
      \caption{\small Extract from \citet*{Takeuchi:2000} reproduced by permission of the AAS. An   Example from \citetalias{Takeuchi:2000} (Case I)  of the effects from applying the estimators on a homogenous MC sample (left) and a clustered MC sample (right). On the left hand plots we observe all four estimators are consistent with each other with no indication of intrinsic bias. However, the clustered sample on the right-hand panel shows how the expected bias of $1/V_{\max}$ affects determination of the faint end.  From this panel, there is no evidence of clustering intrinsically biasing the remaining estimators.  The input luminosity function in both scenarios for this example was a uniform distribution.}        
      \label{fig:TYI1}
    \end{center}
  \end{figure*}
 \begin{figure*}\label{fig:TYI2}
    \begin{center}
      \includegraphics[angle=90,width=1.\textwidth]{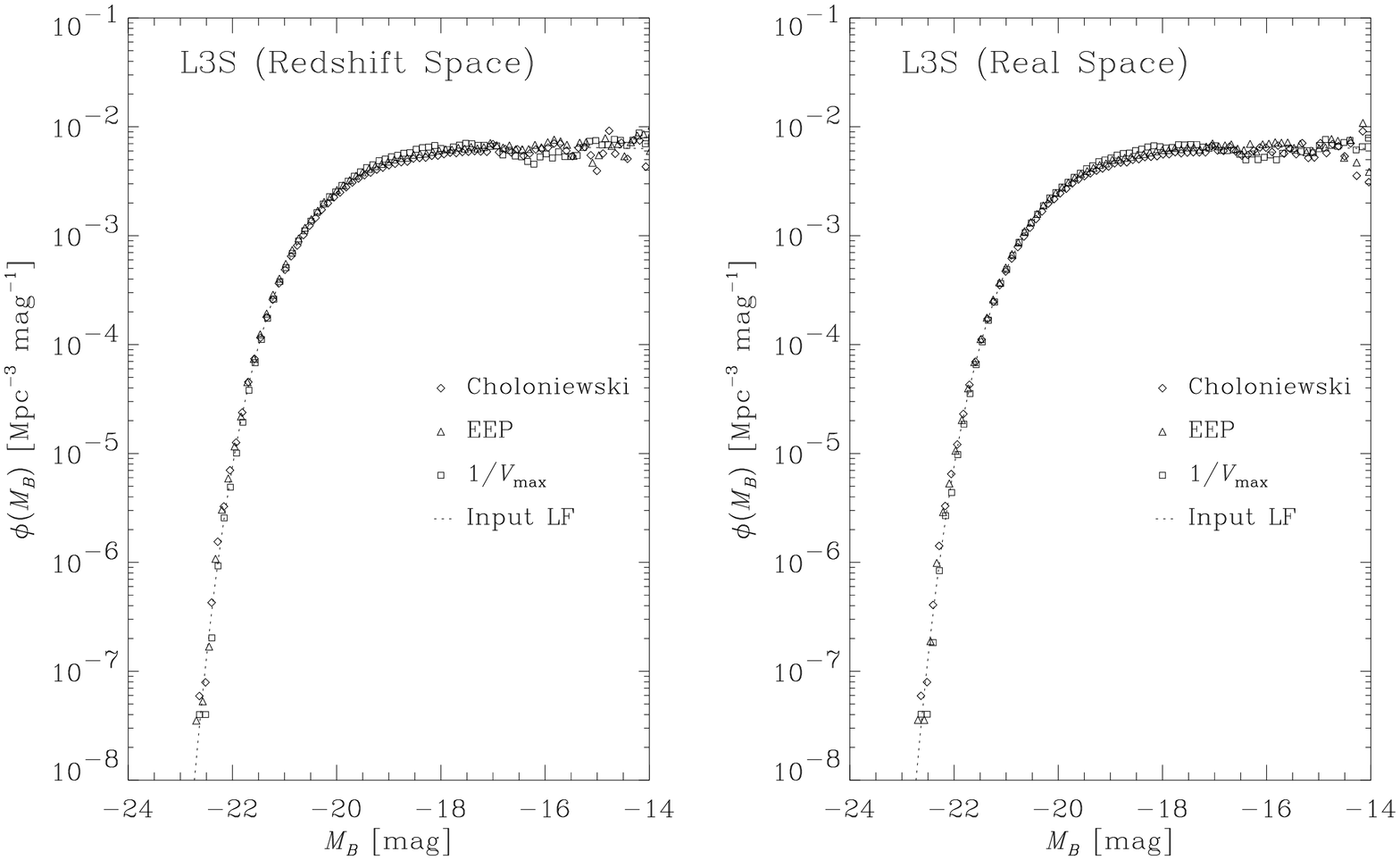}\\
      \vspace{5mm}
        \includegraphics[trim= 0cm 17cm 0cm 0cm, clip=true,width=1.\textwidth]{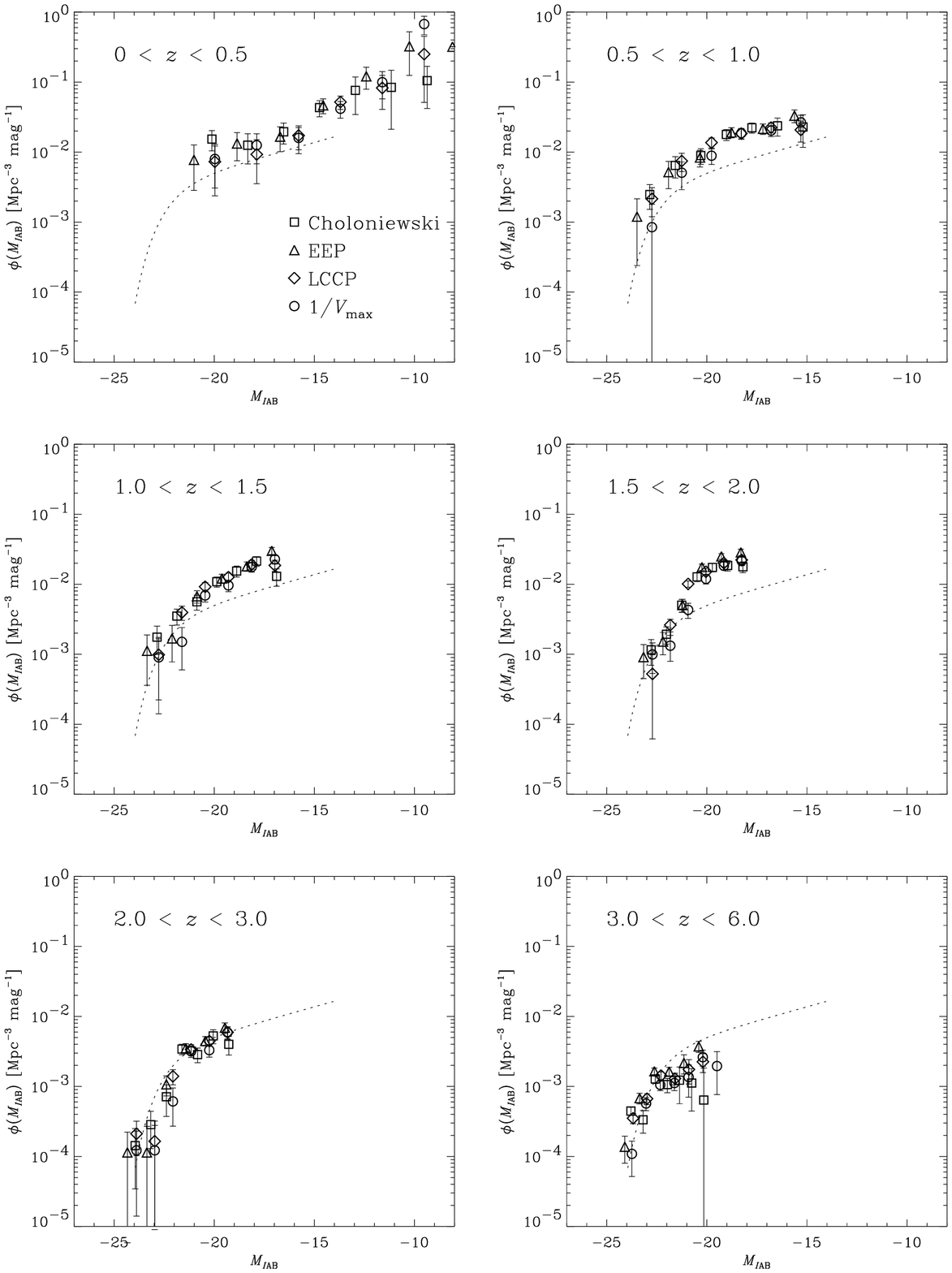}
      \caption{\small Extracts from  \citet*{Takeuchi:2000} reproduced by permission of the AAS.  Top panels:  results from Case II where 2dFGRS MC samples were tested.  They showed that considering the samples in redshift space (left) and real space (right) had little adverse effect on determining the LF.   Both panels show consistency between the estimators under test compared to the input LF.  Bottom panels: example of results from Case III where the estimators were applied the HDF data. In the context of this review it can be seen that in the redshift ranges considered in both panels, once again, all estimators seem to show overall consistency with each other.  }        
      \label{fig:TYI2}
    \end{center}
  \end{figure*}
An MC catalogue was then generated with sample sizes of 100 and 1000 to observe the effects of Poisson noise when one is sampling from small data-sets.
\\
\\
\noindent
The results from this part of the study revealed that in the homogeneous case all estimators gave results consistent with each other, thus contradicting earlier claims by \citetalias{Willmer:1997}, where they found a bias in their results when applying the {\it original} Schmidt $1/V_{\max}$.  

In contrast, the clustered MC sample showed that $1/V_{\max}$ was heavily biased producing an overestimation in the recovered LF  - as would be expected.  Moreover, the results showed the other three estimators were not adversely affected by this inhomogeneity and demonstrated good agreement with all the input LFs.  However, larger error bars were observed when applying the Choloniewski method which implied shot noise was beginning to dominate for the smaller of the sample sizes.  An excerpt of these results in Figure~\ref{fig:TYI1}.
\subsubsection{Case II - 2dFGRS MC catalogue} In this case they apply three of the four  LF estimators to a MC 2dFGRS sample prepared by \cite{Cole:1998MNRAS.300..945C} - it is unclear why the {\it LCCP} method was excluded from this part of the analysis.  In general they found that all estimators gave consistent results compared with the input LF (see the top panels in Figure~\ref{fig:TYI2}).  They did, however, report slight deviations in the $1/V_{\max}$ estimator which were considered to be due to the clustering in the sample.
\subsubsection{Case III - Hubble Deep Field (HDF)}  All four estimators were applied to the photometric redshift HDF  \citep{Fernandez-Soto:1999,Williams:1996}  to probe evolution of the LF shape.   The overall sample size under test consisted of 946 galaxies.
Whilst they provide extensive details of the  intrinsic evolutionary properties found in this study, overall they reported that all four 4 estimators gave consistent results with a distinct evolutionary trend in the LF with redshift.  An example of these results is shown in the bottom panels of Figure~\ref{fig:TYI2}.  It has been noted that rudimentary determination of $K$-corrections adopted by \citetalias{Takeuchi:2000}  in this analysis may have adversely affected  the overall accuracy   of  LF determination for all of the estimators applied in the  HDF results (Takeuchi, private communication). However, as we shall see in the following section,  \cite{Ilbert:2004}  provided a rigorous  study of the effect of K-correction bias on LF estimators.
\\
\\ \noindent
Finally, an interesting aside in this analysis compares the  algorithms for efficiency. They found that the Cho{\l}oniewski method was the fastest, whereas the \citetalias{Efstathiou:1988} method required more iterations and $V_{\rm max}$ required the maximum volume to be calculated for each galaxy which both  contributed to a slower computational time.
\subsection{\cite{Ilbert:2004} } \label{sec:ilbert_comp}
\begin{figure*}
    \begin{center}
      \includegraphics[width=0.8\textwidth]{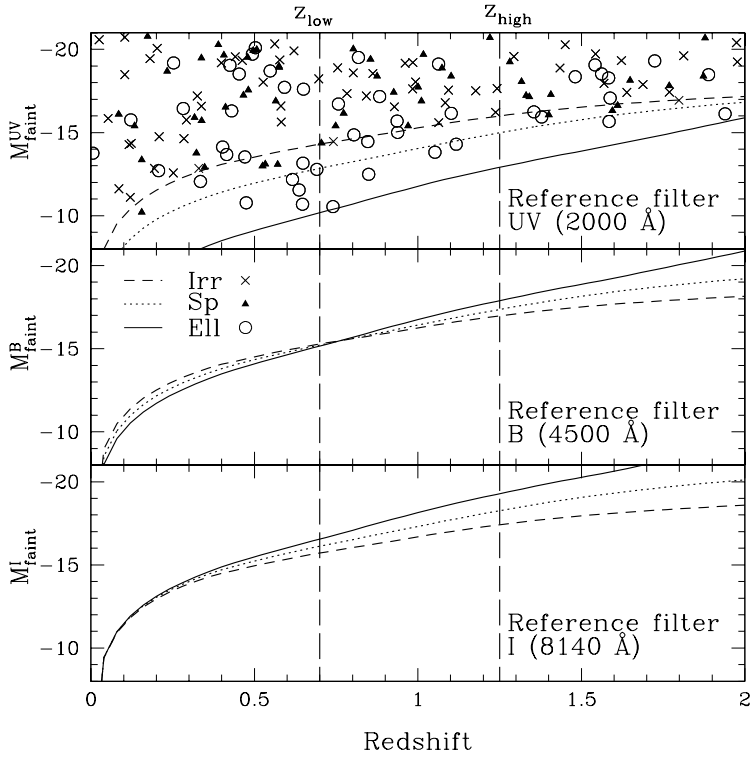}
      \caption{\small Extracts from  \cite{Ilbert:2004} reproduced by permission.  To illustrate how the bias creeps into the global LF consider the three cases in the above panels. In the top panel a galaxy distribution is shown consisting of irregulars (crosses), spirals (solid triangles) and ellipticals (open circles).   For each galaxy type SED,  the corresponding faint observable absolute magnitude limit $M_{\lim}^{\rm faint}$  is shown as a function of redshift.  The vertical dashed lines represent the redshift region $0.7\le z < 1.25$ within which the bias is studied. The section filter is $I$ with $I_{AB}\le 26$~mag and the reference filters are respectively shown from top to bottom as UV FOCA (2000 A), B HST (4500 A), I HST (8140 A).
      }  
      \label{fig:I04_1}
    \end{center}
  \end{figure*}

In contrast to the previous studies examined in this section, \citetalias{Ilbert:2004} explored a specific  intrinsic bias inherent in the estimation of the global LF.    The bias, which is more sensitive  toward the faint end of the LF, arises when the wavelength in the selection filter differs  significantly enough from the  wavelength of the reference filter. For a given data-set comprising different galaxy types having different SEDs will  consequently lead to different absolute magnitude limits within which a given population could  be observed at a given redshift for a fixed apparent magnitude limit. Figure~\ref{fig:I04_1} illustrates such a scenario. 

A relationship can be established that relates the  redshift $z$, the effective wavelength of the selection filter, $\lambda_s$, and the effective wavelength of the reference filter, $\lambda_{Ref}$ which allows us to probe the absolute magnitude limit, $M_{\lim}^{\rm faint}$,  for a given reference filter, as  a function of redshift for different galaxy types. In this example the galaxy types are irregulars, spirals and ellipticals which can be generally referred to ranging, respectively, from the `blue' populations to `red' populations.  The three panels in Figure~\ref{fig:I04_1} represent three different cases  exploring the relationships shown in Equation~\ref{equ:filter_ref_case}  within a given redshift range, $z_{low}\le z <z_{high}$, corresponding to   $0.7\le z < 1.25$. 
\begin{equation}\label{equ:filter_ref_case}
1+z_{low}
\begin{cases}
< \lambda_s/\lambda_{Ref}\\ \\
\sim \lambda_s/\lambda_{Ref}\\ \\
> \lambda_s/\lambda_{Ref}\\
\end{cases}
\end{equation}
Each case  corresponds to observing the three galaxy populations within different reference filters which can be summarised as follows.

\begin{figure*}
    \begin{center}
        \includegraphics[width=1.\textwidth]{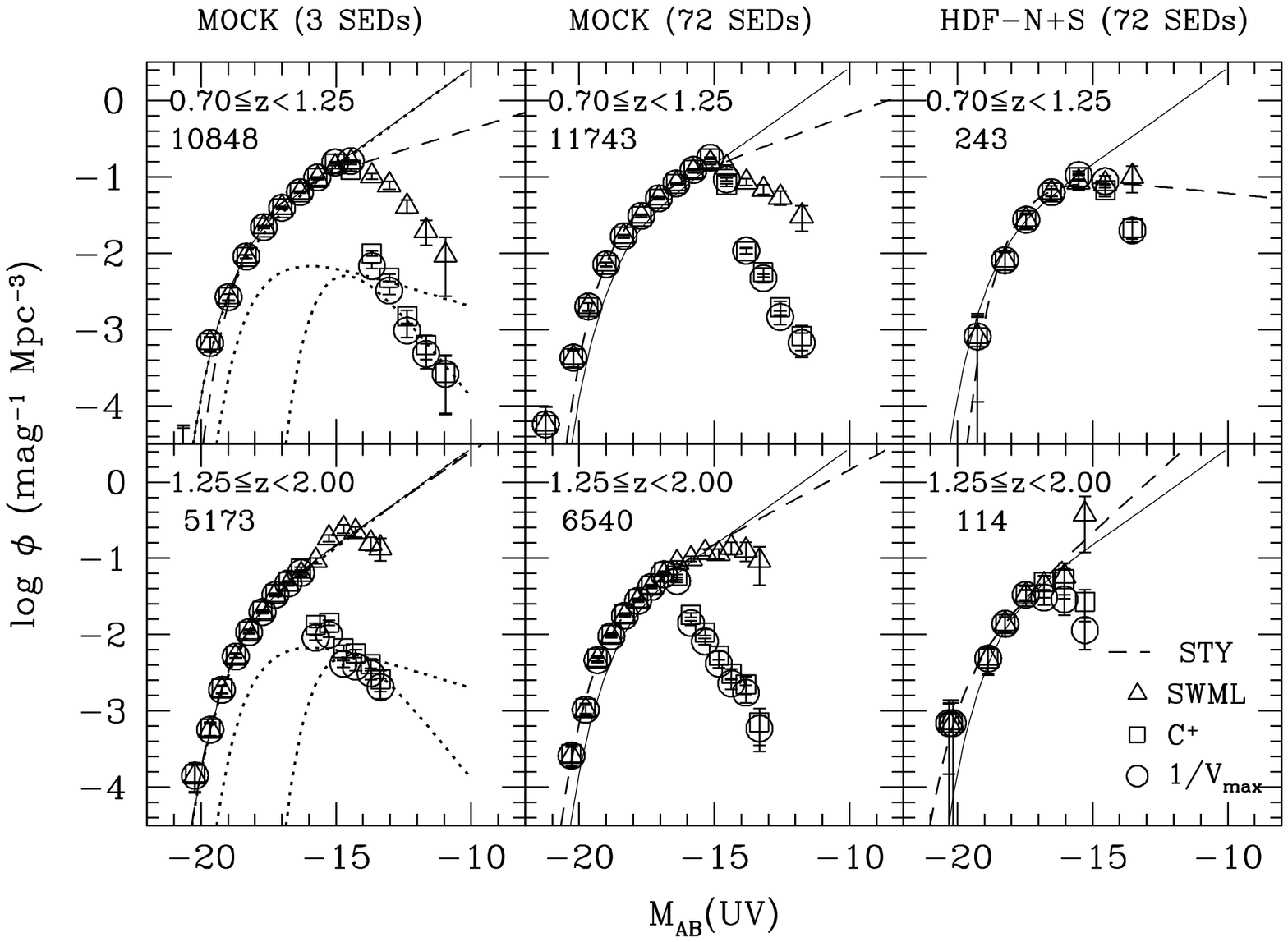}\\\vspace{5mm}
        \includegraphics[width=1.\textwidth]{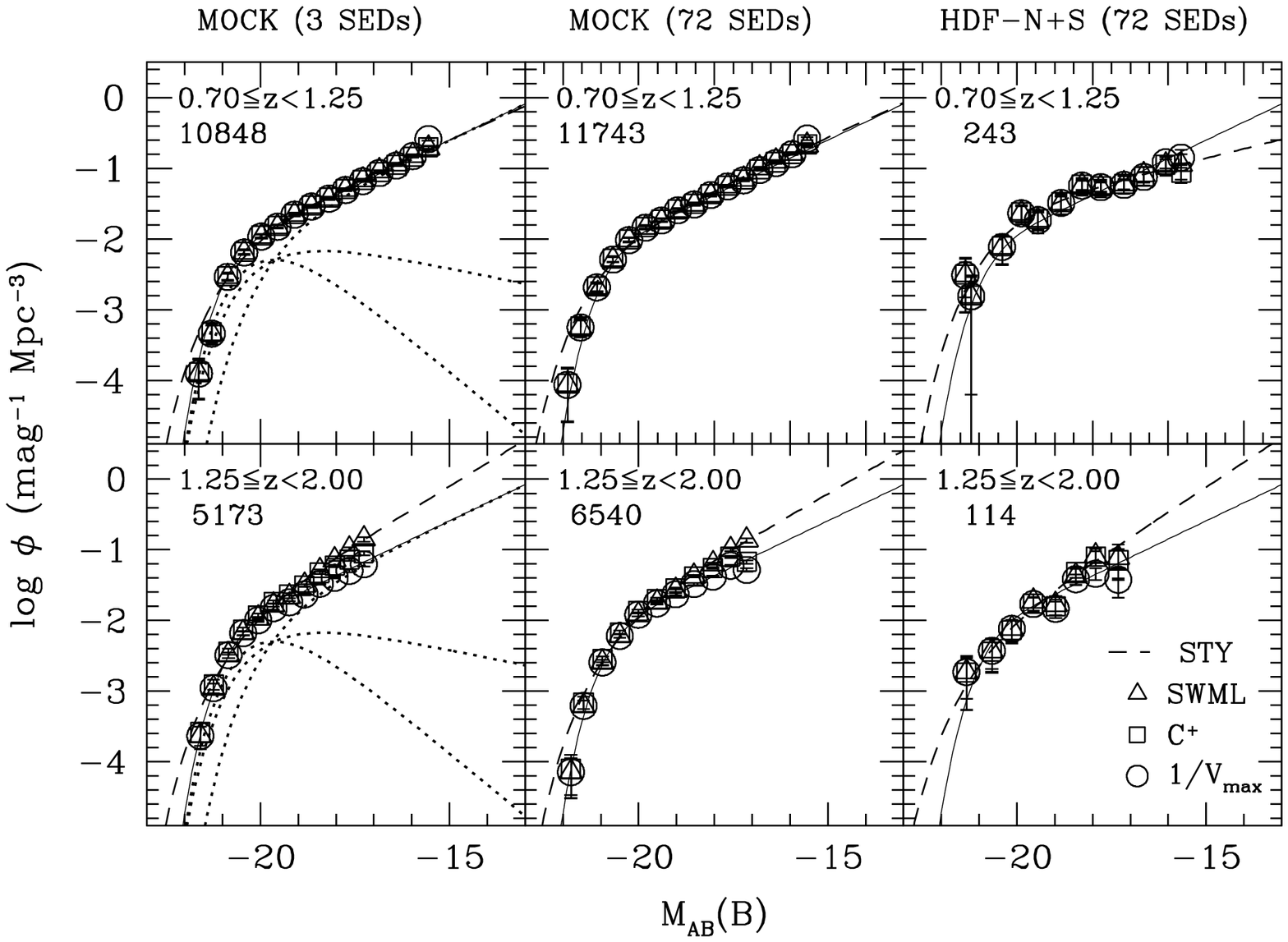}
              \caption{\small Extracts from  \cite{Ilbert:2004} reproduced by permission. Examples of how this form of observational bias affects the recovery of the global LF for different  estimators in two different reference  filters -UV (top panel set) and B-HST [4500A] (bottom panel set).   For each panel set the results for the simulated mock data are shown in the left and middle plots. The LFs from HDF-North and -South are shown in the right-hand plots of each set.  The resulting LF estimates for each estimator are indicated on each panel set.  The LFs corresponding to the three input SEDs used are shown as dotted lines on the left-hand plots of each panel set: from the steepest to the shallowest slope,  irregulars, spirals and ellipticals LF, respectively. The global simulated LF (i.e. the sum of the three input LFs)  is shown as a solid line.}
           \label{fig:I04_2}
    \end{center}
  \end{figure*}

\begin{enumerate}
\item $1+z_{low} < \lambda_{I_{AB}}/\lambda_{\rm UV_{\rm HST}}$:\\ \\
In the top panel of Figure~\ref{fig:I04_1} $M_{\lim}^{\rm faint}$ is defined for the UV {\it Ref} filter and is consequently brighter for blue galaxies than for red.   Thus, in the top panel of the figure the faint irregular SEDs becoming increasingly unobservable within the absolute magnitude range where the  other spiral and elliptical galaxy types can still be detected.  In this filter within the redshift range considered, both  irregular- and spiral-type  galaxies  are not observable beyond  absolute magnitudes fainter than  $M_{\lim}^{\rm faint}=-14.2$.  However, the elliptical galaxies  remain observable out to $M_{\lim}^{\rm faint}=-10.0$.  Therefore, one would expect the global estimation of the LF would become biased between $-10.0\lesssim M_{\lim}^{\rm faint} \lesssim 14.2$. \\

\item $1+z_{low} \sim \lambda_{I_{AB}}/\lambda_{\rm B_{\rm HST}}$: \\\\
In the middle panel of Figure~\ref{fig:I04_1} $M_{\lim}^{\rm faint}$  is now defined for the B  {\it Ref} filter and is  approximately the same for all galaxies for the given B-band reference filter within the redshift range considered.  In this case one would expect minimal bias effect as most galaxy populations would remain observable out to the faintest absolute magnitude limit.\\

\item $1+z_{low} > \lambda_{I_{AB}}/\lambda_{\rm I_{\rm HST}}$:\\\\
Finally, observing galaxies in the I {\it Ref} filter as show in the bottom panel of Figure~\ref{fig:I04_1}  implies that  $M_{\lim}^{\rm faint}$ is now {\it brighter}  for red galaxies than for blue and therefore faint red galaxies are missing from the sample.  Opposite to Case~1, elliptical galaxies are now unobservable within the absolute magnitude range where irregulars would still be detected. However, as can be seen, this bias should be less prevalent than  in Case~1 since the maximum absolute magnitude range this effects is only in order of $\sim1$~mag. Thus, where all SEDs have the same absolute magnitude limit, no bias should be present in the recovered LF. 
\end{enumerate}
\vspace{5mm}
\noindent
To explore the impact of this bias, \citetalias{Ilbert:2004} apply the $1/V_{\max}$, $C^+$, \citetalias{sandage:1979} and the \citetalias{Efstathiou:1988} estimators to both real and simulated data across these reference filters. For this review only results pertaining to the  UV and B filters are summarised which are,  respectively, shown as the top and bottom panel sets in Figure~\ref{fig:I04_2}.
The simulated mocks were drawn from the  multi-colour mock catalogues that are based on an empirical approach using observed LFs to derive redshift and apparent magnitude distributions [see \citetalias{Ilbert:2004} and \cite{Arnouts:1999MNRAS.310..540A} for more details].  In essence, they generate two sets of mock catalogues which have been classified into three main spectral classes: irregulars, spirals and ellipticals.  In the first set of mocks (corresponding to the left-hand panels in each set of  Figure~\ref{fig:I04_2}) they use only one SED per spectral class. This is to compare against the second set of more realistic mocks  where objects have been  drawn from an interpolated set of 72 SEDs (the middle panels in each set of  Figure~\ref{fig:I04_2}).  In this case one spectral class can correspond to multiple SEDs  and therefore the bias may be evident within a single spectral class.  

For the real data sample, they used the photometric catalogue and photometric redshifts  from the Hubble Deep Field (HDF) North and South surveys \citep{Arnouts:1999MNRAS.310..540A,Arnouts:2002MNRAS.329..355A}  (the right-hand panels in each set of  Figure~\ref{fig:I04_2}). For consistency, they have used the exactly the same set of 72 SEDs  as in the mock sample.

Before discussing their findings, the remainder of the setup of  Figure~\ref{fig:I04_2} is discussed first. The upper and lower plots in each panel set  are the recovered LFs in the respective redshift ranges $0.70\leq z<1.25$ and $1.25\leq z<2.00$.  The first of these ranges corresponds to the range shown in Figure~\ref{fig:I04_1}.  The left-hand plots show the input LFs for each spectral type in the mock samples, indicated by the dotted lines. From the steepest to the shallowest slope correspond to irregulars, spirals and ellipticals.  In each plot the global input LF is shown as a solid line. Finally each LF estimator is indicated by a: dashed line (\citetalias{sandage:1979}), triangles (SWML \citetalias{Efstathiou:1988}), squares ($C^+$-method) and circles ($1/V_{\max}$).  In each case the global LF is determined.

\subsubsection{Case 1 analysis - the UV reference filter:}

Due to the nature of how each LF estimator is constructed, the resulting bias affects them in different ways.  Examining firstly the top panel set corresponding to the reference filter, UV, we can see clearly that all LF estimators underestimate the global input LF.  
In the mock data for the first redshift range (top panels), the  $C^+$-method and  $1/V_{\max}$ methods follow the brighter end of the LF well but then show a sharp drop beyond $M\sim-14$, at which point they seem to recover the input LF of the elliptical samples.  A similar trend is observed for the more distant redshift range.  This bias seems entirely consistent with Case~1 described above and shown in the top panel of Figure~\ref{fig:I04_1}.  Whilst the SWML also shows a drop off at this magnitude it is less biased, recovering more the shape of the input elliptical LF as opposed to the LF itself.  Contrastingly, the parametric \citetalias{sandage:1979} method shows only a marginal short-fall in the recovered LF at faint magnitudes, and, moreover, in the left-hand bottom panel it seems to recover the input global LF exactly.

Whilst the HDF data in the right panels do show a drop off for all estimators toward the faint end of the LF, the impact is significantly less than in the mock data.  It was thought that this was due to the very small sample size of the data and thus it was concluded  that in this reference frame the global LF is not wholly recoverable.

\subsubsection{Case 2 analysis - the B reference filter:}

As already discussed and shown in the bottom panel of Figure~\ref{fig:I04_1}, in this reference frame the absolute magnitude limits for the spectral types are very similar across the redshift range indicated.  Thus in the top plots of the middle panel set in Figure~\ref{fig:I04_2} it is clear that all LF estimators recover the global LF extremely well.   In the bottom plots of this set we can observe a slight overestimation in the SWML and \citetalias{sandage:1979}  estimators toward the faint end. This is consistent with the middle panel of Figure~\ref{fig:I04_1} where at this redshift range, the absolute magnitude limits for each spectral type begin to show stronger divergence.

The HDF LF is fairly  recovered in the filter with only slight underestimation in \citetalias{sandage:1979} at $0.70\leq z<1.25$ and slight overestimation at $1.25\leq z<2.00$.

\vspace{10mm}
\noindent
This type of analysis into the robustness of the traditional LF estimators provides a useful tool for probing the deep survey sample where samples may be more prone to biasing from variable magnitude limits resulting from large $K$-corrections across spectral types. The authors conclude that since the \citetalias{sandage:1979} and SWML methods differ in their biasing (beyond Poisson errors) compared to the $1/V_{\max}$  and $C^+$ estimators can be a good indicator that either one has bias in  recovering the global LF.

Several possible ways to reduce this bias entering into LF estimation are offered. These include selecting galaxies subsamples in the closest rest-frame filter to the reference filter as performed by e.g. \cite{Poli:2003ApJ...593L...1P}. However, this requires multi-colour information  to be able to derive the same rest-frame band LF at different redshifts.  Another possible route applies more to large survey data. This would require estimating the global LF within the absolute magnitude range in which {\it all} galaxy types are detected. Of course, this approach would require the cutting of a percentage of the data.  
%
%
% END OF FILE			%section 8
 
 \section{Emerging Generalised Methods}\label{sec:emerging}
Despite the continuing popularity of such methods as $1/V_{\max}$, MLE and the SWML, there has been renewed exploration into  more innovative statistical approaches.  These new methods are largely motivated by the potential hazards and pitfalls inherent with the current traditional approaches.  I now  examine in more detail  three such approaches to estimating the LF that have emerged over the last few years. The first is a semi-parametric approach by \cite{Schafer:2007}, the second is a Bayesian approach by \cite{Kelly:2008}, and the third, by \cite{Takeuchi2010MNRAS.406.1830T}, applies the {\it copula} to construct the bivariate LF.

\subsection{\cite{Schafer:2007} - A semi-parametric approach}
This approach  is statistically rigorous and considers data-sets that are truncated i.e. flux limited. There can be potential advantages  for this method summarised as follows.
\begin{enumerate}
\item{No strict parametric form is assumed for the bivariate density.}
\item{No assumption of  independence between redshift and absolute magnitude is made.}
\item{No binning of data is required.}
\item{A varying selection function can be incorporated.}
\end{enumerate}
By not assuming separability Schafer decomposes the bivariate density $\phi(z,M)$ into,
\begin{equation}\label{equ:schafer}
\log\phi(z,M,\theta)=\mbox{\boldmath$f$}(z)+\mbox{\boldmath$g$}(M)+\mbox{\boldmath$h$}(z,M,\theta),
\end{equation}
where ${\bf{h}}(z,M,\theta)$  has a parametric form that incorporates the dependency between redshift, $z$, absolute magnitude, $M$,  and the real valued parameters, $\theta$,  by folding in, for example, evolutionary models. The functions ${\bf{f}}(z)$ and ${\bf{g}}(M)$ are, however, determined non-parametrically. He then incorporates an extended form of the maximum likelihood approach called the `local' likelihood estimator for the density estimation and applies this to  15,057 quasars from \cite{Richards:2006}. This semi-parametric approach has the advantage of allowing the user to estimate evolution of the LF with redshift without assuming a strict parametric form for the bivariate density.  The only parametric form required is that which models the dependence between redshift and absolute magnitude.  Moreover, it should be noted that this method assumes a {\it complete} data-set. 

\subsubsection{Estimating the local likelihood density:}  
This approach is a non-parametric extension of the MLE where one assumes the data ${\bf X}=(X_1,X_2,...,X_n)$ are observations of independent, identically distributed random variables from a distribution with density $f_0$.  The MLE ($\hat f_{MLE}$) for $f_0$ is defined as the $f \in \mathcal{F}$, where $\mathcal{F}$ denotes the class of candidates for $f_0$, and is maximised as
\begin{equation}
\sum\limits_{j = 1}^n {\log f(X_j ) - \left\{n\left[\int {f(x)dx-1}\right]\right\} } 
\end{equation}
From this, one can localise the likelihood criterion and thus construct  the final local likelihood $\hat f_{LL}$ estimator by smoothing the local estimates giving

\begin{equation}\label{equ:schafer_local}
\hat f_{LL} (x) \equiv {{\left[ {\sum\limits_{u \in \mathcal{G}} {K^* (x,u,\lambda )\hat f_u (x)} } \right]} \mathord{\left/
 {\vphantom {{\left[ {\sum\limits_{u \in \mathcal{G}} {K^* (x,u,\lambda )\hat f_u (x)} } \right]} {\left[ {\sum\limits_{u \in \mathcal{G}} {K^* (x,u,\lambda )} } \right]}}} \right.
 \kern-\nulldelimiterspace} {\left[ {\sum\limits_{u \in \mathcal{G}} {K^* (x,u,\lambda )} } \right]}}
\end{equation}
where $\mathcal{G}$ forms a grid $u\in \mathcal{G}$  of  equally spaced values (between -3 and 3 in the authors example) of a Gaussian density with mean zero and variance of unity. The term, $K^*(x,u,\lambda)$, is therefore a kernel function such that
\begin{equation}
\sum\limits_{u \in G} {K^* (x,u,\lambda )=1} \quad \forall x.
\end{equation}
By making  $\mathcal{G}$ sufficiently large, the amount of smoothing is completely dominated by the kernel function parameter, $\lambda$.
\subsubsection{Extending to flux-limited data}
The local density likelihood is incorporated into the case for flux-limited survey data where one can include the dependence between the redshift, $z$, and absolute magnitude, $M$.  A first order approximation of ${\bf h}$ is made from Equation~\ref{equ:schafer} such that
\begin{equation}
\mbox{\boldmath$h$}(z,M,\theta)=\theta zM.
\end{equation}
After an extensive derivation, a global criterion for the likelihood is found to be given by
\begin{align}
&\mathcal{L}^* (\mbox{\boldmath$f,g,z,M,$}\theta ) \equiv \sum\limits_{j = 1}^n {w_j \left( {\sum\limits_{u \in G} {K^* (z_j ,u,\lambda } )a_u (z_j )} \right.}  \nonumber 
\\ \nonumber
 & + \sum\limits_{u \in \mathcal{G}} {K^* (M_j ,v,\lambda } )\mbox{\boldmath$b_v$}(M_j ) + \mbox{\boldmath$h$}(z_j ,M_j ,\theta ) \\ \nonumber
& - \int\limits_\mathcal{A} {\left\{ {\exp (\mbox{\boldmath$h$}(z,M,\theta ))\left[ {\sum\limits_{u \in G} {K^* (M,v,\lambda } )\exp (\mbox{\boldmath$b_v$} (M))} \right]} \right.} \\
& \times \left. {\left. {\left[ {\sum\limits_{u \in G} {K^* (M,v,\lambda } )\exp (\mbox{\boldmath$a_v$}(M))} \right]dM\,dz} \right\}} \right),
\end{align}
where ${\bf a_u}(z)$ and ${\bf b_v}(M)$ are degree $p$ polynomials which form part of the smoothing term of $K^*$ for local estimates. $\mathcal{A}$ defines the region outside of which the data are truncated on the  $(z,M)$ plane.  The quantity $w_j$  is a weighting to take incompleteness into account and is defined as the inverse of the selection function.

Estimating the LF in this way has the advantage of allowing the user to estimate the evolution of the LF without assuming a strict parametric form of the bivariate density. The method  was  applied to quasar  data from \cite{Richards:2006} and compared to their results which applied the \cite{Page:2000} version of $1/V_{\rm max}$ (see page~\pageref{page2000}). These results shown in Figure~\ref{fig:schaf_LF1} are in good agreement with Richards.
\begin{figure*} \label{fig:schaf_LF1}
    \begin{center}           
     \includegraphics[width=0.9\textwidth]{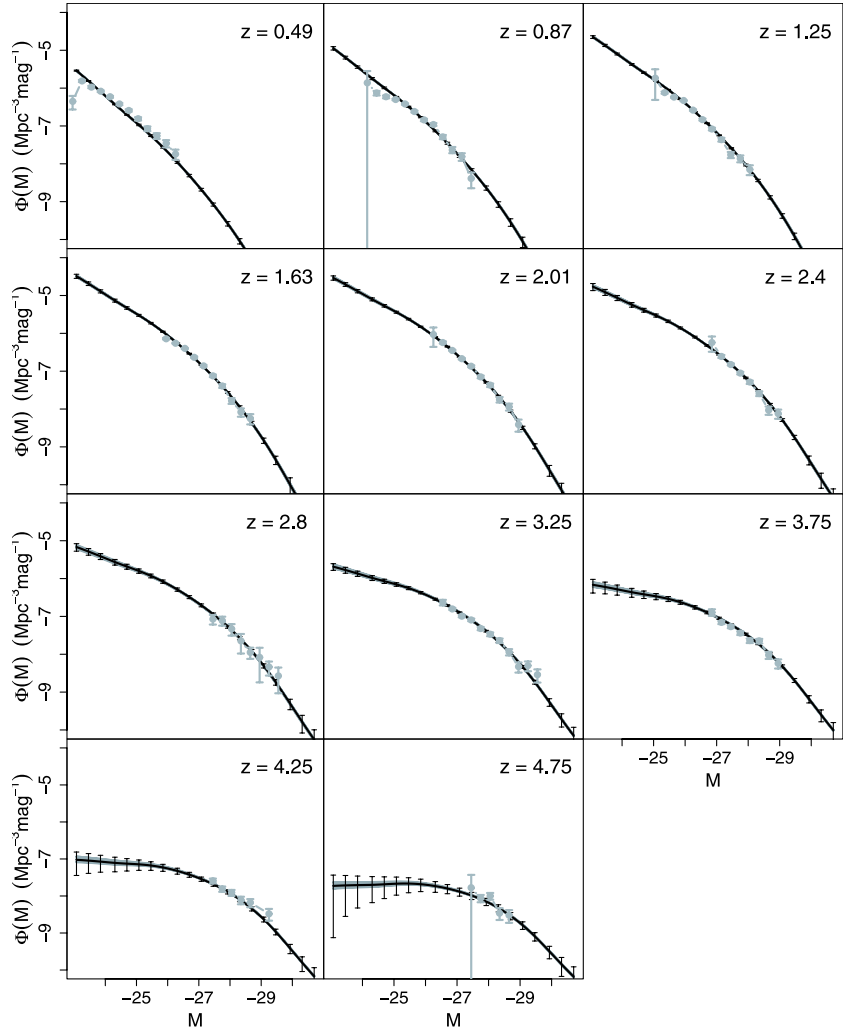}
            \caption{\small Extract from \cite{Schafer:2007} reproduced by permission of the AAS. Estimates of the luminosity function at different redshifts (black solid curves and error bars), compared with estimates from  \cite{Richards:2006}  (light grey solid curves and error bars). Error bars represent one standard error and account for statistical errors only.}
      \label{fig:schaf_LF1}
    \end{center}
  \end{figure*}
\subsection{\cite{Kelly:2008} -- A Bayesian approach}
In \cite{Andreon:2006MNRAS.369..969A} a Bayesian approach to constraining the LF is described by considering the color-magnitude distribution for both cluster and background galaxies (see also the end of \S~\ref{sec:MLE}). This section  now considers a  second Bayesian technique to estimate LF developed by \cite{Kelly:2008}.  In this paper they derive a  likelihood function of the LF  that relates observed data to the true LF (assuming some parametric form).  They then use  a Bayesian framework  to estimate the LF and the posterior probability  distribution of the LF parameters via a mixture of Gaussian functions.  By modelling the LF using  Gaussian functions, they circumvent the problem of having to assume a parametric form and allow for a flexible fitting method.
\subsubsection{Estimating the LF likelihood:}The form of the likelihood function that they adopt for the LF estimation is derived from a binomial distribution.   Whilst they highlight that the traditional approach of using a Poisson distribution is incorrect, they show that as long as the survey's detection probability is small, both approaches yield the same results.  We recall the relation of the LF to the probability density of ($L$,$z$) can be written in the following separable form:
\begin{equation}\label{equ:LF_kelly}
p(L,z)=\frac{1}{N}\phi(L,z)\rho(z),
\end{equation}
where $L$ is the luminosity, $z$ is redshift and $N$ is the normalisation set as  the total number of objects in the observable Universe.  From this starting point, the authors assume a parametric form for $\phi$($L$,$z$), with parameters $\theta$ and show that the observed data likelihood function is given by,
\begin{equation}\label{equ:likeho}
p(L_{obs} ,z_{obs} ,{\mathbf{I}}|\theta ,N) \propto C_n^N [p(I = 0|\theta )]^{N - n} \prod\limits_{i \in \mathcal{A}_{obs} } {p(L_i ,z_i |\theta )},
\end{equation}
where
\begin{equation}
p(L,z|\theta ) = \prod\limits_{i = 1}^N {p(L_i ,z_i |\theta )} 
\end{equation}
is the likelihood function for all $N$ sources in the universe.  $L_{obs}$ and $z_{obs}$  denote the sources observed in a given survey.  By adding in sample selection, the probability that the survey misses a source, given by the parameters $\theta$, is
\begin{equation}
p(I = 0|\theta ) = \iint {p(I = 0|L,z)p(L,z|\theta )dL\;dz}
\end{equation}
In Equation~\ref{equ:likeho}, $\mathbf{I}$ is a vector of size $N$ taking on values
\begin{equation}
I_i
\begin{cases}
1\ \ \ \ \mbox{if  $i$th source is included in survey}\\\nonumber \\
0\ \ \ \ \mbox{otherwise}
\end{cases}
\end{equation}
Finally the term $C^N_n=N!/n!(N-n)!$ is the binomial coefficient and $\mathcal{A}_{obs}$ is the set of $n$ included sources.  Equation~\ref{equ:likeho} thus represents observed data likelihood function given an assumed LF that is parameterised by $\theta$.
\subsubsection{The posterior probability  distribution}
It is  then shown that, for a  given set of  observed data,  the joint posterior probability distribution of $\theta$ and $N$ for the LF is given by
\begin{equation}
p(\theta, N |L_{obs} ,z_{obs} ) \propto p(N|\theta, n)p(\theta |L_{obs} ,z_{obs}), 
\end{equation}

\subsubsection{Model for the LF} 
\begin{figure*} \label{fig:kelly1}
    \begin{center}           
     \includegraphics[angle=90, width=0.49\textwidth]{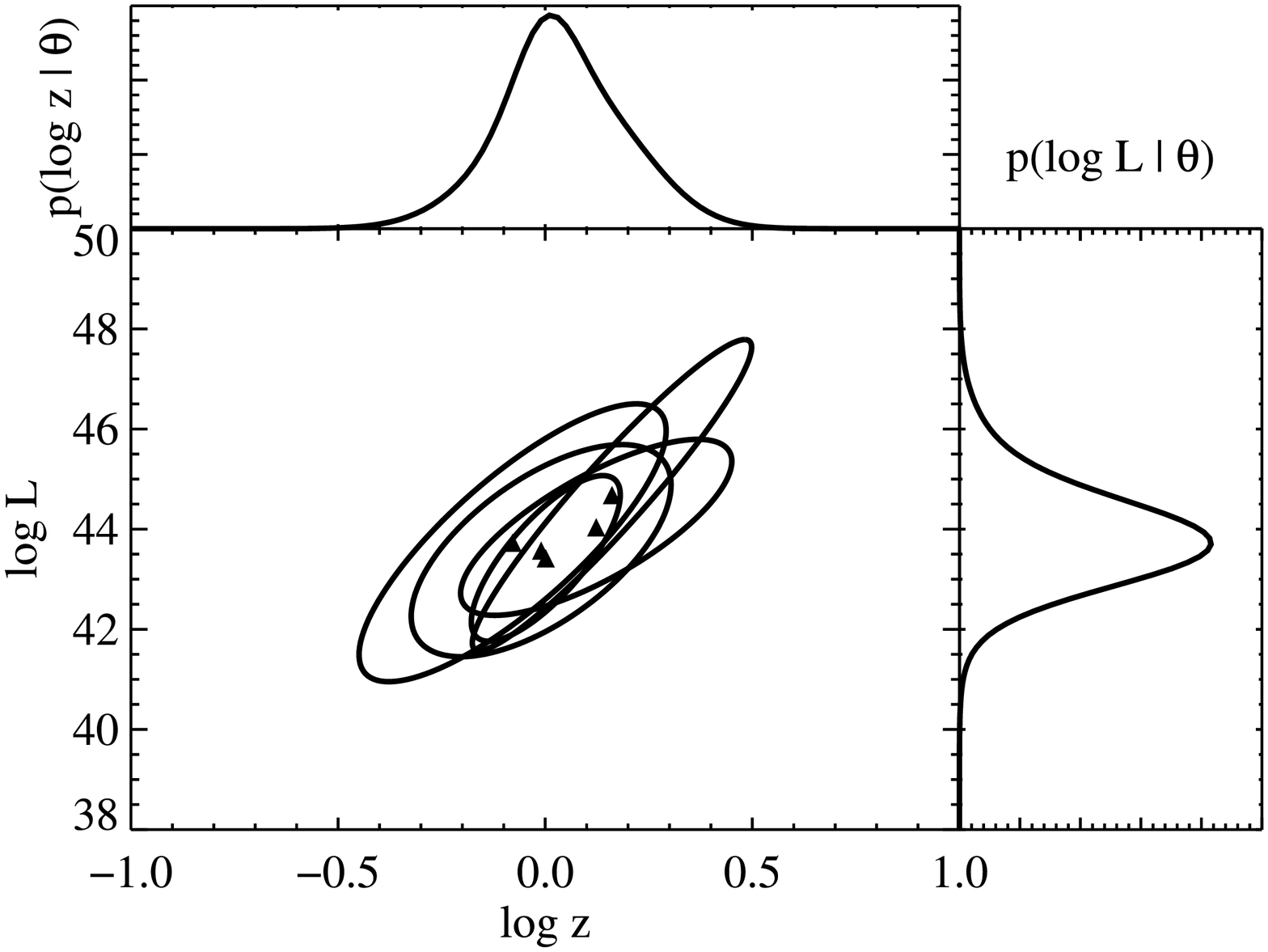}\hfill
          \includegraphics[angle=90, width=0.49\textwidth]{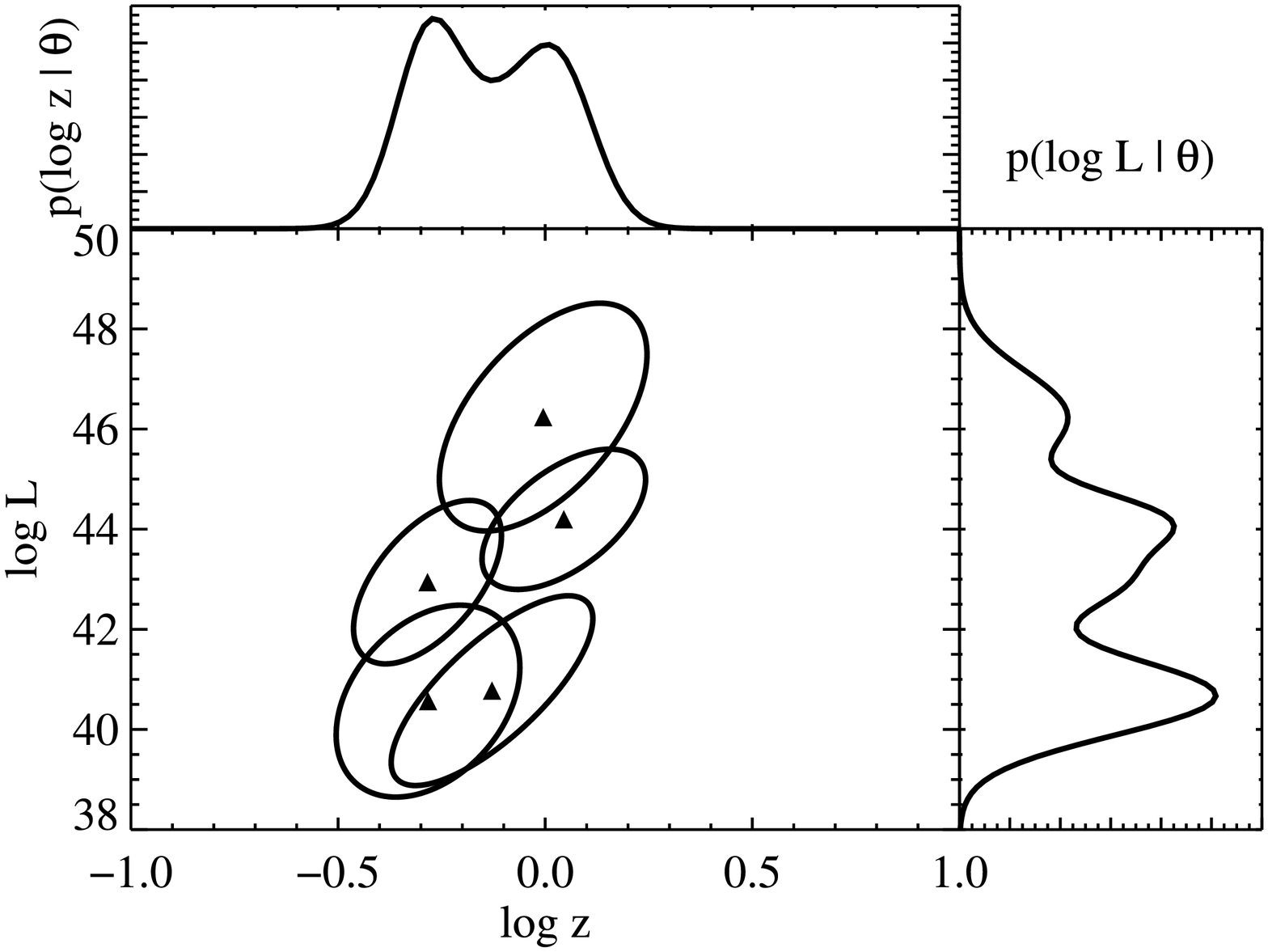}
            \caption{\small From \cite{Kelly:2008}: Illustration of the prior distribution for the mixture Gaussian function (GF) approach to describing the LF with $K=5$ GFs.  For the two cases shown, each has the resulting marginal distributions of $\log z$ and $\log L$ on the respective top and right side of each plot.   The left-hand panel illustrates  unimodal prior assumed in the paper where the GFs are close together.  When the GFs are further apart (right-hand panel) the resulting LF is multimodal. }
      \label{fig:kelly1}
    \end{center}
  \end{figure*}
To model the LF they adopt a similar approach as applied by \cite{Blanton:2003ApJ...592..819B}, where a   mixture of Gaussian functions were used to accurately estimate the LF.   In this case, to minimise the number of Gaussian functions (GFs) required to describe the LF,  they consider log of the  joint distribution of $L$ and $z$.  Equation~\ref{equ:LF_kelly} is now re-written as
\begin{equation}
p(L,z)=\frac{p(\log L,\log z)}{Lz(\ln 10)^2},
\end{equation}
Moreover, unlike  \cite{Blanton:2003ApJ...592..819B}, they allow the widths of GFs to vary and do not fix the centroids to like on a grid of values.  This also adds to minimising the number of GFs required.  The mixture of $K$ Gaussian functions can be written as
\begin{align}
 p(\log L_i ,\log z_i |\pi ,\mu ,\Sigma ) & = \sum\limits_{k = 1}^K  \nonumber
 \frac{{\pi _k }}{{2\pi \left| {\Sigma _k } \right|^{1/2} }} \\ 
&\times \exp \left[ { - \frac{1}{2}({\mathbf x_i}  - \mu _k )^T \sum\nolimits_k^{ - 1} {({\mathbf x_i}   - \mu _k )} } \right], 
 \end{align} 
where $\theta  = (\pi ,\mu ,\Sigma)$, ${\sum}_k^{K}\pi_k=1$, $\mathbf{x_i}=(\log L_i,\log z_i)$,  $\mu_k$ is the 2-element mean position vector for the $k$th Gaussian, $\sum_k$ is the 2 $\times$ 2 covariance matrix for the $k$th Gaussian, and $\mathbf{x}^T$ denotes the transpose of $\mathbf{x}$.  The model for the LF  is given by
\begin{align}
\phi (L,z|\theta ,N) &= \frac{N}
{{Lz(\ln 10)^2 }}\left( {\frac{{dV}}
{{dz}}} \right)^{ - 1} \\ \nonumber 
&\times\sum\limits_{k = 1}^K {\frac{{\pi _k }}
{{2\pi |\sum _k |^{1/2} }}\exp \left[ { - \frac{1}
{2}({\mathbf{x}} - \mu _{\mathbf{k}} )^T {\sum}_k^{ - 1} ({\mathbf{x}} - \mu _{\mathbf{k}} )} \right]} 
\end{align}
Another crucial aspect to using the mixture model is the form of its prior distribution. Essentially, it is assumed, in general,  the LF will be unimodal i.e.  it does not show distinct peaks (see Figure~\ref{fig:kelly1}). Consequently, only the prior distribution has a parametric form allowing the GF widths to be  closer together and thus aiding the convergence of the Markov Chain Monte Carlo (MCMC).
The form of this prior is given by
\begin{align}
p(\pi ,\mu ,\Sigma ,\mu _0 ,A) &\propto \prod\limits_{k = 1}^K {p(\mu _k } |\mu _0 ,\Sigma )p(\Sigma _k |A)  \nonumber  \\
&\propto \prod\limits_{k = 1}^K {{\rm{Cauchy}}_{\rm{2}} (} \mu _k |\mu _0 ,T){\rm{Inv - Wishart}}_{\rm{1}} (\Sigma _k |A),
\end{align}
where ${{\rm{Cauchy}}_{\rm{2}} (} \mu _k |\mu _0 ,T)$ refers to a two-dimensional Cauchy distribution as a function of $\mu_k$, with mean vector $\mu_0$ and scale matrix $T$. The ${\rm{Inv - Wishart}}_{\rm{1}} (\Sigma _k |A)$ is the inverse Wishart density as a function of $\Sigma_k$, with 1 degree of freedom and scale matrix $A$.

Finally, the total joint posterior distribution is shown to be given by
\begin{align}
p(\theta ,N,\mu _0 ,A|\log L_{obs} ,\log z_{obs} ) \propto p(N|\theta n)p(\theta ,\mu _0 ,A|\log L_{obs} ,\log z_{obs} )
\end{align}
MCMC is then applied using the  Metropolis-Hastings algorithm (MHA) \citep{Metropolis:1949,Metropolis:1953JChPh..21.1087M,hastings:1970} for obtaining random draws of the LF from the posterior distribution.  Given a suitably large enough number of Gaussian functions it is flexible enough to give an accurate estimate of any smooth and continuous LF.  They found that $K\sim$ 3 to 6 was generally a suitable number.

\begin{figure*} \label{fig:kellyLF}
    \begin{center}           
     \includegraphics[angle=90, width=1.\textwidth]{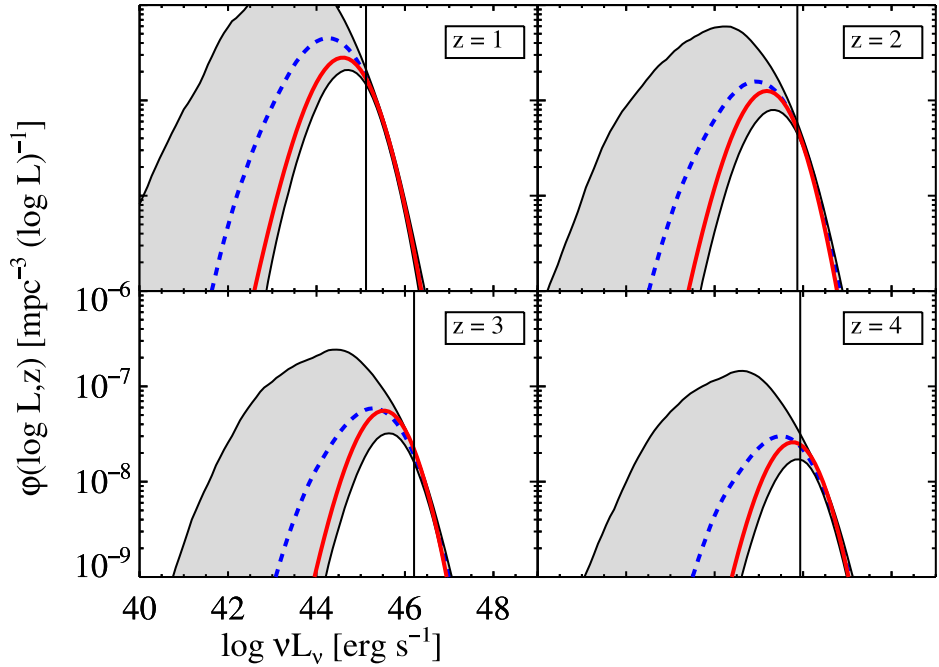}
            \caption{\small Extract from  \cite{Kelly:2008} reproduced by permission of the AAS.  The input LF derived from a simulated sample is shown as a solid red line. The dashed line represents the  posterior median estimate of the LF based on the mixture of Gaussian functions model. The shaded region is 90$\%$ of the posterior probability. The solid vertical  line shows the flux limit of the sample at each redshift.}
      \label{fig:kellyLF}
    \end{center}
  \end{figure*}

This approach was applied to MC simulated quasars derived from a Schechter function and separate simulated SDSS-DR3 quasar data based on  \cite{Richards:2006}. Figure~\ref{fig:kellyLF} shows the resulting LF at different redshifts for the latter sample. For the Schechter MC samples they concluded  that the mixture model proved to be  flexible enough to approximate the {\it true} Schechter function form, but was not as accurate as if one had fitted  the correct parametric  model.   Moreover, they showed that one could just as easily substitute the mixture model for any parametric form of the LF into the likelihood function and posterior distribution. The Bayesian mixture of Gaussian functions model is able to accurately constrain the LF, even below the survey detection limit.  
\subsection{\cite{Takeuchi2010MNRAS.406.1830T} -- Copula and correlation for the Bivariate LF}
Takeuchi adopts a  technique for BLF construction that uses the $copula$  function  to connect  two distribution functions and defines nonparametric measures of their dependence.  To demonstrate this technique the FUV-FIR\footnote{FUV-FIR refers to the far ultraviolet  and the far infrared} BLF is constructed by adopting the \cite{Saunders:1990} LF shown  in  Equation~\ref{equ:LFSaunders} for the IR, and a Schechter function shown in  Equation~\ref{Eq:LF_schechter} for the UV.  However, to summarise this technique some of the  framework  underpinning the copula definition is firstly  introduced.

\subsubsection{Sklar's theorem}
\cite{sklar:1959} created a new class of functions which he called copulae\footnote{Copula is Latin for  ``a link" or ``tie"}.  In the most general case, Sklar's theorem states that if $x_1,x_2,...,x_n$  variables of $H$, an $n$-dimensional distribution function with marginal distribution functions (or  `marginals')  $F_1,F_2,...,F_n$, then there exists an $n$-copula $C$ such that
\begin{equation}
H(x_1,x_2,...,x_n)=C[F_1(x_1),F_2(x_2),...,F_n(x_n)]
\end{equation}
The above relation has the property that if the marginal distributions $F$  are continuous then the resulting copula $C$ is unique. However, if this is not the case then the copula is unique on the range of values of the marginal distributions on Range $F_1$ $\times$ Range $F_2$  $\times....\times$ Range $F_n$, where Range $F_i$ is the range of the function $F_i$.

The implementation of the copula in this work reduces the above to a 2-dimensional bivariate case where $H$ is defined as
\begin{equation}
H(x_1,x_2)=C[F_1(x_1),F_2(x_2)].
\end{equation}
This theorem gives the basis that any bivariate distribution function with given marginals can be expressed in this form. Therefore, in essence, the copulae connect the joint distribution functions to their one-dimensional margins.

\subsubsection{Bivariate copula definition}
The copula itself is defined on the unit $n$-cube[0,1]$^n$ having uniformly distributed marginal distributions.  For  the 2-dimensional case, the copula is a function $C: [0,1]^2\rightarrow [0,1]$ which  has the following properties:
\begin{enumerate}
\item{$C$ is grounded: that is, $\forall \ u,v \in[0,1]$, $C(u,0)=0$ and $C(0,v)=0$}
\item{$C$ is 2-increasing: thus, $\forall \ u_1,u_2,v_1,v_2 \in [0,1]$ such that $u_1\leq u_2$ and  $v_1\leq v_2$, yielding,
\begin{equation}
	C(u_2,v_2)-C(u_2,v_1)-C(u_1,v_2)+C(u_1,v_1)\geq 0 \nonumber
\end{equation}
}
\item{$\forall \  u,v \in[0,1]$, 
\begin{equation}
C(u,1)=u \quad \mbox{and} \quad C(1,v)=v
\end{equation}
}
\end{enumerate}

\subsubsection{Boundaries of the copula}
Lastly,  the Fr\'{e}chet--Hoeffding lower and upper bounds that are defined for every copula $C$ and every (u,v)$\in[0,1]^2$ are such that
\begin{equation}
\max(u+v-1,0)\leq C(u,v) \leq \min(u,v)
\end{equation}

\begin{figure*} \label{fig:T2010}
    \begin{center}           
    	 \includegraphics[width=0.5\textwidth]{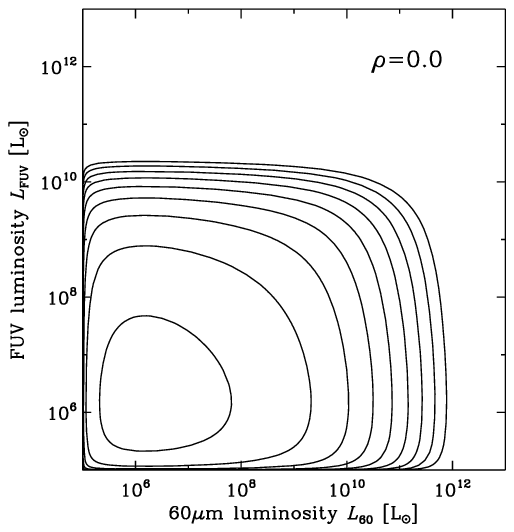}\hfill
     	 \includegraphics[width=0.5\textwidth]{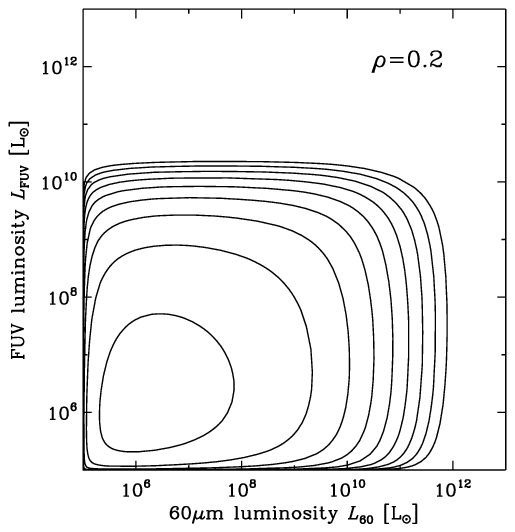}\\
    	 \includegraphics[width=0.5\textwidth]{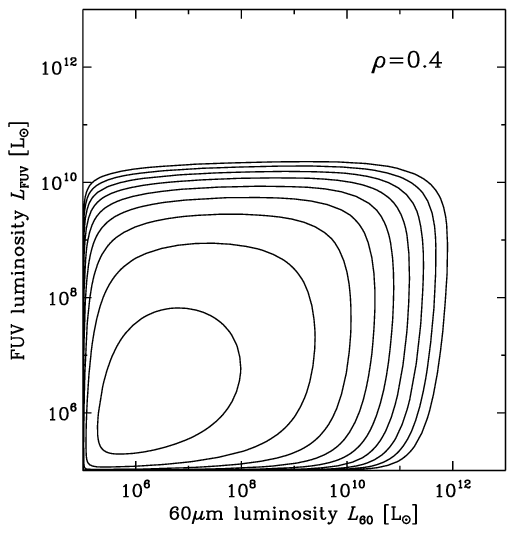}\hfill
     	 \includegraphics[width=0.5\textwidth]{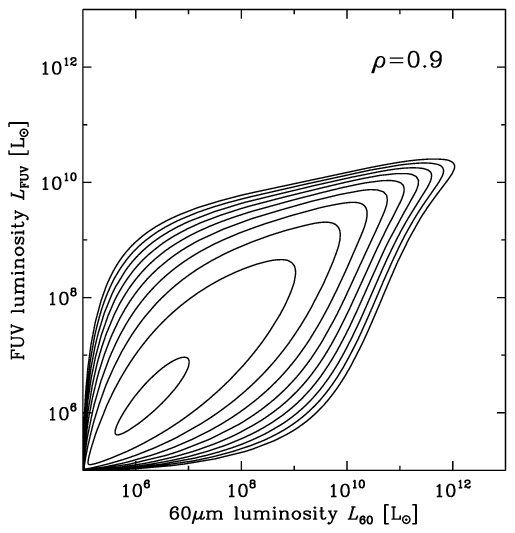}
          \caption{\small Extract from \cite{Takeuchi2010MNRAS.406.1830T}: The analytical BLF constructed with the Gaussian copula with model LFs of UV- and IR-selected galaxies. The BLFs are normalised so that integrating over the whole ranges of L1 and L2 gives 1. The linear correlation coefficient $\rho$ varies from 0.0 - 0.9 from top left to bottom right. The contours are logarithmic with an interval $\Delta\log\phi^{(2)} = 0.5$ drawn from the peak probability. }
      \label{fig:T2010}
    \end{center}
  \end{figure*}

\subsubsection{Constructing the copula for the BLF}
There are a number of forms the copula can take and Takeuchi explores two of them: the Farlie-Gumbel-Morgenstern (FGM) copula and a Gaussian copula.  For illustration of the method and simplicity  only the Gaussian form is considered here.  To measure the dependence between the two distribution functions $F_1(x_1)$ and $F_2(x_2)$ the  linear  Pearson product-moment correlation function is applied. However, it is noted that to explore the  non-linear dependencies, the Kendal $\tau$  \citep{kendall:1938}  or Spearman's $\rho_s$ \citep{spearman:1904} estimators would be preferred.  

The Gaussian copula $C^G$ is defined as
\begin{equation}
C^G(u_1,u_2;\rho)=\Phi_2[\Phi_1^{-1}(u_1),\Phi_1^{-1}(u_2);\rho]
\end{equation}
where $\Phi_1$ is the standard normal CDF (with mean zero and variance of unity) and $\Phi_2$ is the bivariate CDF with correlation $\rho$.  By differentiating $C^G$ w.r.t $u_1$ and $u_2$ one obtains the $C^G$ density function shown to be given by
\begin{equation}
c^G (u_1 ,u_2 ;\rho ) = \frac{1}{{\sqrt {\det \Sigma } }}\exp \left\{ { - \frac{1}{2}\left[ {\Psi ^{ - 1^T } (\Sigma ^{ - 1}  - I)\Psi ^{ - 1} } \right]} \right\}
\end{equation}
where $\Phi^{-1}\equiv [\Phi^{-1}(u_1),\Phi^{-1}(u_2)]^T$ and $I$ is the identity matrix. $\Sigma$ is the covariance matrix given by
\begin{equation}
\Sigma  = \left( {\begin{array}{*{20}c}
   1 & \rho   \\
   \rho  & 1  \\
\end{array}} \right)
\end{equation}
Thus if we now denote the univariate LFs as $\phi_1^{(1)}(L_1)$ and $\phi_2^{(1)}(L_2)$ then the bivariate PDF $\phi^{(2)}(L_1,L_2)$ is described by a differential copula as\footnote{It should be noted that Equation~\ref{phicop}  is the correct  form for the expression of $\phi^{(2)}\left( L_1, L_2 \right)$. The equivalent expression given by  Equation~32 in \cite{Takeuchi2010MNRAS.406.1830T} contained a typographical error.}
\begin{equation}\label{phicop}
\phi^{(2)}\left( L_1, L_2 \right) \equiv c\left[ \phi^{(1)}_1 \left( L_1 \right), \phi^{(1)}_2 \left( L_2 \right)\right] \phi^{(1)}_1 \left( L_1 \right) \phi^{(1)}_2 \left( L_2 \right)
\end{equation}
Thus, under the Gaussian copula the BLF is finally given by
\begin{equation}
\phi ^{(2)} (L_1 ,L_2 ;\rho ) = \frac{1}{{\sqrt {\det \Sigma } }}\exp \left\{ { - \frac{1}{2}\left[ {\Psi ^{ - 1^T } (\Sigma ^{ - 1}  - I)\Psi ^{ - 1} } \right]} \right\} \times \phi _1 ^{(1)} (L_1 )\phi _2^{(1)} (L_2 )
\end{equation}
where
\begin{equation}
\Phi^{-1}\equiv \left[\Phi^{-1}\left(\phi_1^{(1)}(L_1)\right),\Phi^{-1}\left(\phi_2^{(1)}(L_2)\right)\right]^T
\end{equation}
Figure~\ref{fig:T2010} shows a summary of results from the above application of the method using the Gaussian copula to construct the FUV-FIR BLF.  The marginal distributions in this application are given as a standard Schechter LF (Equation~\ref{Eq:LF_schechter} on page~\pageref{Eq:LF_schechter}) for the FIR LF and the log-Gaussian LF (Equation~\ref{equ:LFSaunders} on page~\pageref{equ:LFSaunders}) as defined by  \cite{Saunders:1990} for the FIR LF.  It is demonstrated that if the linear correlation between two variables is strong the Gaussian copula is a useful way in which to construct this type of distribution function since it connects two marginal distributions and is directly related to the linear correlation coefficient between the two variables.
%
%
 %END OF FILE			%section 9
 \section{From Tests of Independence to Completeness Estimators}\label{sec:indep}
This review has had a thorough look at many of  the non-parametric and parametric methods used to determine LFs.  However,  as discussed \S~\ref{sec:intro} the fundamental assumption for the majority of  techniques is the separability of the density function and the LF that allows us to write the probability distribution of observables as
 \begin{equation}\label{Eq:sep}
 \Phi(M,Z)=\phi(M)\rho(Z)
 \end{equation}
The quantity $Z$ now refers to the distance modulus which is used instead of redshift throughout this section. The distance modulus is related to redshift and magnitude in the following way,
\begin{equation}
Z = m-M = 5\log_{10}[d_L(z,\Omega_m,\Omega_\Lambda,H_0)] + 25
\end{equation}

This is a fundamental and crucial assumption that is generally accepted over small redshift bins (or shallow redshift surveys).  However, what is sometimes overlooked is whether the assumption of separability is valid. That is, after all corrections have been made for e.g. evolution, $K$-correction etc.,  are the observables absolute magnitude $M$  and distance modulus $Z$, of Equation~\ref{Eq:sep} statistically independent?
\subsection{Efron and Petrosian independence test}
In order to test this assumption of independence (or separability) between two variables, \cite{Efron:1992} developed a simple ranked-based non-parametric test statistic for data-sets with a single truncation.  Their method is, in principle, an extension of the $C^-$ method.  For this technique Efron and Petrosian construct the Kendall \citep{kendall:1938} test statistic, $\tau$, based on the rejection of independence between the random variables $M$ and $Z$ under test.   Although they give a very detailed and formal  explanation, only the main points are  summarised here.  
\begin{figure} \label{fig:efron}   
   \begin{center}
      \includegraphics[width=1.\textwidth]{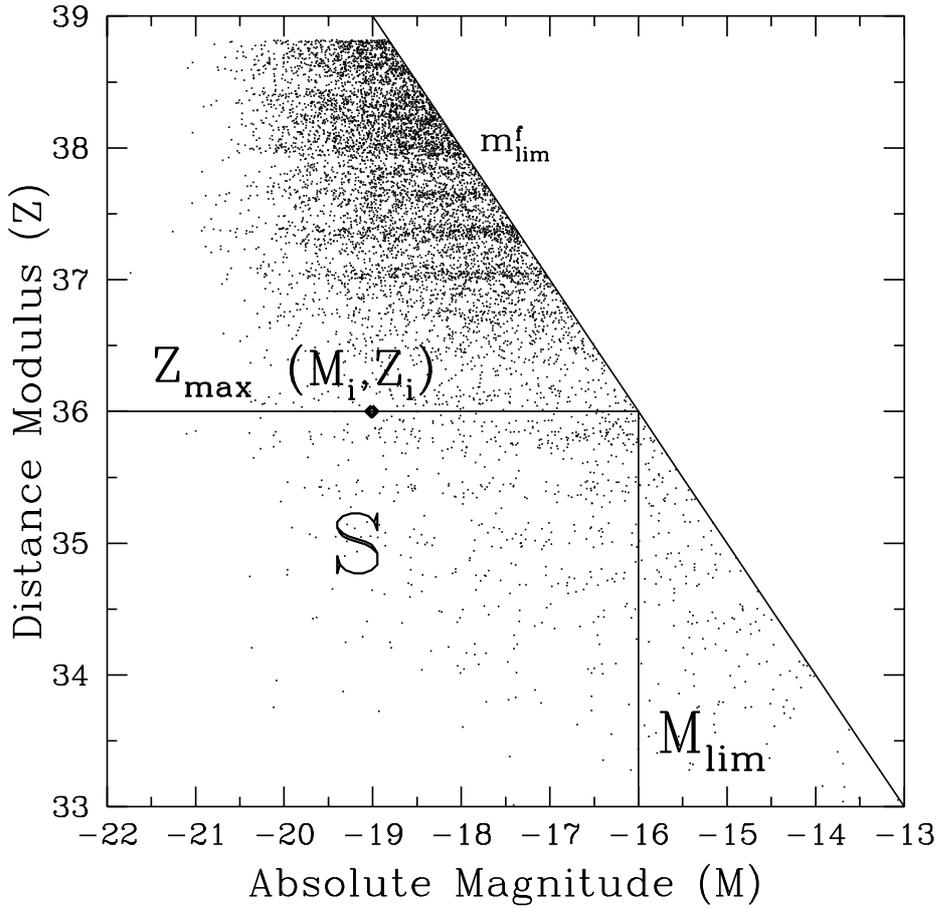}
              \caption[Schematic illustrating the construction of the Efron and Petrosian (1992)  test of independence.]{\small Schematic illustrating the construction of the Efron and Petrosian (1992)  test of independence. In order for this test to be applied correctly it is crucial to account for the presence of the faint apparent magnitude limit,  $m_{\lim}^{\rm f}$ which introduces a correlation between $M$ and $Z$.  Thus, drawing from the basic ideas of the  $C^-$ method, it is possible to construct a separable region, $S$, for each galaxy located at ($M_i,Z_i$) within which one can now can construct the $\tau$ statistic of independence. }
       \label{fig:efron}  
    \end{center}
  \end{figure}
% %
To illustrate this test  refer back to Figure~\ref{fig:cminus} on page~\pageref{fig:cminus}, which schematically shows the construction of the $C^-$ method. 
 
If, for the moment, we imagine the most simplest scenario where we are not hampered by an apparent magnitude limit $m_{\lim}^{\rm f}$,  then $M$ and $Z$ would be independent assuming we have a complete sample. Consequently,  the rank, $R_i$, of $M_i$, would be uniformly distributed between 1 and the number of galaxies in the set, $N_{\rm gal}$,  with respected expectation value and variance,
 \begin{equation}\label{Eq:variance}
 E=\frac{1}{2}(N+1),  \ \ \ \ \ V=\frac{1}{12}(N^2-1).
 \end{equation}
 The quantity $T$ for each galaxy is then defined as
 \begin{equation}
 T_i=\frac{(R_i-E)}{V},
 \end{equation}
such that  $R_i$ is now normalised to have mean of zero and a variance of unity.  The hypothesis of independence is then rejected or accepted depending on the value of $T_i$.   One then constructs confidence levels of rejection by combining $T_i$ into the single statistic $\tau$ such that
\begin{equation}
\tau  = \frac{{\sum\limits_i {(R_i  - E)} }}{{\sqrt {\sum\limits_i V } }}
\end{equation}
where, by definition, a $\tau$ of 1 indicates a 1-$\sigma$ correlation and conversely, a $\tau$ of 0 indicates that the variables are completely uncorrelated.

Now, if the magnitude limit $m_{\lim}^{\rm f}$ is  re-introduced   as in Figure~\ref{fig:efron}, the above  method has to be modified to correctly account for the magnitude limit breaking the separability between $M$ and $Z$.  This requires the construction of  subsets for each galaxy located at ($M_i,Z_i$) to correctly calculate the ranks $R_i$ of the entire set of observables.  Therefore, within each set, all objects which could have been observed up to $m_{\lim}^{\rm f}$  limit are counted such that  for each galaxy  an area $S$ is defined:
\[
S\equiv
\begin{cases}
 \infty  < M_i < M_{\lim} \\ 
 \mbox{and} \\
 0<Z_i<Z_{\rm max}(M_i),
 \end{cases}
 \]
as illustrated in Figure~\ref{fig:efron}.  Within this subset $R_i$ is uniformly distributed between 1 and $N_i(S)$, where $N_i(S)$ is the number of objects in each subset.  With this amendment to the method, the construction of the $\tau$ statistic remains the same with $\tau$ once again being defined as,
\begin{equation}
\tau  = \frac{{\sum\limits_i {(R_i  - E)} }}{{\sqrt {\sum\limits_i V } }}
\end{equation}
where the expectation value $E$ and variance $V$ remain unchanged from Equation~\ref{Eq:variance}.  The statistic $\tau$ thus gives us an unbiased measure of  the correlation of the data-set whilst properly accounting for the apparent magnitude limit.

In \cite{Efron:1998} they extend this method for data which have  a double truncation i.e. a survey that has distinct faint $and$ bright flux limits.  The natural progression was to explore cases where the value of $|{\tau}|\ge$1, which implies random variables $M$ and $Z$ cannot be considered separable.  In this case they assume that there is some form of luminosity evolution in the underlying data.  And so by adopting a parametric form for luminosity evolution that varies with an evolutionary parameter, usually referred to as $\beta$, they were able to show that  for  a particular value of $\beta$, $\tau(\beta)$ would equal 0. 
 
\cite{Maloney:1999} apply these methods on various quasar samples to determine the density functions and the luminosity evolution.  These techniques have also been used by  \cite{Hao:2005} for AGN's to test the correlation between the host galaxy and nuclear luminosity. \cite{Kocevski:2006ApJ...642..371K} applied this to constrain luminosity evolutionary models of Gamma Ray Bursters (GRBs).

\subsection{The $T_c$ and $T_v$ completeness estimators}\label{sec:complete}
In \cite{Rauzy:2000} the authors drew upon the ideas from \cite{Efron:1992} to develop a non-parametric method to fit peculiar velocity field models.  This methodology was later extended in \cite{Rauzy:2001} as a completeness test. The motivation for  this statistical method was to introduce a non-parametric test that could determine the {\it true}  completeness limit in apparent magnitude of a magnitude-redshift survey  whilst retaining as few model assumptions as possible and being independent of spatial clustering.

\begin{figure*} \label{fig:rauzy_mz}
    \begin{center}           
     \includegraphics[width=0.49\textwidth]{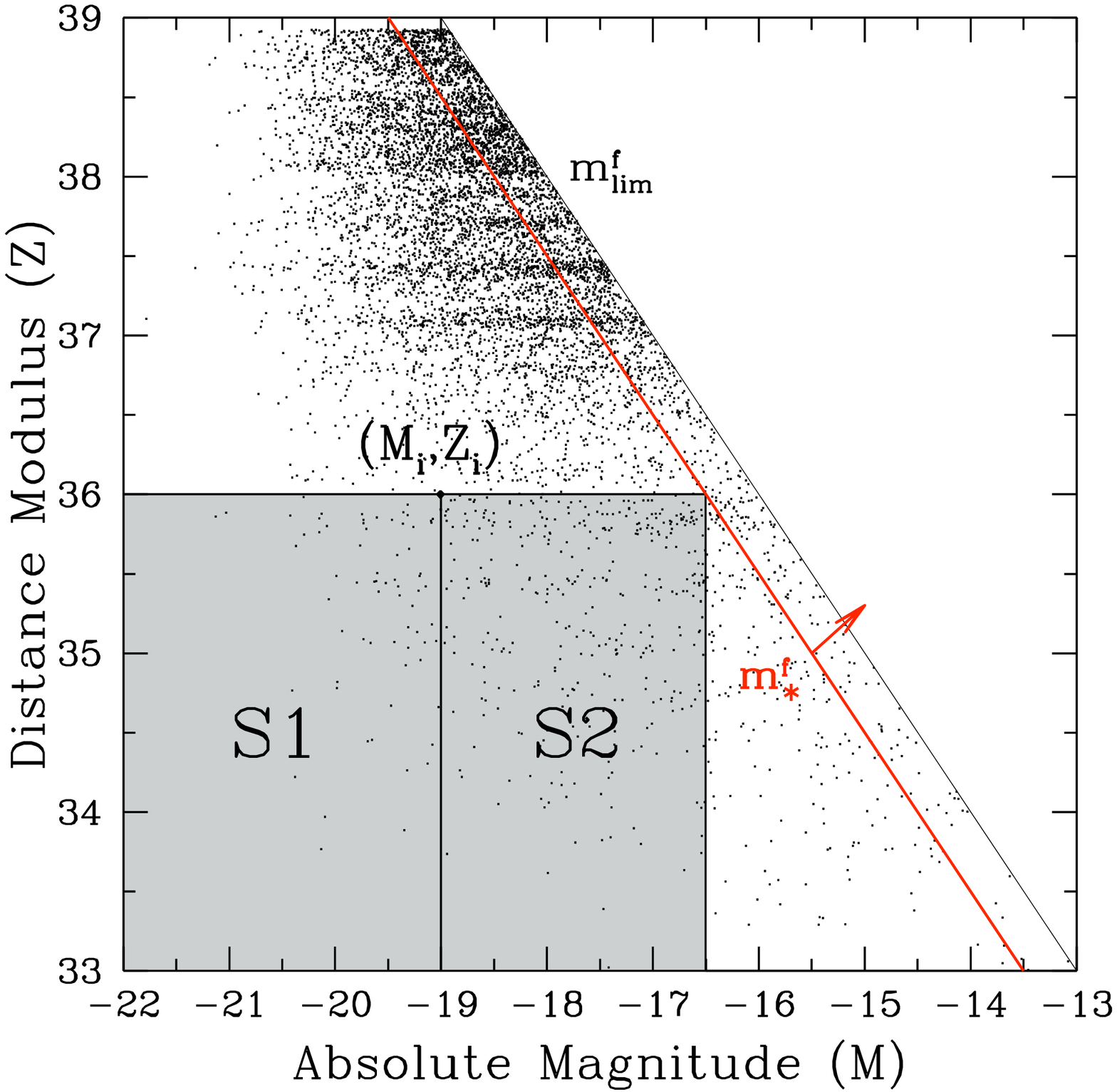}
     \includegraphics[width=0.49\textwidth]{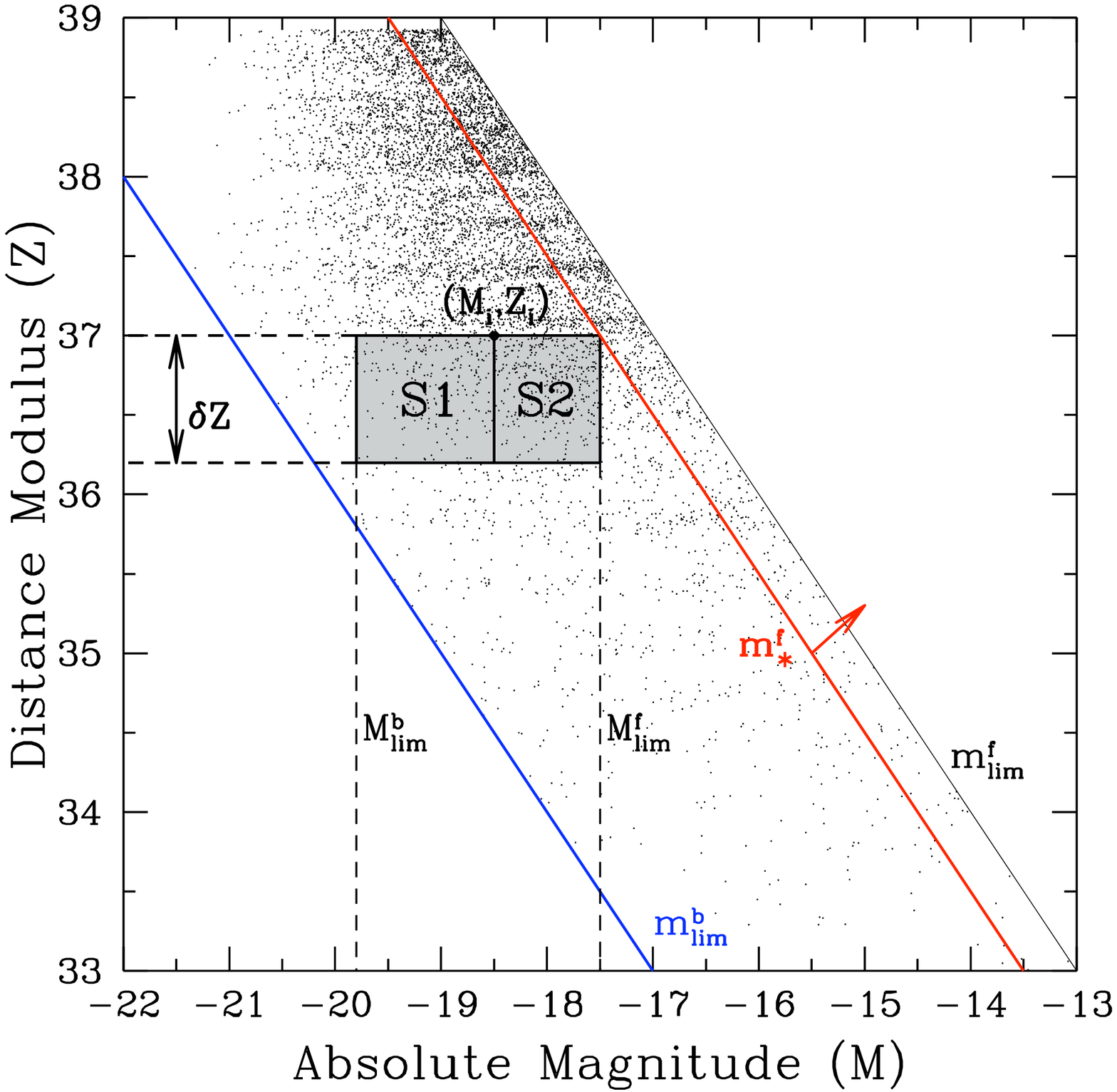}
           \caption{\small Schematic  construction the random variable, $\zeta$, for the \cite{Rauzy:2001} completeness test (left-hand panel) and the \citet*{Johnston:2007}  extension (right-hand panel).  In the left-hand plot the constructed regions $S_1$ and $S_2$ are defined for `trial' faint apparent magnitude limit $m^{\rm f}_*$ for a typical galaxy at $(M_i,Z_i)$. Also shown is the true faint apparent magnitude limit $m^{\rm f}_{\rm lim}$, within which the rectangular regions  $S_1$ and $S_2$ contain a joint distribution of $M$ and $Z$ that is separable. The right-hand plot now shows the extension where one accounts for the presence of a secondary bright apparent magnitude limit. The $S_1$ and $S_2$ regions are uniquely defined  for a slice of fixed width, $\delta Z$, in corrected distance modulus, and for `trial'  faint apparent magnitude limit $m^{\rm f}_*$.  Also shown are the true bright and faint apparent magnitude limits $m^{\rm b}_{\rm lim}$ and $m^{\rm f}_{\rm lim}$, within which the rectangular regions  $S_1$ and $S_2$ once again contain a joint distribution of $M$ and $Z$ that is separable.}
      \label{fig:rauzy_mz}
    \end{center}
  \end{figure*}
The fundamental approach of the Rauzy completeness is the same as Efron and Petrosian i.e assumption of separability between the absolute magnitude, $M$, data and the distance modulus $Z$ data.  As demonstrated in the previous section, the presence of a faint apparent magnitude limit introduces a correlation between the variables $M$ and $Z$ for observable galaxies.  Thus to retain the assumption of separability and construct a completeness statistic, the author defines a separable region defined by the random variable $\zeta$.   The left-hand panel of Figure~\ref{fig:rauzy_mz} illustrates how  the variable $\zeta$ is constructed such that it satisfies

\begin{equation}
\zeta  = \frac{{F(M)}}{{F[M_{\lim } (Z)]}}\equiv\frac{S_1}{S_1\cup S_2},
\end{equation}
where $F(M)$ is the Cumulative Luminosity Function (CLF) defined as
\begin{equation}
F(M) = \int_{ - \infty }^M {f(x)dx}.
\end{equation}
It follows immediately from its definition that  $\zeta$ has the property of being uniformly distributed between 0 and 1.  The random variable $\zeta$ can be estimated directly from the data, without prior knowledge of the functional form of the CLF and is independent of the spatial distribution of galaxies.
\begin{figure*}
    \begin{center}
     \includegraphics[width=0.49\textwidth]{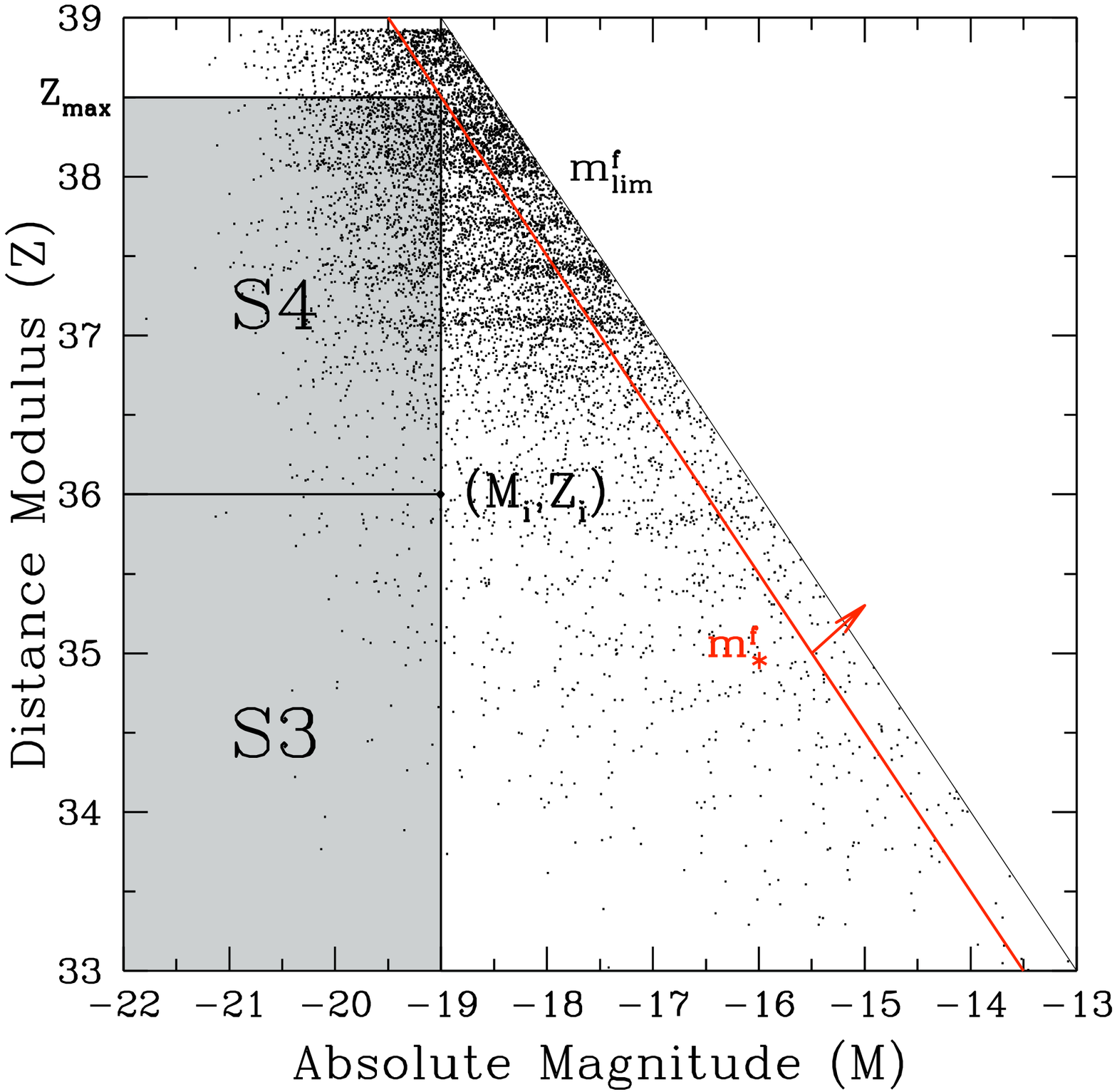}    
     \includegraphics[width=0.49\textwidth]{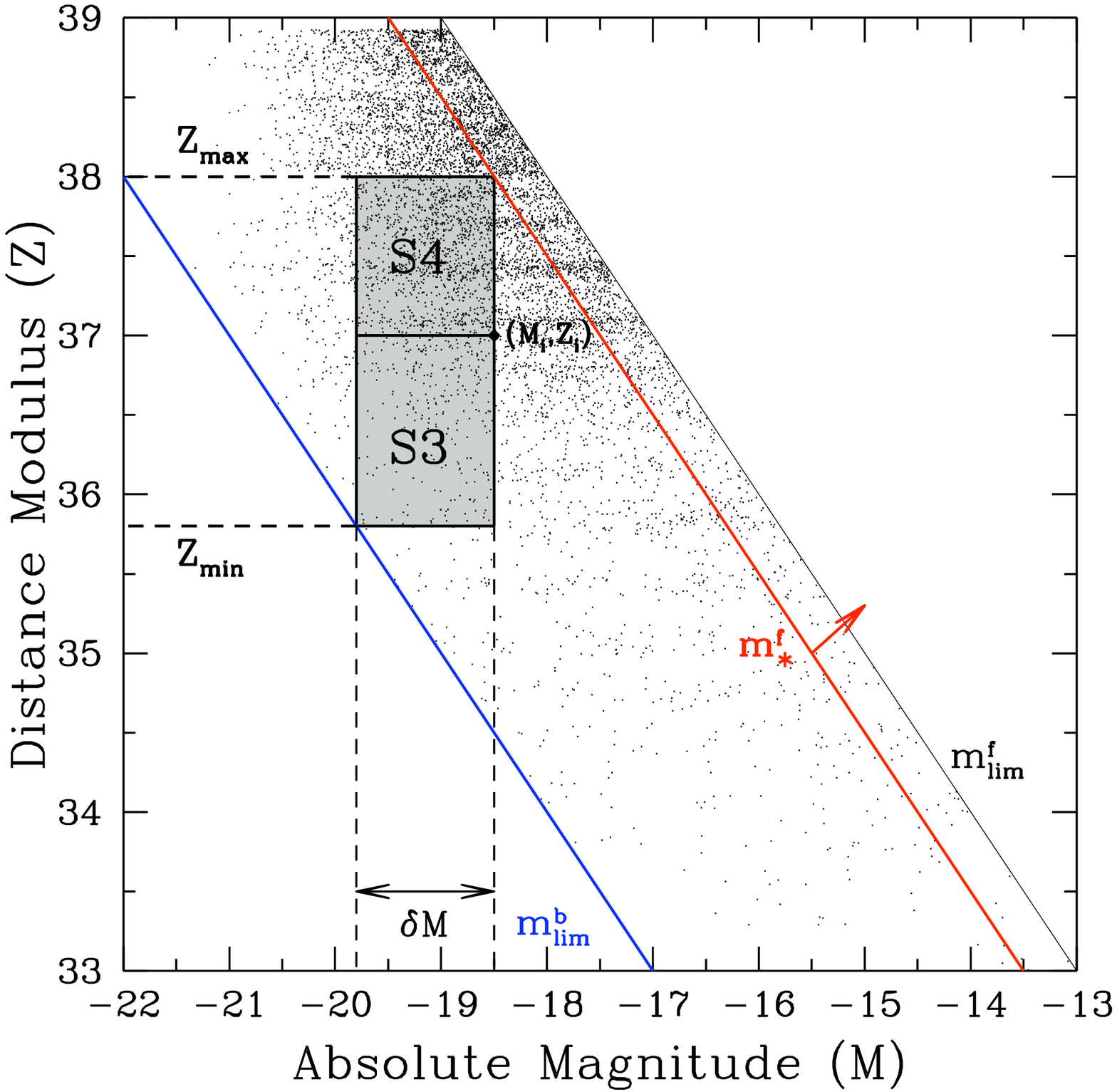}
      \caption{\small Schematic construction of the rectangular regions $S_3$ and $S_4$, defined for a typical galaxy at $(M_i,Z_i)$, which feature in the estimation of the variant completeness test statistic, $T_v$.  The left-hand panel illustrates how $S_3$ and $S_4$ are constructed for a survey with only a faint magnitude limit $m^{\rm f}_{\rm lim}$, and are shown for a trial faint limit $m^{\rm f}_\star$.  The right-hand panel shows the case where the survey also has a true bright limit $m^{\rm b}_{\rm lim}$ (which we assume for simplicity is known), and the rectangles are constructed for trial bright and faint limits $m^{\rm b}_{\rm lim}$ and $m^{\rm f}_{\rm lim}$,  respectively. Note that the construction of the rectangles is unique for a `slice' of fixed width, $\delta M$, in absolute magnitude.
}
      \label{fig:mz2}
    \end{center}
  \end{figure*}
Therefore, the estimator ${\hat \zeta_i}$ can be combined for each observed galaxy into a single statistic, $T_c$, which can then be used to test the magnitude completeness of a given sample for adopted trial magnitude limits $m^{\rm f}_*$.  $T_c$ is defined as
\begin{equation}\label{equ:T_c}
T_c =  {\displaystyle
 \sum_{i=1}^{N_{\rm gal}}
\left ( {\hat \zeta_i} - \frac{1}{2} \right ) }\Bigg/ { {\displaystyle
\left (\sum_{i=1}^{N_{\rm gal}} V_i \right )^{\frac{1}{2}} } },
\end{equation}
where $V_i$ is the variance.  If the sample is complete in apparent magnitude, for a given  trial magnitude limit, then $T_c$ should be normally distributed with mean zero and variance unity.  If, on the other hand, the trial faint  magnitude limit is fainter than the true limit, $T_c$ will become systematically negative, due to the systematic departure of the $\hat \zeta_i$ distribution from uniformity on the interval $[0,1]$.  The  random variable $\zeta$, was applied by \citet*{RHM:2001} in conjunction with the Kolmogrov-Smirnov test   and the correlation coefficient $\rho$, to devise a method to fit luminosity function models to observables.  This method was applied to the Southern Sky Redshift Survey (SSRS2) sample of \cite{Costa:1988}.  

The Rauzy completeness  test has since been applied to a wide variety of data exploring e.g.  the completeness limits of the HI mass function in the HIPASS survey  \citep{Zwaan:2004MNRAS.350.1210Z},  the HI flux completeness of ALFALFA survey data \citep{Toribio:2011arXiv1103.0990T}, and for nearby dwarf galaxies in the local volume \citep{Lee:2009}.  A recent study by \cite{Devereux:2009} adopted the Rauzy method to determine the completeness of multi-wavelength selected data.  

In \citet*{Johnston:2007} they extended its usefulness to survey data that are characterised either by two distinct  faint and bright flux limits and/or the case where the presence of a bright limit is harder to detect.  This is illustrated in the right-hand panel of Figure~\ref{fig:rauzy_mz}.  In addition to this they constructed a new statistic called $T_v$ which was constructed instead from the cumulative distribution function  of the corrected distance modulus for observable galaxies and denoted by $H(Z)$, such that
\begin{equation}\label{CDF}
 H(Z)=\int_{-\infty}^Z \, \overline{h}(Z') \, dZ'.
\end{equation}
Thus, a new random variable, $\tau$, was defined that mirrored the $\zeta$ variable and was constructed for the two cases,
\begin{align}\label{tau_faint}
\mbox{{\bf Case I :  }}\ \ \
&\tau =\frac{H(Z)}{H[Z_{\rm max}(M)]}\equiv\frac{S_3}{S_3\cup S_4}.
\\  \nonumber
\\
\mbox{\bf{Case II :   }}\ \ \
 &\tau =\frac{H(Z) - H[Z_{\min}(M-\delta M)]}{H[Z_{\max}(M)] - H[Z_{\min}(M-\delta M)]}\equiv\frac{S_3}{S_3\cup S_4},\label{equ:deltaM}
\end{align}
as illustrated, respectively, on the left- and right-hand panels in Figure~\ref{fig:mz2}. As with the $T_c$ statistic, one can combine the estimator ${\hat \tau_i}$ for each observed galaxy into a single statistic, $T_v$, which we can use to test the magnitude completeness of our sample for an adopted trial magnitude limit $m^{\rm f}_*$. $T_c$ is defined as
\begin{equation}\label{T_v}
T_v =  {\displaystyle
 \sum_{i=1}^{N_{\rm gal}}
\left ( {\hat \tau_i} - \frac{1}{2} \right ) }\Bigg/{ {\displaystyle
\left (\sum_{i=1}^{N_{\rm gal}} V_i \right )^{\frac{1}{2}} } }.
\end{equation} 

$T_v$ has the exact properties as $T_c$ such that if the sample is complete in apparent magnitude, for a given trial magnitude limit $m_*$,  $T_v$ should be normally distributed with mean zero and variance unity.  %
\begin{figure*}
    \begin{center}
     \includegraphics[width=1.\textwidth]{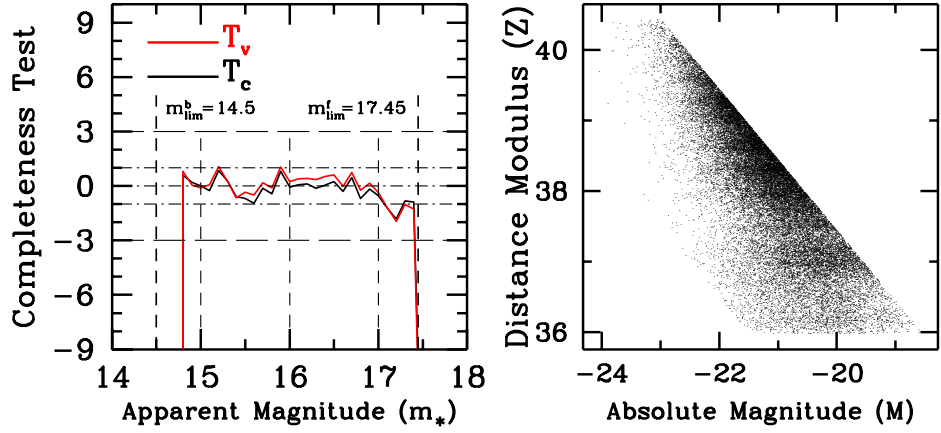}    
      \caption{\small The \citet*{Johnston:2007} $T_c$ and $T_v$ completeness test statistics applied to a SDSS Early Type galaxy sample consisting of 30,000 objects.  Right panel:  shows the distribution of sources in $M$ and $Z$ which is well defined by the respective  faint and bright apparent magnitude limits $m_{\lim}^{\rm b}$=14.5 and  $m_{\lim}^{\rm f}$=17.45~mag.  Left panel: beginning the test at the bright magnitude limit, $T_c$ and $T_v$ are estimated for each trial apparent magnitude limit, $m_*$,  moving toward and beyond the faint limit of the survey in  increments of 0.1.  The $y$-axis indicates the completeness limits, where, for a complete sample, it is expected to fluctuate about zero. The true limit of the survey will be indicated by a systematic drop in  $T_c$ or  $T_v$ below pre-determined limits, in this case $-3\sigma$. The vertical dashed lines indicate the respective limiting magnitudes of the survey sample,  $m_{\lim}^{\rm b}$ and  $m_{\lim}^{\rm f}$.
}
      \label{fig:tcsdss}
    \end{center}
  \end{figure*}

Figure~\ref{fig:tcsdss} illustrates how the test works in practice when applied to real survey data.
The sample under test comprises of early-type galaxies  selected from  the SDSS as described in \cite{Bernardi:2003a,Bernardi:2005}.  To summarise, the galaxies have been selected within the magnitude range $14.5 < m_r <17.75$ within a redshift range $0.05 < z < 0.3$. The resulting distribution in absolute magnitude $M$ and distance modulus $Z$ is shown on the right-hand panel of Figure~\ref{fig:tcsdss}. It should be noted that the magnitudes have been K-corrected.  The left-hand panel shows the resulting completeness test for $T_c$ (black solid line) and $T_v$ (red solid line).  Since this survey sample is well defined by a faint {\it and} bright apparent magnitude limit, the \citet*{Johnston:2007} estimators are applied, adopting a respective fixed $\delta Z$ and $\delta M$ width of 0.2 for $T_c$ and $T_v$. For both estimators the test begins with a trial magnitude limit equal to the bright limit of the sample $m_*=14.5$.  $T_c$ and $T_v$ are then estimated for increasing values of $m_*$ in increments of 0.1.  Note that the test does not register a result  for either estimator until there are enough objects to be counted for the $\zeta$ and $\tau$ random variables.  In this example this corresponds to an $m_*=14.8$.  As the figure shows, both $T_c$ and $T_v$ fluctuate about zero within approximately the $|1|\sigma$ limits. Both test statistics then systematically drop sharply below $-3\sigma$  at the actual magnitude limit of the sample. For this sample under test the completeness test indicates that there are no residual systematics between $14.5 < m_r <17.75$ and that the data are therefore complete in apparent magnitude up to the published survey limit.

In \citet*{Teodoro:2009} extensions to the method were made to  generalise the approach.  This extension based the completeness calculation on a minimum (or constant) signal-to-noise level computed directly from the $\zeta$ (and/or $\tau$) defined regions.  With the introduction of a secondary bright limit (Figure~\ref{fig:rauzy_mz}-right)  one is now restricted with the volume subsample size for each galaxy.   In studying the distribution of galaxies in the ($M ,Z$) plane one is  trying to understand the underlying luminosity function of a given population of galaxies as well as the way that such a function is sampled.  To do so one uses a finite number of galaxies to estimate whatever measurements. This makes them prone to shot noise and also to the existence of some spurious features of the global properties of the data-set.   
%
%
%END OF FILE			%section 10
\section{Applications}\label{sec:applications}
This review has focussed primarily on the statistical methodology surrounding the  luminosity function in the context  galaxy populations.  This section now focuses on selected areas  within astronomy where the LF has had a strong impact.  

It should firstly be noted that the LF  has not been  confined just to the study of  galaxy populations. The   stellar LF  has had an equally long history \citep[see e.g.][for early studies]{Salpeter:1955ApJ...121..161S,Sandage:1957ApJ...125..422S,Schwarzschild:1958ApJ...128..348S,Limber:1960ApJ...131..168L,Simoda:1968ApJ...151..133S,Miller:1979ApJS...41..513M}. In particular, the stellar LF has proven useful when constraining stellar evolution rates  \citep[e.g.][]{Sandquist:1996ApJ...470..910S,Zoccali:2000AA...358..943Z,Hargis:2004ApJ...608..243H}, age and distance determination to globular clusters as well as  constraining stellar evolution processes \citep[see e.g.][]{Bahcall:1972ApJ...177..647B,Renzini:1988ARAA..26..199R}. More recently the Century Survey Galactic Halo project utilised the  spectroscopic redshift surveys, SDSS \citep{York:2000} and 2MASS \citep{Cutri:2003tmc..book.....C},  to survey blue horizontal branch (BHB) stars within the Galactic halo \citep{Brown:2003AJ....126.1362B}. In \cite{Brown:2005AJ....130.1097B} they applied the \citetalias{Efstathiou:1988} step-wise LF estimator to construct the BHB LF and compare with globular cluster  data.  Other studies by e.g. \citet{Covey:2008AJ....136.1778C} explores the LF and mass functions of low mass stars extracting data from the same surveys.  The Hubble Space Telescope has also been pivotal for the study of young star clusters within the Milky Way, and their corresponding LF studies have proved useful in probing the underlying mass function \citep[e.g.][]{1999ApJ...527L..81Z,2006A&A...450..129G,2008AJ....135.2095D,2010ARA&A..48..431P}. The white dwarf luminosity function (WDLF) is another galactic source which provides useful constraints on star-formation rate and history of the Galactic disk \citep[e.g.][]{Liebert:1988ApJ...332..891L,Wood:1992ApJ...386..539W,Harris:2006AJ....131..571H}.

In terms of extragalactic sources, there are extensive LF studies  covering a  broad  range wavelengths which include:  UV selected Lyman break galaxies (LBGs) to probe the very high redshift Universe out to $z\sim7$ \citep[e.g.][]{Steidel:1996ApJ...462L..17S,Steidel:1999ApJ...519....1S,Madau:1996MNRAS.283.1388M,Sawicki:2006ApJ...642..653S,Bouwens2008ApJ...686..230B,Stanway:2008MNRAS.386..370S};   HI selected galaxies \citep[e.g.][]{2001MNRAS.327.1249Z}; and the radio LF \citep[e.g.][]{2001MNRAS.322..536W,1977A&A....57...41A,2007MNRAS.381.1548K}; mid- and  far-infrared \citep{Hacking:1987ApJ...316L..15H,Saunders:1990,Pozzi:2004ApJ...609..122P,Caputi:2007ApJ...660...97C,Rodighiero:2010AA...515A...8R};  sub-mm \citep[e.g][]{Dunne:2000MNRAS.315..115D,Vaccari:2010AA...518L..20V};  X-ray \citep[e.g.][]{Ueda:2003ApJ...598..886U,Hasinger:2005AA...441..417H,Brusa:2009ApJ...693....8B,Aird:2010MNRAS.401.2531A}; and   gamma ray bursters (GRBs) \citep[e.g.][]{2001ApJ...552...36S,2010MNRAS.406.1944W}.

The remainder of this section discusses in more detail some of the recent highlights resulting from optical LF studies at low and intermediate redshifts as well  the LF contribution to probing the evolutionary processes of  QSOs.
\subsection{The optical luminosity function}
The review by \citetalias{Binggeli:1988} gave a very comprehensive overview of the impact of galaxy  LF studies up to the end of the 1980s. This section now examines the work carried out by various groups from the 1990s to present day, and summaries the  developments in probing the optical LF from  low redshift ranges ($z\lesssim0.2$) up to the more intermediate redshifts $z\lesssim1.5$.
\subsubsection{The local universe}
Prior to the emergence of large redshift surveys such as the Two Degree Field Galaxy Redshift Survey (2dFGRS) and the Sloan Digital Sky Survey (SDSS), obtaining large complete samples of the local Universe was largely limited to the technological constraints of obtaining spectroscopic redshifts.   Nevertheless, throughout  the 1990s surveys such as Stromlo-APM Redshift Survey (S-APM) \citep{Loveday:1992ApJ...390..338L}, Las Campanas (LCRS)  \citep{Lin:1996ApJ...464...60L} and the ESO Slice Project (ESP) \citep{Zucca:1997AA...326..477Z} made significant progress to constrain the LF to $0.05<z<0.2$ of field galaxies.  
\begin{figure}
    \begin{center}
      \includegraphics[width=0.7\textwidth]{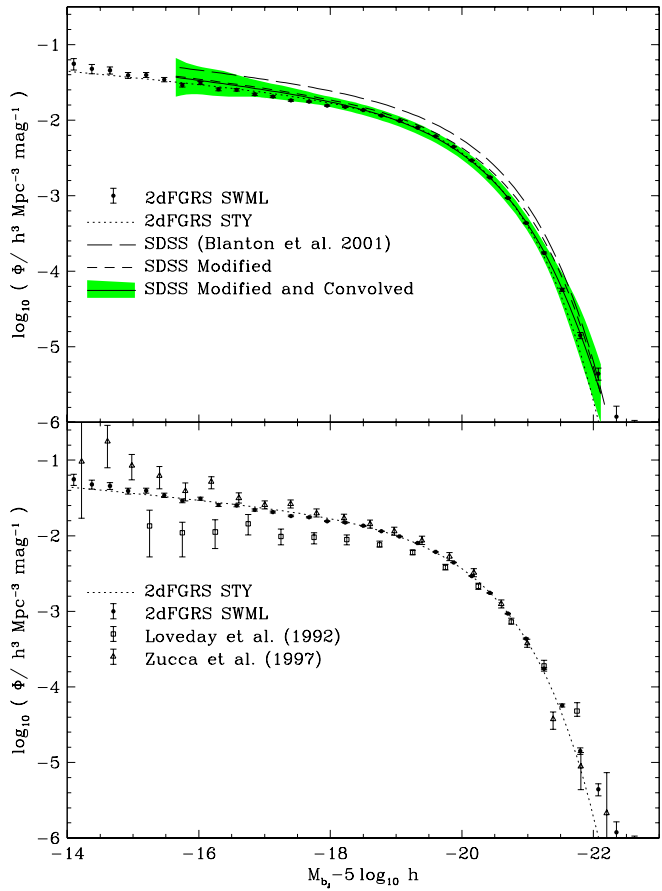}
          \caption{\small  Reproduced from \cite{Norberg:2002b} by permission.  The top panel shows the 2dFGRS LF results using the non-parametric SWML and  the parametric MLE estimators and compared to the SDSS LF results from \cite{Blanton:2001AJ....121.2358B}. The bottom panel shows comparisons with  \cite{Zucca:1997AA...326..477Z} and  \cite{Loveday:1992ApJ...390..338L}.}
          
      \label{fig:2dflf}
    \end{center}
  \end{figure}

In 1992 the S-APM survey measured $b_J$ magnitudes of 1769 galaxies out to $b_J=17.15$~mags at a redshift $z\sim1.0$.  They applied the MLEs of \citetalias{sandage:1979} and \citetalias{Efstathiou:1988} to estimate the LF and constrained Schechter parameters  $M^*=-19.50\pm0.13$, a faint end slope of $\alpha=-0.97\pm0.15$ and a normalisation $\phi^*=(1.4\pm0.17)\times10^{-2}$  Mpc$^{-3}$.  LF results from the CfA survey were then published by \cite{Marzke:1994ApJ...428...43M} using a sample of 9063 galaxies out to $m_z\le 15.5$ and within a range $-13\le M_z\le -22$.  By applying the  \citetalias{sandage:1979}  and the \citetalias{Efstathiou:1988} methods they determined an overall   $M^*=-18.8\pm0.3$,  $\alpha=-1.0\pm0.2$ and a normalisation of $\phi^*=(4.0\pm1.0)\times10^{-2}$  Mpc$^{-3}$ \citep[see also][for their extended analysis to morphological types]{Marzke:1994AJ....108..437M}.  By 1996, the LCRS had imaged 23,690 objects from which 18,678 galaxies were selected out to a limiting $r$-band magnitude of $r\lesssim17.5$ and an average redshift of $z=0.1$. It was one of the first to use multi-fibre spectroscopy, allowing between 50 and 112 redshifts to be measured simultaneously.  They too adopted the same LF estimators to the data and measured LF parameters roughly consistent with S-APM survey. Table~4 shows a summary of the results to compare to other redshift surveys. 

 One year later, the ESP survey published LF results that pushed approximately  2 magnitudes fainter, out to $b_J=19.4$~mag but with a smaller sample of  3342 galaxies out to  a median redshift at $z\sim1.0$.  To estimate the LF they use the \citetalias{sandage:1979} MLE and the modified non-parametric Lynden-Bell $C^-$ estimator called, `$C^+$', as described by Equation~\ref{equ:zucca} on page~\pageref{equ:zucca}. Whilst the $M^*$ and $\phi^*$ values were consistent with the previous two surveys, the faint-end slope was considerably different, yielding a slope of  $\alpha=-1.22$.  The discrepancy in the faint-end slopes was thought to be more an effect of incompleteness and sample variance toward the faint end in LCRS and S-APM. 

However, as multi-fibre technology improved surveys like the 2dFGRS \citep{Colless:1998} marked a new era of mapping the local Universe. With a  final catalogue of $\sim220,000$ measured spectroscopic redshifts out to a $z_{\max}\sim0.3$ and a limiting magnitude of $b_J \lesssim19.45$~mag, the 2dFGRS was, at this point, the largest redshift survey compiled.  Initial LF analysis was performed by \cite{Folkes:1999MNRAS.308..459F} using a preliminary sample of 5869 galaxies in the range $0.01\le z\le0.2$ to $b_J=19.45$~mag. They applied the standard Schmidt $V_{\max}$ and \citetalias{sandage:1979} estimators and constrained overall values of  $M^*=-19.7$ and $\alpha=-1.3$, which showed a much closer consistency with the ESP LF parameters. As the survey reached its conclusion,   later studies by \cite{Madgwick:2002MNRAS.333..133M} (examining the LF as a function of spectral type) and  \cite{Norberg:2002b}, using respective sample sizes of 75,000 and  110,500 galaxies, would  confirm these original results with global LF parameters of $M^*=-19.66$, $\alpha=-1.21$ and , $\phi^*=(1.21\pm0.03)\times10^{-2}$  Mpc$^{-3}$  and a luminosity density of  $j_{b_J}=$ ($1.82\pm0.17$) $\times10^8h$$L_\odot$Mpc$^{-3}$ from \cite{Norberg:2002b}. 

Finally, the advent of SDSS \citep{York:2000} brought, for the first time, a single survey imaging the sky in five bands ($u'$,$g'$,r$'$,$i'$,$z'$ centred at 3540, 4770, 6230, 7630 and 9130 $\AA$ respectively). It overtook the 2dFGRS in terms of the number of objects and remains, for the time being, the largest photometric and spectroscopic survey compiled to date.  With  the final catalogue containing $\sim1,000,000$ spectroscopically confirmed galaxy redshifts, SDSS represents the most accurate map of the nearby Universe out to $z\lesssim0.3$.  Particular to this survey was the use of non-standard filters \citep[see][for more details]{Fukugita:1996AJ....111.1748F} which required converting to make direct comparisons to existing surveys. The first LF analysis reported by SDSS  was by \cite{Blanton:2001AJ....121.2358B} using the a sample 11,275 galaxies of the commissioning data.  As well as computing LFs  all the SDSS bands, they were able to use  the five-band colour information to convert their magnitudes into previous survey bands to make an accurate comparison of previous LF results.  They converted to  LCRS  $r$-band  and $b_J$ (to compare to 2dF).  These results are also shown on Table~4. Generally,  their measured LF parameters were closer to that of 2dF than LCRS. However, they noted their luminosity densities were 1.4 times  higher to that of 2dF, as measured by \cite{Folkes:1999MNRAS.308..459F} reporting  $j_{r}=$ ($2.6\pm0.3$) $\times10^8h$$L_\odot$Mpc$^{-3}$, concluding that previous surveys had missed a considerable fraction of luminosity density in the Universe. Figure~\ref{fig:2dflf} shows LF comparisons of the  S-APM, ESP and the SDSS \cite{Blanton:2001AJ....121.2358B} survey samples taken from the  \cite{Norberg:2002b} paper.

Around two years later, \cite{Blanton:2003ApJ...592..819B} compiled a  sampled from the second data release SDSS-DR2 consisting of 147,986 galaxies to measure the LF at z=0.1 in all SDSS bands.  The main difference in their analysis with the previous paper was the incorporation of evolutionary model into the LF.  This resulted in an  $r^*$-band Schechter function fit with $M^*=-20.44\pm0.01$ and $\alpha=-1.05\pm0.01$, and a newly computed  luminosity density  of $j_{r}=$ ($1.84\pm0.04$) $\times10^8h$$L_\odot$Mpc$^{-3}$ which was now consistent with the 2dF.

The Millennium Galaxy Catalogue (MGC) \citep{Liske:2003} used $B$-band imaging on the Wide Field Camera on the 2.5m Issac Newton Telescope to produce higher quality photometry compared to 2dF and SDSS.  The survey was designed to be fully contained within the 2dF and SDSS-DR1 fields and imaged 10,095 galaxies out to a limiting magnitude $B_{\rm MGC}<20.0$~mag  within a redshift range $0.013<z<0.18$.  In \cite{Driver:2005} they  applied their bivariate Step-Wise LF estimator (as detailed in \S~\ref{sec:biv_surfdens}) and  constrained Schechter LF parameters $M_*$=19.60 $\pm$ 0.04, $\alpha= -1.13$ $\pm$ 0.02 and a normalisation of $\phi_*=$  1.77 $\pm$ 0.15.	Whilst their $M_*$ value was in line with 2dF, the recovered faint-end slope was flatter.  Their resulting luminosity density of $j_{b_J}=$ ($1.99\pm0.17$) $\times10^8h$$L_\odot$Mpc$^{-3}$ was similar to that of 2dF.

A more recent analysis by   \cite{Montero:2009MNRAS.399.1106M} has included  516,891 galaxies in the $r^*$-band  from SDSS-DR6,  achieving a much  greater redshift completeness than the previous analyses.   Their results made accurate constraints on both the bright and faint ends of the LF with $r^*$-band Schechter fits of  $M^*=-20.73\pm0.04$ and $\alpha=-1.23\pm0.02$.  They reported notable differences in the bright-end of the best-fit  LF of an excess at the $M_u\eqsim-20.5$. The excess then weakens moving in the $g$-band and fades away toward the $z$-band.  It was concluded that this excess may be the result of QSOs.
\subsubsection{Probing evolution at intermediate redshifts}\label{sec:appintz}
Our recent ability to probe toward higher redshifts (z$\lesssim1.0$) with significant enough number statistics, has opened a new window to the evolutionary processes of different galaxy populations through the study of their LFs.  However, it should be noted that  making comparisons between various surveys has been a tricky process largely due to the ways in which  galaxy types have been defined e.g. by spectral type or by morphology \citep[see e.g.][for a discussion]{Ilbert:2006AA...453..809I}.  Nevertheless, over the last 15-20 years, an overall consistent picture of galaxy evolution at this redshift range seems to be reaching convergence \citep[for reviews see][]{koo:1992ARAA..30..613K,Ellis:1997ARAA..35..389E,Faber:2007}.

Through the mid to late 1990s work in this area was largely led by groups such as the Canada-France Redshift Survey (CFRS) \citep{Lilly:1995ApJ...455..108L}, Autofib I \& II \citep{Ellis:1996MNRAS.280..235E,Heyl:1997}, the Hawaii Deep Fields \citep{Cowie:1996AJ....112..839C}, the Canadian Network for Observational Cosmology survey (CNOC1 \& 2) \citep{Lin:1997ApJ...475..494L,Lin:1999}  and the Norris Survey of the Corona Borealis Supercluster \citep{Small:1997ApJ...487..512S,Lin:1999}.  Whilst these survey samples  typically ranged from  hundreds to just under 2000 galaxies, a clear picture of evolution between  `red' early-type galaxies and   `blue' star-forming late-type galaxies had emerged in which  star-forming galaxies were observed to have  more rapid and stronger evolutionary properties compared to early-type galaxies.  Table~5 provides a summary these  and other surveys related to this study.

The CFRS, using a sample of 730 $I$-band selected galaxies, published results that seemed to lend support to  the monolithic collapse scenario \citep[e.g.][]{Eggen:1962,Searle:1978ApJ...225..357S} in which galaxies originate from large regions of primordial baryonic gas which then collapses  to form  stars within the central  region of a dark matter halo, thus allowing  the most massive galaxies to form first.   They found that the LF of red galaxies showed no evidence for either number density or luminosity evolution over the range $0<z<1.0$. However, their results for the blue population of galaxies indicated evolution in the form of a steepening in the faint-end slope of the LF.

The Autofib Redshift Survey explored both the global evolution properties of the B-band LF \citep[Autofib I][]{Ellis:1996MNRAS.280..235E} and evolution by spectral type  \citep[Autofib II][]{Heyl:1997} out to $z\sim0.75$ and covering a range in magnitude $11.5<b_J<24.0$. It was also this paper that saw extensions made to both  the SWML method of \citetalias{Efstathiou:1988} and the MLE of \citetalias{sandage:1979} for the combining of multiple samples (see e.g.  \S~\ref{sec:swml} on page~\pageref{sec:swml}).  The galaxy population were divided into early-type E/S0s, early spirals (Sab/Sbc) and late spirals (Scd/Sdm/starburst).  For the early-types, they found  no real evidence for evolution out to $z\sim0.5$.  The early-type spirals showed some evidence for evolution with slight steepening of the faint end slope of the LF as a function of redshift. Overall there seemed no significant change in both luminosity and number density for the early-type spiral galaxies.  However, the late-type spirals showed significant evolution over the same redshift range in both luminosity and increased number density. Moreover, using [O$_{\rm II}$] emission as an indicator of star-formation, they also found a steep increase in such emission implying a rapid increase in star formation rate as a function of redshift. 

At around the same time, the  CNOC  redshift surveys were being carried out, which by the end of the 1990s utilised a sample of $\sim2000$ galaxies to probe evolution of the LF between $0.12<z<0.55$ within the magnitude range $17.0<R_C<21.5$ \citep[][respectively known as CNOC1 and CNOC2]{Lin:1997ApJ...475..494L,Lin:1999}.  At this time it was the largest intermediate-redshift galaxy survey.  They classified galaxies into early-types, intermediate-types and late-types,  estimating rest-frame $B$, $R_C$, and $U$ band LFs.  Applying both the \citetalias{sandage:1979} and  \citetalias{Efstathiou:1988} estimators they found that when looking back in redshift they observed their early-type sample showing an increasing luminosity evolution but with a reduced number density. Their intermediate-types showed no evidence for number evolution but indicated a slight increase in luminosity. The late-type sample showed strong evidence for number evolution with  stronger star formation at higher redshifts. Overall, they reported distinct separation of luminosity (early-types) and density evolution (late-types).  It was noted that due to differences in both  the way galaxies were classified and evolutionary models adopted, direct comparison of CNOC2 and Autofib was difficult. However, in general the CNOC2 group found their results  qualitatively consistent. 

The  COMBO-17 \citep{Wolf:2003AA...401...73W} survey marked a significant increase in the number of galaxies surveyed from $0.2<z<1.2$, totalling 25,431 measured photometric redshifts out to a limiting magnitude $R<24.0$. They provided a thorough study  of evolution as a function of  LF and comoving luminosity density at different rest-wavelengths with comparisons made to the CFRS, CNOC2 and CADIS \citep{Fried:2001AA...367..788F} surveys.  The galaxies were split into four types based on templates from \cite{Kinney:1996ApJ...467...38K} where Types 1,2,3 \& 4 were, respectively, categorised as E-Sa, Sa-Sbc, Sbc-SB6 \& SB6-SB1. After accounting for the differences in galaxy-type classification, they found their results to  be consistent with CNOC2.  They reported  evolution of the LF was strongly SED type dependent  in both density normalisation and characteristic luminosity.  For their early-type sample (Type 1) they found a faint-end slope of $\alpha=0.5\pm0.2$ which then steepened to $\alpha=1.5\pm0.06$ for bluer galaxies (Type 4). For the latest-type galaxies they found no real change in the density normalisation, $\phi^*$ over their redshift range, but instead, found a progressively faint $M_*$.  Conversely, their sample of early-type galaxies they found an order magnitude increase in $\phi^*$ from $z=1.2$ to present day. $M_*$, however, did become progressively fainter, albeit to a smaller degree compared to the latest-type galaxies. 

\cite{Bell:2004ApJ...608..752B} described a model independent method for segregating galaxies into early- and late-types by identifying  the bimodal rest-frame color distributions of galaxies out to $z\sim1$.  This would prove an invaluable approach for future studies. 

The VIMOS-VLT Deep Survey  (VVDS) \citep[e.g.][]{Ilbert:2005} combined spectroscopic redshifts in the  Chandra Deep Field South and photometric redshifts from COMBO-17 and multi-colour images from the HST/ACS Great Observatories Origin Deep Survey (GOODS) to probe evolution out to $z\sim1.5$.  Analysis of this survey was performed in two different ways: in \cite{Zucca:2006AA...455..879Z} they classified galaxies according to spectral type, in  \cite{Ilbert:2006AA...453..809I} they instead look at galaxy morphology for classification.

As with COMBO-17 \cite{Zucca:2006AA...455..879Z} split their sample population of 7713 galaxies into four spectral types, but based their classifications on different templates using CCW  \cite{Coleman:1980ApJS...43..393C}. In this way the four types were classified as:  E/S0 (Type 1), early spirals (Type 2), late spirals (Type 3) and irregulars (Type 4).  For their LF studies they applied their ALF algorithm and  compared their results with previous surveys such as CNOC2 and COMBO-17.  The comparisons indicated general agreement with CNOC2 in terms of negative density evolution for early-types and a strong positive density evolution for late-type galaxies.  However, unlike CNOC2, they did not observe luminosity evolution for early-types.  Their comparison to COMBO-17 did yield notable differences in the shapes and normalisation of the LFs. For example, the main discrepancy that was reported concerned the evolution of the Type 1 galaxies.  In COMBO-17 they saw very strong density evolution which decreased with increasing redshift. Such evolution was not observed in the VVDS sample.  It was thought this may be due to differences in galaxy classification. As a test they consolidated their Type 1 and 2 populations and re-estimated their LFs. Whilst this seemed to reduce the discrepancy the differences in the slopes and normalisation of the LFs remained.

In  \cite{Ilbert:2006AA...453..809I}  their classification scheme of the VVDS data was morphological, based on structural features of  galaxies within which three main populations were established - disk-dominated populations, red bulge-dominated populations and blue bulge-dominated populations. As in \cite{Zucca:2006AA...455..879Z} they applied their ALF algorithm to compute the LFs and presented results from the rest-frame $B$-band LF out to $z=1.2$. In general they observed a strong dependency on the shape of the LF according to morphological type. At  redshift range $0.4<z<0.8$ the red bulge-dominated population LFs that indicated a shallow faint-end slope of $\alpha=0.55\pm0.21$. This was  was in stark contrast to the disk population which showed a much steeper slope of $\alpha=-1.19\pm0.07$.  Due to the irregular galaxies being included in the disk population sample, it was expected that strong evolution would be observed. Instead, the opposite was found. It was thought the effects such as cosmic variance and the domination of spiral galaxies  in this sample were possible reasons for this effect. With the red bulge-dominated population they found that in general there was a distinct number density evolution with the age of the Universe consistent with the hierarchical scenario, and indicating that E/S0 galaxies are already established at $z\sim1$. Finally the blue bulge-dominated population was found to have strong evolution with a brightening of $\sim0.7$ mag between $0.6\lesssim z \lesssim1.0$. 

At around the same time as VDDS the Deep Extragalactic Evolutionary  Probe 2 (DEEP2) \citep[e.g.][]{Willmer:2006ApJ...647..853W} survey team reported their $B$-band  LF results of $\sim 11,000$ galaxies out to $z\sim1$. As well as examining the global LF they also provide analysis for subdivided blue and red populations (using colour bimodality). As with previous studies their findings  showed red and blue galaxies evolving differently. In general, the blue galaxies show strong luminosity evolution but little number evolution, whereas the red galaxies show strong number evolution with little change in luminosity over the redshift range. Their results showed similar trends to previous surveys such as COMBO-17.

\cite{Faber:2007} combined data from DEEP2 and COMBO-17 to produce a catalogue of 39,000 galaxies which helped reduce effects from  cosmic variance and Poisson noise by 7\% and 15\% per redshift bin. Using the colour bimodality approach as first employed by \cite{Bell:2004ApJ...608..752B}, galaxies were classified into blue and red sub samples. They found that for a fixed number density moving toward higher redshifts, the blue population showed a brightening in magnitude. The red galaxies, however, showed almost no evolution for a fixed absolute magnitude. Their Schechter LF fits showed good agreement with the  original analysis in both DEEP2 and COMBO-17 at all redshifts.  When combining their distant Schechter LF parameters with local estimates they concluded that the number density of blue galaxies has remained the same since $z=1$, whilst the red galaxies have shown an increase in number density by a factor of $\sim2$ since $z\sim1$.  This result is restricted to galaxies near $L_B^*$  and does not extend to more luminous galaxies.

Finally, the more recent and ongoing COSMOS survey \citep{Scoville:2007ApJS..172....1S} and the redshift follow-up zCOSMOS \citep{Lilly:2007,Zucca:2009} have published LF studies have using  the 10k bright sample which, so far, consists of 10, 644 objects  \citep{Zucca:2009}. This paper estimates both the  global LF (gLF) in the range $0.01 <z <1.3$ and the LF as function of both spectrophotometric  and morphological type.   In terms of spectrophotometric typing, galaxies were divided into similar categories as in \cite{Zucca:2006AA...455..879Z} with Type 1 = E/S0, Type 2 = early spirals, Type 3  = late spirals and Type 4 = irregular and starburst galaxies.  As in previous studies, the ALF algorithm was applied to estimate all LFs. To allow for better constraints on luminosity evolution, the parameter $\alpha$ was fixed to the value obtained in the redshift range $0.30<z<0.80$.  Overall, they found that evolutionary properties in both luminosity and number density was consistent with previous VVDS studies. To better constrain evolutionary trends their Type 3 and 4 populations were combined.  With this the Type 1 galaxies showed both luminosity and number evolution with a brightening of $M_*$ by $\sim0.6$ mag and a decrease in $\phi_*$ from the lowest redshift bin [0.1 to 0.35]  to the highest [0.75 to 1.00]. Whilst the Type 2 population showed an overall brightening of $\sim0.25$ mag, there was no significant number evolution observed. Finally, the Type 3 + 4 galaxies showed evolution in both number and luminosity with a brightening  of $\sim0.5$ mag and an increase of $\phi_*$ with redshift.

Their morphological studies saw galaxies classified into three groups: early-types, spirals and irregulars, and were analysed over the same redshift range as the spectrophotometric  types. The summary of these results showed that early-type galaxies dominate the bright-end of the LF at low redshifts ($z<0.35$), whilst spirals tended to dominate at the faint-end. Their sample showed that at the redshift range $0.35<z<0.75$ a positive luminosity evolution was observed in both the spiral and early-type populations. At the highest redshift range, $z>0.75$, the irregulars were observed to show strong increase in  number evolution toward high redshifts. 
\subsection{The QSO LF}
Quasi-stellar objects (QSO) were first identified  in 1963 by Maarten Schmidt and early pioneering work by e.g. \cite{Schmidt:1963Natur.197.1040S,Schmidt:1968,Schmidt:1972,Schmidt:1983ApJ...269..352S} confirmed that they had strong evolutionary properties with a steep increase in space densities with increasing redshift.  Central to understanding their evolutionary process is estimating the  quasar luminosity function (QLF) as a function of redshift. 

Research in this field  has garnered renewed interest in recent times due, in part, to surveys such as the  2dF and the SDSS which increased the number of QSO detections by orders of magnitude, vastly improving on number statistics, providing more comprehensive catalogues to fainter limiting magnitudes. Furthermore, recent observational evidence by \citet{Kormendy:1995ARAA..33..581K,Magorrian:1998AJ....115.2285M,Ferrarese:2000ApJ...539L...9F} has established a relationship between the evolution of galaxies with their central supermassive black holes.  Thus, the QSO luminosity function (QLF) can potentially provide strong constraints on SMBH formation and evolution  \citep{Haiman:2001ApJ...552..459H,Yu:2002MNRAS.335..965Y}, constraints on structure formation models  coupled with  the environmental  processes such as AGN feedback \citep[e.g.][]{Small:1992MNRAS.259..725S,Haehnelt:1993MNRAS.263..168H,Efstathiou:1988MNRAS.230P...5E,Kauffmann:2000MNRAS.311..576K}.   

Work by  \cite{Boyle:1988ASPC....2....1B,Boyle:1988MNRAS.235..935B} originally showed that at low redshifts the QLF can be well described by a broken double power-law of the form
\begin{equation}\label{equ_LFQSO}
\Phi (M,z) = \frac{{\Phi ^* }}{{10^{0.4[M - M^*](\alpha  + 1)}  + 10^{0.4[M - M^*](\beta  + 1)} }}
\end{equation}
with a break at $M^*$ and a bright-end slope  $\alpha$ steeper than  the faint-end slope  $\beta$.

The work performed during the 1980s and early 1990s using parametric fits to data had built a picture for the  QSO LF at redshifts $z\lesssim2.2$ which  was described by a steep bright end slope ($\alpha\sim-3.6$) and  a very flat faint-end slope ($\beta\sim-1.2$)  and showed strong evolutionary properties that could be described by fitting pure luminosity evolution models  (PLE ; see \S~\ref{sec:evolution}) with a peak in space density at $z\sim$ 2 - 3 \citep[e.g.][]{Osmer:1982ApJ...253...28O,Marshall:1983,Marshall:1985ApJ...299..109M,Koo:1988ApJ...325...92K,Boyle:1990,Hartwick:1990ARAA..28..437H,Warren:1994ApJ...421..412W}.    The understanding of how the QLF  evolves  with redshift  was put on a more solid  footing with the advent of surveys that brought the number of detections into the  tens of thousands allowing accurate determination of the QLF at $z<2.5$.

The AGN LF  has also been studied extensively in the X-ray.  Whilst early measurements of the X-ray AGN LF by \cite{Boyle:1993MNRAS.260...49B} showed evidence that evolution was following the standard  PLE  model, more recent results have also reported the AGN downsizing. In this scenario lower luminosity AGN peak in their comoving space density at lower redshifts ($z\lesssim1$), whilst higher luminosity AGN peak at higher redshifts $z\sim2$ \citep[e.g.][]{Ueda:2003ApJ...598..886U,Barger:2005AJ....129..578B,Hasinger:2005AA...441..417H,Babic:2007AA...474..755B,Brusa:2009ApJ...693....8B}

The advent of the The 2-degree Field QSO redshift survey  (2QZ) \citep{Boyle2000MNRAS.317.1014B,Croom:2001MNRAS.322L..29C,Croom:2004MNRAS.349.1397C} provided a substantial increase in the number of QSOs cataloguing a total of 23,338 within the magnitude range $18.25<b_J<20.85$ out to $0.4<z<2.1$.  However, it was not until the 2dF-SDSS LRG and QSO (2SLAQ) survey \citep{Richards:2005MNRAS.360..839R,Croom:2009MNRAS.392...19C} that evidence of downsizing was also observed. The 2SLAQ survey combined  photometry from SDSS-DR1 \citep[e.g.][]{Gunn:1998AJ.116.3040G,Stoughton:2002AJ.123.485S,Abazajian:2003AJ.126.2081A} to produce a catalogue of 10,637 QSOs. 2SLAQ provided a  sample 1 magnitude deeper than 2QZ  allowing to probe the faint end of the QLF out to $b_J<21.85$~mag and to a redshift of $z<2.6$.  By applying a similar approach to construct the QLF as in \cite{Croom:2004MNRAS.349.1397C}, they provided constraints on both the faint and bright end of the QLF, finding evidence for number evolution in the form of  downsizing which had not been observed in earlier studies by e.g. \cite{Richards:2005MNRAS.360..839R} (using early results from 2SLAQ) and \cite{Wolf:2003AA...401...73W}  (using the COMBO-17 survey).  Using this catalogue,  \cite{Croom:2009MNRAS.399.1755C} showed how the activity in low-luminosity active galactic nuclei (AGNs) peaks at a lower redshift than that of more luminous AGNs. 

Accurately  constraining  the QLF at  $z\ge3$ is much more  of a challenge where low numbers of low-luminosity objects  at high redshift  have meant faint end of the QLF is much more poorly understood. As such, whether or not downsizing is observed at earlier epochs remains a pertinent question.   However, there have been made a lot of efforts to build larger catalogues   \citep[e.g.][]{Fan:2000AJ....120.1167F,Fan:2001AJ....122.2833F,Fan:2003AJ....125.1649F,Fan:2004AJ....128..515F,Richards:2006,Goto:2006MNRAS.371..769G,Siana:2008,Glikman:2010ApJ...710.1498G,Willott:2010AJ....139..906W}.  There have been a number  of studies of the faint-end slope at $z\sim3$  \citep[e.g.][]{Hunt:2004ApJ...605..625H,Bongiorno:2007AA...472..443B,Fontanot:2007AA...461...39F,Siana:2008}. However, recent work by \cite{Ikeda:2010arXiv1011.2280I} and initial studies by \citet{Glikman:2010ApJ...710.1498G}   have both provided constraints on the QLF at $z\sim 4$ that, at first, yielded conflicting results. 

 \citet{Glikman:2010ApJ...710.1498G} using both Deep Lens Survey \cite{Wittman:2002SPIE.4836...73W} and the  NOAO Deep Wide-Field Survey \citep{Jannuzi:1999ASPC..193..258J} found 23 QSOs candidates between $3.74 < z< 5.06$ down to a limiting $R\leq23$ magnitude.   To construct the QLF they applied the standard $V_{\max}$ estimator and also use the \citetalias{sandage:1979} MLE to constrain the LF parameters of Equation~\ref{equ_LFQSO}. They found that when combining their QLF data with that of  the  \cite{Richards:2006} SDSS data,  the QLF at $z\sim4$ was best described by a single power-law with  a $\beta=-2.3\pm0.2$. However, \cite{Ikeda:2010arXiv1011.2280I}, using the COSMOS survey, identified eight low-luminosity QSOs at this redshift within a magnitude range $22<i'<24$ using the FOCAS instrument on the Subaru Telescope. By applying a least-squares weighted fit to the QLF of Equation~\ref{equ_LFQSO},  their results yielded a much shallower faint-end slope of $\beta=-1.67^{+0.11}_{-0.17}$, which they found to be more consistent with the downsizing scenario of AGN. Ikeda  et al.  suggested potential contamination of the Glikman et al. sample being the cause for the large discrepancy. However, follow-up  work by \cite{Glikman:2011ApJ...728L..26G} examined an increased  sample  of a further 40 candidates. Whilst they have now  found a $\beta=-1.6^{+0.8}_{-0.6}$  being in agreement with Ikeda, their normalisation  is higher by a factor of 4. It will be interesting to see how these results will converge as the number statistics improve in the future.
  
Work continues exploring the most distant QLFs out to  $z\sim 6$ using  $i$-band dropout techniques \citep[see e.g.][]{Fan:2003AJ....125.1649F,Fan:2004AJ....128..515F,Richards:2004AJ....127.1305R,Fan:2006AJ....132..117F,Jiang:2009AJ....138..305J,Willott:2010AJ....139..906W}. The SDSS and the Canada-France High-$z$ Quasar Survey (CFHQS)  \citep[e.g.][]{Willott:2009AJ....137.3541W} have increased the number of high redshift quasars at $z\sim6$ to $\sim60$. Work by  \citet{Fan:2003AJ....125.1649F,Fan:2004AJ....128..515F} and \cite{Richards:2004AJ....127.1305R} indicated a bright-end slope reaching a value close to $\beta\sim-4$.  Using SDSS samples,  \cite{Jiang:2009AJ....138..305J} found a significantly shallower slope at $\beta=-2.6\pm0.3$.   It was suggested by   \cite{Willott:2010AJ....139..906W}  that there may be a break in the QLF at $M_{1450}\approx-26$ which could account for this discrepancy; a feature that they also found in their analysis using data from the  Canada-France High-z Quasar Survey (CFHQS) combined with SDSS that  identify 40 QSOs candidates at redshift range $5.74 <z<6.42$.  In their analysis, they found a bright-end slope of $\beta=-2.81$.

 At the more intermediate redshift ranges, the QSO LF seems very well constrained and significant progress has already been  made in probing the high redshift QSO LF. However, it is clear that, in general, many more QSOs identifications are required to effectively constrain QLFs at earlier epochs if they are to be used in constraining AGN and  SMBH formation evolutionary models.

\begin{landscape}
\begin{deluxetable}{lcccccccc}
\small
\tablewidth{0pc}
\tablecaption{Summary of LF parameters from selected redshift surveys at low redshifts.  The columns are as follows: (1) the paper from which the LF parameters were computed, 
(2) the limiting apparent magnitude for the survey, (3) number of galaxies in the sample, (4) the  redshift the LF was computed at, (5) (6) \& (7) are the LF parameters, and (8) lists the LF estimators used: 
\citetalias{sandage:1979} is the parametric maximum likelihood estimator, \citetalias{Efstathiou:1988}  is the Step Wise MLE and $C^+$ is the modified $C^-$ estimator.  }
%\label{table:lf_params}
\tablehead{
\\
\colhead{Survey Data} 	&
\colhead{$m_{\lim}$}&
\colhead{${N_{\rm gal}}$} &
\colhead{Redshift}    &
\colhead{$M^*$ }      &
\colhead{${\alpha}$}  	&
\colhead{${\phi^*}$}   	&
\colhead{LF Method} 	&
\\ 
& & &$(z)$ & &($\times$10$^{-2}$$h^3$Mpc$^{-3}$) &  
}
\startdata 
\\
{\bf LF studies out to ${\bf z\lesssim0.2}$} 
\\
\\
 S-APM   \citep{Loveday:1992ApJ...390..338L}   		&$b_J=17.15$		&1769        & $z\sim0.1$		&-19.50 $\pm$ 0.13  	&  -0.97 $\pm$ 0.150  	&   1.40 $\pm$ 0.13 	 	&\citetalias{sandage:1979}, \citetalias{Efstathiou:1988} &\nl
 CfA   \citep{Marzke:1994ApJ...428...43M}   			&$m_z=15.55$		&9063      & $z\sim0.05$		&-18.80 $\pm$ 0.3  		&  -1.00 $\pm$ 0.2	  	&   4.00 $\pm$ 1.0 	 	&\citetalias{sandage:1979}, \citetalias{Efstathiou:1988} &\nl
LCRS    \citep{Lin:1996ApJ...464...60L}            			&$r\lesssim17.50$	&18,678     &$z\sim0.1$        	& -20.29 $\pm$ 0.02 	&  -0.70 $\pm$ 0.05   	&   1.90 $\pm$ 0.10      	&\citetalias{sandage:1979}, \citetalias{Efstathiou:1988} &\nl
ESP   \citep{Zucca:1997AA...326..477Z}      			&$b_J=19.40$		& 3342       &$z\lesssim0.2$      	& -19.16			  	&      -1.22 	             	&   	2.00				&$C^+$, \citetalias{sandage:1979}   & \nl
2dFGRS \citep{Folkes:1999MNRAS.308..459F}        		&$b_J=19.45$		& 5869	 &$z\sim0.2$		& -19.70   			  	&      -1.30    		     	&    1.70				&\citetalias{sandage:1979}, $1/V_{\max}$ &  \nl
2dFGRS \citep{Madgwick:2002MNRAS.333..133M}		&$b_J=19.45$		&75,589	 &$z\le0.15$		& -19.79 $\pm$ 0.04		& -1.19 $\pm$ 0.01  		&  1.59 $\pm$ 0.1		&\citetalias{sandage:1979}, \citetalias{Efstathiou:1988} &\\
2dFGRS \citep{Norberg:2002b}                           		 	&$b_J=19.45$		&110,500	&$z<0.2$			& -19.66 $\pm$ 0.07  	&  -1.21 $\pm$ 0.03  		&    1.61 $\pm$ 0.08		&\citetalias{sandage:1979}, \citetalias{Efstathiou:1988}&  \\\\
SDSS-DR1  \citep{Blanton:2001AJ....121.2358B}   	 	&$r^*=17.60$		&11,275	&$z\lesssim0.2$	& -20.83 $\pm$ 0.03	  	&    -1.20 $\pm$ 0.03    	&1.46 $\pm$ 0.12		&\citetalias{sandage:1979}, \citetalias{Efstathiou:1988}, $1/V_{\max}$& \nl
\quad\quad $b_J$ converted SDSS Petrosian			&$b_J=17.80$		&-		&-				&-19.70 $\pm$ 0.04		& -1.22 $\pm$ 0.05    	& 2.69 $\pm$ 0.34 		& -&\\
\quad\quad $r$ converted SDSS Petrosian			&$r=17.40$		&-		&-				&-20.80 $\pm$ 0.03		& -1.17 $\pm$ 0.03    	& 1.92 $\pm$ 0.23		&- &\\\\

SDSS-DR2 \citep{Blanton:2003ApJ...592..819B}        	&$r^*=17.79$		&147,986	&$z=0.1$			& -20.44 $\pm$ 0.01  	&  -1.05 $\pm$ 0.01  	&    1.49 $\pm$ 0.04			& based on  \citetalias{Efstathiou:1988} &  \nl
MGC		\citep{Liske:2003}						&$B=20.0$ &10,095	&$z<0.18$		 &-19.60 $\pm$ 0.04      & -1.13 $\pm$ 0.02  & 1.77 $\pm$ 0.15			& Bivariate SWML &\\
SDSS-DR6      									&$r^*=17.77$		&516,891	&$0.02<z<0.2$		& -20.73 $\pm$ 0.04  	&  -1.23 $\pm$ 0.02  	&    0.90 $\pm$ 0.07			&\citetalias{Efstathiou:1988}  &  \nl
\citep{Montero:2009MNRAS.399.1106M}  \\
\\
\enddata
\end{deluxetable}

\begin{deluxetable}{lccccc}
\small
\tablewidth{0pc}
\tablecaption{Summary of galaxy LF evolution studies at intermediate redshifts. }
%\label{table:lf_params}
\tablehead{
\\
\colhead{Survey Data} 	&
\colhead{${m_{\lim}}$}&
\colhead{${N_{\rm gal}}$} &
\colhead{Redshift}    &
\colhead{LF Method} 	&
\\ 
& & &$(z)$ & &

}
\startdata 
\\
{\bf LF studies out to ${\bf z\lesssim1.5}$} 
\\
\\
CFRS  \citep{Lilly:1995ApJ...455..108L}				&$17.5<I_{\rm AB}<22.50$	&730			&$0<z<1.0$        			 				&\citetalias{sandage:1979}, \citetalias{Efstathiou:1988}  \\
Autofib  I \citep{Ellis:1996MNRAS.280..235E}			&$11.5<b_J<24.0$			&$1700$	&$0<z\lesssim0.75$						&$1/V_{\max}$, \citetalias{Efstathiou:1988}   \\
Hawaii Deep Fields \citep{Cowie:1996AJ....112..839C}	&$K=20.0$, $I=23$, $B=24.5$ & 393		&$0.2<z\lesssim1.7$							&	$1/V_{\max}$ \\\
CNOC1  \citep{Lin:1997ApJ...475..494L}				&variable					&$389$		&$0.2<z<0.6$         					&\citetalias{sandage:1979}, \citetalias{Efstathiou:1988}  \\
Autofib II \citep{Heyl:1997}			&$11.5<b_J<24.0$			&1700		&$0<z\lesssim0.75$						&\citetalias{sandage:1979}, \citetalias{Efstathiou:1988}  \\	
NORRIS \citep{Small:1997ApJ...487..512S}			&$14\lesssim r\lesssim 20.0$	&493		&$0<z\le 0.5$							&\citetalias{Efstathiou:1988} \\
HST imaging of CFRS \& Autofib					&						&341		&$z\lesssim1.0$						&$1/V_{\max}$ \\
 \citep{Brinchmann:1998ApJ...499..112B} \\
CNOC2  \citep{Lin:1999}							&$17.0<R_C<21.5$			&$\sim 2000$	&$0.12<z<0.55$         					&\citetalias{sandage:1979}, \citetalias{Efstathiou:1988}  \\
CADIS \citep{Fried:2001AA...367..788F}				&$I_{815}=30.0$			&2779 		&$0.1<z\lesssim 1.0$					&$1/V_{\max}$, \citetalias{sandage:1979} \\
CFGRS \citep{Cohen:2002ApJ...567..672C}			&$R<24.0$				&553		&$0.01<z<1.5$							&\citetalias{sandage:1979} \\
DEEP Groth Strip Survey \citep{Im:2002ApJ...571..136I}	&$16.5<I<22.0$			&145		&$z\lesssim1.0$						&$1/V_{\max}$, \citetalias{sandage:1979} \\
COMBO-17 \citep{Wolf:2003AA...401...73W}			&$R\le24.0$				&$\sim25,000$	&$0.2<z<1.2$								&$1/V_{\max}$, \citetalias{sandage:1979} \\
ESO-S  \citep{Lapparent:2003AA...404..831D}   		&$R_C\le20.5$				&617 		&$0.1<z<0.5$								&\citetalias{sandage:1979}, \citetalias{Efstathiou:1988} 	\\		
VIMOS VLT Deep   \citep{Ilbert:2006AA...453..809I} 		&$I_{\rm AB}=24.0$			&$\sim4160$	&$0.4<z<1.0$ 								&$1/V_{\max}$, $C^+$, \citetalias{sandage:1979}, \citetalias{Efstathiou:1988}  \\
VIMOS VLT Deep  \citep{Zucca:2006AA...455..879Z}  	 &$I_{\rm AB}=24.0$			&7713		&$0.05\lesssim z \lesssim1.5$ 	    				&$1/V_{\max}$, $C^+$, \citetalias{sandage:1979}, \citetalias{Efstathiou:1988}    \nl
DEEP2 \citep{Willmer:2006ApJ...647..853W}  			&$R_{\rm AB}\sim25.5$		&$\sim10,000$	&$z\lesssim1.0$			    				&$1/V_{\max}$ \citep{Eales:1993}, \citetalias{sandage:1979}   \nl
zCOSMOS \citep{Zucca:2009}						&$15\le I \le 22.5$						&8478		&$z\lesssim1.0$				&$1/V_{\max}$, $C^+$, \citetalias{sandage:1979}, \citetalias{Efstathiou:1988} \\ \\
\enddata
\end{deluxetable}

\end{landscape}
%section 11
 \section{Concluding Remarks}\label{sec:conclusions}
\begin{figure} 
   \begin{center}
	 \includegraphics[width=1.0\textwidth]{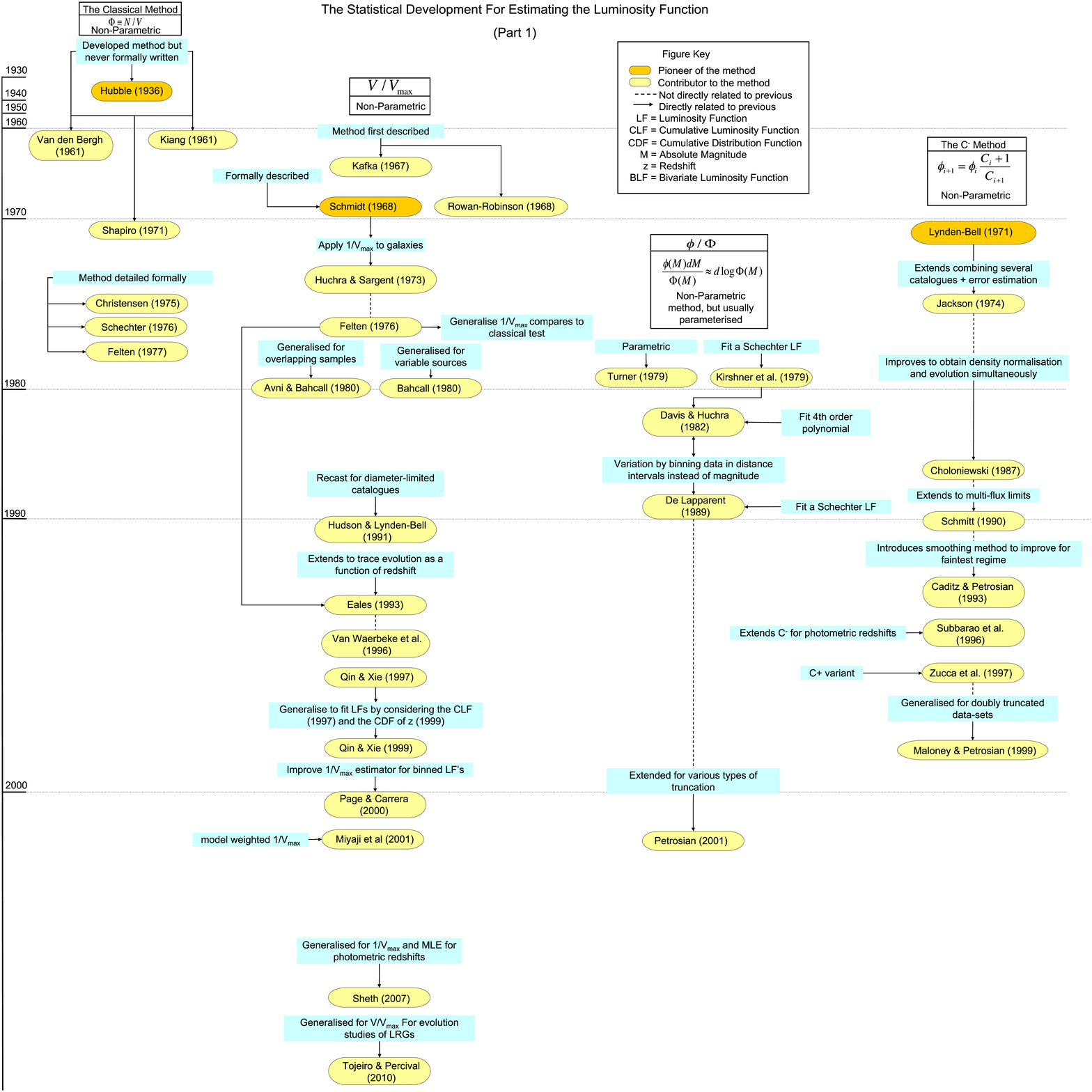}
              \caption {\small Part 1 of a schematic charting the development of all the major statistical methods that estimate the galaxy LF.  A larger high resolution chart is  available on request. Fig. 23 shows Part 2. }
       \label{fig:sumchart1}  
    \end{center}
  \end{figure}

\begin{figure} 
   \begin{center}
	 \includegraphics[width=1.0\textwidth]{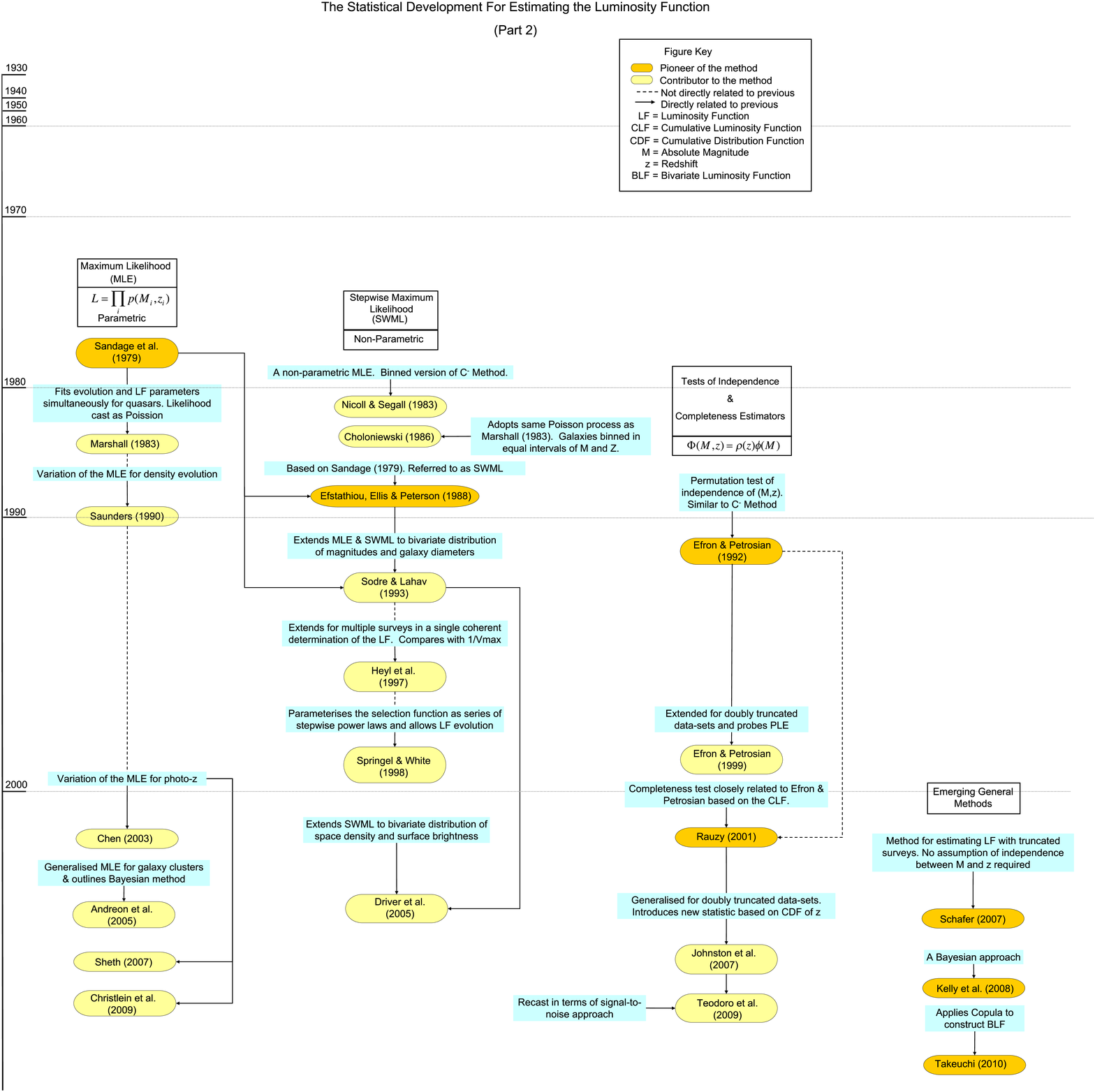}
              \caption {\small Part 2 of a schematic charting the development of all the major statistical methods that estimate the galaxy LF.  A larger high resolution chart is  available on request.  }
       \label{fig:sumchart2}  
    \end{center}
  \end{figure}

This review is an attempt to consolidate all the most innovative statistics developed for estimating the LF from their early beginnings 75 years ago to present day.  Figures~\ref{fig:sumchart1} and \ref{fig:sumchart2} show a time-line diagram which charts the genealogical progress of all these estimators (a larger high-resolution version is available on request).  

Within the non-parametric regime it was discussed that whilst the traditional number count {\it classical} approach is straightforward in its construction, it is limited by its assumption of spatial homogeneity - a limitation also shared with the $1/V_{\max}$ estimator.   This intrinsic bias  of $V_{\max}$ has been demonstrated by \citetalias{Willmer:1997} and \citetalias{Takeuchi:2000} when direct comparisons  to other LF estimators showed overestimation of the LF particularly at the faint-end slope. 

In terms of the predictive power of  the ${V}/{V}_{\rm max}$ counterpart I would add a cautionary note.   Probing evolution in data can pose degenerate qualities making it difficult to determine whether a significant departure from the expectation value of $V/V_{\rm max}$ ($\approx 1/2$) is  due to inhomogeneity effects (introduced by clustering) or evolution,  or simply an indication of underlying incompleteness of the catalogue from other sources of contamination.   Nevertheless, this seems not to have  deterred its steady growth in popularity within the astronomical  community  as is evident from its numerous extensions to multiple surveys with varying flux limits, diameter-limited surveys, fitting generic LFs and adaptation to photometric redshifts etc...   In fact,  $1/V_{\max}$ and  $V/V_{\rm max}$  remain one of the most widely applied non-parametric estimators for analysing the statistical properties of extragalactic sources.  This is most likely due to the ease with which it can be applied.  

The construction of the $\phi/\Phi$ estimator  and, in particular, the $C^-$ method  offered a way to effectively circumvent the assumption of homogeneity by the cancellation of the density terms within their construction.  The $C^-$ method could be considered as a breakthrough since the method required sampling the CLF and therefore no binning of the data was required \citep[see][for an interesting discussion on the pitfalls of binning]{Andreon:2005MNRAS.360..727A}.   And yet, despite the fact that it was pioneered just three years after Schmidt's estimator and eight years before $\phi/\Phi$,  it did not grow in popularity as other tests did.   \cite{Petrosian:1992} also demonstrated that all non-parametric methods are essentially variations of the $C^-$ method under the limit of having one galaxy per bin.  It is therefore a slight mystery as to why this approach has not been applied more often. Perhaps this is due to other maximum likelihood estimators (MLE) that  developed as a result.

Probably the most notable of the MLEs is the so-called \citetalias{sandage:1979} estimator named after its developers, Sandage, Tammann and Yahil in 1979.  Its main appeal is that one must adopt a analytical form for the LF and  estimate its parameters via a maximum likelihood process.  The form of the LF one chooses is rather arbitrary and in general  the Schechter function \citep{Schechter:1976} is favoured due to its robustness to various survey data.   Approaching the problem this way avoids many of the problems associated with binning techniques and also avoids issues of density inhomogeneities.  However, one needs to assess the goodness-of-fit and determine the normalisation  via independent means. 

\citetalias{Efstathiou:1988} extended the ideas of Lynden-Bell's $C^-$ method  and the  \citetalias{sandage:1979} MLE by developing the  non-parametric case of the MLE  by replacing the analytical form of the LF with a series of step functions.  This step-wise maximum likelihood (SWML) method  has become an equally favoured estimator in recent times. It, like $V_{\max}$ has seen various extensions for e.g. bivariate distributions and combining multiple surveys to improve its usage.

Also examined in detail  were the more recent LF estimators that have emerged within the last few years that depart from the traditional approaches. The first one, developed by \cite{Schafer:2007}, states key potential advantages for its construction  that include making no strict assumption of separability between the probability densities $\rho(z)$ and $\Phi(M)$ and incorporating a varying selection function into the method. The second by \cite{Kelly:2008} takes a Bayesian approach adopting a mixture of Gaussian functions for the prior distribution. Finally, the more recent \cite{Takeuchi2010MNRAS.406.1830T} bivariate construction of the LF applied the copula which connects two distribution functions. This parametric technique was applied to the FUV-FIR BLF.

The final part of the review shifted focus  to one of the basic assumptions of most LF estimators - the assumption of separability between the luminosity function $\phi(M)$ and the density function $\rho(z)$.  By building on Lynden Bell's $C^-$ method, \cite{Efron:1992} constructed test statistic that can serve as a test of correlation between the assumed independent variables $M$ and $z$ and furthermore be used a to constrain pure luminosity models, where any such evolution would introduce  correlations in the ($M,z$) distribution.  It was then demonstrated how \cite{Rauzy:2001} built upon the ideas in EP92 and turned them into a simple but powerful robust non-parametric test of magnitude completeness.  This completeness  test does not require any  modelling of the LF and unlike $V/V_{\max}$ is independent of the spatial distribution of sources.  Moreover, it returns a differential measure of completeness where one can assess the level of incompleteness over a range of magnitudes by applying  successive trial apparent magnitude limits.

What has probably become obvious through this review is that choosing the {\it correct} LF estimator for your data is not a straight forward task. However, recent statistical advances mean that there are a myriad of options available. This decision usually comes down to taste and experience, and perhaps more crucially, how far one is willing to extend their analysis beyond simple applications to obtain rigorous, if more computationally challenging, results. For now, the practice of applying several complementary LF estimators to a given data-set provides a reasonable, if time consuming, approach.

In closing, the notion  that we have entered into an era of {\it precision cosmology} has been cited for more than a decade \citep{Turner:1998} and can perhaps be justified when applied to CMBr measurements \citep[e.g.][]{Spergel:2003,Larson:2010}. However, in the study of galaxy redshift surveys, the era of precision cosmology  would appear to be approaching but will require improvement in both the quality and size of our data-sets and, crucially, in our statistical toolbox before we can claim that is has truly arrived.  Whether the more recent methods will become  as popular as the traditional ones remains to be seen, since shifts towards seemingly more complex  techniques can take time to catch on.  Nevertheless, it is encouraging that  this branch of  astronomy continues to develop and embrace new statistical methods.  With  with the  next generation of galaxy redshift surveys on the horizon,  it will be interesting to see how the current approaches to constraining  the  luminosity function will adapt and which new methods will come to prominence.			%section 12
\begin{acknowledgements}
There are a number of people who deserve a lot of  thanks. Firstly, I would like to thank David Valls-Gabaud for  encouraging me to  write this review and also taking the time to read the several drafts along the way. The following people also deserve my gratitude for  reading various parts of the review and/or  providing useful comments and criticisms (in alphabetical order): Steve Crawford, Nick Cross, Andreas Faltenbacher, Martin Hendry, Chris Koen, Roy Maartens (for his title suggestion!),  Prina Patel, Mat Smith, Tsutomu Takeuchi, Luis Teodoro and Christopher Willmer.  I would like to extend special thanks  to  Jeremy Heyl, Olivier Ilbert, Brandon Kelly, Peder Norberg, Chad Schafer, Tsutomu Takeuchi, Rita Tojeiro and Christopher Willmer   for being kind enough to  allow me to include some of their figures - it saved me a lot of time and helped bring more clarity to this review.  I  would also like to acknowledge the current  support of the National Research Foundation (South Africa) and the funding body Engineering and Physical Sciences Research Council (EPSRC) whilst at Glasgow University (UK).  Finally, thank you to Joel Bregman at A\&AR for his guidance and the anonymous referee for his/her helpful comments.
\end{acknowledgements}	         	%acknowledgments
%
%%SUMMARY CHART FIGURE INPUT HERE.
%%THIS IS THE FINAL FIGURE OF THE DOCUMENT
\clearpage

\clearpage
%%

%BIBLIOGRAPHY CALLED HERE
\bibliographystyle{aa}
\bibliography{bibliography_CIII}

\begin{thebibliography}{399}
\expandafter\ifx\csname natexlab\endcsname\relax\def\natexlab#1{#1}\fi

\bibitem[{{Aarseth}(1963)}]{Aarseth:1963MNRAS.126..223A}
{Aarseth}, S.~J. 1963, \mnras, 126, 223

\bibitem[{{Aarseth} {et~al.}(1979){Aarseth}, {Turner}, \&
  {Gott}}]{Aarseth:1979ApJ...228..664A}
{Aarseth}, S.~J., {Turner}, E.~L., \& {Gott}, III, J.~R. 1979, \apj, 228, 664

\bibitem[{{Abazajian} {et~al.}(2003){Abazajian}, {Adelman-McCarthy},
  {Ag{\"u}eros}, {Allam}, {Anderson}, {Annis}, {Bahcall}, {Baldry}, {Bastian},
  {Berlind}, {Bernardi}, {Blanton}, {Blythe}, {Bochanski}, {Boroski},
  {Brewington}, {Briggs}, {Brinkmann}, {Brunner}, {Budav{\'a}ri}, {Carey},
  {Carr}, {Castander}, {Chiu}, {Collinge}, {Connolly}, {Covey}, {Csabai},
  {Dalcanton}, {Dodelson}, {Doi}, {Dong}, {Eisenstein}, {Evans}, {Fan},
  {Feldman}, {Finkbeiner}, {Friedman}, {Frieman}, {Fukugita}, {Gal},
  {Gillespie}, {Glazebrook}, {Gonzalez}, {Gray}, {Grebel}, {Grodnicki}, {Gunn},
  {Gurbani}, {Hall}, {Hao}, {Harbeck}, {Harris}, {Harris}, {Harvanek},
  {Hawley}, {Heckman}, {Helmboldt}, {Hendry}, {Hennessy}, {Hindsley}, {Hogg},
  {Holmgren}, {Holtzman}, {Homer}, {Hui}, {Ichikawa}, {Ichikawa}, {Inkmann},
  {Ivezi{\'c}}, {Jester}, {Johnston}, {Jordan}, {Jordan}, {Jorgensen},
  {Juri{\'c}}, {Kauffmann}, {Kent}, {Kleinman}, {Knapp}, {Kniazev}, {Kron},
  {Krzesi{\'n}ski}, {Kunszt}, {Kuropatkin}, {Lamb}, {Lampeitl}, {Laubscher},
  {Lee}, {Leger}, {Li}, {Lidz}, {Lin}, {Loh}, {Long}, {Loveday}, {Lupton},
  {Malik}, {Margon}, {McGehee}, {McKay}, {Meiksin}, {Miknaitis}, {Moorthy},
  {Munn}, {Murphy}, {Nakajima}, {Narayanan}, {Nash}, {Neilsen}, {Newberg},
  {Newman}, {Nichol}, {Nicinski}, {Nieto-Santisteban}, {Nitta}, {Odenkirchen},
  {Okamura}, {Ostriker}, {Owen}, {Padmanabhan}, {Peoples}, {Pier}, {Pindor},
  {Pope}, {Quinn}, {Rafikov}, {Raymond}, {Richards}, {Richmond}, {Rix},
  {Rockosi}, {Schaye}, {Schlegel}, {Schneider}, {Schroeder}, {Scranton},
  {Sekiguchi}, {Seljak}, {Sergey}, {Sesar}, {Sheldon}, {Shimasaku}, {Siegmund},
  {Silvestri}, {Sinisgalli}, {Sirko}, {Smith}, {Smol{\v c}i{\'c}}, {Snedden},
  {Stebbins}, {Steinhardt}, {Stinson}, {Stoughton}, {Strateva}, {Strauss},
  {SubbaRao}, {Szalay}, {Szapudi}, {Szkody}, {Tasca}, {Tegmark}, {Thakar},
  {Tremonti}, {Tucker}, {Uomoto}, {Vanden Berk}, {Vandenberg}, {Vogeley},
  {Voges}, {Vogt}, {Walkowicz}, {Weinberg}, {West}, {White}, {Wilhite},
  {Willman}, {Xu}, {Yanny}, {Yarger}, {Yasuda}, {Yip}, {Yocum}, {York},
  {Zakamska}, {Zehavi}, {Zheng}, {Zibetti}, \&
  {Zucker}}]{Abazajian:2003AJ.126.2081A}
{Abazajian}, K., {Adelman-McCarthy}, J.~K., {Ag{\"u}eros}, M.~A., {et~al.}
  2003, \aj, 126, 2081

\bibitem[{{Abazajian} {et~al.}(2009){Abazajian}, {Adelman-McCarthy},
  {Ag{\"u}eros}, {Allam}, {Allende Prieto}, {An}, {Anderson}, {Anderson},
  {Annis}, {Bahcall}, \& et~al.}]{Abazajian:2009ApJS..182..543A}
{Abazajian}, K.~N., {Adelman-McCarthy}, J.~K., {Ag{\"u}eros}, M.~A., {et~al.}
  2009, \apjs, 182, 543

\bibitem[{{Abramowitz} \& {Stegun}(1964)}]{Abramowitz:1964}
{Abramowitz}, M. \& {Stegun}, I. 1964, {Handbook of Mathematical Functions with
  Formulas, Graphs, and Mathematical Tables} (Dover Publications)

\bibitem[{{Adelman-McCarthy} {et~al.}(2007){Adelman-McCarthy}, {Ag{\"u}eros},
  {Allam}, {Anderson}, {Anderson}, {Annis}, {Bahcall}, {Bailer-Jones},
  {Baldry}, {Barentine}, {Beers}, {Belokurov}, {Berlind}, {Bernardi},
  {Blanton}, {Bochanski}, {Boroski}, {Bramich}, {Brewington}, {Brinchmann},
  {Brinkmann}, {Brunner}, {Budav{\'a}ri}, {Carey}, {Carliles}, {Carr},
  {Castander}, {Connolly}, {Cool}, {Cunha}, {Csabai}, {Dalcanton}, {Doi},
  {Eisenstein}, {Evans}, {Evans}, {Fan}, {Finkbeiner}, {Friedman}, {Frieman},
  {Fukugita}, {Gillespie}, {Gilmore}, {Glazebrook}, {Gray}, {Grebel}, {Gunn},
  {de Haas}, {Hall}, {Harvanek}, {Hawley}, {Hayes}, {Heckman}, {Hendry},
  {Hennessy}, {Hindsley}, {Hirata}, {Hogan}, {Hogg}, {Holtzman}, {Ichikawa},
  {Ichikawa}, {Ivezi{\'c}}, {Jester}, {Johnston}, {Jorgensen}, {Juri{\'c}},
  {Kauffmann}, {Kent}, {Kleinman}, {Knapp}, {Kniazev}, {Kron}, {Krzesinski},
  {Kuropatkin}, {Lamb}, {Lampeitl}, {Lee}, {Leger}, {Lima}, {Lin}, {Long},
  {Loveday}, {Lupton}, {Mandelbaum}, {Margon}, {Mart{\'{\i}}nez-Delgado},
  {Matsubara}, {McGehee}, {McKay}, {Meiksin}, {Munn}, {Nakajima}, {Nash},
  {Neilsen}, {Newberg}, {Nichol}, {Nieto-Santisteban}, {Nitta}, {Oyaizu},
  {Okamura}, {Ostriker}, {Padmanabhan}, {Park}, {Peoples}, {Pier}, {Pope},
  {Pourbaix}, {Quinn}, {Raddick}, {Re Fiorentin}, {Richards}, {Richmond},
  {Rix}, {Rockosi}, {Schlegel}, {Schneider}, {Scranton}, {Seljak}, {Sheldon},
  {Shimasaku}, {Silvestri}, {Smith}, {Smol{\v c}i{\'c}}, {Snedden}, {Stebbins},
  {Stoughton}, {Strauss}, {SubbaRao}, {Suto}, {Szalay}, {Szapudi}, {Szkody},
  {Tegmark}, {Thakar}, {Tremonti}, {Tucker}, {Uomoto}, {Vanden Berk},
  {Vandenberg}, {Vidrih}, {Vogeley}, {Voges}, {Vogt}, {Weinberg}, {West},
  {White}, {Wilhite}, {Yanny}, {Yocum}, {York}, {Zehavi}, {Zibetti}, \&
  {Zucker}}]{Adelman-McCarthy:2007}
{Adelman-McCarthy}, J.~K., {Ag{\"u}eros}, M.~A., {Allam}, S.~S., {et~al.} 2007,
  \apjs, 172, 634

\bibitem[{{Adelman-McCarthy} {et~al.}(2006){Adelman-McCarthy}, {Ag{\"u}eros},
  {Allam}, {Anderson}, {Anderson}, {Annis}, {Bahcall}, {Baldry}, {Barentine},
  {Berlind}, {Bernardi}, {Blanton}, {Boroski}, {Brewington}, {Brinchmann},
  {Brinkmann}, {Brunner}, {Budav{\'a}ri}, {Carey}, {Carr}, {Castander},
  {Connolly}, {Csabai}, {Czarapata}, {Dalcanton}, {Doi}, {Dong}, {Eisenstein},
  {Evans}, {Fan}, {Finkbeiner}, {Friedman}, {Frieman}, {Fukugita}, {Gillespie},
  {Glazebrook}, {Gray}, {Grebel}, {Gunn}, {Gurbani}, {de Haas}, {Hall},
  {Harris}, {Harvanek}, {Hawley}, {Hayes}, {Hendry}, {Hennessy}, {Hindsley},
  {Hirata}, {Hogan}, {Hogg}, {Holmgren}, {Holtzman}, {Ichikawa}, {Ivezi{\'c}},
  {Jester}, {Johnston}, {Jorgensen}, {Juri{\'c}}, {Kent}, {Kleinman}, {Knapp},
  {Kniazev}, {Kron}, {Krzesinski}, {Kuropatkin}, {Lamb}, {Lampeitl}, {Lee},
  {Leger}, {Lin}, {Long}, {Loveday}, {Lupton}, {Margon},
  {Mart{\'{\i}}nez-Delgado}, {Mandelbaum}, {Matsubara}, {McGehee}, {McKay},
  {Meiksin}, {Munn}, {Nakajima}, {Nash}, {Neilsen}, {Newberg}, {Newman},
  {Nichol}, {Nicinski}, {Nieto-Santisteban}, {Nitta}, {O'Mullane}, {Okamura},
  {Owen}, {Padmanabhan}, {Pauls}, {Peoples}, {Pier}, {Pope}, {Pourbaix},
  {Quinn}, {Richards}, {Richmond}, {Rockosi}, {Schlegel}, {Schneider},
  {Schroeder}, {Scranton}, {Seljak}, {Sheldon}, {Shimasaku}, {Smith}, {Smol{\v
  c}i{\'c}}, {Snedden}, {Stoughton}, {Strauss}, {SubbaRao}, {Szalay},
  {Szapudi}, {Szkody}, {Tegmark}, {Thakar}, {Tucker}, {Uomoto}, {Vanden Berk},
  {Vandenberg}, {Vogeley}, {Voges}, {Vogt}, {Walkowicz}, {Weinberg}, {West},
  {White}, {Xu}, {Yanny}, {Yocum}, {York}, {Zehavi}, {Zibetti}, \&
  {Zucker}}]{Adelman-McCarthy:2006}
{Adelman-McCarthy}, J.~K., {Ag{\"u}eros}, M.~A., {Allam}, S.~S., {et~al.} 2006,
  \apjs, 162, 38

\bibitem[{{Aird} {et~al.}(2010){Aird}, {Nandra}, {Laird}, {Georgakakis},
  {Ashby}, {Barmby}, {Coil}, {Huang}, {Koekemoer}, {Steidel}, \&
  {Willmer}}]{Aird:2010MNRAS.401.2531A}
{Aird}, J., {Nandra}, K., {Laird}, E.~S., {et~al.} 2010, \mnras, 401, 2531

\bibitem[{{Aldrich}(1997)}]{Aldrich:1997}
{Aldrich}, J. 1997, Statistical Science, 12, 162

\bibitem[{{Andreon}(2006)}]{Andreon:2006MNRAS.369..969A}
{Andreon}, S. 2006, \mnras, 369, 969

\bibitem[{{Andreon}(2010)}]{Andreon:2010MNRAS.407..263A}
{Andreon}, S. 2010, \mnras, 407, 263

\bibitem[{{Andreon} {et~al.}(2006){Andreon}, {Cuillandre}, {Puddu}, \&
  {Mellier}}]{Andreon:2006MNRAS.372...60A}
{Andreon}, S., {Cuillandre}, J.-C., {Puddu}, E., \& {Mellier}, Y. 2006, \mnras,
  372, 60

\bibitem[{{Andreon} {et~al.}(2008){Andreon}, {Puddu}, {de Propris}, \&
  {Cuillandre}}]{Andreon:2008MNRAS.385..979A}
{Andreon}, S., {Puddu}, E., {de Propris}, R., \& {Cuillandre}, J.-C. 2008,
  \mnras, 385, 979

\bibitem[{{Andreon} {et~al.}(2005){Andreon}, {Punzi}, \&
  {Grado}}]{Andreon:2005MNRAS.360..727A}
{Andreon}, S., {Punzi}, G., \& {Grado}, A. 2005, \mnras, 360, 727

\bibitem[{{Arnouts} {et~al.}(1999){Arnouts}, {Cristiani}, {Moscardini},
  {Matarrese}, {Lucchin}, {Fontana}, \&
  {Giallongo}}]{Arnouts:1999MNRAS.310..540A}
{Arnouts}, S., {Cristiani}, S., {Moscardini}, L., {et~al.} 1999, \mnras, 310,
  540

\bibitem[{{Arnouts} {et~al.}(2002){Arnouts}, {Moscardini}, {Vanzella},
  {Colombi}, {Cristiani}, {Fontana}, {Giallongo}, {Matarrese}, \&
  {Saracco}}]{Arnouts:2002MNRAS.329..355A}
{Arnouts}, S., {Moscardini}, L., {Vanzella}, E., {et~al.} 2002, \mnras, 329,
  355

\bibitem[{{Auriemma} {et~al.}(1977){Auriemma}, {Perola}, {Ekers}, {Fanti},
  {Lari}, {Jaffe}, \& {Ulrich}}]{1977A&A....57...41A}
{Auriemma}, C., {Perola}, G.~C., {Ekers}, R.~D., {et~al.} 1977, \aap, 57, 41

\bibitem[{{Avni} \& {Bahcall}(1980)}]{avni:1980}
{Avni}, Y. \& {Bahcall}, J.~N. 1980, \apj, 235, 694

\bibitem[{{Babi{\'c}} {et~al.}(2007){Babi{\'c}}, {Miller}, {Jarvis}, {Turner},
  {Alexander}, \& {Croom}}]{Babic:2007AA...474..755B}
{Babi{\'c}}, A., {Miller}, L., {Jarvis}, M.~J., {et~al.} 2007, \aap, 474, 755

\bibitem[{{Bahcall} \& {Yahil}(1972)}]{Bahcall:1972ApJ...177..647B}
{Bahcall}, J.~N. \& {Yahil}, A. 1972, \apj, 177, 647

\bibitem[{{Ball} {et~al.}(2008){Ball}, {Brunner}, {Myers}, {Strand}, {Alberts},
  \& {Tcheng}}]{Ball:2008}
{Ball}, N.~M., {Brunner}, R.~J., {Myers}, A.~D., {et~al.} 2008, \apj, 683, 12

\bibitem[{{Ball} {et~al.}(2006){Ball}, {Loveday}, {Brunner}, {Baldry}, \&
  {Brinkmann}}]{Ball2006MNRAS.373..845B}
{Ball}, N.~M., {Loveday}, J., {Brunner}, R.~J., {Baldry}, I.~K., \&
  {Brinkmann}, J. 2006, \mnras, 373, 845

\bibitem[{{Banerji} {et~al.}(2008){Banerji}, {Abdalla}, {Lahav}, \&
  {Lin}}]{Banerji:2008MNRAS.386.1219B}
{Banerji}, M., {Abdalla}, F.~B., {Lahav}, O., \& {Lin}, H. 2008, \mnras, 386,
  1219

\bibitem[{{Barger} {et~al.}(2005){Barger}, {Cowie}, {Mushotzky}, {Yang},
  {Wang}, {Steffen}, \& {Capak}}]{Barger:2005AJ....129..578B}
{Barger}, A.~J., {Cowie}, L.~L., {Mushotzky}, R.~F., {et~al.} 2005, \aj, 129,
  578

\bibitem[{{Baum}(1962)}]{Baum:1962}
{Baum}, W.~A. 1962, in IAU Symposium, Vol.~15, Problems of Extra-Galactic
  Research, ed. G.~C. {McVittie}, 390--+

\bibitem[{{Beckwith} {et~al.}(2006){Beckwith}, {Stiavelli}, {Koekemoer},
  {Caldwell}, {Ferguson}, {Hook}, {Lucas}, {Bergeron}, {Corbin}, {Jogee},
  {Panagia}, {Robberto}, {Royle}, {Somerville}, \& {Sosey}}]{Beckwith:2006}
{Beckwith}, S.~V.~W., {Stiavelli}, M., {Koekemoer}, A.~M., {et~al.} 2006, \aj,
  132, 1729

\bibitem[{{Bell} {et~al.}(2004){Bell}, {Wolf}, {Meisenheimer}, {Rix}, {Borch},
  {Dye}, {Kleinheinrich}, {Wisotzki}, \& {McIntosh}}]{Bell:2004ApJ...608..752B}
{Bell}, E.~F., {Wolf}, C., {Meisenheimer}, K., {et~al.} 2004, \apj, 608, 752

\bibitem[{{Ben{\'{\i}}tez}(2000)}]{Benitez:2000}
{Ben{\'{\i}}tez}, N. 2000, \apj, 536, 571

\bibitem[{{Bernardi} {et~al.}(2003{\natexlab{a}}){Bernardi}, {Sheth}, {Annis},
  {Burles}, {Eisenstein}, {Finkbeiner}, {Hogg}, {Lupton}, {Schlegel},
  {SubbaRao}, {Bahcall}, {Blakeslee}, {Brinkmann}, {Castander}, {Connolly},
  {Csabai}, {Doi}, {Fukugita}, {Frieman}, {Heckman}, {Hennessy}, {Ivezi{\'c}},
  {Knapp}, {Lamb}, {McKay}, {Munn}, {Nichol}, {Okamura}, {Schneider}, {Thakar},
  \& {York}}]{Bernardi:2003b}
{Bernardi}, M., {Sheth}, R.~K., {Annis}, J., {et~al.} 2003{\natexlab{a}}, \aj,
  125, 1849

\bibitem[{{Bernardi} {et~al.}(2003{\natexlab{b}}){Bernardi}, {Sheth}, {Annis},
  {Burles}, {Eisenstein}, {Finkbeiner}, {Hogg}, {Lupton}, {Schlegel},
  {SubbaRao}, {Bahcall}, {Blakeslee}, {Brinkmann}, {Castander}, {Connolly},
  {Csabai}, {Doi}, {Fukugita}, {Frieman}, {Heckman}, {Hennessy}, {Ivezi{\'c}},
  {Knapp}, {Lamb}, {McKay}, {Munn}, {Nichol}, {Okamura}, {Schneider}, {Thakar},
  \& {York}}]{Bernardi:2003a}
{Bernardi}, M., {Sheth}, R.~K., {Annis}, J., {et~al.} 2003{\natexlab{b}}, \aj,
  125, 1817

\bibitem[{{Bernardi} {et~al.}(2005){Bernardi}, {Sheth}, {Nichol}, {Schneider},
  \& {Brinkmann}}]{Bernardi:2005}
{Bernardi}, M., {Sheth}, R.~K., {Nichol}, R.~C., {Schneider}, D.~P., \&
  {Brinkmann}, J. 2005, \aj, 129, 61

\bibitem[{Bernoulli(1769)}]{Bernoulli:1769}
Bernoulli, D. 1769, Dijudicatio maxime probabilis plurium observationum
  discrepantium atque verisimillima inductio inde formanda., Bernoulli MSS,
  f.299--305

\bibitem[{Bernoulli(1778)}]{Bernoulli:1778}
Bernoulli, D. 1778, Dijudicatio maxime probabilis plurium observationum
  discrepantium atque verisimillima inductio inde formanda, {\it Acta Academiae
  Scientiarum Imperialis Petropolitanae} for 1777, pars prior 3--23 (English
  translation in {\cite{Kendall:1961}} 3--13, reprinted 1970 in Pearson, Egon
  S. and Kendall, M. G. (eds.))

\bibitem[{{Bertone} {et~al.}(2007){Bertone}, {De Lucia}, \&
  {Thomas}}]{Bertone:2007MNRAS.379.1143B}
{Bertone}, S., {De Lucia}, G., \& {Thomas}, P.~A. 2007, \mnras, 379, 1143

\bibitem[{{Binggeli} {et~al.}(1988){Binggeli}, {Sandage}, \&
  {Tammann}}]{Binggeli:1988}
{Binggeli}, B., {Sandage}, A., \& {Tammann}, G.~A. 1988, \araa, 26, 509

\bibitem[{{Blake} \& {Bridle}(2005)}]{Blake:2005MNRAS.363.1329B}
{Blake}, C. \& {Bridle}, S. 2005, \mnras, 363, 1329

\bibitem[{{Blanton} {et~al.}(2001){Blanton}, {Dalcanton}, {Eisenstein},
  {Loveday}, {Strauss}, {SubbaRao}, {Weinberg}, {Anderson}, {Annis}, {Bahcall},
  {Bernardi}, {Brinkmann}, {Brunner}, {Burles}, {Carey}, {Castander},
  {Connolly}, {Csabai}, {Doi}, {Finkbeiner}, {Friedman}, {Frieman}, {Fukugita},
  {Gunn}, {Hennessy}, {Hindsley}, {Hogg}, {Ichikawa}, {Ivezi{\'c}}, {Kent},
  {Knapp}, {Lamb}, {Leger}, {Long}, {Lupton}, {McKay}, {Meiksin}, {Merelli},
  {Munn}, {Narayanan}, {Newcomb}, {Nichol}, {Okamura}, {Owen}, {Pier}, {Pope},
  {Postman}, {Quinn}, {Rockosi}, {Schlegel}, {Schneider}, {Shimasaku},
  {Siegmund}, {Smee}, {Snir}, {Stoughton}, {Stubbs}, {Szalay}, {Szokoly},
  {Thakar}, {Tremonti}, {Tucker}, {Uomoto}, {Vanden Berk}, {Vogeley},
  {Waddell}, {Yanny}, {Yasuda}, \& {York}}]{Blanton:2001AJ....121.2358B}
{Blanton}, M.~R., {Dalcanton}, J., {Eisenstein}, D., {et~al.} 2001, \aj, 121,
  2358

\bibitem[{{Blanton} {et~al.}(2003{\natexlab{a}}){Blanton}, {Hogg}, {Bahcall},
  {Baldry}, {Brinkmann}, {Csabai}, {Eisenstein}, {Fukugita}, {Gunn},
  {Ivezi{\'c}}, {Lamb}, {Lupton}, {Loveday}, {Munn}, {Nichol}, {Okamura},
  {Schlegel}, {Shimasaku}, {Strauss}, {Vogeley}, \&
  {Weinberg}}]{Blanton:2003ApJ...594..186B}
{Blanton}, M.~R., {Hogg}, D.~W., {Bahcall}, N.~A., {et~al.} 2003{\natexlab{a}},
  \apj, 594, 186

\bibitem[{{Blanton} {et~al.}(2003{\natexlab{b}}){Blanton}, {Hogg}, {Bahcall},
  {Brinkmann}, {Britton}, {Connolly}, {Csabai}, {Fukugita}, {Loveday},
  {Meiksin}, {Munn}, {Nichol}, {Okamura}, {Quinn}, {Schneider}, {Shimasaku},
  {Strauss}, {Tegmark}, {Vogeley}, \& {Weinberg}}]{Blanton:2003ApJ...592..819B}
{Blanton}, M.~R., {Hogg}, D.~W., {Bahcall}, N.~A., {et~al.} 2003{\natexlab{b}},
  \apj, 592, 819

\bibitem[{{Blanton} {et~al.}(2003{\natexlab{c}}){Blanton}, {Lin}, {Lupton},
  {Maley}, {Young}, {Zehavi}, \& {Loveday}}]{Blanton:2003AJ.125.2276B}
{Blanton}, M.~R., {Lin}, H., {Lupton}, R.~H., {et~al.} 2003{\natexlab{c}}, \aj,
  125, 2276

\bibitem[{{Blumenthal} {et~al.}(1984){Blumenthal}, {Faber}, {Primack}, \&
  {Rees}}]{Blumenthal:1984Natur.311..517B}
{Blumenthal}, G.~R., {Faber}, S.~M., {Primack}, J.~R., \& {Rees}, M.~J. 1984,
  \nat, 311, 517

\bibitem[{{Bolzonella} {et~al.}(2000){Bolzonella}, {Miralles}, \&
  {Pell{\'o}}}]{Bolzonella:2000}
{Bolzonella}, M., {Miralles}, J.-M., \& {Pell{\'o}}, R. 2000, \aap, 363, 476

\bibitem[{{Bonfield} {et~al.}(2010){Bonfield}, {Sun}, {Davey}, {Jarvis},
  {Abdalla}, {Banerji}, \& {Adams}}]{Bonfield:2010MNRAS.405..987B}
{Bonfield}, D.~G., {Sun}, Y., {Davey}, N., {et~al.} 2010, \mnras, 405, 987

\bibitem[{{Bongiorno} {et~al.}(2007){Bongiorno}, {Zamorani}, {Gavignaud},
  {Marano}, {Paltani}, {Mathez}, {M{\o}ller}, {Picat}, {Cirasuolo},
  {Lamareille}, {Bottini}, {Garilli}, {Le Brun}, {Le F{\`e}vre}, {Maccagni},
  {Scaramella}, {Scodeggio}, {Tresse}, {Vettolani}, {Zanichelli}, {Adami},
  {Arnouts}, {Bardelli}, {Bolzonella}, {Cappi}, {Charlot}, {Ciliegi},
  {Contini}, {Foucaud}, {Franzetti}, {Guzzo}, {Ilbert}, {Iovino}, {McCracken},
  {Marinoni}, {Mazure}, {Meneux}, {Merighi}, {Pell{\`o}}, {Pollo}, {Pozzetti},
  {Radovich}, {Zucca}, {Hatziminaoglou}, {Polletta}, {Bondi}, {Brinchmann},
  {Cucciati}, {de la Torre}, {Gregorini}, {Mellier}, {Merluzzi}, {Temporin},
  {Vergani}, \& {Walcher}}]{Bongiorno:2007AA...472..443B}
{Bongiorno}, A., {Zamorani}, G., {Gavignaud}, I., {et~al.} 2007, \aap, 472, 443

\bibitem[{{Bouwens} {et~al.}(2009){Bouwens}, {Illingworth}, {Bradley}, {Ford},
  {Franx}, {Zheng}, {Broadhurst}, {Coe}, \&
  {Jee}}]{Bouwens:2009ApJ...690.1764B}
{Bouwens}, R.~J., {Illingworth}, G.~D., {Bradley}, L.~D., {et~al.} 2009, \apj,
  690, 1764

\bibitem[{{Bouwens} {et~al.}(2008{\natexlab{a}}){Bouwens}, {Illingworth},
  {Franx}, \& {Ford}}]{Bouwens:2008}
{Bouwens}, R.~J., {Illingworth}, G.~D., {Franx}, M., \& {Ford}, H.
  2008{\natexlab{a}}, \apj, 686, 230

\bibitem[{{Bouwens} {et~al.}(2008{\natexlab{b}}){Bouwens}, {Illingworth},
  {Franx}, \& {Ford}}]{Bouwens2008ApJ...686..230B}
{Bouwens}, R.~J., {Illingworth}, G.~D., {Franx}, M., \& {Ford}, H.
  2008{\natexlab{b}}, \apj, 686, 230

\bibitem[{{Bower} {et~al.}(2006){Bower}, {Benson}, {Malbon}, {Helly}, {Frenk},
  {Baugh}, {Cole}, \& {Lacey}}]{Bower:2006MNRAS.370..645B}
{Bower}, R.~G., {Benson}, A.~J., {Malbon}, R., {et~al.} 2006, \mnras, 370, 645

\bibitem[{{Boyle} {et~al.}(1990){Boyle}, {Fong}, {Shanks}, \&
  {Peterson}}]{Boyle:1990}
{Boyle}, B.~J., {Fong}, R., {Shanks}, T., \& {Peterson}, B.~A. 1990, \mnras,
  243, 1

\bibitem[{{Boyle} {et~al.}(1993){Boyle}, {Griffiths}, {Shanks}, {Stewart}, \&
  {Georgantopoulos}}]{Boyle:1993MNRAS.260...49B}
{Boyle}, B.~J., {Griffiths}, R.~E., {Shanks}, T., {Stewart}, G.~C., \&
  {Georgantopoulos}, I. 1993, \mnras, 260, 49

\bibitem[{{Boyle} {et~al.}(2000){Boyle}, {Shanks}, {Croom}, {Smith}, {Miller},
  {Loaring}, \& {Heymans}}]{Boyle2000MNRAS.317.1014B}
{Boyle}, B.~J., {Shanks}, T., {Croom}, S.~M., {et~al.} 2000, \mnras, 317, 1014

\bibitem[{{Boyle} {et~al.}(1988{\natexlab{a}}){Boyle}, {Shanks}, \&
  {Peterson}}]{Boyle:1988ASPC....2....1B}
{Boyle}, B.~J., {Shanks}, T., \& {Peterson}, B.~A. 1988{\natexlab{a}}, in
  Astronomical Society of the Pacific Conference Series, Vol.~2, Optical
  Surveys for Quasars, ed. {P.~Osmer \& M.~M.~Phillips}, 1--+

\bibitem[{{Boyle} {et~al.}(1988{\natexlab{b}}){Boyle}, {Shanks}, \&
  {Peterson}}]{Boyle:1988MNRAS.235..935B}
{Boyle}, B.~J., {Shanks}, T., \& {Peterson}, B.~A. 1988{\natexlab{b}}, \mnras,
  235, 935

\bibitem[{{Brinchmann} {et~al.}(1998){Brinchmann}, {Abraham}, {Schade},
  {Tresse}, {Ellis}, {Lilly}, {Le Fevre}, {Glazebrook}, {Hammer}, {Colless},
  {Crampton}, \& {Broadhurst}}]{Brinchmann:1998ApJ...499..112B}
{Brinchmann}, J., {Abraham}, R., {Schade}, D., {et~al.} 1998, \apj, 499, 112

\bibitem[{{Brown} {et~al.}(2003){Brown}, {Allende Prieto}, {Beers}, {Wilhelm},
  {Geller}, {Kenyon}, \& {Kurtz}}]{Brown:2003AJ....126.1362B}
{Brown}, W.~R., {Allende Prieto}, C., {Beers}, T.~C., {et~al.} 2003, \aj, 126,
  1362

\bibitem[{{Brown} {et~al.}(2005){Brown}, {Geller}, {Kenyon}, {Kurtz}, {Allende
  Prieto}, {Beers}, \& {Wilhelm}}]{Brown:2005AJ....130.1097B}
{Brown}, W.~R., {Geller}, M.~J., {Kenyon}, S.~J., {et~al.} 2005, \aj, 130, 1097

\bibitem[{{Brunner} {et~al.}(1997){Brunner}, {Connolly}, {Szalay}, \&
  {Bershady}}]{Brunner:1997}
{Brunner}, R.~J., {Connolly}, A.~J., {Szalay}, A.~S., \& {Bershady}, M.~A.
  1997, \apjl, 482, L21+

\bibitem[{{Brusa} {et~al.}(2009){Brusa}, {Comastri}, {Gilli}, {Hasinger},
  {Iwasawa}, {Mainieri}, {Mignoli}, {Salvato}, {Zamorani}, {Bongiorno},
  {Cappelluti}, {Civano}, {Fiore}, {Merloni}, {Silverman}, {Trump}, {Vignali},
  {Capak}, {Elvis}, {Ilbert}, {Impey}, \& {Lilly}}]{Brusa:2009ApJ...693....8B}
{Brusa}, M., {Comastri}, A., {Gilli}, R., {et~al.} 2009, \apj, 693, 8

\bibitem[{{Budav{\'a}ri}(2009)}]{Budavri:2009}
{Budav{\'a}ri}, T. 2009, \apj, 695, 747

\bibitem[{{Caditz} \& {Petrosian}(1993)}]{Caditz:1993}
{Caditz}, D. \& {Petrosian}, V. 1993, \apj, 416, 450

\bibitem[{{Caputi} {et~al.}(2007){Caputi}, {Lagache}, {Yan}, {Dole},
  {Bavouzet}, {Le Floc'h}, {Choi}, {Helou}, \&
  {Reddy}}]{Caputi:2007ApJ...660...97C}
{Caputi}, K.~I., {Lagache}, G., {Yan}, L., {et~al.} 2007, \apj, 660, 97

\bibitem[{{Cen} \& {Ostriker}(1992)}]{Cen:1992ApJ...399L.113C}
{Cen}, R. \& {Ostriker}, J.~P. 1992, \apjl, 399, L113

\bibitem[{{Chapin} {et~al.}(2009){Chapin}, {Hughes}, \&
  {Aretxaga}}]{Chapin2009MNRAS.393..653C}
{Chapin}, E.~L., {Hughes}, D.~H., \& {Aretxaga}, I. 2009, \mnras, 393, 653

\bibitem[{{Chapman} {et~al.}(2003){Chapman}, {Helou}, {Lewis}, \&
  {Dale}}]{Chapman2003ApJ...588..186C}
{Chapman}, S.~C., {Helou}, G., {Lewis}, G.~F., \& {Dale}, D.~A. 2003, \apj,
  588, 186

\bibitem[{{Charlier, C. V. L.}(1922)}]{Charlier:1922}
{Charlier, C. V. L.} 1922, Arkiv. f$\ddot{o}$r. Astron. Fys.,, 16, 1

\bibitem[{{Chen} {et~al.}(2003){Chen}, {Marzke}, {McCarthy}, {Martini},
  {Carlberg}, {Persson}, {Bunker}, {Bridge}, \&
  {Abraham}}]{Chen:2003ApJ...586..745C}
{Chen}, H., {Marzke}, R.~O., {McCarthy}, P.~J., {et~al.} 2003, \apj, 586, 745

\bibitem[{{Chen} {et~al.}(1995){Chen}, {Chao}, \& {Lo}}]{chen:1995}
{Chen}, K., {Chao}, M., \& {Lo}, S. 1995, {\aos}, 23, 440

\bibitem[{{Cho{\l}oniewski}(1985)}]{Choloniewski1985MNRAS.214..197C}
{Cho{\l}oniewski}, J. 1985, \mnras, 214, 197

\bibitem[{{Cho{\l}oniewski}(1986)}]{Choloniewski:1986}
{Cho{\l}oniewski}, J. 1986, \mnras, 223, 1

\bibitem[{{Cho{\l}oniewski}(1987)}]{Choloniewski:1987}
{Cho{\l}oniewski}, J. 1987, \mnras, 226, 273

\bibitem[{{Christensen}(1975)}]{Christensen:1975}
{Christensen}, C.~G. 1975, \aj, 80, 282

\bibitem[{{Christlein} {et~al.}(2009){Christlein}, {Gawiser}, {Marchesini}, \&
  {Padilla}}]{Christlein2009MNRAS.400..429C}
{Christlein}, D., {Gawiser}, E., {Marchesini}, D., \& {Padilla}, N. 2009,
  \mnras, 400, 429

\bibitem[{{Cohen}(2002)}]{Cohen:2002ApJ...567..672C}
{Cohen}, J.~G. 2002, \apj, 567, 672

\bibitem[{{Cole}(2011)}]{Cole:2011arXiv1104.0009C}
{Cole}, S. 2011, ArXiv e-prints, astro-ph/1104.0009

\bibitem[{{Cole} {et~al.}(1994){Cole}, {Aragon-Salamanca}, {Frenk}, {Navarro},
  \& {Zepf}}]{Cole:1994MNRAS.271..781C}
{Cole}, S., {Aragon-Salamanca}, A., {Frenk}, C.~S., {Navarro}, J.~F., \&
  {Zepf}, S.~E. 1994, \mnras, 271, 781

\bibitem[{{Cole} {et~al.}(1998){Cole}, {Hatton}, {Weinberg}, \&
  {Frenk}}]{Cole:1998MNRAS.300..945C}
{Cole}, S., {Hatton}, S., {Weinberg}, D.~H., \& {Frenk}, C.~S. 1998, \mnras,
  300, 945

\bibitem[{{Coleman} {et~al.}(1980){Coleman}, {Wu}, \&
  {Weedman}}]{Coleman:1980ApJS...43..393C}
{Coleman}, G.~D., {Wu}, C., \& {Weedman}, D.~W. 1980, \apjs, 43, 393

\bibitem[{{Colless}(1998)}]{Colless:1998}
{Colless}, M. 1998, in Wide Field Surveys in Cosmology, 14th IAP meeting held
  May 26-30, 1998, Paris. Publisher: Editions Frontieres. ISBN: 2-8 6332-241-9,
  p. 77., ed. S.~{Colombi}, Y.~{Mellier}, \& B.~{Raban}, 77--+

\bibitem[{{Collister} {et~al.}(2007){Collister}, {Lahav}, {Blake}, {Cannon},
  {Croom}, {Drinkwater}, {Edge}, {Eisenstein}, {Loveday}, {Nichol}, {Pimbblet},
  {de Propris}, {Roseboom}, {Ross}, {Schneider}, {Shanks}, \&
  {Wake}}]{Collister:2007MNRAS.375...68C}
{Collister}, A., {Lahav}, O., {Blake}, C., {et~al.} 2007, \mnras, 375, 68

\bibitem[{{Condon}(1989)}]{Condon:1989ApJ...338...13C}
{Condon}, J.~J. 1989, \apj, 338, 13

\bibitem[{{Connolly} {et~al.}(1995){Connolly}, {Csabai}, {Szalay}, {Koo},
  {Kron}, \& {Munn}}]{Connolly:1995}
{Connolly}, A.~J., {Csabai}, I., {Szalay}, A.~S., {et~al.} 1995, \aj, 110, 2655

\bibitem[{{Covey} {et~al.}(2008){Covey}, {Hawley}, {Bochanski}, {West}, {Reid},
  {Golimowski}, {Davenport}, {Henry}, {Uomoto}, \&
  {Holtzman}}]{Covey:2008AJ....136.1778C}
{Covey}, K.~R., {Hawley}, S.~L., {Bochanski}, J.~J., {et~al.} 2008, \aj, 136,
  1778

\bibitem[{{Cowie} {et~al.}(1996){Cowie}, {Songaila}, {Hu}, \&
  {Cohen}}]{Cowie:1996AJ....112..839C}
{Cowie}, L.~L., {Songaila}, A., {Hu}, E.~M., \& {Cohen}, J.~G. 1996, \aj, 112,
  839

\bibitem[{{Crawford} {et~al.}(2009){Crawford}, {Bershady}, \&
  {Hoessel}}]{Crawford:2009}
{Crawford}, S.~M., {Bershady}, M.~A., \& {Hoessel}, J.~G. 2009, \apj, 690, 1158

\bibitem[{{Croom} {et~al.}(2009{\natexlab{a}}){Croom}, {Richards}, {Shanks},
  {Boyle}, {Sharp}, {Bland-Hawthorn}, {Bridges}, {Brunner}, {Cannon}, {Carson},
  {Chiu}, {Colless}, {Couch}, {de Propris}, {Drinkwater}, {Edge}, {Fine},
  {Loveday}, {Miller}, {Myers}, {Nichol}, {Outram}, {Pimbblet}, {Roseboom},
  {Ross}, {Schneider}, {Smith}, {Stoughton}, {Strauss}, \&
  {Wake}}]{Croom:2009MNRAS.392...19C}
{Croom}, S.~M., {Richards}, G.~T., {Shanks}, T., {et~al.} 2009{\natexlab{a}},
  \mnras, 392, 19

\bibitem[{{Croom} {et~al.}(2009{\natexlab{b}}){Croom}, {Richards}, {Shanks},
  {Boyle}, {Strauss}, {Myers}, {Nichol}, {Pimbblet}, {Ross}, {Schneider},
  {Sharp}, \& {Wake}}]{Croom:2009MNRAS.399.1755C}
{Croom}, S.~M., {Richards}, G.~T., {Shanks}, T., {et~al.} 2009{\natexlab{b}},
  \mnras, 399, 1755

\bibitem[{{Croom} {et~al.}(2001){Croom}, {Smith}, {Boyle}, {Shanks}, {Loaring},
  {Miller}, \& {Lewis}}]{Croom:2001MNRAS.322L..29C}
{Croom}, S.~M., {Smith}, R.~J., {Boyle}, B.~J., {et~al.} 2001, \mnras, 322, L29

\bibitem[{{Croom} {et~al.}(2004){Croom}, {Smith}, {Boyle}, {Shanks}, {Miller},
  {Outram}, \& {Loaring}}]{Croom:2004MNRAS.349.1397C}
{Croom}, S.~M., {Smith}, R.~J., {Boyle}, B.~J., {et~al.} 2004, \mnras, 349,
  1397

\bibitem[{{Cross} \& {Driver}(2002)}]{Cross2002MNRAS.329..579C}
{Cross}, N. \& {Driver}, S.~P. 2002, \mnras, 329, 579

\bibitem[{{Cuesta-Bolao} \& {Serna}(2003)}]{Cuesta:2003AA...405..917C}
{Cuesta-Bolao}, M.~J. \& {Serna}, A. 2003, \aap, 405, 917

\bibitem[{{Cutri} {et~al.}(2003){Cutri}, {Skrutskie}, {van Dyk}, {Beichman},
  {Carpenter}, {Chester}, {Cambresy}, {Evans}, {Fowler}, {Gizis}, {Howard},
  {Huchra}, {Jarrett}, {Kopan}, {Kirkpatrick}, {Light}, {Marsh}, {McCallon},
  {Schneider}, {Stiening}, {Sykes}, {Weinberg}, {Wheaton}, {Wheelock}, \&
  {Zacarias}}]{Cutri:2003tmc..book.....C}
{Cutri}, R.~M., {Skrutskie}, M.~F., {van Dyk}, S., {et~al.} 2003, {The IRSA
  2MASS All-Sky Point Source Catalog, NASA/IPAC Infrared Science Archive.}

\bibitem[{{da Costa} {et~al.}(1994){da Costa}, {Geller}, {Pellegrini},
  {Latham}, {Fairall}, {Marzke}, {Willmer}, {Huchra}, {Calderon}, {Ramella}, \&
  {Kurtz}}]{daCosta:1994ApJ...424L...1D}
{da Costa}, L.~N., {Geller}, M.~J., {Pellegrini}, P.~S., {et~al.} 1994, \apjl,
  424, L1

\bibitem[{{da Costa} {et~al.}(1988){da Costa}, {Pellegrini}, {Sargent},
  {Tonry}, {Davis}, {Meiksin}, {Latham}, {Menzies}, \& {Coulson}}]{Costa:1988}
{da Costa}, L.~N., {Pellegrini}, P.~S., {Sargent}, W.~L.~W., {et~al.} 1988,
  \apj, 327, 544

\bibitem[{{da Costa} {et~al.}(1998){da Costa}, {Willmer}, {Pellegrini},
  {Chaves}, {Rit{\'e}}, {Maia}, {Geller}, {Latham}, {Kurtz}, {Huchra},
  {Ramella}, {Fairall}, {Smith}, \&
  {L{\'{\i}}pari}}]{Costa:1998AJ....116....1D}
{da Costa}, L.~N., {Willmer}, C.~N.~A., {Pellegrini}, P.~S., {et~al.} 1998,
  \aj, 116, 1

\bibitem[{{Dav{\'e}} {et~al.}(2008){Dav{\'e}}, {Oppenheimer}, \&
  {Sivanandam}}]{Dav:2008MNRAS.391..110D}
{Dav{\'e}}, R., {Oppenheimer}, B.~D., \& {Sivanandam}, S. 2008, \mnras, 391,
  110

\bibitem[{{Davis} {et~al.}(1985){Davis}, {Efstathiou}, {Frenk}, \&
  {White}}]{Davis:1985ApJ...292..371D}
{Davis}, M., {Efstathiou}, G., {Frenk}, C.~S., \& {White}, S.~D.~M. 1985, \apj,
  292, 371

\bibitem[{{Davis} \& {Huchra}(1982)}]{Davis:1982}
{Davis}, M. \& {Huchra}, J. 1982, \apj, 254, 437

\bibitem[{{Davis} {et~al.}(1980){Davis}, {Tonry}, {Huchra}, \&
  {Latham}}]{Davis:1980}
{Davis}, M., {Tonry}, J., {Huchra}, J., \& {Latham}, D.~W. 1980, \apjl, 238,
  L113

\bibitem[{{de Blok}(2010)}]{2010AdAst2010E...5D}
{de Blok}, W.~J.~G. 2010, Advances in Astronomy, 2010

\bibitem[{{de Lapparent} {et~al.}(2003){de Lapparent}, {Galaz}, {Bardelli}, \&
  {Arnouts}}]{Lapparent:2003AA...404..831D}
{de Lapparent}, V., {Galaz}, G., {Bardelli}, S., \& {Arnouts}, S. 2003, \aap,
  404, 831

\bibitem[{{de Lapparent} {et~al.}(1989){de Lapparent}, {Geller}, \&
  {Huchra}}]{Lapparent:1989}
{de Lapparent}, V., {Geller}, M.~J., \& {Huchra}, J.~P. 1989, \apj, 343, 1

\bibitem[{{De Lucia} \& {Blaizot}(2007)}]{DeLucia:2007MNRAS.375....2D}
{De Lucia}, G. \& {Blaizot}, J. 2007, \mnras, 375, 2

\bibitem[{{Devereux} {et~al.}(2009){Devereux}, {Willner}, {Ashby}, {Willmer},
  \& {Hriljac}}]{Devereux:2009}
{Devereux}, N., {Willner}, S.~P., {Ashby}, M.~L.~N., {Willmer}, C.~N.~A., \&
  {Hriljac}, P. 2009, \apj, 702, 955

\bibitem[{{Devine} {et~al.}(2008){Devine}, {Churchwell}, {Indebetouw},
  {Watson}, \& {Crawford}}]{2008AJ....135.2095D}
{Devine}, K.~E., {Churchwell}, E.~B., {Indebetouw}, R., {Watson}, C., \&
  {Crawford}, S.~M. 2008, \aj, 135, 2095

\bibitem[{{Driver} {et~al.}(2005){Driver}, {Liske}, {Cross}, {De Propris}, \&
  {Allen}}]{Driver:2005}
{Driver}, S.~P., {Liske}, J., {Cross}, N.~J.~G., {De Propris}, R., \& {Allen},
  P.~D. 2005, \mnras, 360, 81

\bibitem[{{Dunne} {et~al.}(2000){Dunne}, {Eales}, {Edmunds}, {Ivison},
  {Alexander}, \& {Clements}}]{Dunne:2000MNRAS.315..115D}
{Dunne}, L., {Eales}, S., {Edmunds}, M., {et~al.} 2000, \mnras, 315, 115

\bibitem[{{Dye} \& {Eales}(2010)}]{Dye:2010MNRAS.tmp..610D}
{Dye}, S. \& {Eales}, S.~A. 2010, \mnras, 610

\bibitem[{{Eadie} {et~al.}(1971){Eadie}, {Drijard}, \&
  {James}}]{Eadie:1971smep.book.....E}
{Eadie}, W.~T., {Drijard}, D., \& {James}, F.~E. 1971, {Statistical methods in
  experimental physics} (Amsterdam: North-Holland)

\bibitem[{{Eales}(1993)}]{Eales:1993}
{Eales}, S. 1993, \apj, 404, 51

\bibitem[{{Eales} {et~al.}(2010){Eales}, {Raymond}, {Roseboom}, {Altieri},
  {Amblard}, {Arumugam}, {Auld}, {Aussel}, {Babbedge}, {Blain}, {Bock},
  {Boselli}, {Brisbin}, {Buat}, {Burgarella}, {Castro-Rodriguez}, {Cava},
  {Chanial}, {Clements}, {Conley}, {Conversi}, {Cooray}, {Dowell}, {Dwek},
  {Dye}, {Elbaz}, {Farrah}, {Fox}, {Franceschini}, {Gear}, {Glenn},
  {Gonzalez\~{}Solares}, {Griffin}, {Harwit}, {Hatziminaoglou}, {Huang},
  {Ibar}, {Isaak}, {Ivison}, {Lagache}, {Levenson}, {Lonsdale}, {Lu}, {Madden},
  {Maffei}, {Mainetti}, {Marchetti}, {Morrison}, {Mortier}, {Nguyen},
  {O'Halloran}, {Oliver}, {Omont}, {Owen}, {Page}, {Pannella}, {Panuzzo},
  {Papageorgiou}, {Pearson}, {Perez-Fournon}, {Pohlen}, {Rawlings},
  {Rigopoulou}, {Rizzo}, {Rowan-Robinson}, {Sanchez Portal}, {Schulz}, {Scott},
  {Seymour}, {Shupe}, {Smith}, {Stevens}, {Strazzullo}, {Symeonidis},
  {Trichas}, {Tugwell}, {Vaccari}, {Valtchanov}, {Vigroux}, {Wang}, {Ward},
  {Wright}, {Xu}, \& {Zemcov}}]{Eales:2010arXiv1005.2189E}
{Eales}, S., {Raymond}, G., {Roseboom}, I.~G., {et~al.} 2010, ArXiv e-prints

\bibitem[{{Efron} \& {Petrosian}(1992)}]{Efron:1992}
{Efron}, B. \& {Petrosian}, V. 1992, \apj, 399, 345

\bibitem[{{Efron} \& {Petrosian}(1999)}]{Efron:1998}
{Efron}, B. \& {Petrosian}, V. 1999, Journal of the American Statistical
  Association, 94(447), 824

\bibitem[{{Efstathiou} {et~al.}(1988){Efstathiou}, {Ellis}, \&
  {Peterson}}]{Efstathiou:1988}
{Efstathiou}, G., {Ellis}, R.~S., \& {Peterson}, B.~A. 1988, \mnras, 232, 431

\bibitem[{{Efstathiou} \& {Rees}(1988)}]{Efstathiou:1988MNRAS.230P...5E}
{Efstathiou}, G. \& {Rees}, M.~J. 1988, \mnras, 230, 5P

\bibitem[{{Eggen} {et~al.}(1962){Eggen}, {Lynden-Bell}, \&
  {Sandage}}]{Eggen:1962}
{Eggen}, O.~J., {Lynden-Bell}, D., \& {Sandage}, A.~R. 1962, \apj, 136, 748

\bibitem[{{Eisenstein} {et~al.}(2001){Eisenstein}, {Annis}, {Gunn}, {Szalay},
  {Connolly}, {Nichol}, {Bahcall}, {Bernardi}, {Burles}, {Castander},
  {Fukugita}, {Hogg}, {Ivezi{\'c}}, {Knapp}, {Lupton}, {Narayanan}, {Postman},
  {Reichart}, {Richmond}, {Schneider}, {Schlegel}, {Strauss}, {SubbaRao},
  {Tucker}, {Vanden Berk}, {Vogeley}, {Weinberg}, \&
  {Yanny}}]{Eisenstein:2001AJ.122.2267E}
{Eisenstein}, D.~J., {Annis}, J., {Gunn}, J.~E., {et~al.} 2001, \aj, 122, 2267

\bibitem[{{Eisenstein} {et~al.}(2005){Eisenstein}, {Zehavi}, {Hogg},
  {Scoccimarro}, {Blanton}, {Nichol}, {Scranton}, {Seo}, {Tegmark}, {Zheng},
  {Anderson}, {Annis}, {Bahcall}, {Brinkmann}, {Burles}, {Castander},
  {Connolly}, {Csabai}, {Doi}, {Fukugita}, {Frieman}, {Glazebrook}, {Gunn},
  {Hendry}, {Hennessy}, {Ivezi{\'c}}, {Kent}, {Knapp}, {Lin}, {Loh}, {Lupton},
  {Margon}, {McKay}, {Meiksin}, {Munn}, {Pope}, {Richmond}, {Schlegel},
  {Schneider}, {Shimasaku}, {Stoughton}, {Strauss}, {SubbaRao}, {Szalay},
  {Szapudi}, {Tucker}, {Yanny}, \& {York}}]{Eisenstein:2005ApJ...633..560E}
{Eisenstein}, D.~J., {Zehavi}, I., {Hogg}, D.~W., {et~al.} 2005, \apj, 633, 560

\bibitem[{{Ellis}(1997)}]{Ellis:1997ARAA..35..389E}
{Ellis}, R.~S. 1997, \araa, 35, 389

\bibitem[{{Ellis} {et~al.}(1996){Ellis}, {Colless}, {Broadhurst}, {Heyl}, \&
  {Glazebrook}}]{Ellis:1996MNRAS.280..235E}
{Ellis}, R.~S., {Colless}, M., {Broadhurst}, T., {Heyl}, J., \& {Glazebrook},
  K. 1996, \mnras, 280, 235

\bibitem[{{Faber} {et~al.}(2007){Faber}, {Willmer}, {Wolf}, {Koo}, {Weiner},
  {Newman}, {Im}, {Coil}, {Conroy}, {Cooper}, {Davis}, {Finkbeiner}, {Gerke},
  {Gebhardt}, {Groth}, {Guhathakurta}, {Harker}, {Kaiser}, {Kassin},
  {Kleinheinrich}, {Konidaris}, {Kron}, {Lin}, {Luppino}, {Madgwick},
  {Meisenheimer}, {Noeske}, {Phillips}, {Sarajedini}, {Schiavon}, {Simard},
  {Szalay}, {Vogt}, \& {Yan}}]{Faber:2007}
{Faber}, S.~M., {Willmer}, C.~N.~A., {Wolf}, C., {et~al.} 2007, \apj, 665, 265

\bibitem[{{Falco} {et~al.}(1999){Falco}, {Kurtz}, {Geller}, {Huchra}, {Peters},
  {Berlind}, {Mink}, {Tokarz}, \& {Elwell}}]{Falco:1999}
{Falco}, E.~E., {Kurtz}, M.~J., {Geller}, M.~J., {et~al.} 1999, \pasp, 111, 438

\bibitem[{{Fall} \& {Efstathiou}(1980)}]{Fall:1980MNRAS.193..189F}
{Fall}, S.~M. \& {Efstathiou}, G. 1980, \mnras, 193, 189

\bibitem[{{Fan} {et~al.}(2004){Fan}, {Hennawi}, {Richards}, {Strauss},
  {Schneider}, {Donley}, {Young}, {Annis}, {Lin}, {Lampeitl}, {Lupton}, {Gunn},
  {Knapp}, {Brandt}, {Anderson}, {Bahcall}, {Brinkmann}, {Brunner}, {Fukugita},
  {Szalay}, {Szokoly}, \& {York}}]{Fan:2004AJ....128..515F}
{Fan}, X., {Hennawi}, J.~F., {Richards}, G.~T., {et~al.} 2004, \aj, 128, 515

\bibitem[{{Fan} {et~al.}(2001){Fan}, {Narayanan}, {Lupton}, {Strauss}, {Knapp},
  {Becker}, {White}, {Pentericci}, {Leggett}, {Haiman}, {Gunn}, {Ivezi{\'c}},
  {Schneider}, {Anderson}, {Brinkmann}, {Bahcall}, {Connolly}, {Csabai}, {Doi},
  {Fukugita}, {Geballe}, {Grebel}, {Harbeck}, {Hennessy}, {Lamb}, {Miknaitis},
  {Munn}, {Nichol}, {Okamura}, {Pier}, {Prada}, {Richards}, {Szalay}, \&
  {York}}]{Fan:2001AJ....122.2833F}
{Fan}, X., {Narayanan}, V.~K., {Lupton}, R.~H., {et~al.} 2001, \aj, 122, 2833

\bibitem[{{Fan} {et~al.}(2006){Fan}, {Strauss}, {Becker}, {White}, {Gunn},
  {Knapp}, {Richards}, {Schneider}, {Brinkmann}, \&
  {Fukugita}}]{Fan:2006AJ....132..117F}
{Fan}, X., {Strauss}, M.~A., {Becker}, R.~H., {et~al.} 2006, \aj, 132, 117

\bibitem[{{Fan} {et~al.}(2003){Fan}, {Strauss}, {Schneider}, {Becker}, {White},
  {Haiman}, {Gregg}, {Pentericci}, {Grebel}, {Narayanan}, {Loh}, {Richards},
  {Gunn}, {Lupton}, {Knapp}, {Ivezi{\'c}}, {Brandt}, {Collinge}, {Hao},
  {Harbeck}, {Prada}, {Schaye}, {Strateva}, {Zakamska}, {Anderson},
  {Brinkmann}, {Bahcall}, {Lamb}, {Okamura}, {Szalay}, \&
  {York}}]{Fan:2003AJ....125.1649F}
{Fan}, X., {Strauss}, M.~A., {Schneider}, D.~P., {et~al.} 2003, \aj, 125, 1649

\bibitem[{{Fan} {et~al.}(2000){Fan}, {White}, {Davis}, {Becker}, {Strauss},
  {Haiman}, {Schneider}, {Gregg}, {Gunn}, {Knapp}, {Lupton}, {Anderson},
  {Anderson}, {Annis}, {Bahcall}, {Boroski}, {Brunner}, {Chen}, {Connolly},
  {Csabai}, {Doi}, {Fukugita}, {Hennessy}, {Hindsley}, {Ichikawa},
  {Ivezi{\'c}}, {Loveday}, {Meiksin}, {McKay}, {Munn}, {Newberg}, {Nichol},
  {Okamura}, {Pier}, {Sekiguchi}, {Shimasaku}, {Stoughton}, {Szalay},
  {Szokoly}, {Thakar}, {Vogeley}, \& {York}}]{Fan:2000AJ....120.1167F}
{Fan}, X., {White}, R.~L., {Davis}, M., {et~al.} 2000, \aj, 120, 1167

\bibitem[{{Felten}(1976)}]{Felten:1976}
{Felten}, J.~E. 1976, \apj, 207, 700

\bibitem[{{Felten}(1977)}]{Felten:1977}
{Felten}, J.~E. 1977, \aj, 82, 861

\bibitem[{{Ferguson} {et~al.}(2000){Ferguson}, {Dickinson}, \&
  {Williams}}]{Ferguson:2000ARAA..38..667F}
{Ferguson}, H.~C., {Dickinson}, M., \& {Williams}, R. 2000, \araa, 38, 667

\bibitem[{{Fern{\'a}ndez-Soto} {et~al.}(1999){Fern{\'a}ndez-Soto}, {Lanzetta},
  \& {Yahil}}]{Fernandez-Soto:1999}
{Fern{\'a}ndez-Soto}, A., {Lanzetta}, K.~M., \& {Yahil}, A. 1999, \apj, 513, 34

\bibitem[{{Ferrarese} \& {Merritt}(2000)}]{Ferrarese:2000ApJ...539L...9F}
{Ferrarese}, L. \& {Merritt}, D. 2000, \apjl, 539, L9

\bibitem[{{Finkelstein} {et~al.}(2010){Finkelstein}, {Papovich}, {Giavalisco},
  {Reddy}, {Ferguson}, {Koekemoer}, \&
  {Dickinson}}]{Finkelstein:2010ApJ...719.1250F}
{Finkelstein}, S.~L., {Papovich}, C., {Giavalisco}, M., {et~al.} 2010, \apj,
  719, 1250

\bibitem[{{Firth} {et~al.}(2003){Firth}, {Lahav}, \&
  {Somerville}}]{Firth:2003MNRAS.339.1195F}
{Firth}, A.~E., {Lahav}, O., \& {Somerville}, R.~S. 2003, \mnras, 339, 1195

\bibitem[{Fisher {et~al.}(1995)Fisher, Huchra, Strauss, Davis, Yahil, \&
  Schlegel}]{Fisher:1995.APJS.100.69}
Fisher, K., Huchra, J., Strauss, M., {et~al.} 1995, \apjs, 100, 69

\bibitem[{{Fisher}(1912)}]{Fisher:1912}
{Fisher}, R.~A. 1912, Messenger of Mathematics, 41, 155

\bibitem[{{Fisher}(1922)}]{Fisher:1922}
{Fisher}, R.~A. 1922, Philos. Trans. Roy. Soc. London Ser. A, 222, 309

\bibitem[{{Folkes} {et~al.}(1999){Folkes}, {Ronen}, {Price}, {Lahav},
  {Colless}, {Maddox}, {Deeley}, {Glazebrook}, {Bland-Hawthorn}, {Cannon},
  {Cole}, {Collins}, {Couch}, {Driver}, {Dalton}, {Efstathiou}, {Ellis},
  {Frenk}, {Kaiser}, {Lewis}, {Lumsden}, {Peacock}, {Peterson}, {Sutherland},
  \& {Taylor}}]{Folkes:1999MNRAS.308..459F}
{Folkes}, S., {Ronen}, S., {Price}, I., {et~al.} 1999, \mnras, 308, 459

\bibitem[{{Fontanot} {et~al.}(2007){Fontanot}, {Cristiani}, {Monaco}, {Nonino},
  {Vanzella}, {Brandt}, {Grazian}, \& {Mao}}]{Fontanot:2007AA...461...39F}
{Fontanot}, F., {Cristiani}, S., {Monaco}, P., {et~al.} 2007, \aap, 461, 39

\bibitem[{{Fried} {et~al.}(2001){Fried}, {von Kuhlmann}, {Meisenheimer}, {Rix},
  {Wolf}, {Hippelein}, {K{\"u}mmel}, {Phleps}, {R{\"o}ser}, {Thierring}, \&
  {Maier}}]{Fried:2001AA...367..788F}
{Fried}, J.~W., {von Kuhlmann}, B., {Meisenheimer}, K., {et~al.} 2001, \aap,
  367, 788

\bibitem[{{Fukugita} {et~al.}(1996){Fukugita}, {Ichikawa}, {Gunn}, {Doi},
  {Shimasaku}, \& {Schneider}}]{Fukugita:1996AJ....111.1748F}
{Fukugita}, M., {Ichikawa}, T., {Gunn}, J.~E., {et~al.} 1996, \aj, 111, 1748

\bibitem[{{Geller} \& {Huchra}(1989)}]{Geller:1989}
{Geller}, M.~J. \& {Huchra}, J.~P. 1989, Science, 246, 897

\bibitem[{{Gerdes} {et~al.}(2010){Gerdes}, {Sypniewski}, {McKay}, {Hao},
  {Weis}, {Wechsler}, \& {Busha}}]{Gerdes:2010ApJ...715..823G}
{Gerdes}, D.~W., {Sypniewski}, A.~J., {McKay}, T.~A., {et~al.} 2010, \apj, 715,
  823

\bibitem[{{Giannantonio} {et~al.}(2008){Giannantonio}, {Scranton},
  {Crittenden}, {Nichol}, {Boughn}, {Myers}, \&
  {Richards}}]{Giannantonio:2008PhRvD..77l3520G}
{Giannantonio}, T., {Scranton}, R., {Crittenden}, R.~G., {et~al.} 2008, \prd,
  77, 123520

\bibitem[{{Gieles} {et~al.}(2006){Gieles}, {Larsen}, {Bastian}, \&
  {Stein}}]{2006A&A...450..129G}
{Gieles}, M., {Larsen}, S.~S., {Bastian}, N., \& {Stein}, I.~T. 2006, \aap,
  450, 129

\bibitem[{{Gingold} \& {Monaghan}(1977)}]{Gingold:1977MNRAS.181..375G}
{Gingold}, R.~A. \& {Monaghan}, J.~J. 1977, \mnras, 181, 375

\bibitem[{{Glikman} {et~al.}(2010){Glikman}, {Bogosavljevi{\'c}}, {Djorgovski},
  {Stern}, {Dey}, {Jannuzi}, \& {Mahabal}}]{Glikman:2010ApJ...710.1498G}
{Glikman}, E., {Bogosavljevi{\'c}}, M., {Djorgovski}, S.~G., {et~al.} 2010,
  \apj, 710, 1498

\bibitem[{{Glikman} {et~al.}(2011){Glikman}, {Djorgovski}, {Stern}, {Dey},
  {Jannuzi}, \& {Lee}}]{Glikman:2011ApJ...728L..26G}
{Glikman}, E., {Djorgovski}, S.~G., {Stern}, D., {et~al.} 2011, \apjl, 728,
  L26+

\bibitem[{{Gnedin} \& {Zhao}(2002)}]{Gnedin:2002MNRAS.333..299G}
{Gnedin}, O.~Y. \& {Zhao}, H. 2002, \mnras, 333, 299

\bibitem[{{Goto}(2006)}]{Goto:2006MNRAS.371..769G}
{Goto}, T. 2006, \mnras, 371, 769

\bibitem[{{Governato} {et~al.}(2004){Governato}, {Mayer}, {Wadsley}, {Gardner},
  {Willman}, {Hayashi}, {Quinn}, {Stadel}, \&
  {Lake}}]{Governato:2004ApJ...607..688G}
{Governato}, F., {Mayer}, L., {Wadsley}, J., {et~al.} 2004, \apj, 607, 688

\bibitem[{{Governato} {et~al.}(2007){Governato}, {Willman}, {Mayer}, {Brooks},
  {Stinson}, {Valenzuela}, {Wadsley}, \& {Quinn}}]{2007MNRAS.374.1479G}
{Governato}, F., {Willman}, B., {Mayer}, L., {et~al.} 2007, \mnras, 374, 1479

\bibitem[{{Gunn} {et~al.}(1998){Gunn}, {Carr}, {Rockosi}, {Sekiguchi}, {Berry},
  {Elms}, {de Haas}, {Ivezi{\'c}}, {Knapp}, {Lupton}, {Pauls}, {Simcoe},
  {Hirsch}, {Sanford}, {Wang}, {York}, {Harris}, {Annis}, {Bartozek},
  {Boroski}, {Bakken}, {Haldeman}, {Kent}, {Holm}, {Holmgren}, {Petravick},
  {Prosapio}, {Rechenmacher}, {Doi}, {Fukugita}, {Shimasaku}, {Okada}, {Hull},
  {Siegmund}, {Mannery}, {Blouke}, {Heidtman}, {Schneider}, {Lucinio}, \&
  {Brinkman}}]{Gunn:1998AJ.116.3040G}
{Gunn}, J.~E., {Carr}, M., {Rockosi}, C., {et~al.} 1998, \aj, 116, 3040

\bibitem[{{Gunn} \& {Gott}(1972)}]{Gunn:1972ApJ...176....1G}
{Gunn}, J.~E. \& {Gott}, III, J.~R. 1972, \apj, 176, 1

\bibitem[{{Haberzettl} {et~al.}(2009){Haberzettl}, {Williger}, {Lauroesch},
  {Haines}, {Valls-Gabaud}, {Harris}, {Koekemoer}, {Loveday}, {Campusano},
  {Clowes}, {Dav{\'e}}, {Graham}, \& {S{\"o}chting}}]{Haberzettl:2009}
{Haberzettl}, L., {Williger}, G.~M., {Lauroesch}, J.~T., {et~al.} 2009, \apj,
  702, 506

\bibitem[{{Hacking} {et~al.}(1987){Hacking}, {Houck}, \&
  {Condon}}]{Hacking:1987ApJ...316L..15H}
{Hacking}, P., {Houck}, J.~R., \& {Condon}, J.~J. 1987, \apjl, 316, L15

\bibitem[{{Haehnelt} \& {Rees}(1993)}]{Haehnelt:1993MNRAS.263..168H}
{Haehnelt}, M.~G. \& {Rees}, M.~J. 1993, \mnras, 263, 168

\bibitem[{{Haiman} \& {Loeb}(2001)}]{Haiman:2001ApJ...552..459H}
{Haiman}, Z. \& {Loeb}, A. 2001, \apj, 552, 459

\bibitem[{{Hao} {et~al.}(2005){Hao}, {Strauss}, {Tremonti}, {Schlegel},
  {Heckman}, {Kauffmann}, {Blanton}, {Fan}, {Gunn}, {Hall}, {Ivezi{\'c}},
  {Knapp}, {Krolik}, {Lupton}, {Richards}, {Schneider}, {Strateva}, {Zakamska},
  {Brinkmann}, {Brunner}, \& {Szokoly}}]{Hao:2005}
{Hao}, L., {Strauss}, M.~A., {Tremonti}, C.~A., {et~al.} 2005, \aj, 129, 1783

\bibitem[{{Hargis} {et~al.}(2004){Hargis}, {Sandquist}, \&
  {Bolte}}]{Hargis:2004ApJ...608..243H}
{Hargis}, J.~R., {Sandquist}, E.~L., \& {Bolte}, M. 2004, \apj, 608, 243

\bibitem[{{Harris} {et~al.}(2006){Harris}, {Munn}, {Kilic}, {Liebert},
  {Williams}, {von Hippel}, {Levine}, {Monet}, {Eisenstein}, {Kleinman},
  {Metcalfe}, {Nitta}, {Winget}, {Brinkmann}, {Fukugita}, {Knapp}, {Lupton},
  {Smith}, \& {Schneider}}]{Harris:2006AJ....131..571H}
{Harris}, H.~C., {Munn}, J.~A., {Kilic}, M., {et~al.} 2006, \aj, 131, 571

\bibitem[{{Hartwick} \& {Schade}(1990)}]{Hartwick:1990ARAA..28..437H}
{Hartwick}, F.~D.~A. \& {Schade}, D. 1990, \araa, 28, 437

\bibitem[{{Hasinger} {et~al.}(2005){Hasinger}, {Miyaji}, \&
  {Schmidt}}]{Hasinger:2005AA...441..417H}
{Hasinger}, G., {Miyaji}, T., \& {Schmidt}, M. 2005, \aap, 441, 417

\bibitem[{{Hastings}(1970)}]{hastings:1970}
{Hastings}, W.~K. 1970, Biometrika, 57, 97

\bibitem[{{Hendry} \& {Simmons}(1990)}]{Hendry:1990A&A...237..275H}
{Hendry}, M.~A. \& {Simmons}, J.~F.~L. 1990, \aap, 237, 275

\bibitem[{{Heyl} {et~al.}(1997){Heyl}, {Colless}, {Ellis}, \&
  {Broadhurst}}]{Heyl:1997}
{Heyl}, J., {Colless}, M., {Ellis}, R.~S., \& {Broadhurst}, T. 1997, \mnras,
  285, 613

\bibitem[{{Hill} {et~al.}(2010){Hill}, {Driver}, {Cameron}, {Cross}, {Liske},
  \& {Robotham}}]{Hill:2010MNRAS.404.1215H}
{Hill}, D.~T., {Driver}, S.~P., {Cameron}, E., {et~al.} 2010, \mnras, 404, 1215

\bibitem[{{Hogg} {et~al.}(2002){Hogg}, {Baldry}, {Blanton}, \&
  {Eisenstein}}]{Hogg:2002}
{Hogg}, D.~W., {Baldry}, I.~K., {Blanton}, M.~R., \& {Eisenstein}, D.~J. 2002,
  ArXiv Astrophysics e-prints, astro-ph/0210394

\bibitem[{{Hubble}(1936{\natexlab{a}})}]{Hubble:1936b}
{Hubble}, E. 1936{\natexlab{a}}, \apj, 84, 517

\bibitem[{{Hubble}(1936{\natexlab{b}})}]{Hubble:1936}
{Hubble}, E.~P. 1936{\natexlab{b}}, {Realm of the Nebulae} (Realm of the
  Nebulae, by E.P.~Hubble.~ New Haven: Yale University Press, 1936)

\bibitem[{{Huchra} {et~al.}(1983){Huchra}, {Davis}, {Latham}, \&
  {Tonry}}]{Huchra:1983}
{Huchra}, J., {Davis}, M., {Latham}, D., \& {Tonry}, J. 1983, \apjs, 52, 89

\bibitem[{{Huchra} \& {Sargent}(1973)}]{Huchra:1973}
{Huchra}, J. \& {Sargent}, W.~L.~W. 1973, \apj, 186, 433

\bibitem[{{Hudson} \& {Lynden-Bell}(1991)}]{Hudson:1991}
{Hudson}, M.~J. \& {Lynden-Bell}, D. 1991, \mnras, 252, 219

\bibitem[{{Humason} {et~al.}(1956){Humason}, {Mayall}, \&
  {Sandage}}]{Humason:1956}
{Humason}, M.~L., {Mayall}, N.~U., \& {Sandage}, A.~R. 1956, \aj, 61, 97

\bibitem[{{Hunt} {et~al.}(2004){Hunt}, {Steidel}, {Adelberger}, \&
  {Shapley}}]{Hunt:2004ApJ...605..625H}
{Hunt}, M.~P., {Steidel}, C.~C., {Adelberger}, K.~L., \& {Shapley}, A.~E. 2004,
  \apj, 605, 625

\bibitem[{{Ikeda} {et~al.}(2011){Ikeda}, {Nagao}, {Matsuoka}, {Taniguchi},
  {Shioya}, {Trump}, {Capak}, {Comastri}, {Enoki}, {Ideue}, {Kakazu},
  {Koekemoer}, {Morokuma}, {Murayama}, {Saito}, {Salvato}, {Schinnerer},
  {Scoville}, \& {Silverman}}]{Ikeda:2010arXiv1011.2280I}
{Ikeda}, H., {Nagao}, T., {Matsuoka}, K., {et~al.} 2011, \apjl, 728, L25+

\bibitem[{{Ilbert} {et~al.}(2006){Ilbert}, {Lauger}, {Tresse}, {Buat},
  {Arnouts}, {Le F{\`e}vre}, {Burgarella}, {Zucca}, {Bardelli}, {Zamorani},
  {Bottini}, {Garilli}, {Le Brun}, {Maccagni}, {Picat}, {Scaramella},
  {Scodeggio}, {Vettolani}, {Zanichelli}, {Adami}, {Arnaboldi}, {Bolzonella},
  {Cappi}, {Charlot}, {Contini}, {Foucaud}, {Franzetti}, {Gavignaud}, {Guzzo},
  {Iovino}, {McCracken}, {Marano}, {Marinoni}, {Mathez}, {Mazure}, {Meneux},
  {Merighi}, {Paltani}, {Pello}, {Pollo}, {Pozzetti}, {Radovich}, {Bondi},
  {Bongiorno}, {Busarello}, {Ciliegi}, {Mellier}, {Merluzzi}, {Ripepi}, \&
  {Rizzo}}]{Ilbert:2006AA...453..809I}
{Ilbert}, O., {Lauger}, S., {Tresse}, L., {et~al.} 2006, \aap, 453, 809

\bibitem[{{Ilbert} {et~al.}(2004){Ilbert}, {Tresse}, {Arnouts}, {Zucca},
  {Bardelli}, {Zamorani}, {Adami}, {Cappi}, {Garilli}, {Le F{\`e}vre},
  {Maccagni}, {Meneux}, {Scaramella}, {Scodeggio}, {Vettolani}, \&
  {Zanichelli}}]{Ilbert:2004}
{Ilbert}, O., {Tresse}, L., {Arnouts}, S., {et~al.} 2004, \mnras, 351, 541

\bibitem[{{Ilbert} {et~al.}(2005){Ilbert}, {Tresse}, {Zucca}, {Bardelli},
  {Arnouts}, {Zamorani}, {Pozzetti}, {Bottini}, {Garilli}, {Le Brun}, {Le
  F{\`e}vre}, {Maccagni}, {Picat}, {Scaramella}, {Scodeggio}, {Vettolani},
  {Zanichelli}, {Adami}, {Arnaboldi}, {Bolzonella}, {Cappi}, {Charlot},
  {Contini}, {Foucaud}, {Franzetti}, {Gavignaud}, {Guzzo}, {Iovino},
  {McCracken}, {Marano}, {Marinoni}, {Mathez}, {Mazure}, {Meneux}, {Merighi},
  {Paltani}, {Pello}, {Pollo}, {Radovich}, {Bondi}, {Bongiorno}, {Busarello},
  {Ciliegi}, {Lamareille}, {Mellier}, {Merluzzi}, {Ripepi}, \&
  {Rizzo}}]{Ilbert:2005}
{Ilbert}, O., {Tresse}, L., {Zucca}, E., {et~al.} 2005, \aap, 439, 863

\bibitem[{{Im} {et~al.}(2002){Im}, {Simard}, {Faber}, {Koo}, {Gebhardt},
  {Willmer}, {Phillips}, {Illingworth}, {Vogt}, \&
  {Sarajedini}}]{Im:2002ApJ...571..136I}
{Im}, M., {Simard}, L., {Faber}, S.~M., {et~al.} 2002, \apj, 571, 136

\bibitem[{{Jackson}(1974)}]{Jackson:1974}
{Jackson}, J.~C. 1974, \mnras, 166, 281

\bibitem[{{Jannuzi} \& {Dey}(1999)}]{Jannuzi:1999ASPC..193..258J}
{Jannuzi}, B.~T. \& {Dey}, A. 1999, in Astronomical Society of the Pacific
  Conference Series, Vol. 193, The Hy-Redshift Universe: Galaxy Formation and
  Evolution at High Redshift, ed. {A.~J.~Bunker \& W.~J.~M.~van Breugel},
  258--+

\bibitem[{{Jiang} {et~al.}(2009){Jiang}, {Fan}, {Bian}, {Annis}, {Chiu},
  {Jester}, {Lin}, {Lupton}, {Richards}, {Strauss}, {Malanushenko},
  {Malanushenko}, \& {Schneider}}]{Jiang:2009AJ....138..305J}
{Jiang}, L., {Fan}, X., {Bian}, F., {et~al.} 2009, \aj, 138, 305

\bibitem[{{Johnston} {et~al.}(2007){Johnston}, {Teodoro}, \&
  {Hendry}}]{Johnston:2007}
{Johnston}, R., {Teodoro}, L., \& {Hendry}, M. 2007, \mnras, 376, 1757

\bibitem[{{Kafka}(1967)}]{Kafka:1967}
{Kafka}, P. 1967, \nat, 213, 346

\bibitem[{{Kaiser} \& {Best}(2007)}]{2007MNRAS.381.1548K}
{Kaiser}, C.~R. \& {Best}, P.~N. 2007, \mnras, 381, 1548

\bibitem[{{Katz} \& {Gunn}(1991)}]{Katz:1991ApJ...377..365K}
{Katz}, N. \& {Gunn}, J.~E. 1991, \apj, 377, 365

\bibitem[{{Katz} \& {White}(1993)}]{Katz:1993ApJ...412..455K}
{Katz}, N. \& {White}, S.~D.~M. 1993, \apj, 412, 455

\bibitem[{{Kauffmann} \& {Haehnelt}(2000)}]{Kauffmann:2000MNRAS.311..576K}
{Kauffmann}, G. \& {Haehnelt}, M. 2000, \mnras, 311, 576

\bibitem[{{Kelly} {et~al.}(2008){Kelly}, {Fan}, \& {Vestergaard}}]{Kelly:2008}
{Kelly}, B.~C., {Fan}, X., \& {Vestergaard}, M. 2008, \apj, 682, 874

\bibitem[{{Kendall}(1938)}]{kendall:1938}
{Kendall}, M. 1938, Biometrika, 56, 231

\bibitem[{Kendall(1961)}]{Kendall:1961}
Kendall, M.~G. 1961, Biometrika, 48, 1

\bibitem[{{Kere{\v s}} {et~al.}(2009){Kere{\v s}}, {Katz}, {Fardal},
  {Dav{\'e}}, \& {Weinberg}}]{2009MNRAS.395..160K}
{Kere{\v s}}, D., {Katz}, N., {Fardal}, M., {Dav{\'e}}, R., \& {Weinberg},
  D.~H. 2009, \mnras, 395, 160

\bibitem[{{Kere{\v s}} {et~al.}(2005){Kere{\v s}}, {Katz}, {Weinberg}, \&
  {Dav{\'e}}}]{Kere:2005MNRAS.363....2K}
{Kere{\v s}}, D., {Katz}, N., {Weinberg}, D.~H., \& {Dav{\'e}}, R. 2005,
  \mnras, 363, 2

\bibitem[{{Kiang}(1961)}]{Kiang:1961}
{Kiang}, T. 1961, \mnras, 122, 263

\bibitem[{{Kinney} {et~al.}(1996){Kinney}, {Calzetti}, {Bohlin}, {McQuade},
  {Storchi-Bergmann}, \& {Schmitt}}]{Kinney:1996ApJ...467...38K}
{Kinney}, A.~L., {Calzetti}, D., {Bohlin}, R.~C., {et~al.} 1996, \apj, 467, 38

\bibitem[{{Kirshner} {et~al.}(1979){Kirshner}, {Oemler}, \&
  {Schechter}}]{Kirshner:1979}
{Kirshner}, R.~P., {Oemler}, Jr., A., \& {Schechter}, P.~L. 1979, \aj, 84, 951

\bibitem[{{Klypin} {et~al.}(1999){Klypin}, {Kravtsov}, {Valenzuela}, \&
  {Prada}}]{1999ApJ...522...82K}
{Klypin}, A., {Kravtsov}, A.~V., {Valenzuela}, O., \& {Prada}, F. 1999, \apj,
  522, 82

\bibitem[{{Kneib} {et~al.}(2004){Kneib}, {Ellis}, {Santos}, \&
  {Richard}}]{Kneib:2004ApJ...607..697K}
{Kneib}, J., {Ellis}, R.~S., {Santos}, M.~R., \& {Richard}, J. 2004, \apj, 607,
  697

\bibitem[{{Kocevski} \& {Liang}(2006)}]{Kocevski:2006ApJ...642..371K}
{Kocevski}, D. \& {Liang}, E. 2006, \apj, 642, 371

\bibitem[{{Kodama} {et~al.}(1999){Kodama}, {Bell}, \& {Bower}}]{Kodama:1999}
{Kodama}, T., {Bell}, E.~F., \& {Bower}, R.~G. 1999, \mnras, 302, 152

\bibitem[{{Koo} \& {Kron}(1988)}]{Koo:1988ApJ...325...92K}
{Koo}, D.~C. \& {Kron}, R.~G. 1988, \apj, 325, 92

\bibitem[{{Koo} \& {Kron}(1992)}]{koo:1992ARAA..30..613K}
{Koo}, D.~C. \& {Kron}, R.~G. 1992, \araa, 30, 613

\bibitem[{{Koranyi} \& {Strauss}(1997)}]{1997ApJ...477...36K}
{Koranyi}, D.~M. \& {Strauss}, M.~A. 1997, \apj, 477, 36

\bibitem[{{Kormendy} \& {Richstone}(1995)}]{Kormendy:1995ARAA..33..581K}
{Kormendy}, J. \& {Richstone}, D. 1995, \araa, 33, 581

\bibitem[{{Kravtsov}(2010)}]{2010AdAst2010E...8K}
{Kravtsov}, A. 2010, Advances in Astronomy, 2010

\bibitem[{{Lacey} \& {Cole}(1994)}]{Lacey:1994MNRAS.271..676L}
{Lacey}, C. \& {Cole}, S. 1994, \mnras, 271, 676

\bibitem[{{Lanzetta} {et~al.}(2002){Lanzetta}, {Yahata}, {Pascarelle}, {Chen},
  \& {Fern{\'a}ndez-Soto}}]{Lanzetta:2002ApJ...570..492L}
{Lanzetta}, K.~M., {Yahata}, N., {Pascarelle}, S., {Chen}, H., \&
  {Fern{\'a}ndez-Soto}, A. 2002, \apj, 570, 492

\bibitem[{{Larson} {et~al.}(2011){Larson}, {Dunkley}, {Hinshaw}, {Komatsu},
  {Nolta}, {Bennett}, {Gold}, {Halpern}, {Hill}, {Jarosik}, {Kogut}, {Limon},
  {Meyer}, {Odegard}, {Page}, {Smith}, {Spergel}, {Tucker}, {Weiland},
  {Wollack}, \& {Wright}}]{Larson:2010}
{Larson}, D., {Dunkley}, J., {Hinshaw}, G., {et~al.} 2011, \apjs, 192, 16

\bibitem[{{Lawrence} {et~al.}(1986){Lawrence}, {Walker}, {Rowan-Robinson},
  {Leech}, \& {Penston}}]{Lawrence:1986}
{Lawrence}, A., {Walker}, D., {Rowan-Robinson}, M., {Leech}, K.~J., \&
  {Penston}, M.~V. 1986, \mnras, 219, 687

\bibitem[{{Lawrence} {et~al.}(2007){Lawrence}, {Warren}, {Almaini}, {Edge},
  {Hambly}, {Jameson}, {Lucas}, {Casali}, {Adamson}, {Dye}, {Emerson},
  {Foucaud}, {Hewett}, {Hirst}, {Hodgkin}, {Irwin}, {Lodieu}, {McMahon},
  {Simpson}, {Smail}, {Mortlock}, \& {Folger}}]{Lawrence:2007}
{Lawrence}, A., {Warren}, S.~J., {Almaini}, O., {et~al.} 2007, \mnras, 379,
  1599

\bibitem[{{Le Borgne} {et~al.}(2009){Le Borgne}, {Elbaz}, {Ocvirk}, \&
  {Pichon}}]{Borgne:2009AA...504..727L}
{Le Borgne}, D., {Elbaz}, D., {Ocvirk}, P., \& {Pichon}, C. 2009, \aap, 504,
  727

\bibitem[{{Lee} {et~al.}(2009){Lee}, {Kennicutt}, {Jos{\'e} G.~Funes}, {Sakai},
  \& {Akiyama}}]{Lee:2009}
{Lee}, J.~C., {Kennicutt}, R.~C., {Jos{\'e} G.~Funes}, S.~J., {Sakai}, S., \&
  {Akiyama}, S. 2009, \apj, 692, 1305

\bibitem[{{Liebert} {et~al.}(1988){Liebert}, {Dahn}, \&
  {Monet}}]{Liebert:1988ApJ...332..891L}
{Liebert}, J., {Dahn}, C.~C., \& {Monet}, D.~G. 1988, \apj, 332, 891

\bibitem[{{Lilly}(2007)}]{Lilly:2007}
{Lilly}, S.~J.~{\etal}. 2007, \apjs, 172, 70

\bibitem[{{Lilly} {et~al.}(1995){Lilly}, {Tresse}, {Hammer}, {Crampton}, \& {Le
  Fevre}}]{Lilly:1995ApJ...455..108L}
{Lilly}, S.~J., {Tresse}, L., {Hammer}, F., {Crampton}, D., \& {Le Fevre}, O.
  1995, \apj, 455, 108

\bibitem[{{Lima} {et~al.}(2008){Lima}, {Cunha}, {Oyaizu}, {Frieman}, {Lin}, \&
  {Sheldon}}]{Lima:2008}
{Lima}, M., {Cunha}, C.~E., {Oyaizu}, H., {et~al.} 2008, \mnras, 390, 118

\bibitem[{{Limber}(1960)}]{Limber:1960ApJ...131..168L}
{Limber}, D.~N. 1960, \apj, 131, 168

\bibitem[{{Lin} {et~al.}(1996){Lin}, {Kirshner}, {Shectman}, {Landy}, {Oemler},
  {Tucker}, \& {Schechter}}]{Lin:1996ApJ...464...60L}
{Lin}, H., {Kirshner}, R.~P., {Shectman}, S.~A., {et~al.} 1996, \apj, 464, 60

\bibitem[{{Lin} {et~al.}(1997){Lin}, {Yee}, {Carlberg}, \&
  {Ellingson}}]{Lin:1997ApJ...475..494L}
{Lin}, H., {Yee}, H.~K.~C., {Carlberg}, R.~G., \& {Ellingson}, E. 1997, \apj,
  475, 494

\bibitem[{{Lin} {et~al.}(1999){Lin}, {Yee}, {Carlberg}, {Morris}, {Sawicki},
  {Patton}, {Wirth}, \& {Shepherd}}]{Lin:1999}
{Lin}, H., {Yee}, H.~K.~C., {Carlberg}, R.~G., {et~al.} 1999, \apj, 518, 533

\bibitem[{{Liske} {et~al.}(2003){Liske}, {Lemon}, {Driver}, {Cross}, \&
  {Couch}}]{Liske:2003}
{Liske}, J., {Lemon}, D.~J., {Driver}, S.~P., {Cross}, N.~J.~G., \& {Couch},
  W.~J. 2003, \mnras, 344, 307

\bibitem[{{Loh} \& {Spillar}(1986)}]{Loh:1986a}
{Loh}, E.~D. \& {Spillar}, E.~J. 1986, \apj, 303, 154

\bibitem[{{Loveday} {et~al.}(1992){Loveday}, {Peterson}, {Efstathiou}, \&
  {Maddox}}]{Loveday:1992ApJ...390..338L}
{Loveday}, J., {Peterson}, B.~A., {Efstathiou}, G., \& {Maddox}, S.~J. 1992,
  \apj, 390, 338

\bibitem[{{Lucy}(1974)}]{Lucy:1974AJ.....79..745L}
{Lucy}, L.~B. 1974, \aj, 79, 745

\bibitem[{{Lucy}(1977)}]{Lucy:1977AJ.....82.1013L}
{Lucy}, L.~B. 1977, \aj, 82, 1013

\bibitem[{{Lynden-Bell}(1971)}]{lynden:1971}
{Lynden-Bell}, D. 1971, \mnras, 155, 95

\bibitem[{{Maccacaro} {et~al.}(1991){Maccacaro}, {della Ceca}, {Gioia},
  {Morris}, {Stocke}, \& {Wolter}}]{Maccacaro:1991}
{Maccacaro}, T., {della Ceca}, R., {Gioia}, I.~M., {et~al.} 1991, \apj, 374,
  117

\bibitem[{{Madau} {et~al.}(1996){Madau}, {Ferguson}, {Dickinson}, {Giavalisco},
  {Steidel}, \& {Fruchter}}]{Madau:1996MNRAS.283.1388M}
{Madau}, P., {Ferguson}, H.~C., {Dickinson}, M.~E., {et~al.} 1996, \mnras, 283,
  1388

\bibitem[{{Madgwick} {et~al.}(2002){Madgwick}, {Lahav}, {Baldry}, {Baugh},
  {Bland-Hawthorn}, {Bridges}, {Cannon}, {Cole}, {Colless}, {Collins}, {Couch},
  {Dalton}, {De Propris}, {Driver}, {Efstathiou}, {Ellis}, {Frenk},
  {Glazebrook}, {Jackson}, {Lewis}, {Lumsden}, {Maddox}, {Norberg}, {Peacock},
  {Peterson}, {Sutherland}, \& {Taylor}}]{Madgwick:2002MNRAS.333..133M}
{Madgwick}, D.~S., {Lahav}, O., {Baldry}, I.~K., {et~al.} 2002, \mnras, 333,
  133

\bibitem[{{Magnelli} {et~al.}(2009){Magnelli}, {Elbaz}, {Chary}, {Dickinson},
  {Le Borgne}, {Frayer}, \& {Willmer}}]{Magnelli:2009AA...496...57M}
{Magnelli}, B., {Elbaz}, D., {Chary}, R.~R., {et~al.} 2009, \aap, 496, 57

\bibitem[{{Magnelli} {et~al.}(2011){Magnelli}, {Elbaz}, {Chary}, {Dickinson},
  {Le Borgne}, {Frayer}, \& {Willmer}}]{Magnelli:2011AA...528A..35M}
{Magnelli}, B., {Elbaz}, D., {Chary}, R.~R., {et~al.} 2011, \aap, 528, A35+

\bibitem[{{Magorrian} {et~al.}(1998){Magorrian}, {Tremaine}, {Richstone},
  {Bender}, {Bower}, {Dressler}, {Faber}, {Gebhardt}, {Green}, {Grillmair},
  {Kormendy}, \& {Lauer}}]{Magorrian:1998AJ....115.2285M}
{Magorrian}, J., {Tremaine}, S., {Richstone}, D., {et~al.} 1998, \aj, 115, 2285

\bibitem[{{Maloney} \& {Petrosian}(1999)}]{Maloney:1999}
{Maloney}, A. \& {Petrosian}, V. 1999, \apj, 518, 32

\bibitem[{{Markarian}(1967)}]{Markarian:1967}
{Markarian}, B.~E. 1967, Astrofizika, 3, 55

\bibitem[{{Markarian}(1969{\natexlab{a}})}]{Markarian:1969a}
{Markarian}, B.~E. 1969{\natexlab{a}}, Astrofizika, 5, 443

\bibitem[{{Markarian}(1969{\natexlab{b}})}]{Markarian:1969b}
{Markarian}, B.~E. 1969{\natexlab{b}}, Astrofizika, 5, 581

\bibitem[{{Markarian} {et~al.}(1971){Markarian}, {Lipovetskij}, \&
  {Lipovetsky}}]{Markarian:1971}
{Markarian}, B.~E., {Lipovetskij}, V.~A., \& {Lipovetsky}, V.~A. 1971,
  Astrofizika, 7, 511

\bibitem[{{Marsden} {et~al.}(2010){Marsden}, {Chapin}, {Halpern}, {Patanchon},
  {Scott}, {Truch}, {Valiante}, {Viero}, \&
  {Wiebe}}]{Marsden2010arXiv1010.1176M}
{Marsden}, G., {Chapin}, E.~L., {Halpern}, M., {et~al.} 2010, ArXiv e-prints,
  astro-ph/10101176

\bibitem[{{Marshall}(1985)}]{Marshall:1985ApJ...299..109M}
{Marshall}, H.~L. 1985, \apj, 299, 109

\bibitem[{{Marshall} {et~al.}(1983){Marshall}, {Tananbaum}, {Avni}, \&
  {Zamorani}}]{Marshall:1983}
{Marshall}, H.~L., {Tananbaum}, H., {Avni}, Y., \& {Zamorani}, G. 1983, \apj,
  269, 35

\bibitem[{{Marzke} {et~al.}(1994{\natexlab{a}}){Marzke}, {Geller}, {Huchra}, \&
  {Corwin}}]{Marzke:1994AJ....108..437M}
{Marzke}, R.~O., {Geller}, M.~J., {Huchra}, J.~P., \& {Corwin}, Jr., H.~G.
  1994{\natexlab{a}}, \aj, 108, 437

\bibitem[{{Marzke} {et~al.}(1994{\natexlab{b}}){Marzke}, {Huchra}, \&
  {Geller}}]{Marzke:1994ApJ...428...43M}
{Marzke}, R.~O., {Huchra}, J.~P., \& {Geller}, M.~J. 1994{\natexlab{b}}, \apj,
  428, 43

\bibitem[{{Massarotti} {et~al.}(2001){Massarotti}, {Iovino}, {Buzzoni}, \&
  {Valls-Gabaud}}]{Massarotti:2001AA...380..425M}
{Massarotti}, M., {Iovino}, A., {Buzzoni}, A., \& {Valls-Gabaud}, D. 2001,
  \aap, 380, 425

\bibitem[{{McConnachie} {et~al.}(2009){McConnachie}, {Irwin}, {Ibata},
  {Dubinski}, {Widrow}, {Martin}, {C{\^o}t{\'e}}, {Dotter}, {Navarro},
  {Ferguson}, {Puzia}, {Lewis}, {Babul}, {Barmby}, {Bienaym{\'e}}, {Chapman},
  {Cockcroft}, {Collins}, {Fardal}, {Harris}, {Huxor}, {Mackey},
  {Pe{\~n}arrubia}, {Rich}, {Richer}, {Siebert}, {Tanvir}, {Valls-Gabaud}, \&
  {Venn}}]{2009Natur.461...66M}
{McConnachie}, A.~W., {Irwin}, M.~J., {Ibata}, R.~A., {et~al.} 2009, \nat, 461,
  66

\bibitem[{{McLure} {et~al.}(2010){McLure}, {Dunlop}, {Cirasuolo}, {Koekemoer},
  {Sabbi}, {Stark}, {Targett}, \& {Ellis}}]{McLure:2010MNRAS.403..960M}
{McLure}, R.~J., {Dunlop}, J.~S., {Cirasuolo}, M., {et~al.} 2010, \mnras, 403,
  960

\bibitem[{{Metropolis} {et~al.}(1953){Metropolis}, {Rosenbluth}, {Rosenbluth},
  {Teller}, \& {Teller}}]{Metropolis:1953JChPh..21.1087M}
{Metropolis}, N., {Rosenbluth}, A.~W., {Rosenbluth}, M.~N., {Teller}, A.~H., \&
  {Teller}, E. 1953, \jcp, 21, 1087

\bibitem[{{Metropolis} \& {Ulam}(1949)}]{Metropolis:1949}
{Metropolis}, N. \& {Ulam}, S. 1949, \jasa, 44, 335

\bibitem[{{Miller} \& {Scalo}(1979)}]{Miller:1979ApJS...41..513M}
{Miller}, G.~E. \& {Scalo}, J.~M. 1979, \apjs, 41, 513

\bibitem[{{Miyaji} {et~al.}(2001){Miyaji}, {Hasinger}, \&
  {Schmidt}}]{Miyaji:2001AA...369...49M}
{Miyaji}, T., {Hasinger}, G., \& {Schmidt}, M. 2001, \aap, 369, 49

\bibitem[{{Montero-Dorta} \& {Prada}(2009)}]{Montero:2009MNRAS.399.1106M}
{Montero-Dorta}, A.~D. \& {Prada}, F. 2009, \mnras, 399, 1106

\bibitem[{{Moore} {et~al.}(1999){Moore}, {Ghigna}, {Governato}, {Lake},
  {Quinn}, {Stadel}, \& {Tozzi}}]{1999ApJ...524L..19M}
{Moore}, B., {Ghigna}, S., {Governato}, F., {et~al.} 1999, \apjl, 524, L19

\bibitem[{{Navarro} \& {Benz}(1991)}]{Navarro:1991ApJ...380..320N}
{Navarro}, J.~F. \& {Benz}, W. 1991, \apj, 380, 320

\bibitem[{{Navarro} {et~al.}(1994){Navarro}, {Frenk}, \&
  {White}}]{Navarro:1994MNRAS.267L...1N}
{Navarro}, J.~F., {Frenk}, C.~S., \& {White}, S.~D.~M. 1994, \mnras, 267, L1+

\bibitem[{{Navarro} {et~al.}(2004){Navarro}, {Hayashi}, {Power}, {Jenkins},
  {Frenk}, {White}, {Springel}, {Stadel}, \&
  {Quinn}}]{Navarro:2004MNRAS.349.1039N}
{Navarro}, J.~F., {Hayashi}, E., {Power}, C., {et~al.} 2004, \mnras, 349, 1039

\bibitem[{{Nicoll} \& {Segal}(1983)}]{Nicoll:1983}
{Nicoll}, J.~F. \& {Segal}, I.~E. 1983, \aap, 118, 180

\bibitem[{{Norberg} {et~al.}(2002){Norberg}, {Cole}, {Baugh}, {Frenk},
  {Baldry}, {Bland-Hawthorn}, {Bridges}, {Cannon}, {Colless}, {Collins},
  {Couch}, {Cross}, {Dalton}, {De Propris}, {Driver}, {Efstathiou}, {Ellis},
  {Glazebrook}, {Jackson}, {Lahav}, {Lewis}, {Lumsden}, {Maddox}, {Madgwick},
  {Peacock}, {Peterson}, {Sutherland}, \& {Taylor}}]{Norberg:2002b}
{Norberg}, P., {Cole}, S., {Baugh}, C.~M., {et~al.} 2002, \mnras, 336, 907

\bibitem[{{Oesch} {et~al.}(2010){Oesch}, {Bouwens}, {Carollo}, {Illingworth},
  {Magee}, {Trenti}, {Stiavelli}, {Franx}, {Labb{\'e}}, \& {van
  Dokkum}}]{Oesch:2010ApJ...725L.150O}
{Oesch}, P.~A., {Bouwens}, R.~J., {Carollo}, C.~M., {et~al.} 2010, \apjl, 725,
  L150

\bibitem[{{Oke} \& {Sandage}(1968)}]{Oke:1968ApJ...154...21O}
{Oke}, J.~B. \& {Sandage}, A. 1968, \apj, 154, 21

\bibitem[{{Osmer}(1982)}]{Osmer:1982ApJ...253...28O}
{Osmer}, P.~S. 1982, \apj, 253, 28

\bibitem[{{Oyaizu} {et~al.}(2008){Oyaizu}, {Lima}, {Cunha}, {Lin}, {Frieman},
  \& {Sheldon}}]{Oyaizu:2008ApJ...674..768O}
{Oyaizu}, H., {Lima}, M., {Cunha}, C.~E., {et~al.} 2008, \apj, 674, 768

\bibitem[{{Page} \& {Carrera}(2000)}]{Page:2000}
{Page}, M.~J. \& {Carrera}, F.~J. 2000, \mnras, 311, 433

\bibitem[{{Peacock}(1985)}]{Peacock:1985MNRAS.217..601P}
{Peacock}, J.~A. 1985, \mnras, 217, 601

\bibitem[{{Peacock} \& {Gull}(1981)}]{Peacock:1981MNRAS.196..611P}
{Peacock}, J.~A. \& {Gull}, S.~F. 1981, \mnras, 196, 611

\bibitem[{Peebles(1980)}]{Peebles:1980}
Peebles, P. 1980, The Large Scale Structure of the Universe ((Princeton:
  Princeton Univ. Press))

\bibitem[{{Peebles}(1970)}]{Peebles:1970AJ.....75...13P}
{Peebles}, P.~J.~E. 1970, \aj, 75, 13

\bibitem[{{Peebles}(1973)}]{Peebles:1973PASJ...25..291P}
{Peebles}, P.~J.~E. 1973, \pasj, 25, 291

\bibitem[{{Peebles}(1974)}]{Peebles:1974ApJ...189L..51P}
{Peebles}, P.~J.~E. 1974, \apjl, 189, L51+

\bibitem[{{Peebles} \& {Ratra}(1988)}]{Peebles:1988ApJ...325L..17P}
{Peebles}, P.~J.~E. \& {Ratra}, B. 1988, \apjl, 325, L17

\bibitem[{{Perlmutter} {et~al.}(1999){Perlmutter}, {Turner}, \&
  {White}}]{Perlmutter:1999}
{Perlmutter}, S., {Turner}, M.~S., \& {White}, M. 1999, Physical Review
  Letters, 83, 670

\bibitem[{{Petrosian}(1992)}]{Petrosian:1992}
{Petrosian}, V. 1992, in Statistical Challenges in Modern Astronomy, 173--200

\bibitem[{{Phillipps} \& {Driver}(1995)}]{Phillipps:1995MNRAS.274..832P}
{Phillipps}, S. \& {Driver}, S. 1995, \mnras, 274, 832

\bibitem[{{Poli} {et~al.}(2003){Poli}, {Giallongo}, {Fontana}, {Menci},
  {Zamorani}, {Nonino}, {Saracco}, {Vanzella}, {Donnarumma}, {Salimbeni},
  {Cimatti}, {Cristiani}, {Daddi}, {D'Odorico}, {Mignoli}, {Pozzetti}, \&
  {Renzini}}]{Poli:2003ApJ...593L...1P}
{Poli}, F., {Giallongo}, E., {Fontana}, A., {et~al.} 2003, \apjl, 593, L1

\bibitem[{{Portegies Zwart} {et~al.}(2010){Portegies Zwart}, {McMillan}, \&
  {Gieles}}]{2010ARA&A..48..431P}
{Portegies Zwart}, S.~F., {McMillan}, S.~L.~W., \& {Gieles}, M. 2010, \araa,
  48, 431

\bibitem[{{Pozzi} {et~al.}(2004){Pozzi}, {Gruppioni}, {Oliver}, {Matute}, {La
  Franca}, {Lari}, {Zamorani}, {Serjeant}, {Franceschini}, \&
  {Rowan-Robinson}}]{Pozzi:2004ApJ...609..122P}
{Pozzi}, F., {Gruppioni}, C., {Oliver}, S., {et~al.} 2004, \apj, 609, 122

\bibitem[{{Prescott} {et~al.}(2009){Prescott}, {Baldry}, \&
  {James}}]{Prescott:2009MNRAS.397...90P}
{Prescott}, M., {Baldry}, I.~K., \& {James}, P.~A. 2009, \mnras, 397, 90

\bibitem[{{Press} \& {Schechter}(1974)}]{Press:1974}
{Press}, W.~H. \& {Schechter}, P. 1974, \apj, 187, 425

\bibitem[{{Qin} \& {Xie}(1997)}]{Qin:1997}
{Qin}, Y.~P. \& {Xie}, G.~Z. 1997, \apj, 486, 100

\bibitem[{{Qin} \& {Xie}(1999)}]{Qin:1999}
{Qin}, Y.-P. \& {Xie}, G.-Z. 1999, \aap, 341, 693

\bibitem[{{Rauzy}(2001)}]{Rauzy:2001}
{Rauzy}, S. 2001, \mnras, 324, 51

\bibitem[{{Rauzy} \& {Hendry}(2000)}]{Rauzy:2000}
{Rauzy}, S. \& {Hendry}, M.~A. 2000, \mnras, 316, 621

\bibitem[{{Rauzy} {et~al.}(2001){Rauzy}, {Hendry}, \& {D'Mellow}}]{RHM:2001}
{Rauzy}, S., {Hendry}, M.~A., \& {D'Mellow}, K. 2001, \mnras, 328, 1016

\bibitem[{{Renzini} \& {Fusi Pecci}(1988)}]{Renzini:1988ARAA..26..199R}
{Renzini}, A. \& {Fusi Pecci}, F. 1988, \araa, 26, 199

\bibitem[{{Richards} {et~al.}(2005){Richards}, {Croom}, {Anderson},
  {Bland-Hawthorn}, {Boyle}, {De Propris}, {Drinkwater}, {Fan}, {Gunn},
  {Ivezi{\'c}}, {Jester}, {Loveday}, {Meiksin}, {Miller}, {Myers}, {Nichol},
  {Outram}, {Pimbblet}, {Roseboom}, {Ross}, {Schneider}, {Shanks}, {Sharp},
  {Stoughton}, {Strauss}, {Szalay}, {Vanden Berk}, \&
  {York}}]{Richards:2005MNRAS.360..839R}
{Richards}, G.~T., {Croom}, S.~M., {Anderson}, S.~F., {et~al.} 2005, \mnras,
  360, 839

\bibitem[{{Richards} {et~al.}(2006){Richards}, {Strauss}, {Fan}, {Hall},
  {Jester}, {Schneider}, {Vanden Berk}, {Stoughton}, {Anderson}, {Brunner},
  {Gray}, {Gunn}, {Ivezi{\'c}}, {Kirkland}, {Knapp}, {Loveday}, {Meiksin},
  {Pope}, {Szalay}, {Thakar}, {Yanny}, {York}, {Barentine}, {Brewington},
  {Brinkmann}, {Fukugita}, {Harvanek}, {Kent}, {Kleinman}, {Krzesi{\'n}ski},
  {Long}, {Lupton}, {Nash}, {Neilsen}, {Nitta}, {Schlegel}, \&
  {Snedden}}]{Richards:2006}
{Richards}, G.~T., {Strauss}, M.~A., {Fan}, X., {et~al.} 2006, \aj, 131, 2766

\bibitem[{{Richards} {et~al.}(2004){Richards}, {Strauss}, {Pindor}, {Haiman},
  {Fan}, {Eisenstein}, {Schneider}, {Bahcall}, {Brinkmann}, \&
  {Brunner}}]{Richards:2004AJ....127.1305R}
{Richards}, G.~T., {Strauss}, M.~A., {Pindor}, B., {et~al.} 2004, \aj, 127,
  1305

\bibitem[{{Riess} {et~al.}(1998){Riess}, {Filippenko}, {Challis},
  {Clocchiatti}, {Diercks}, {Garnavich}, {Gilliland}, {Hogan}, {Jha},
  {Kirshner}, {Leibundgut}, {Phillips}, {Reiss}, {Schmidt}, {Schommer},
  {Smith}, {Spyromilio}, {Stubbs}, {Suntzeff}, \& {Tonry}}]{Riess:1998}
{Riess}, A.~G., {Filippenko}, A.~V., {Challis}, P., {et~al.} 1998, \aj, 116,
  1009

\bibitem[{{Robertson} {et~al.}(2004){Robertson}, {Yoshida}, {Springel}, \&
  {Hernquist}}]{Robertson:2004ApJ...606...32R}
{Robertson}, B., {Yoshida}, N., {Springel}, V., \& {Hernquist}, L. 2004, \apj,
  606, 32

\bibitem[{{Robertson}(1978)}]{Robertson:1978MNRAS.182..617R}
{Robertson}, J.~G. 1978, \mnras, 182, 617

\bibitem[{{Robertson}(1980)}]{Robertson:1980MNRAS.190..143R}
{Robertson}, J.~G. 1980, \mnras, 190, 143

\bibitem[{{Rodighiero} {et~al.}(2010){Rodighiero}, {Vaccari}, {Franceschini},
  {Tresse}, {Le Fevre}, {Le Brun}, {Mancini}, {Matute}, {Cimatti}, {Marchetti},
  {Ilbert}, {Arnouts}, {Bolzonella}, {Zucca}, {Bardelli}, {Lonsdale}, {Shupe},
  {Surace}, {Rowan-Robinson}, {Garilli}, {Zamorani}, {Pozzetti}, {Bondi}, {de
  la Torre}, {Vergani}, {Santini}, {Grazian}, \&
  {Fontana}}]{Rodighiero:2010AA...515A...8R}
{Rodighiero}, G., {Vaccari}, M., {Franceschini}, A., {et~al.} 2010, \aap, 515,
  A8+

\bibitem[{{Rossi} \& {Sheth}(2008)}]{Rossi:2008}
{Rossi}, G. \& {Sheth}, R.~K. 2008, \mnras, 387, 735

\bibitem[{{Rossi} {et~al.}(2010){Rossi}, {Sheth}, \&
  {Park}}]{Rossi2010MNRAS.401..666R}
{Rossi}, G., {Sheth}, R.~K., \& {Park}, C. 2010, \mnras, 401, 666

\bibitem[{{Rowan-Robinson}(1968)}]{Rowan-Robinson1968MNRAS.138..445R}
{Rowan-Robinson}, M. 1968, \mnras, 138, 445

\bibitem[{{Rowan-Robinson} {et~al.}(1993){Rowan-Robinson}, {Benn}, {Lawrence},
  {McMahon}, \& {Broadhurst}}]{1993MNRAS.263..123R}
{Rowan-Robinson}, M., {Benn}, C.~R., {Lawrence}, A., {McMahon}, R.~G., \&
  {Broadhurst}, T.~J. 1993, \mnras, 263, 123

\bibitem[{{Rubin} {et~al.}(1980){Rubin}, {Ford}, \&
  {.~Thonnard}}]{Rubin:1980ApJ...238..471R}
{Rubin}, V.~C., {Ford}, W.~K.~J., \& {.~Thonnard}, N. 1980, \apj, 238, 471

\bibitem[{{Salpeter}(1955)}]{Salpeter:1955ApJ...121..161S}
{Salpeter}, E.~E. 1955, \apj, 121, 161

\bibitem[{{Sandage}(1957)}]{Sandage:1957ApJ...125..422S}
{Sandage}, A. 1957, \apj, 125, 422

\bibitem[{{Sandage} {et~al.}(1979){Sandage}, {Tammann}, \&
  {Yahil}}]{sandage:1979}
{Sandage}, A., {Tammann}, G.~A., \& {Yahil}, A. 1979, \apj, 232, 352

\bibitem[{{Sanders} {et~al.}(2003){Sanders}, {Mazzarella}, {Kim}, {Surace}, \&
  {Soifer}}]{Sanders:2003AJ....126.1607S}
{Sanders}, D.~B., {Mazzarella}, J.~M., {Kim}, D.-C., {Surace}, J.~A., \&
  {Soifer}, B.~T. 2003, \aj, 126, 1607

\bibitem[{{Sandquist} {et~al.}(1996){Sandquist}, {Bolte}, {Stetson}, \&
  {Hesser}}]{Sandquist:1996ApJ...470..910S}
{Sandquist}, E.~L., {Bolte}, M., {Stetson}, P.~B., \& {Hesser}, J.~E. 1996,
  \apj, 470, 910

\bibitem[{Santiago {et~al.}(1996)Santiago, Strauss, Lahav, Davis, Dressler, \&
  Huchra}]{Santiago:1996}
Santiago, B., Strauss, M., Lahav, O., {et~al.} 1996, \apj, 461, 38

\bibitem[{{Saunders} {et~al.}(1990){Saunders}, {Rowan-Robinson}, {Lawrence},
  {Efstathiou}, {Kaiser}, {Ellis}, \& {Frenk}}]{Saunders:1990}
{Saunders}, W., {Rowan-Robinson}, M., {Lawrence}, A., {et~al.} 1990, \mnras,
  242, 318

\bibitem[{{Saunders} {et~al.}(2000){Saunders}, {Sutherland}, {Maddox},
  {Keeble}, {Oliver}, {Rowan-Robinson}, {McMahon}, {Efstathiou}, {Tadros},
  {White}, {Frenk}, {Carrami{\~n}ana}, \& {Hawkins}}]{Saunders:2000}
{Saunders}, W., {Sutherland}, W.~J., {Maddox}, S.~J., {et~al.} 2000, \mnras,
  317, 55

\bibitem[{{Sawicki} \& {Thompson}(2006)}]{Sawicki:2006ApJ...642..653S}
{Sawicki}, M. \& {Thompson}, D. 2006, \apj, 642, 653

\bibitem[{{Schaeffer} \& {Silk}(1988)}]{Schaeffer:1988AA...203..273S}
{Schaeffer}, R. \& {Silk}, J. 1988, \aap, 203, 273

\bibitem[{{Schafer}(2007)}]{Schafer:2007}
{Schafer}, C.~M. 2007, \apj, 661, 703

\bibitem[{{Schechter}(1976)}]{Schechter:1976}
{Schechter}, P. 1976, \apj, 203, 297

\bibitem[{{Schmidt}(1963)}]{Schmidt:1963Natur.197.1040S}
{Schmidt}, M. 1963, \nat, 197, 1040

\bibitem[{{Schmidt}(1968)}]{Schmidt:1968}
{Schmidt}, M. 1968, \apj, 151, 393

\bibitem[{{Schmidt}(1972)}]{Schmidt:1972}
{Schmidt}, M. 1972, \apj, 176, 303

\bibitem[{{Schmidt}(1976)}]{Schmidt:1976}
{Schmidt}, M. 1976, \apjl, 209, L55+

\bibitem[{{Schmidt}(2001)}]{2001ApJ...552...36S}
{Schmidt}, M. 2001, \apj, 552, 36

\bibitem[{{Schmidt} \& {Green}(1983)}]{Schmidt:1983ApJ...269..352S}
{Schmidt}, M. \& {Green}, R.~F. 1983, \apj, 269, 352

\bibitem[{{Schmitt}(1990)}]{Schmitt:1990}
{Schmitt}, J.~H.~M.~M. 1990, \aap, 240, 556

\bibitem[{{Schrabback} {et~al.}(2010){Schrabback}, {Hartlap}, {Joachimi},
  {Kilbinger}, {Simon}, {Benabed}, {Brada{\v c}}, {Eifler}, {Erben},
  {Fassnacht}, {High}, {Hilbert}, {Hildebrandt}, {Hoekstra}, {Kuijken},
  {Marshall}, {Mellier}, {Morganson}, {Schneider}, {Semboloni}, {van Waerbeke},
  \& {Velander}}]{Schrabback:2010AA...516A..63S}
{Schrabback}, T., {Hartlap}, J., {Joachimi}, B., {et~al.} 2010, \aap, 516, A63+

\bibitem[{{Schwarzschild} \&
  {H{\"a}rm}(1958)}]{Schwarzschild:1958ApJ...128..348S}
{Schwarzschild}, M. \& {H{\"a}rm}, R. 1958, \apj, 128, 348

\bibitem[{{Scoville} {et~al.}(2007){Scoville}, {Aussel}, {Brusa}, {Capak},
  {Carollo}, {Elvis}, {Giavalisco}, {Guzzo}, {Hasinger}, {Impey}, {Kneib},
  {LeFevre}, {Lilly}, {Mobasher}, {Renzini}, {Rich}, {Sanders}, {Schinnerer},
  {Schminovich}, {Shopbell}, {Taniguchi}, \&
  {Tyson}}]{Scoville:2007ApJS..172....1S}
{Scoville}, N., {Aussel}, H., {Brusa}, M., {et~al.} 2007, \apjs, 172, 1

\bibitem[{{Searle} \& {Zinn}(1978)}]{Searle:1978ApJ...225..357S}
{Searle}, L. \& {Zinn}, R. 1978, \apj, 225, 357

\bibitem[{{Shapiro}(1971)}]{Shapiro:1971}
{Shapiro}, S.~L. 1971, \aj, 76, 291

\bibitem[{{Sheth}(2007)}]{Sheth:2007}
{Sheth}, R.~K. 2007, \mnras, 378, 709

\bibitem[{{Siana} {et~al.}(2008){Siana}, {Polletta}, {Smith}, {Lonsdale},
  {Gonzalez-Solares}, {Farrah}, {Babbedge}, {Rowan-Robinson}, {Surace},
  {Shupe}, {Fang}, {Franceschini}, \& {Oliver}}]{Siana:2008}
{Siana}, B., {Polletta}, M.~d.~C., {Smith}, H.~E., {et~al.} 2008, \apj, 675, 49

\bibitem[{{Simoda} \& {Kimura}(1968)}]{Simoda:1968ApJ...151..133S}
{Simoda}, M. \& {Kimura}, H. 1968, \apj, 151, 133

\bibitem[{{Simon} \& {Geha}(2007)}]{Simon:2007ApJ...670..313S}
{Simon}, J.~D. \& {Geha}, M. 2007, \apj, 670, 313

\bibitem[{{Sklar}(1959)}]{sklar:1959}
{Sklar}, A. 1959, \pisup, 8, 229

\bibitem[{{Skrutskie} {et~al.}(2006){Skrutskie}, {Cutri}, {Stiening},
  {Weinberg}, {Schneider}, {Carpenter}, {Beichman}, {Capps}, {Chester},
  {Elias}, {Huchra}, {Liebert}, {Lonsdale}, {Monet}, {Price}, {Seitzer},
  {Jarrett}, {Kirkpatrick}, {Gizis}, {Howard}, {Evans}, {Fowler}, {Fullmer},
  {Hurt}, {Light}, {Kopan}, {Marsh}, {McCallon}, {Tam}, {Van Dyk}, \&
  {Wheelock}}]{Skrutskie:2006AJ....131.1163S}
{Skrutskie}, M.~F., {Cutri}, R.~M., {Stiening}, R., {et~al.} 2006, \aj, 131,
  1163

\bibitem[{{Small} \& {Blandford}(1992)}]{Small:1992MNRAS.259..725S}
{Small}, T.~A. \& {Blandford}, R.~D. 1992, \mnras, 259, 725

\bibitem[{{Small} {et~al.}(1997){Small}, {Sargent}, \&
  {Hamilton}}]{Small:1997ApJ...487..512S}
{Small}, T.~A., {Sargent}, W.~L.~W., \& {Hamilton}, D. 1997, \apj, 487, 512

\bibitem[{{Sodre} \& {Lahav}(1993)}]{Sodre:1993}
{Sodre}, L.~J. \& {Lahav}, O. 1993, \mnras, 260, 285

\bibitem[{{Sofue} \& {Rubin}(2001)}]{2001ARA&A..39..137S}
{Sofue}, Y. \& {Rubin}, V. 2001, \araa, 39, 137

\bibitem[{{Spearman}(1904)}]{spearman:1904}
{Spearman}, C. 1904, \ajp, 15, 72

\bibitem[{{Spergel} {et~al.}(2003){Spergel}, {Verde}, {Peiris}, {Komatsu},
  {Nolta}, {Bennett}, {Halpern}, {Hinshaw}, {Jarosik}, {Kogut}, {Limon},
  {Meyer}, {Page}, {Tucker}, {Weiland}, {Wollack}, \& {Wright}}]{Spergel:2003}
{Spergel}, D.~N., {Verde}, L., {Peiris}, H.~V., {et~al.} 2003, \apjs, 148, 175

\bibitem[{{Springel}(2005)}]{Springel:2005MNRAS.364.1105S}
{Springel}, V. 2005, \mnras, 364, 1105

\bibitem[{{Springel} \& {White}(1998)}]{Springel:1998}
{Springel}, V. \& {White}, S.~D.~M. 1998, \mnras, 298, 143

\bibitem[{{Springel} {et~al.}(2001){Springel}, {Yoshida}, \&
  {White}}]{Springel:2001NewA....6...79S}
{Springel}, V., {Yoshida}, N., \& {White}, S.~D.~M. 2001, \na, 6, 79

\bibitem[{{Stabenau} {et~al.}(2008){Stabenau}, {Connolly}, \&
  {Jain}}]{Stabenau:2008}
{Stabenau}, H.~F., {Connolly}, A., \& {Jain}, B. 2008, \mnras, 387, 1215

\bibitem[{{Stanway} {et~al.}(2008){Stanway}, {Bremer}, {Squitieri}, {Douglas},
  \& {Lehnert}}]{Stanway:2008MNRAS.386..370S}
{Stanway}, E.~R., {Bremer}, M.~N., {Squitieri}, V., {Douglas}, L.~S., \&
  {Lehnert}, M.~D. 2008, \mnras, 386, 370

\bibitem[{{Steidel} {et~al.}(1999){Steidel}, {Adelberger}, {Giavalisco},
  {Dickinson}, \& {Pettini}}]{Steidel:1999ApJ...519....1S}
{Steidel}, C.~C., {Adelberger}, K.~L., {Giavalisco}, M., {Dickinson}, M., \&
  {Pettini}, M. 1999, \apj, 519, 1

\bibitem[{{Steidel} {et~al.}(1996){Steidel}, {Giavalisco}, {Pettini},
  {Dickinson}, \& {Adelberger}}]{Steidel:1996ApJ...462L..17S}
{Steidel}, C.~C., {Giavalisco}, M., {Pettini}, M., {Dickinson}, M., \&
  {Adelberger}, K.~L. 1996, \apjl, 462, L17+

\bibitem[{{Steinhardt} {et~al.}(1999){Steinhardt}, {Wang}, \&
  {Zlatev}}]{Steinhardt:1999PhRvD..59l3504S}
{Steinhardt}, P.~J., {Wang}, L., \& {Zlatev}, I. 1999, \prd, 59, 123504

\bibitem[{{Stigler}(2008)}]{Stigler2008arXiv0804.2996S}
{Stigler}, S.~M. 2008, {\SSc}, 22, 598

\bibitem[{{Stoughton} {et~al.}(2002){Stoughton}, {Lupton}, {Bernardi},
  {Blanton}, {Burles}, {Castander}, {Connolly}, {Eisenstein}, {Frieman},
  {Hennessy}, {Hindsley}, {Ivezi{\'c}}, {Kent}, {Kunszt}, {Lee}, {Meiksin},
  {Munn}, {Newberg}, {Nichol}, {Nicinski}, {Pier}, {Richards}, {Richmond},
  {Schlegel}, {Smith}, {Strauss}, {SubbaRao}, {Szalay}, {Thakar}, {Tucker},
  {Vanden Berk}, {Yanny}, {Adelman}, {Anderson}, {Anderson}, {Annis},
  {Bahcall}, {Bakken}, {Bartelmann}, {Bastian}, {Bauer}, {Berman},
  {B{\"o}hringer}, {Boroski}, {Bracker}, {Briegel}, {Briggs}, {Brinkmann},
  {Brunner}, {Carey}, {Carr}, {Chen}, {Christian}, {Colestock}, {Crocker},
  {Csabai}, {Czarapata}, {Dalcanton}, {Davidsen}, {Davis}, {Dehnen},
  {Dodelson}, {Doi}, {Dombeck}, {Donahue}, {Ellman}, {Elms}, {Evans}, {Eyer},
  {Fan}, {Federwitz}, {Friedman}, {Fukugita}, {Gal}, {Gillespie}, {Glazebrook},
  {Gray}, {Grebel}, {Greenawalt}, {Greene}, {Gunn}, {de Haas}, {Haiman},
  {Haldeman}, {Hall}, {Hamabe}, {Hansen}, {Harris}, {Harris}, {Harvanek},
  {Hawley}, {Hayes}, {Heckman}, {Helmi}, {Henden}, {Hogan}, {Hogg}, {Holmgren},
  {Holtzman}, {Huang}, {Hull}, {Ichikawa}, {Ichikawa}, {Johnston}, {Kauffmann},
  {Kim}, {Kimball}, {Kinney}, {Klaene}, {Kleinman}, {Klypin}, {Knapp},
  {Korienek}, {Krolik}, {Kron}, {Krzesi{\'n}ski}, {Lamb}, {Leger},
  {Limmongkol}, {Lindenmeyer}, {Long}, {Loomis}, {Loveday}, {MacKinnon},
  {Mannery}, {Mantsch}, {Margon}, {McGehee}, {McKay}, {McLean}, {Menou},
  {Merelli}, {Mo}, {Monet}, {Nakamura}, {Narayanan}, {Nash}, {Neilsen},
  {Newman}, {Nitta}, {Odenkirchen}, {Okada}, {Okamura}, {Ostriker}, {Owen},
  {Pauls}, {Peoples}, {Peterson}, {Petravick}, {Pope}, {Pordes}, {Postman},
  {Prosapio}, {Quinn}, {Rechenmacher}, {Rivetta}, {Rix}, {Rockosi}, {Rosner},
  {Ruthmansdorfer}, {Sandford}, {Schneider}, {Scranton}, {Sekiguchi}, {Sergey},
  {Sheth}, {Shimasaku}, {Smee}, {Snedden}, {Stebbins}, {Stubbs}, {Szapudi},
  {Szkody}, {Szokoly}, {Tabachnik}, {Tsvetanov}, {Uomoto}, {Vogeley}, {Voges},
  {Waddell}, {Walterbos}, {Wang}, {Watanabe}, {Weinberg}, {White}, {White},
  {Wilhite}, {Wolfe}, {Yasuda}, {York}, {Zehavi}, \&
  {Zheng}}]{Stoughton:2002AJ.123.485S}
{Stoughton}, C., {Lupton}, R.~H., {Bernardi}, M., {et~al.} 2002, \aj, 123, 485

\bibitem[{{Strauss} \& {Willick}(1995)}]{Strauss:1995PhR...261..271S}
{Strauss}, M.~A. \& {Willick}, J.~A. 1995, \physrep, 261, 271

\bibitem[{{Subbarao} {et~al.}(1996){Subbarao}, {Connolly}, {Szalay}, \&
  {Koo}}]{Subbarao1996AJ....112..929S}
{Subbarao}, M.~U., {Connolly}, A.~J., {Szalay}, A.~S., \& {Koo}, D.~C. 1996,
  \aj, 112, 929

\bibitem[{{Takeuchi}(2010)}]{Takeuchi2010MNRAS.406.1830T}
{Takeuchi}, T.~T. 2010, \mnras, 406, 1830

\bibitem[{{Takeuchi} {et~al.}(2006){Takeuchi}, {Ishii}, {Dole}, {Dennefeld},
  {Lagache}, \& {Puget}}]{Takeuchi:2006AA...448..525T}
{Takeuchi}, T.~T., {Ishii}, T.~T., {Dole}, H., {et~al.} 2006, \aap, 448, 525

\bibitem[{{Takeuchi} {et~al.}(2000){Takeuchi}, {Yoshikawa}, \&
  {Ishii}}]{Takeuchi:2000}
{Takeuchi}, T.~T., {Yoshikawa}, K., \& {Ishii}, T.~T. 2000, \apjs, 129, 1

\bibitem[{{Teodoro} {et~al.}(2010){Teodoro}, {Johnston}, \&
  {Hendry}}]{Teodoro:2009}
{Teodoro}, L., {Johnston}, R., \& {Hendry}, M. 2010, \mnras

\bibitem[{{Tilvi} {et~al.}(2010){Tilvi}, {Rhoads}, {Hibon}, {Malhotra}, {Wang},
  {Veilleux}, {Swaters}, {Probst}, {Krug}, {Finkelstein}, \&
  {Dickinson}}]{Tilvi:2010ApJ...721.1853T}
{Tilvi}, V., {Rhoads}, J.~E., {Hibon}, P., {et~al.} 2010, \apj, 721, 1853

\bibitem[{{Tojeiro} \& {Percival}(2010)}]{Tojeiro2010MNRAS.405.2534T}
{Tojeiro}, R. \& {Percival}, W.~J. 2010, \mnras, 405, 2534

\bibitem[{{Toribio} {et~al.}(2011){Toribio}, {Solanes}, {Giovanelli}, {Haynes},
  \& {Martin}}]{Toribio:2011arXiv1103.0990T}
{Toribio}, M.~C., {Solanes}, J.~M., {Giovanelli}, R., {Haynes}, M.~P., \&
  {Martin}, A. 2011, ArXiv e-prints

\bibitem[{{Turner}(1979)}]{Turner:1979}
{Turner}, E.~L. 1979, \apj, 231, 645

\bibitem[{{Turner}(1998)}]{Turner:1998}
{Turner}, M.~S. 1998, APS Meeting Abstracts, 101

\bibitem[{{Ueda} {et~al.}(2003){Ueda}, {Akiyama}, {Ohta}, \&
  {Miyaji}}]{Ueda:2003ApJ...598..886U}
{Ueda}, Y., {Akiyama}, M., {Ohta}, K., \& {Miyaji}, T. 2003, \apj, 598, 886

\bibitem[{{Vaccari} {et~al.}(2010){Vaccari}, {Marchetti}, {Franceschini},
  {Altieri}, {Amblard}, {Arumugam}, {Auld}, {Aussel}, {Babbedge}, {Blain},
  {Bock}, {Boselli}, {Buat}, {Burgarella}, {Castro-Rodr{\'{\i}}guez}, {Cava},
  {Chanial}, {Clements}, {Conley}, {Conversi}, {Cooray}, {Dowell}, {Dwek},
  {Dye}, {Eales}, {Elbaz}, {Farrah}, {Fox}, {Gear}, {Glenn}, {Gonz{\'a}lez
  Solares}, {Griffin}, {Halpern}, {Hatziminaoglou}, {Huang}, {Ibar}, {Isaak},
  {Ivison}, {Lagache}, {Levenson}, {Lu}, {Madden}, {Maffei}, {Mainetti},
  {Mortier}, {Nguyen}, {O'Halloran}, {Oliver}, {Omont}, {Page}, {Panuzzo},
  {Papageorgiou}, {Pearson}, {P{\'e}rez-Fournon}, {Pohlen}, {Rawlings},
  {Raymond}, {Rigopoulou}, {Rizzo}, {Rodighiero}, {Roseboom}, {Rowan-Robinson},
  {S{\'a}nchez Portal}, {Schulz}, {Scott}, {Seymour}, {Shupe}, {Smith},
  {Stevens}, {Symeonidis}, {Trichas}, {Tugwell}, {Valiante}, {Valtchanov},
  {Vigroux}, {Wang}, {Ward}, {Wright}, {Xu}, \&
  {Zemcov}}]{Vaccari:2010AA...518L..20V}
{Vaccari}, M., {Marchetti}, L., {Franceschini}, A., {et~al.} 2010, \aap, 518,
  L20+

\bibitem[{{van den Bergh}(1961)}]{VandenBergh:1961}
{van den Bergh}, S. 1961, Zeitschrift fur Astrophysik, 53, 219

\bibitem[{{van Waerbeke} {et~al.}(1996){van Waerbeke}, {Mathez}, {Mellier},
  {Bonnet}, \& {Lachieze-Rey}}]{Waerbeke:1996}
{van Waerbeke}, L., {Mathez}, G., {Mellier}, Y., {Bonnet}, H., \&
  {Lachieze-Rey}, M. 1996, \aap, 316, 1

\bibitem[{{Vanzella} {et~al.}(2004){Vanzella}, {Cristiani}, {Fontana},
  {Nonino}, {Arnouts}, {Giallongo}, {Grazian}, {Fasano}, {Popesso}, {Saracco},
  \& {Zaggia}}]{Vanzella:2004AA...423..761V}
{Vanzella}, E., {Cristiani}, S., {Fontana}, A., {et~al.} 2004, \aap, 423, 761

\bibitem[{{Wake} {et~al.}(2006){Wake}, {Nichol}, {Eisenstein}, {Loveday},
  {Edge}, {Cannon}, {Smail}, {Schneider}, {Scranton}, {Carson}, {Ross},
  {Brunner}, {Colless}, {Couch}, {Croom}, {Driver}, {da {\^A}ngela}, {Jester},
  {de Propris}, {Drinkwater}, {Bland-Hawthorn}, {Pimbblet}, {Roseboom},
  {Shanks}, {Sharp}, \& {Brinkmann}}]{Wake:2006MNRAS.372..537W}
{Wake}, D.~A., {Nichol}, R.~C., {Eisenstein}, D.~J., {et~al.} 2006, \mnras,
  372, 537

\bibitem[{{Wall} \& {Jenkins}(2003)}]{wall:2003}
{Wall}, J.~V. \& {Jenkins}, C.~R. 2003, Practical Statistics fo Astronomers,
  1st edn. (Cambridge)

\bibitem[{{Wall} {et~al.}(1980){Wall}, {Pearson}, \&
  {Longair}}]{Wall:1980MNRAS.193..683W}
{Wall}, J.~V., {Pearson}, T.~J., \& {Longair}, M.~S. 1980, \mnras, 193, 683

\bibitem[{{Wall} {et~al.}(2008){Wall}, {Pope}, \&
  {Scott}}]{Wall:2008MNRAS.383..435W}
{Wall}, J.~V., {Pope}, A., \& {Scott}, D. 2008, \mnras, 383, 435

\bibitem[{{Wanderman} \& {Piran}(2010)}]{2010MNRAS.406.1944W}
{Wanderman}, D. \& {Piran}, T. 2010, \mnras, 406, 1944

\bibitem[{{Wang} {et~al.}(1998){Wang}, {Bahcall}, \& {Turner}}]{Wang:1998}
{Wang}, Y., {Bahcall}, N., \& {Turner}, E.~L. 1998, \aj, 116, 2081

\bibitem[{{Warren} {et~al.}(2007){Warren}, {Hambly}, {Dye}, {Almaini}, {Cross},
  {Edge}, {Foucaud}, {Hewett}, {Hodgkin}, {Irwin}, {Jameson}, {Lawrence},
  {Lucas}, {Adamson}, {Bandyopadhyay}, {Bryant}, {Collins}, {Davis}, {Dunlop},
  {Emerson}, {Evans}, {Gonzales-Solares}, {Hirst}, {Jarvis}, {Kendall}, {Kerr},
  {Leggett}, {Lewis}, {Mann}, {McLure}, {McMahon}, {Mortlock}, {Rawlings},
  {Read}, {Riello}, {Simpson}, {Smith}, {Sutorius}, {Targett}, \&
  {Varricatt}}]{Warren:2007MNRAS.375..213W}
{Warren}, S.~J., {Hambly}, N.~C., {Dye}, S., {et~al.} 2007, \mnras, 375, 213

\bibitem[{{Warren} {et~al.}(1988){Warren}, {Hewett}, \&
  {Osmer}}]{Warren:1988ASPC....2...96W}
{Warren}, S.~J., {Hewett}, P.~C., \& {Osmer}, P.~S. 1988, in Astronomical
  Society of the Pacific Conference Series, Vol.~2, Optical Surveys for
  Quasars, ed. {P.~Osmer \& M.~M.~Phillips}, 96--+

\bibitem[{{Warren} {et~al.}(1994){Warren}, {Hewett}, \&
  {Osmer}}]{Warren:1994ApJ...421..412W}
{Warren}, S.~J., {Hewett}, P.~C., \& {Osmer}, P.~S. 1994, \apj, 421, 412

\bibitem[{{Weinberg} {et~al.}(2008){Weinberg}, {Colombi}, {Dav{\'e}}, \&
  {Katz}}]{Weinberg:2008ApJ...678....6W}
{Weinberg}, D.~H., {Colombi}, S., {Dav{\'e}}, R., \& {Katz}, N. 2008, \apj,
  678, 6

\bibitem[{{White} {et~al.}(1987){White}, {Davis}, {Efstathiou}, \&
  {Frenk}}]{White:1987Natur.330..451W}
{White}, S.~D.~M., {Davis}, M., {Efstathiou}, G., \& {Frenk}, C.~S. 1987, \nat,
  330, 451

\bibitem[{{White} \& {Frenk}(1991)}]{White:1991ApJ...379...52W}
{White}, S.~D.~M. \& {Frenk}, C.~S. 1991, \apj, 379, 52

\bibitem[{{White} \& {Rees}(1978)}]{White:1978MNRAS.183..341W}
{White}, S.~D.~M. \& {Rees}, M.~J. 1978, \mnras, 183, 341

\bibitem[{{Williams} {et~al.}(1996){Williams}, {Blacker}, {Dickinson}, {Dixon},
  {Ferguson}, {Fruchter}, {Giavalisco}, {Gilliland}, {Heyer}, {Katsanis},
  {Levay}, {Lucas}, {McElroy}, {Petro}, {Postman}, {Adorf}, \&
  {Hook}}]{Williams:1996}
{Williams}, R.~E., {Blacker}, B., {Dickinson}, M., {et~al.} 1996, \aj, 112,
  1335

\bibitem[{{Willman} {et~al.}(2004){Willman}, {Governato}, {Dalcanton}, {Reed},
  \& {Quinn}}]{2004MNRAS.353..639W}
{Willman}, B., {Governato}, F., {Dalcanton}, J.~J., {Reed}, D., \& {Quinn}, T.
  2004, \mnras, 353, 639

\bibitem[{{Willmer}(1997)}]{Willmer:1997}
{Willmer}, C.~N.~A. 1997, \aj, 114, 898

\bibitem[{{Willmer} {et~al.}(2006){Willmer}, {Faber}, {Koo}, {Weiner},
  {Newman}, {Coil}, {Connolly}, {Conroy}, {Cooper}, {Davis}, {Finkbeiner},
  {Gerke}, {Guhathakurta}, {Harker}, {Kaiser}, {Kassin}, {Konidaris}, {Lin},
  {Luppino}, {Madgwick}, {Noeske}, {Phillips}, \&
  {Yan}}]{Willmer:2006ApJ...647..853W}
{Willmer}, C.~N.~A., {Faber}, S.~M., {Koo}, D.~C., {et~al.} 2006, \apj, 647,
  853

\bibitem[{{Willott} {et~al.}(2009){Willott}, {Delorme}, {Reyl{\'e}}, {Albert},
  {Bergeron}, {Crampton}, {Delfosse}, {Forveille}, {Hutchings}, {McLure},
  {Omont}, \& {Schade}}]{Willott:2009AJ....137.3541W}
{Willott}, C.~J., {Delorme}, P., {Reyl{\'e}}, C., {et~al.} 2009, \aj, 137, 3541

\bibitem[{{Willott} {et~al.}(2010){Willott}, {Delorme}, {Reyl{\'e}}, {Albert},
  {Bergeron}, {Crampton}, {Delfosse}, {Forveille}, {Hutchings}, {McLure},
  {Omont}, \& {Schade}}]{Willott:2010AJ....139..906W}
{Willott}, C.~J., {Delorme}, P., {Reyl{\'e}}, C., {et~al.} 2010, \aj, 139, 906

\bibitem[{{Willott} {et~al.}(2001){Willott}, {Rawlings}, {Blundell}, {Lacy}, \&
  {Eales}}]{2001MNRAS.322..536W}
{Willott}, C.~J., {Rawlings}, S., {Blundell}, K.~M., {Lacy}, M., \& {Eales},
  S.~A. 2001, \mnras, 322, 536

\bibitem[{{Wittman} {et~al.}(2002){Wittman}, {Tyson}, {Dell'Antonio}, {Becker},
  {Margoniner}, {Cohen}, {Norman}, {Loomba}, {Squires}, {Wilson}, {Stubbs},
  {Hennawi}, {Spergel}, {Boeshaar}, {Clocchiatti}, {Hamuy}, {Bernstein},
  {Gonzalez}, {Guhathakurta}, {Hu}, {Seljak}, \&
  {Zaritsky}}]{Wittman:2002SPIE.4836...73W}
{Wittman}, D.~M., {Tyson}, J.~A., {Dell'Antonio}, I.~P., {et~al.} 2002, in
  Presented at the Society of Photo-Optical Instrumentation Engineers (SPIE)
  Conference, Vol. 4836, Society of Photo-Optical Instrumentation Engineers
  (SPIE) Conference Series, ed. {J.~A.~Tyson \& S.~Wolff}, 73--82

\bibitem[{{Wolf}(2009)}]{Wolf:2009}
{Wolf}, C. 2009, \mnras, 397, 520

\bibitem[{{Wolf} {et~al.}(2004){Wolf}, {Meisenheimer}, {Kleinheinrich},
  {Borch}, {Dye}, {Gray}, {Wisotzki}, {Bell}, {Rix}, {Cimatti}, {Hasinger}, \&
  {Szokoly}}]{Wolf:2004AA...421..913W}
{Wolf}, C., {Meisenheimer}, K., {Kleinheinrich}, M., {et~al.} 2004, \aap, 421,
  913

\bibitem[{{Wolf} {et~al.}(2003){Wolf}, {Meisenheimer}, {Rix}, {Borch}, {Dye},
  \& {Kleinheinrich}}]{Wolf:2003AA...401...73W}
{Wolf}, C., {Meisenheimer}, K., {Rix}, H., {et~al.} 2003, \aap, 401, 73

\bibitem[{{Wood}(1992)}]{Wood:1992ApJ...386..539W}
{Wood}, M.~A. 1992, \apj, 386, 539

\bibitem[{{Woodroofe}(1985)}]{woodroofe:1985}
{Woodroofe}, M. 1985, {\aos}, 13, 163

\bibitem[{{Y{\`e}che} {et~al.}(2010){Y{\`e}che}, {Petitjean}, {Rich},
  {Aubourg}, {Busca}, {Hamilton}, {Le Goff}, {Paris}, {Peirani}, {Pichon},
  {Rollinde}, \& {Vargas-Maga{\~n}a}}]{Yeche:2010AA...523A..14Y}
{Y{\`e}che}, C., {Petitjean}, P., {Rich}, J., {et~al.} 2010, \aap, 523, A14+

\bibitem[{{York} {et~al.}(2000){York}, {Adelman}, {Anderson}, {Anderson},
  {Annis}, {Bahcall}, {Bakken}, {Barkhouser}, {Bastian}, {Berman}, {Boroski},
  {Bracker}, {Briegel}, {Briggs}, {Brinkmann}, {Brunner}, {Burles}, {Carey},
  {Carr}, {Castander}, {Chen}, {Colestock}, {Connolly}, {Crocker}, {Csabai},
  {Czarapata}, {Davis}, {Doi}, {Dombeck}, {Eisenstein}, {Ellman}, {Elms},
  {Evans}, {Fan}, {Federwitz}, {Fiscelli}, {Friedman}, {Frieman}, {Fukugita},
  {Gillespie}, {Gunn}, {Gurbani}, {de Haas}, {Haldeman}, {Harris}, {Hayes},
  {Heckman}, {Hennessy}, {Hindsley}, {Holm}, {Holmgren}, {Huang}, {Hull},
  {Husby}, {Ichikawa}, {Ichikawa}, {Ivezi{\'c}}, {Kent}, {Kim}, {Kinney},
  {Klaene}, {Kleinman}, {Kleinman}, {Knapp}, {Korienek}, {Kron}, {Kunszt},
  {Lamb}, {Lee}, {Leger}, {Limmongkol}, {Lindenmeyer}, {Long}, {Loomis},
  {Loveday}, {Lucinio}, {Lupton}, {MacKinnon}, {Mannery}, {Mantsch}, {Margon},
  {McGehee}, {McKay}, {Meiksin}, {Merelli}, {Monet}, {Munn}, {Narayanan},
  {Nash}, {Neilsen}, {Neswold}, {Newberg}, {Nichol}, {Nicinski}, {Nonino},
  {Okada}, {Okamura}, {Ostriker}, {Owen}, {Pauls}, {Peoples}, {Peterson},
  {Petravick}, {Pier}, {Pope}, {Pordes}, {Prosapio}, {Rechenmacher}, {Quinn},
  {Richards}, {Richmond}, {Rivetta}, {Rockosi}, {Ruthmansdorfer}, {Sandford},
  {Schlegel}, {Schneider}, {Sekiguchi}, {Sergey}, {Shimasaku}, {Siegmund},
  {Smee}, {Smith}, {Snedden}, {Stone}, {Stoughton}, {Strauss}, {Stubbs},
  {SubbaRao}, {Szalay}, {Szapudi}, {Szokoly}, {Thakar}, {Tremonti}, {Tucker},
  {Uomoto}, {Vanden Berk}, {Vogeley}, {Waddell}, {Wang}, {Watanabe},
  {Weinberg}, {Yanny}, \& {Yasuda}}]{York:2000}
{York}, D.~G., {Adelman}, J., {Anderson}, Jr., J.~E., {et~al.} 2000, \aj, 120,
  1579

\bibitem[{{Yu} \& {Tremaine}(2002)}]{Yu:2002MNRAS.335..965Y}
{Yu}, Q. \& {Tremaine}, S. 2002, \mnras, 335, 965

\bibitem[{{Zawislak-Raczka} \&
  {Kumor-Obryk}(1986)}]{Zawislak:1986MNRAS.222..487Z}
{Zawislak-Raczka}, J. \& {Kumor-Obryk}, B. 1986, \mnras, 222, 487

\bibitem[{{Zawislak-Raczka} \&
  {Kumor-Obryk}(1987)}]{Zawislak:1987ApSS.139..305Z}
{Zawislak-Raczka}, J. \& {Kumor-Obryk}, B. 1987, \apss, 139, 305

\bibitem[{{Zawislak-Raczka} \&
  {Kumor-Obryk}(1990)}]{Zawislak:1990ApSS.172...89Z}
{Zawislak-Raczka}, J. \& {Kumor-Obryk}, B. 1990, \apss, 172, 89

\bibitem[{{Zhang} \& {Fall}(1999)}]{1999ApJ...527L..81Z}
{Zhang}, Q. \& {Fall}, S.~M. 1999, \apjl, 527, L81

\bibitem[{{Zheng} {et~al.}(2009){Zheng}, {Bradley}, {Bouwens}, {Ford},
  {Illingworth}, {Ben{\'{\i}}tez}, {Broadhurst}, {Frye}, {Infante}, {Jee},
  {Motta}, {Shu}, \& {Zitrin}}]{Zheng:2009ApJ...697.1907Z}
{Zheng}, W., {Bradley}, L.~D., {Bouwens}, R.~J., {et~al.} 2009, \apj, 697, 1907

\bibitem[{{Zoccali} \& {Piotto}(2000)}]{Zoccali:2000AA...358..943Z}
{Zoccali}, M. \& {Piotto}, G. 2000, \aap, 358, 943

\bibitem[{{Zucca} {et~al.}(2009){Zucca}, {Bardelli}, {Bolzonella}, {Zamorani},
  {Ilbert}, {Pozzetti}, {Mignoli}, {Kova{\v c}}, {Lilly}, {Tresse}, {Tasca},
  {Cassata}, {Halliday}, {Vergani}, {Caputi}, {Carollo}, {Contini}, {Kneib},
  {Le F{\`e}vre}, {Mainieri}, {Renzini}, {Scodeggio}, {Bongiorno}, {Coppa},
  {Cucciati}, {de La Torre}, {de Ravel}, {Franzetti}, {Garilli}, {Iovino},
  {Kampczyk}, {Knobel}, {Lamareille}, {Le Borgne}, {Le Brun}, {Maier},
  {Pell{\`o}}, {Peng}, {Perez-Montero}, {Ricciardelli}, {Silverman}, {Tanaka},
  {Abbas}, {Bottini}, {Cappi}, {Cimatti}, {Guzzo}, {Koekemoer}, {Leauthaud},
  {Maccagni}, {Marinoni}, {McCracken}, {Memeo}, {Meneux}, {Moresco}, {Oesch},
  {Porciani}, {Scaramella}, {Arnouts}, {Aussel}, {Capak}, {Kartaltepe},
  {Salvato}, {Sanders}, {Scoville}, {Taniguchi}, \& {Thompson}}]{Zucca:2009}
{Zucca}, E., {Bardelli}, S., {Bolzonella}, M., {et~al.} 2009, \aap, 508, 1217

\bibitem[{{Zucca} {et~al.}(2006){Zucca}, {Ilbert}, {Bardelli}, {Tresse},
  {Zamorani}, {Arnouts}, {Pozzetti}, {Bolzonella}, {McCracken}, {Bottini},
  {Garilli}, {Le Brun}, {Le F{\`e}vre}, {Maccagni}, {Picat}, {Scaramella},
  {Scodeggio}, {Vettolani}, {Zanichelli}, {Adami}, {Arnaboldi}, {Cappi},
  {Charlot}, {Ciliegi}, {Contini}, {Foucaud}, {Franzetti}, {Gavignaud},
  {Guzzo}, {Iovino}, {Marano}, {Marinoni}, {Mazure}, {Meneux}, {Merighi},
  {Paltani}, {Pell{\`o}}, {Pollo}, {Radovich}, {Bondi}, {Bongiorno},
  {Busarello}, {Cucciati}, {Gregorini}, {Lamareille}, {Mathez}, {Mellier},
  {Merluzzi}, {Ripepi}, \& {Rizzo}}]{Zucca:2006AA...455..879Z}
{Zucca}, E., {Ilbert}, O., {Bardelli}, S., {et~al.} 2006, \aap, 455, 879

\bibitem[{{Zucca} {et~al.}(1997){Zucca}, {Zamorani}, {Vettolani}, {Cappi},
  {Merighi}, {Mignoli}, {Stirpe}, {MacGillivray}, {Collins}, {Balkowski},
  {Cayatte}, {Maurogordato}, {Proust}, {Chincarini}, {Guzzo}, {Maccagni},
  {Scaramella}, {Blanchard}, \& {Ramella}}]{Zucca:1997AA...326..477Z}
{Zucca}, E., {Zamorani}, G., {Vettolani}, G., {et~al.} 1997, \aap, 326, 477

\bibitem[{{Zwaan} {et~al.}(2001){Zwaan}, {Briggs}, \&
  {Sprayberry}}]{2001MNRAS.327.1249Z}
{Zwaan}, M.~A., {Briggs}, F.~H., \& {Sprayberry}, D. 2001, \mnras, 327, 1249

\bibitem[{{Zwaan} {et~al.}(2004){Zwaan}, {Meyer}, {Webster}, {Staveley-Smith},
  {Drinkwater}, {Barnes}, {Bhathal}, {de Blok}, {Disney}, {Ekers}, {Freeman},
  {Garcia}, {Gibson}, {Harnett}, {Henning}, {Howlett}, {Jerjen}, {Kesteven},
  {Kilborn}, {Knezek}, {Koribalski}, {Mader}, {Marquarding}, {Minchin},
  {O'Brien}, {Oosterloo}, {Pierce}, {Price}, {Putman}, {Ryan-Weber}, {Ryder},
  {Sadler}, {Stevens}, {Stewart}, {Stootman}, {Waugh}, \&
  {Wright}}]{Zwaan:2004MNRAS.350.1210Z}
{Zwaan}, M.~A., {Meyer}, M.~J., {Webster}, R.~L., {et~al.} 2004, \mnras, 350,
  1210

\bibitem[{{Zwicky}(1933)}]{Zwicky:1933AcHPh...6..110Z}
{Zwicky}, F. 1933, Helvetica Physica Acta, 6, 110

\end{thebibliography}
\end{document}